\documentclass[prb,aps,amssymb,twocolumn,notitlepage]{revtex4-2}
\usepackage{amsmath}
\usepackage{amssymb}
\usepackage{amsthm}
\usepackage{amsfonts}
\usepackage{listings}
\usepackage{physics}
\usepackage{enumerate}
\usepackage{latexsym}
\usepackage{psfrag}
\usepackage{bm}
\usepackage{graphicx}
\usepackage[caption=false]{subfig}
\usepackage{blkarray}
\usepackage{array}
\usepackage{color}
\usepackage[normalem]{ulem}
\usepackage{hyperref}
\usepackage{cleveref}
\usepackage{babel}
\usepackage{dsfont}
\usepackage{subfig}
\usepackage{tabularx}
\usepackage{multirow}
\usepackage{xfp}
\usepackage{calc}
\usepackage[nomessages]{fp}
\usepackage{chemformula}

\NewExpandableDocumentCommand{\bettersquareroot}{O{16}m}{%
  \fpeval{round(sqrt(#2),#1)}%
}
\hypersetup{
     colorlinks = true,
     linkcolor = magenta,
     citecolor = magenta
}

\def\beq{\begin{equation}}
\def\eeq{\end{equation}}
\def\bal{\begin{aligned}}
\def\eal{\end{aligned}}
\def\sg135{$P4_2/mbc1^\prime$ (\# 135)}
\def\sgb{$P4/ncc1^\prime$ (\# 130) }

\begin{document}

\title{Ground state stability, symmetry, and degeneracy in Mott insulators with long range interactions}

\author{Dmitry Manning-Coe}
\email{dmitry2@illinois.edu}
\affiliation{Department of Physics and Institute for Condensed Matter Theory, University of Illinois at Urbana-Champaign, Urbana IL, 61801-3080, USA}

\author{Barry Bradlyn}
\email{bbradlyn@illinois.edu}
\affiliation{Department of Physics and Institute for Condensed Matter Theory, University of Illinois at Urbana-Champaign, Urbana IL, 61801-3080, USA}


\begin{abstract}
Recently, models with long-range interactions---known as Hatsugai-Kohmoto (HK) models---have emerged as a promising tool to study the emergence of superconductivity and topology in strongly correlated systems. 
Two obstacles, however, have made it difficult to understand the applicability of these models, especially to topological features: they have thermodynamically large ground state degeneracies, and they tacitly assume spin conservation. 
We show that neither are essential to HK models and that both can be avoided by introducing interactions between tight-binding states in the orbital basis, rather than between energy eigenstates. 
To solve these ``orbital’’ models, we introduce a general technique for solving HK models and show that previous models appear as special cases. 
We illustrate our method by exactly solving graphene and the Kane-Mele model with HK interactions. 
Both realize Mott insulating phases with finite magnetic susceptibility; the graphene model has a fourfold degenerate ground state while the Kane-Mele model has a nondegenerate ground state in the presence of interactions. 
Our technique then allows us to study the effect of strong interactions on symmetry-enforced degeneracy in spin-orbit coupled double-Dirac semimetals. 
We show that adding HK interactions to a double Dirac semi-metal leads to a Mott insulating, spin liquid phase. 
We then use a Schrieffer-Wolff transformation to express the low-energy Hamiltonian in terms of the spin degrees of freedom, making the spin-charge separation explicit. 
Finally, we enumerate a broader class of symmetry-preserving HK interactions and show how they can violate insulating filling constraints derived from space group symmetries. 
This suggests that new approaches are needed to study topological order in the presence of long-range interactions of the HK type.

\end{abstract}
\maketitle

\section{Introduction}

The interplay between interactions and topology is one of the major frontiers in condensed matter physics.
On the one hand, strongly correlated topological systems are predicted to host robust ground state degeneracies and quasiparticles with fractional quantum numbers and exotic statistics~\cite{,moore1991nonabelions,sarma2006topological,read2012topological,wen2017colloquium,kitaev2006anyons}. 
Several classification schemes and exactly-solvable models for such ``topologically ordered'' phases have been proposed~\cite{chen2013symmetry,thorngren2018gauging,wen1990topological,kapustin2014symmetry,levin2005string,dijkgraaf1990topological,barkeshli2013classification}, and many properties of topologically ordered phases have been experimentally verified in fractional quantum Hall systems~\cite{de1998direct,banerjee2018observation,nakamura2020direct}. 
Furthermore, several strongly-correlated materials such as $\alpha$-RuCl$_3$ that do not magnetically order at low temperatures are predicted to host topologically-ordered spin liquid ground states~\cite{banerjee2017neutron,zhou2017quantum,norman2016colloquium,savary2016quantum}. 

On the other hand, outside of the fractional quantum Hall effect, the connection between the microscopic Hamiltonian for interacting electrons and the topological order of the ground state remains elusive. 
Topological order is characterized by the absence of a local order parameter and by the lack of adiabatic continuity to the noninteracting ground state. 
Most analytical tools for treating the interacting electron problem, however rely on perturbation theory around a known non-interacting or mean field ground state~\cite{coleman2015introduction,georges1996dynamical}, or else introduce field-theoretic techniques that can obscure the connection to microscopic degrees of freedom~\cite{shankar1994renormalization}. 
While parton-based mean field approximations for topologically ordered systems exist~\cite{read1991large,wen1991mean,2011WenKMH,maciejko2015fractionalized}, they are generally uncontrolled and must be justified a posteriori. 
It is thus desirable to find a class of analytically tractable models for interacting electrons that could be applied to search for topologically ordered phases.

Recently, a class of exactly-solvable models for interacting electrons proposed by Hatsugai and Komohto (HK) in Ref.~\cite{TheOGHK1992} has gained renewed attention. 
These models include a long-range ring exchange interaction between electrons that is diagonal in momentum space, such that crystal momentum remains a good quantum number and the ground state factorizes as a tensor product over states at different crystal momentum. 
This renders these HK models exactly solvable. 
Up to this point, the HK interaction has been considered only for Hamiltonians with a conserved component of spin, with the HK interaction taken to be diagonal in the energy eigenbasis of the non-interacting part of the Hamiltonian. 
Concretely, this ``band-HK'' Hamiltonian takes the form
\begin{equation}
\label{eq:bandHKmodelintro}
H_{\mathrm{band-HK}} = \sum_{\mathbf{k}m\sigma}\epsilon_{m\sigma}(\mathbf{k})\bar{n}_{\mathbf{k}m\sigma} + U_1\sum_{\mathbf{k}m}\bar{n}_{\mathbf{k}m\uparrow}\bar{n}_{\mathbf{k}m\downarrow},
\end{equation}
where $m$ is a band index, $\sigma=\uparrow,\downarrow$ indexes the $z$-component of the electron spin, $\epsilon_{m\sigma}(\mathbf{k})$ is the noninteracting band dispersion, $U_1$ is the HK interaction strength, and
\begin{equation}
\bar{n}_{\mathbf{k}m\sigma} = \bar{c}^\dag_{\mathbf{k}m\sigma}\bar{c}_{\mathbf{k}m\sigma}\label{eq:bandhk}
\end{equation}
counts the number of electrons in band $m$ with crystal momentum $\mathbf{k}$ and spin $\sigma$ (throughout this work, we will use an overbar to emphasize operators that are diagonal in the band basis). 

Although band-HK Hamiltonians are easily solvable, they still capture several key features of strongly-correlated electron systems~\cite{2022PhilipHKsuperconductor,phillips2020exact,leeb2023quantum}. 
First, much like the Hubbard model, the ground state of \Cref{eq:bandHKmodelintro} is a correlated insulator at half-filling for large $U_1$ (large compared to the bandwidth $W$ of the noninteracting dispersion). 
Away from half filling, the system exhibits spectral weight transfer. 
Second, the single-particle self energy at generic filling exhibits divergences, indicating that the system cannot be obtained from perturbation theory around a noninteracting limit. 
Third, it was recently argued that the Mott insulator transition in the HK model is in the same universality class as the metal-insulator transition in the Hubbard model, and that the only instabilities of the ground state are towards either superconductivity or magnetic ordering~\cite{2022Philipuniversalityargument,zhao2023proof}.
Finally, since HK interactions preserve translational symmetry, they can in principle be realized experimentally in momentum space lattices. Recently there has been progress in realizing these lattices in cold atom systems~\cite{gadwaymsl21}.

The utility of the HK model in understanding the correlated electron problem more generally raises the question of whether it can be applied to study topology in interacting electron systems. 
Indeed, prior work in this direction has been quite promising. 
Ref.~\cite{mai2023topological} introduced an exactly-solvable model of a quantum anomalous Hall Mott insulator based on the HK interaction. 
This was extended in Ref.~\cite{2022PhilipSpinHallHK} to a model of a quantum spin Hall Mott insulator. 
Additionally, Ref.~\cite{setty2023symmetry} initiated the application of tools from non-interacting topological band theory to the study of poles in the single-particle self energy, as part of a larger program to study topology in correlated electron systems via single-particle Green functions~\cite{wagner2023mott}.

Despite this progress, the band-HK models present several obstacles to extending these lines of inquiry. 
First, due to the spin degeneracy in the band-HK model, the zero-temperature ground state is extensively degenerate at any filling for sufficiently strong interactions; there is a twofold degeneracy at every $\mathbf{k}$ in the Brillouin zone (BZ) in the Mott insulating phase. 
This means that care must be taken when topological invariants (which are ground state properties) are to be calculated~\cite{zhao2023failure,yanghk2019}. 
Additionally, a consequence of this extensive ground state degeneracy is a strong instability towards ferromagnetic ordering: the zero-temperature magnetic susceptibility of the HK model diverges, and an infinitesimal Zeeman field leads to magnetic ordering~\cite{yang2021exactly}. 
In many interesting cases, the ferromagnetically ordered state is adiabatically connected to a band insulator. 
This raises questions as to what aspects of Mottness are essential features of the HK model, and which are accidental features of the ground state degeneracy.

Second, it is important to remember that the HK interaction is infinitely long-ranged. 
This poses a conceptual difficulty because topological order is conventionally defined in terms of short-ranged Hamiltonians. 
One defining feature of a topologically ordered state is that no short-ranged Hamiltonian can connect or distinguish topologically degenerate ground states. 
Long-ranged Hamiltonians can of course have nonvanishing matrix elements between topologically degenerate states, and hence can lift the topological ground state degeneracy. 
Viewed from a different perspective, we can say that the characteristic feature of topological order is long-range entanglement in the ground state. 
While this is exotic and robust for ground states of short-ranged Hamiltonians, long-range entanglement is rather pedestrian when the Hamiltonian itself has infinite range.

To see this concretely, let us consider symmetry-enriched Lieb-Schultz-Mattis (LSM) type theorems applied to electronic systems with both time-reversal and crystal symmetries.
As shown in Refs.~\cite{watanabe2015filling,LIEB1961407,hastings2004lieb,oshikawa2000commensurability,parameswaran2013topological}, these theorems give us (mostly) tight bounds on what fillings can host gapped, symmetric, topologically trivial ground states in a given space group. 
These filling constraints provide a powerful tool for identifying topologically nontrivial systems that do not rely on computing a complicated invariant: if a system has an energy gap, a symmetric ground state, and violates a filling constraint then it must be topologically nontrivial. 
However, proofs of the LSM theorems (beyond Kramers's theorem and the requirement of integer filling per unit cell) all rely on the short-rangedness of the Hamiltonian in an essential way. 
For instance, the geometric proof in Ref.~\cite{watanabe2015filling} applies Kramers's theorem to the system placed on nontrivial flat manifolds. 
Crucial to the proof is the fact that, since all flat manifold are locally indistinguishable from Euclidean space, deep in the bulk the system is unaffected by ``twists'' in the manifold. 
This assumption breaks down if the Hamiltonian is---like the HK model---infinitely long ranged. 
In concert with earlier questions on the ground state degeneracy of HK models, this raises the question as to what degree LSM theorems can be used as a tool in these systems.

To address these questions, we will in this work introduced a generalized class of HK-like models which we call orbital HK models. 
Orbital-HK models have interactions that are diagonal in momentum space, but crucially are not diagonal in the band basis. 
Instead, we will formulate HK-like interactions in terms of creation and annihilation operators for tight-binding basis states. 
This allows us to formulate exactly-solvable models for interacting electron systems with spin-orbit coupling, even when no component of spin is conserved. 
We show that the orbital HK model maps to an $N$-site Hubbard model at each crystal momentum, where $N$ is the number of orbitals (per spin) in the unit cell; the band-HK models are a limiting case where the Hubbard Hamiltonian is diagonal. 

We will show through several examples that orbital HK models do not suffer from the extensive ground state degeneracy that has plagued the band-HK model. 
Instead, our orbital-HK models have order-one ground state degeneracies, and can even have nondegenerate ground states for certain choices of interaction. 
We show that this removes the tendency towards magnetic ordering, rendering the ground state stable to infinitesimal Zeeman fields. 
Crucially, we show that even when the ground state degeneracy is removed by the orbital HK interaction, signatures of Mottness remain: the self-energy diverges signifying that the orbital HK ground state is distinct from a band insulator. 
The ground state degeneracy of the band-HK model is thus reminiscent of Ref.~\cite{dave2013absence}, where an extensively degenerate toy model was introduced to demonstrate the failure of Luttinger's theorem in Mott insulators; as in Ref.~\cite{dave2013absence}, the extensive degeneracy of the band HK model is not essential to the formation of a Mott insulator.

Finally, to address the role that LSM theorems can play in the study of HK-like models, we will focus on spin-orbit coupled and interacting double-Dirac semimetals in space group \sg135 \footnote{Note that we follow the convention of Ref.~\cite{bradley1972mathematical} and include a $1^\prime$ in the space group symbol to emphasize the presence of time-reversal symmetry}.
Space group $P4_2/mbc1^\prime$ (\# 135) is a particularly interesting case for two reasons.
First, it allows, although it does not require, a different insulating filling constraint in the interacting and non-interacting cases.
As shown in Refs.~\cite{wieder2016double,bradlyn2016dirac}, all single-particle electronic states at the high-symmetry $A$ point of the BZ come in eightfold-degenerate multiplets in this space group, with linear double-Dirac dispersion away from this point. 
This means that non-interacting band insulators in space group $P4_2/mbc1^\prime$ (\# 135) must have filling $\nu=8n$ electrons per unit cell~\cite{watanabe2016fillingenforced}. 
With interactions, however, the situation is less clear: $P4_2/mbc1^\prime$ (\# 135) is one of a handful of space groups where the LSM bound of Ref.~\cite{watanabe2015filling} is less tight than the non-interacting filling bound, admitting the possibility of a gapped, symmetric, topologically trivial insulator in space group $P4_2/mbc1^\prime$ (\# 135) with $\nu=4$ electrons per unit cell in the presence of interactions. 
Space group \sg135 is thus a candidate for a unique kind of symmetric, insulating ground state, which does not have a non-interacting analogue. 

Second, there are a number of experimental candidates for realizing such a material.
Layered ternary borocarbide compounds are invariant under \sg135 and can have significant interactions strengths \cite{wiitkarborocarbide1994}.
No model Hamiltonians, however, for nonmagnetic interacting insulators in this space group have been put forward.\\

\indent In this work, we explore the phases accessible from a double Dirac semimetal in space group $P4_2/mbc1^\prime$ (
\# 135) at half-filling ($\nu=4$) in the presence of orbital-HK interactions. 
We show that the simplest orbital HK interaction leads to a Mott insulator with gapless spin excitations and no magnetic order, i.e. a gapless spin liquid. 
Next, we show that there exist orbital HK interactions which respect the symmetries of the space group and lead to the emergence of a gapped, nondegenerate ground state at half-filling, seemingly realizing the LSM lower bound in this space group. 
We compare this model to similar orbital HK models in space groups $P4/ncc1^\prime$ (\# 130) and $P2_1/c1^\prime$ (\# 14) where the LSM bound is realized by a band insulator, we provide evidence that the gapped, nondegenerate ground states realized by orbital HK models have long-range entanglement due to the infinite-ranged interaction, calling into question the utility of LSM theorems for these systems. Furthermore, our model in space group $P4/ncc1^\prime$ (\# 130) may be useful for shedding light on low temperature phases of the antiferromagnetic Mott insulator Bi$_2$CuO$_4$~\cite{bradlyn2016dirac,disante2017realizing}.

\subsection*{Guide to the Results}

The structure of the paper is as follows: First, in Sec.~\ref{sec:orbital-hk}, we review the HK model in the band basis, and introduce the orbital HK model in general. 
In Sec.~\ref{subsec:graphene} we show through the example of graphene that the simplest orbital HK interaction results in a ground state at half filling that is only fourfold degenerate, as opposed to the thermodynamically large degeneracy of graphene with the band-HK interaction. 
We show that the zero-temperature ground state is a Mott insulator with finite magnetic susceptibility. 
Next, in Sec.~\ref{subsec:HKKM} we extend our results to models with spin-orbit coupling and analyze the Kane-Mele (KM) model with orbital HK interaction. 
We show that at half filling the ground state of the orbital HK-KM model is nondegenerate. 
Our analysis shows how the orbital HK interaction allows us to treat spin-independent and spin-orbit coupled systems on equal footing. 
Furthermore, it reveals that the extensive ground state degeneracy in band-HK models is purely accidental, and is not germane to the underlying Mott physics.

In Sec.~\ref{sec:DDSL} we move to examine a three dimensional model of an interacting double-Dirac spin liquid. 
We consider a spin-orbit coupled double Dirac semimetal at half filling in space group $P4_2/mbc1^\prime$ (\# 135) in the presence of the simplest uniform orbital HK interaction. 
In Sec.~\ref{subsec:ddslmodel} we review the properties of the space group and introduce the microscopic model. 
Next, in Sec.~\ref{subsec:ddslprops} we compute the ground state, neutral excitation spectrum, and single-particle Green function for the model. 
We show that the ground state is a Mott insulator with a sixteen-fold degenerate ground state. 
The neutral excitation spectrum is gapless, consisting of spinon excitations near the $A$ point in the BZ. 
To make this precise, we show in Sec.~\ref{subsec:Schrieffer} how to apply a Schrieffer-Wolff transformation to orbital HK models. 
For our particular case of a double-Dirac semimetal at half filling, this gives an effective long-range spin Hamiltonian that approximates the neutral excitation spectrum of our model. 

Next, in Sec.~\ref{sec:General Interactions} we consider more general orbital HK interactions consistent with the space group symmetries. 
To begin, we show that the sixteen degenerate ground states can be labelled by irreducible representations of the space group. 
By identifying copies of the trivial representation, we are able in Sec.~\ref{subsec:splitting} to construct generalized orbital HK interactions that project onto the trivial representation which yields a nondegenerate, symmetric ground state. 
We compute the single-particle Green function in the nondegenerate ground state to verify that this represents a symmetric Mott insulating ground state that is not adiabatically connected to a band insulator. 
Thus, our system is a candidate for a gapped, symmetric, topologically trivial insulator at $\nu=4$ in this space group, as allowed for by the LSM theorem of Ref.~\cite{watanabe2015filling}.

To further understand the nature of this insulator and its relation to LSM theorems, we construct for comparison an orbital HK model in space group $P4/ncc1^\prime$ (\# 130). 
At $\nu=4$, the LSM theorem forbids the existence of a gapped, nondegenerate, short-range entangled ground state. 
Nevertheless, using a similar procedure as in space group $P4_2/mbc1^\prime$ (\# 135), we are able to construct a generalized orbital HK model with a gapped, symmetric, nondegenerate ground state. 
Taken together, this suggests that the nondegenerate ground states of orbital HK models are long-range entangled, due to the long-ranged interaction in position space. 
Thus, these models evade generalized LSM theorems that rely on locality of the Hamiltonian. 
The ground states take the form of cat states, which are stabilized by the long-ranged interactions.

We conclude in Sec.~\ref{sec:conclusion} by discussing the implications of our results for the use of HK models to understand correlated topological phases. 
We additionally include several appendices with details of derivations and additional supporting results.

\section{Orbital Hatsugai-Kohmoto interactions}\label{sec:orbital-hk}
We consider tight binding models defined on a parent lattice with additional degrees of freedom within the parent unit cell. 
We denote these ``orbital'' degrees of freedom with the Greek letters $\mu,\tau, ...$ and index a collection of them with Latin letters as $i=(\mu,\tau,...)$. 
We let $\sigma=\pm 1$ index the $z$ component of the electron spin, and we denote the chemical potential by $\mu_0$. 

In the simplest case of a non-interacting, spin-independent, tight binding model with no orbital degrees of freedom, we can straightforwardly diagonalize the non-interacting Hamiltonian to obtain the dispersion $\xi(\mathbf{k})$. 
The original Hatsugai-Kohmoto model, first introduced in Ref.~\cite{TheOGHK1992}, is the simplest example of what we earlier called a `band' HK model in \cref{eq:bandHKmodelintro}. 
It adds to the non-interacting model an energy cost $U_{1}$ for doubly occupying each momentum state through the interaction Hamiltonian
\begin{equation}
\label{eq:originalHK}
H_{HK}=\sum_{\mathbf{k}\sigma}\xi(\mathbf{k})\Bar{n}_{\mathbf{k}\sigma}+U_{1}\sum_{\mathbf{k}}\Bar{n}_{\mathbf{k}\uparrow}\Bar{n}_{\mathbf{k}\downarrow}\text{ }.
\end{equation} 
Hatsugai and Kohmoto showed that the model undergoes a metal-insulator transition when the interaction energy $U_1$ exceeds the bandwidth $W$, and that the energy gap in \Cref{eq:originalHK} at half-filling corresponds to the energy gap of the 1D Hubbard model in the large $U_1$ limit. 
This raises the hope that there may be other features of the Hubbard model that can be learned from the HK model.

This program is complicated by two consequences of the original HK model, which are shared by band HK models generally. 
In the insulating phase $U>W$, all possible configuration of spins in the lower band have the same energy. 
This means that in an $N$ particle system, there is an exponentially large $2^N$ ground state degeneracy. 
This makes it difficult to define properties of the HK model that depend on the ground state wavefunction, which is especially important for analyzing the topology of the ground state.

Second, it is implicitly assumed in \Cref{eq:bandHKmodelintro} that the spin projection $S_{z}$ commutes with the Hamiltonian. 
For a generic, spin-orbit coupled single-particle Hamiltonian, it is no longer clear how to write down such a band HK interaction. 
To deal with this problem, \cite{setty2023symmetry} introduced spin-dependent terms perturbatively to a non-spin dependent HK ground state.

Neither the ground state degeneracy nor the presence of spin conservation are necessary consequences of an HK model. 
In this paper, we will instead consider HK-type interactions which impose an energy penalty on doubly-occupied \textit{orbital} degrees of freedom $i$. 
In momentum space, the simplest such interaction is:
\begin{equation}\label{eq:bandhkgeneral}
    H^{1}_{HK}=U_{1}\sum_{\mathbf{k}i\sigma}n_{\mathbf{k}i\sigma}n_{\mathbf{k}i-\sigma}\text{ 
}.
\end{equation}
 We note that a similar interaction emerged in Ref.~\cite{2023Wysokynski}, where a translation-symmetry-breaking potential was added to the one-band HK model \Cref{eq:originalHK}. 
 To warm-up, we show in the next section, ~\ref{subsec:graphene}, that adding the orbital HK interaction \Cref{eq:bandhkgeneral} to a tight binding model of graphene leads to a ground state that is no longer thermodynamically degenerate. 
Instead, the ground state is only twofold degenerate at the $K$ and $K'$ high symmetry points. 
This order one degeneracy comes from the fact that we can map an orbital HK model to a finite site Hubbard model at every $\mathbf{k}$ point. 
From the perspective of the equivalent finite-site Hubbard models, band HK models correspond to the special case where the equivalent Hubbard model has zero hopping, and hence has a ground state degeneracy at each $\mathbf{k}$-point.

We then show in \Cref{subsec:HKKM} that we can easily extend orbital HK models to non-interacting Hamiltonians which depend on spin. 
To illustrate this, we consider adding $H^{1}_{HK}$ to a Kane-Mele model and solve for the resulting spectrum.

\subsection{Lifting the thermodynamic degeneracy: Graphene}
\label{subsec:graphene}
\subsubsection{Band and Orbital HK models}

\begin{figure}[ht]
\includegraphics[scale=0.5]{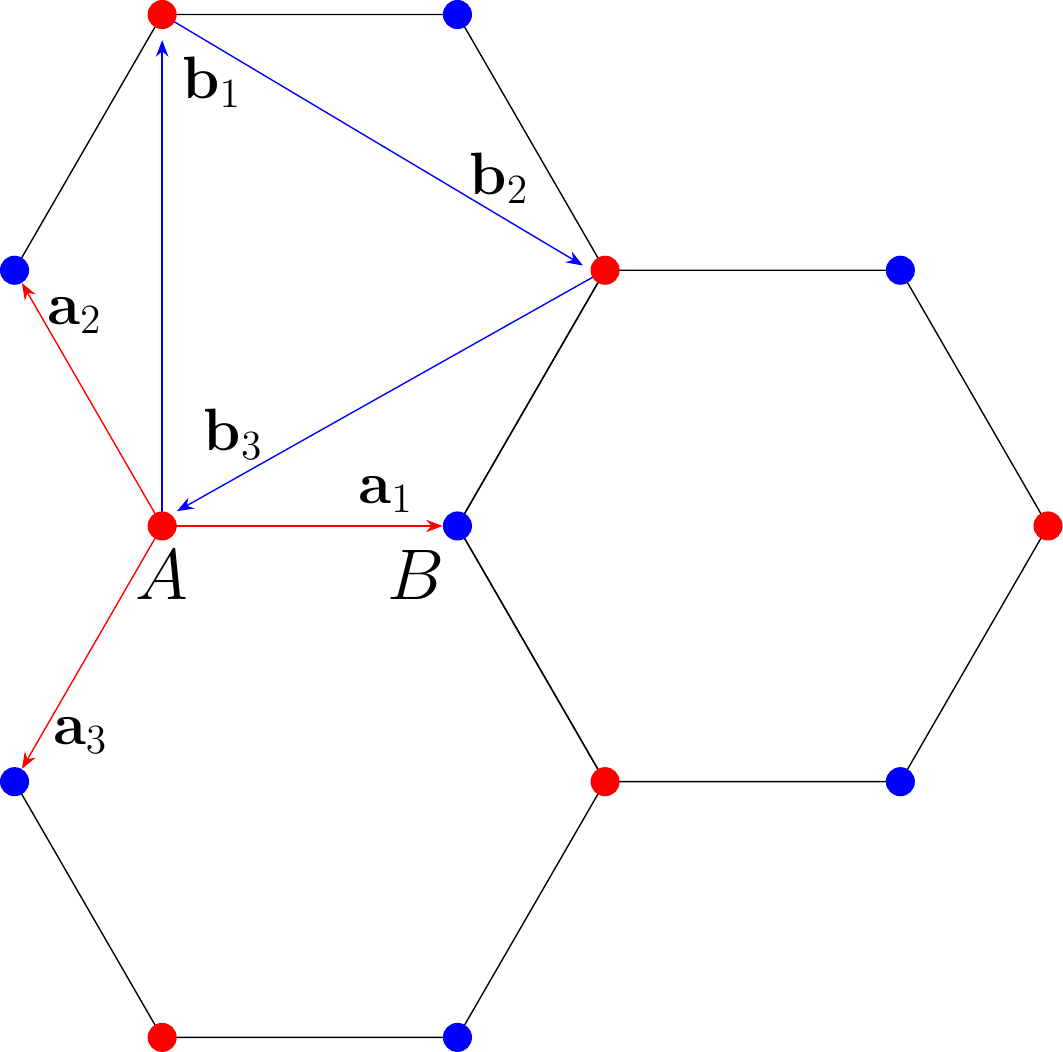}
\caption{Sublattices, nearest neighbor, and next-nearest neighbor vectors for the graphene and Kane-Mele tight binding models on the honeycomb lattice. 
The red arrows represent the $\mathbf{a}_{i}$ nearest neighbor lattice vectors. 
For lattice spacing $a$ the nearest neighbor vectors are given by $\mathbf{a}_1=a(1,0),\mathbf{a}_2=a(-\tfrac{1}{2},\tfrac{\sqrt{3}}{2}),\mathbf{a}_3=a(-\tfrac{1}{2},-\tfrac{\sqrt{3}}{2})$. 
The blue arrows the next-nearest neighbor lattice vectors $\mathbf{b}_i$, given by $\mathbf{b}_1=a(0,\sqrt{3}), \mathbf{b}_2=a(\tfrac{3}{2},-\tfrac{\sqrt{3}}{2}), \mathbf{b}_3=a(-\tfrac{3}{2},-\tfrac{\sqrt{3}}{2})$}
\label{fig:graphene}
\end{figure}

To analyze the ground state of an orbital HK model, we start with the standard non-interacting tight-binding model for graphene with nearest neighbor hopping between different sublattice sites~\cite{semenoff1984condensedmatter}. 
The second-quantized single-particle Hamiltonian is given by
\begin{equation}
H^{0}_{g}=t\sum_{\mathbf{k}\sigma}g(\mathbf{k})c^\dagger_{\mathbf{k}B\sigma}c_{\mathbf{k}A\sigma}+h.c.-\mu_{0}\sum_{\mathbf{k}\mu\sigma}n_{\mathbf{k}\mu\sigma},
\end{equation}
where the hoppings $g(\mathbf{k})$ are:
\begin{align}
g(\mathbf{k})&=\sum_{i}e^{i\mathbf{k}\cdot\mathbf{a}_i},
\end{align}
with $\mathbf{a}_i$ vectors to the nearest neighbor sites, as shown in \Cref{fig:graphene}.

We then add orbital HK interactions $H^{1}_{HK}$ to obtain an orbital HK model of graphene $H_{g}$:
\begin{align}
\label{eq:grapheneH}
H_{g}&=H^{0}_{g}+H^{1}_{HK} \\
&=t\sum_{\mathbf{k}\sigma}g(\mathbf{k})c^\dagger_{\mathbf{k}B\sigma}c_{\mathbf{k}A\sigma}+h.c.+U_{1}\sum_{\mathbf{k}\mu}n_{\mathbf{k}\mu\uparrow}n_{\mathbf{k}\mu\downarrow} \nonumber \\
&-\mu_0\sum_{\mathbf{k}\mu\sigma}n_{\mathbf{k}\mu\sigma}, \nonumber
\end{align}
where $\mu=\pm 1$ indexes the sublattice sites A and B respectively. 
 Since $H_{g}$ is diagonal in momentum space, we can consider a block at given $\mathbf{k}$,  $H_{g}(\mathbf{k})$
\begin{align}\label{eq:graphenekblockham}
H_{g}(\mathbf{k})&=t\sum_{\sigma}g(\mathbf{k})c^\dagger_{\mathbf{k}B\sigma}c_{\mathbf{k}A\sigma}+h.c.+U_{1}\sum_{\mu}n_{\mathbf{k}\mu\uparrow}n_{\mathbf{k}\mu\downarrow} \\
&-\mu_{0}\sum_{\mu\sigma}n_{\mathbf{k}\mu\sigma}\text{ }. \nonumber
\end{align}

The key observation for our technique is that the block $H_{g}(\mathbf{k})$ is equivalent to a two-site Hubbard model with complex hopping coefficients. 
We can see this explicitly by constructing a representation of this Hamiltonian on a basis and solving it by the same exact diagonalization procedure as is used in the Hubbard model \cite{2008JafaritwositeHubbard}.
First we note that $H_{g}(\mathbf{k})$ commutes with the number operator at a given $\mathbf{k}$ and $\sigma$, summed over the sublattice indices $\mu$ :
\begin{equation}
    \left[\sum_{\mu}n_{\mathbf{k},\mu\sigma},H_{g}(\mathbf{k})\right]=0,
\end{equation}
which in turns allows us to further block-diagonalize $H_{g}(\mathbf{k})$ into blocks of fixed particle number at each $\mathbf{k}$ point. 
At half-filling (here two electrons per unit cell), we can construct the matrix elements of $H_g(\mathbf{k})$ in the two-particle sector. 
We first choose an ordering for the six two-particle basis states at each $\mathbf{k}$, defined in \Cref{table:HKKMbasis}.
\begin{table}[ht]
\caption{The six basis states for the two particle sector for our model of graphene with orbital-HK interactions, used to form the matrix $\mathcal{H}_{g}(\mathbf{k})$.} 
\centering 
\renewcommand*{\arraystretch}{1.4}
\begin{tabular}{c c c c c c} 
\hline\hline 
Index & & State & & Label &\\[0.5ex]
\hline 
$1$ & &$c^\dagger_{\mathbf{k}A\uparrow}c^\dagger_{\mathbf{k}B\uparrow}\ket{0}$ & & $\ket{A\uparrow;B\uparrow}$& \\
$2$ & &$c^\dagger_{\mathbf{k}A\uparrow}c^\dagger_{\mathbf{k}A\downarrow}\ket{0}$ & & $\ket{A\uparrow;A\downarrow}$& \\
$3$ & &$c^\dagger_{\mathbf{k}A\downarrow}c^\dagger_{\mathbf{k}B\uparrow}\ket{0}$ & & $\ket{A\downarrow;B\uparrow}$& \\
$4$ & &$c^\dagger_{\mathbf{k}A\uparrow}c^\dagger_{\mathbf{k}B\downarrow}\ket{0}$ & & $\ket{A\uparrow;B\downarrow}$& \\
$5$ & &$c^\dagger_{\mathbf{k}B\uparrow}c^\dagger_{\mathbf{k}B\downarrow}\ket{0}$ & & $\ket{B\uparrow;B\downarrow}$& \\
$6$ & &$c^\dagger_{\mathbf{k}A\downarrow}c^\dagger_{\mathbf{k}B\downarrow}\ket{0}$ & & $\ket{A\downarrow;B\downarrow}$& \\

\hline 
\end{tabular}
\label{table:HKKMbasis} 
\end{table}
The matrix elements $\mathcal{H}_{g,2}(\mathbf{k})$ of $\mathcal{H}_{g}(\mathbf{k})$ in this basis are then given by
\begin{equation}\label{eq:graphene2ptclham}
\mathcal{H}_{g,2}(\mathbf{k})=
\begin{pmatrix}
0&0&0&0&0&0\\
0&U_{1}&-tg^{*}&tg^{*}&0&0\\
0&-tg&0&0&-tg^*&0\\
0&tg&0&0&tg^{*}&0\\
0&0&-tg&tg&U_1&0\\
0&0&0&0&0&0
\end{pmatrix}-2\mu_0\mathbb{I}_{6},
\end{equation}
where $\mathbb{I}_6$ is the $6\times6$ identity matrix. 
\Cref{eq:graphene2ptclham} is the same Hamiltonian as a two-site Hubbard model in the two-electron sector, with a complex hopping coefficient $\tilde{t}\equiv tg(\mathbf{k})$:
\begin{align}
H_{H}&=\tilde{t}\sum_{\sigma}\sum_{<a,b=0>}^{1}c^\dagger_{a\sigma}c_{b\sigma}+h.c.+U_{1}\sum_{i=0}^{1}n_{a\uparrow}n_{a\downarrow} \nonumber \\
&-\mu_{0}\sum_{\sigma}\sum_{i=0}^{1}n_{i\sigma} ,
\end{align}
where $a,b$ are labels that run over the sites $0$ and $1$. 

\begin{figure*}
\FPeval{\overlap}{0.35}
\FPeval{\scalevalue}{round(0.33*(1.075)/\overlap,2)}
\scalebox{\scalevalue}{
\setlength{\tabcolsep}{0pt} 
\def\tabularxcolumn#1{m{#1}}
\hskip-0.75cm\begin{tabularx}{\textwidth}{@{}XXX@{}}
\subfloat[]{\includegraphics[width=\overlap\textwidth]{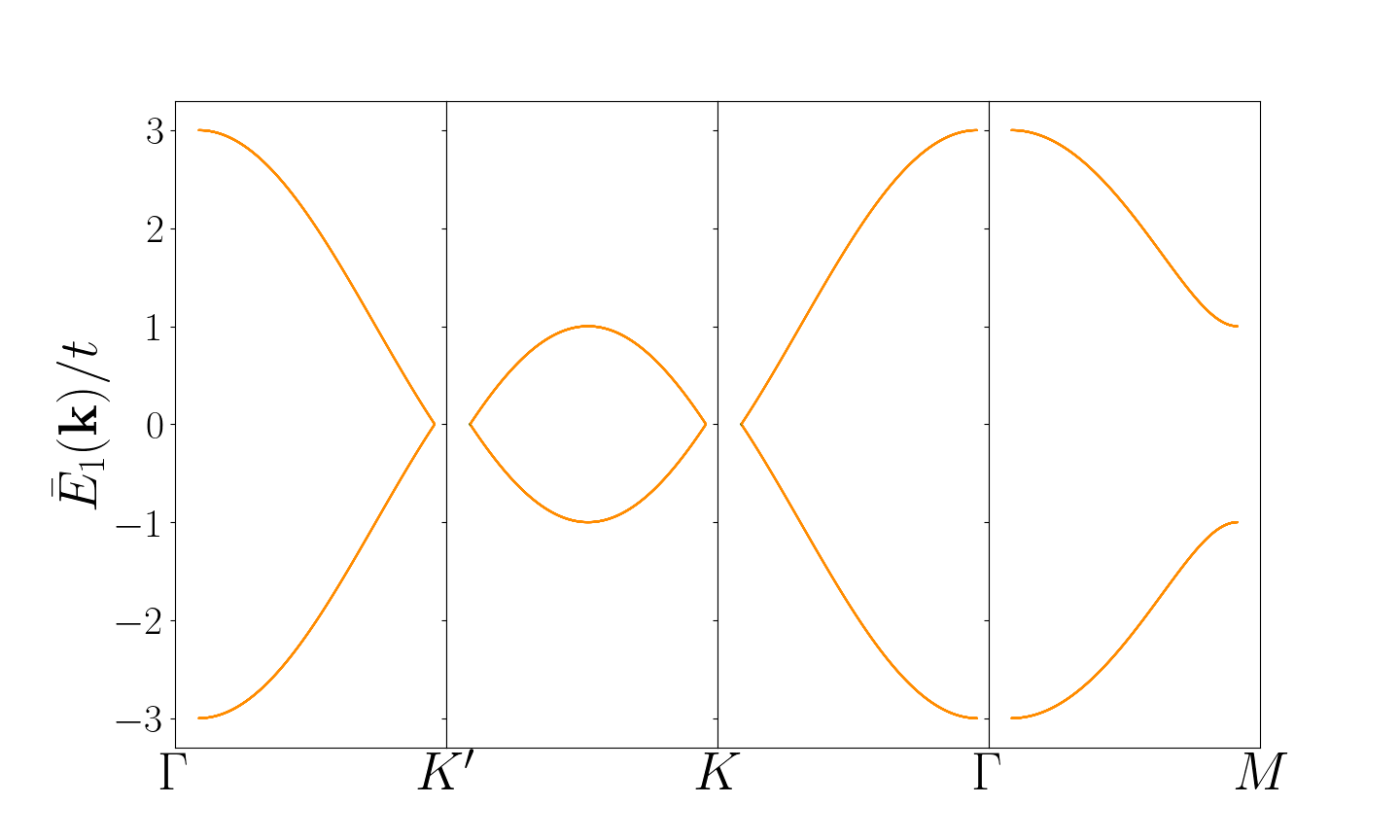}} &
\subfloat[]{\includegraphics[width=\overlap\textwidth]{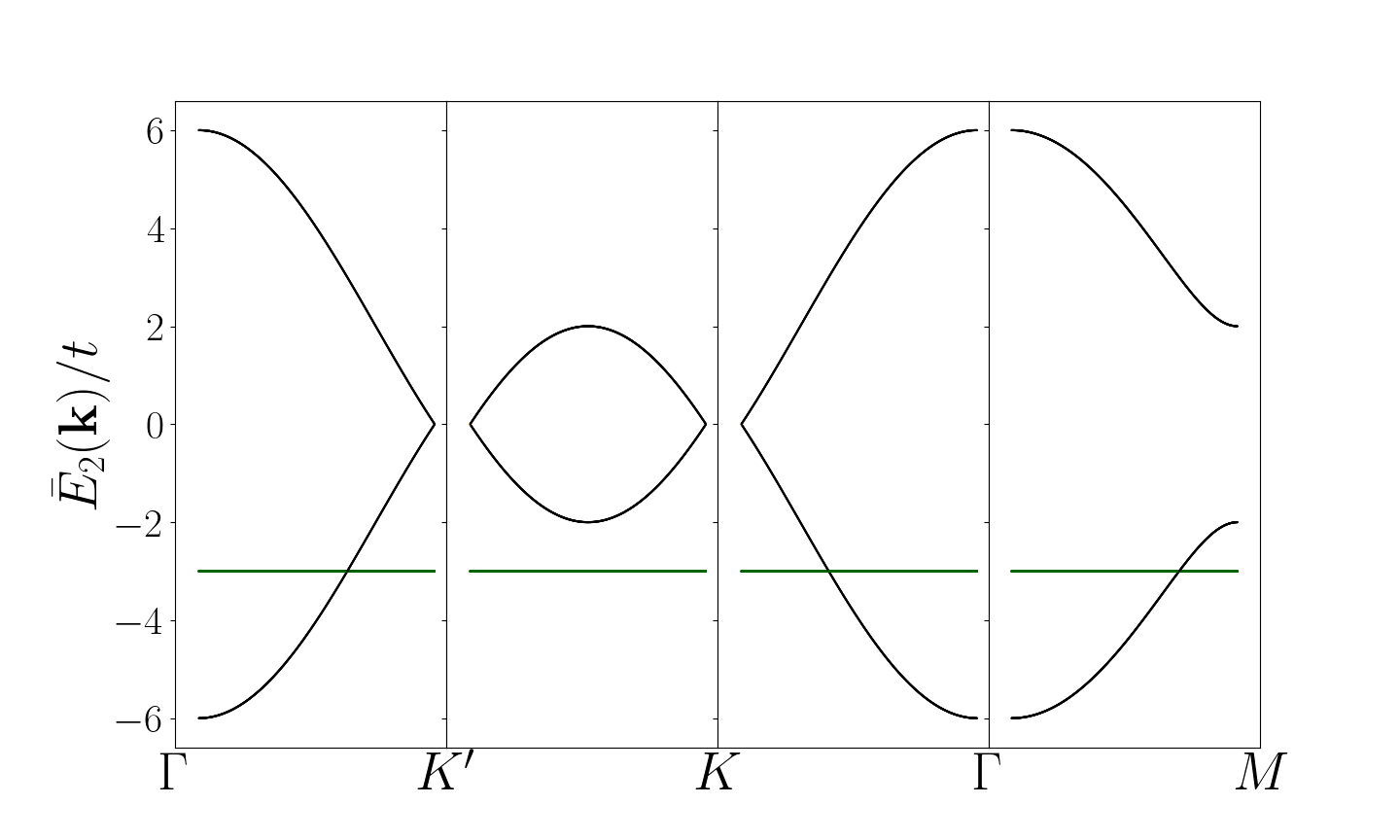}} &
\subfloat[]{\includegraphics[width=\overlap\textwidth]{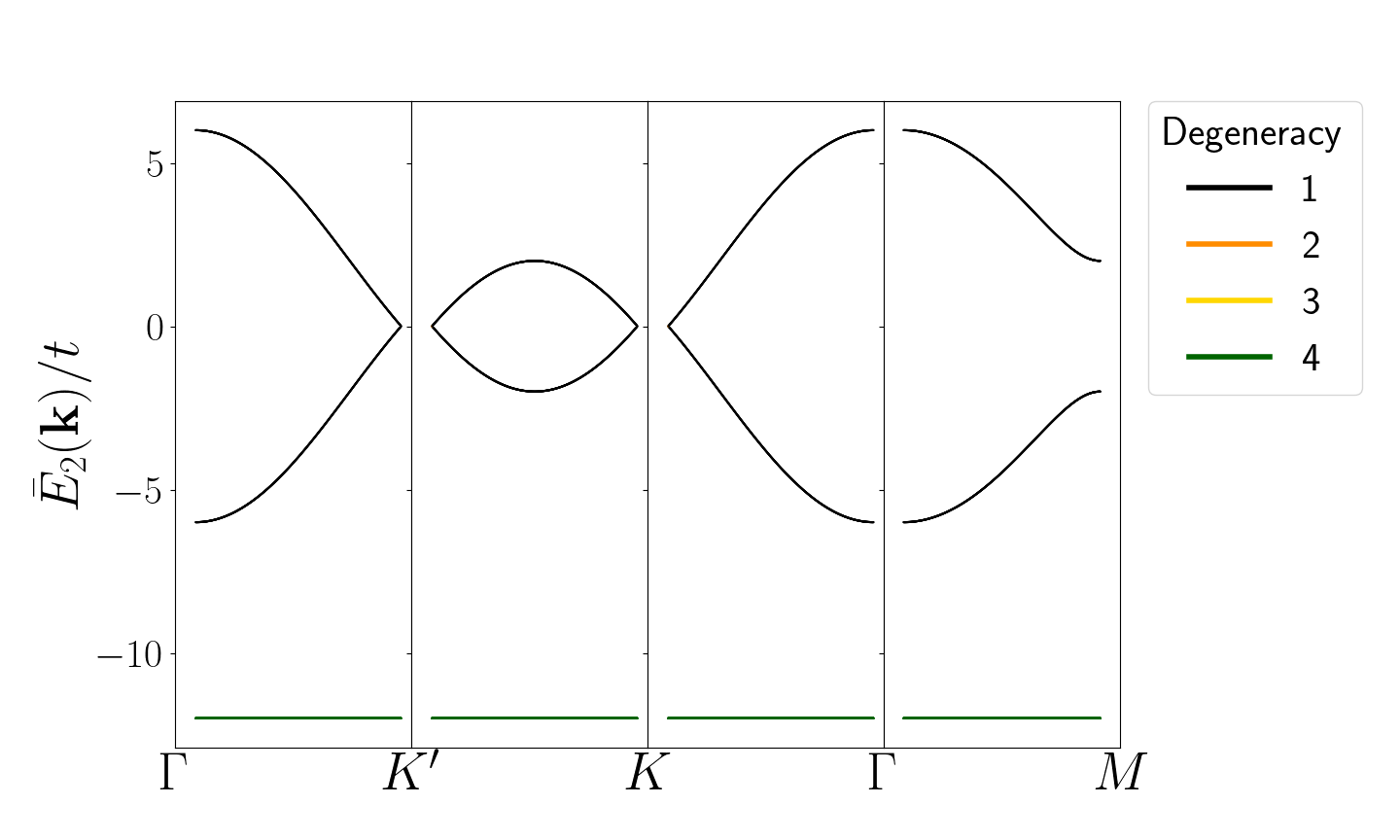}} \\[-3ex]
\subfloat[]{\includegraphics[width=\overlap\textwidth]{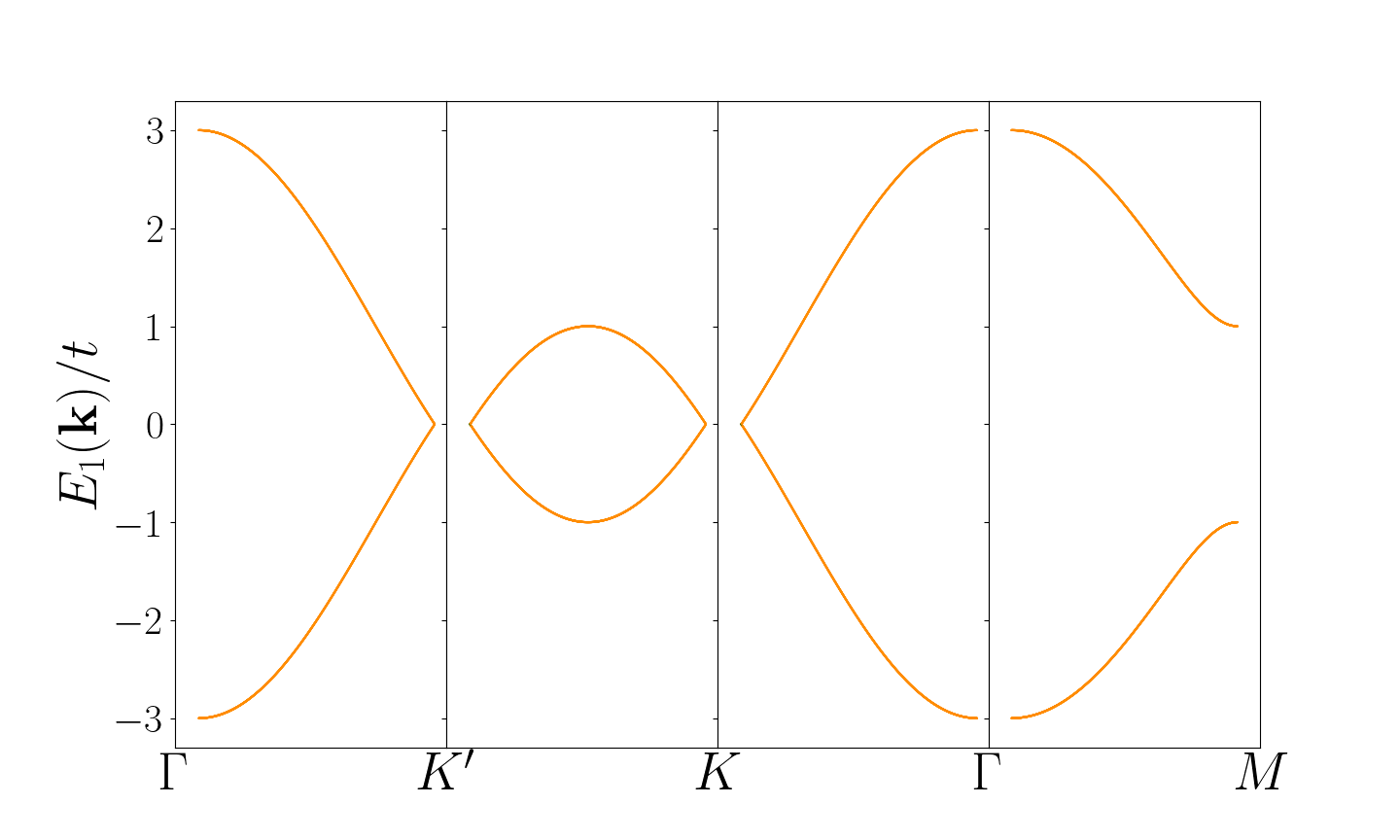}} &
\subfloat[]{\includegraphics[width=\overlap\textwidth]{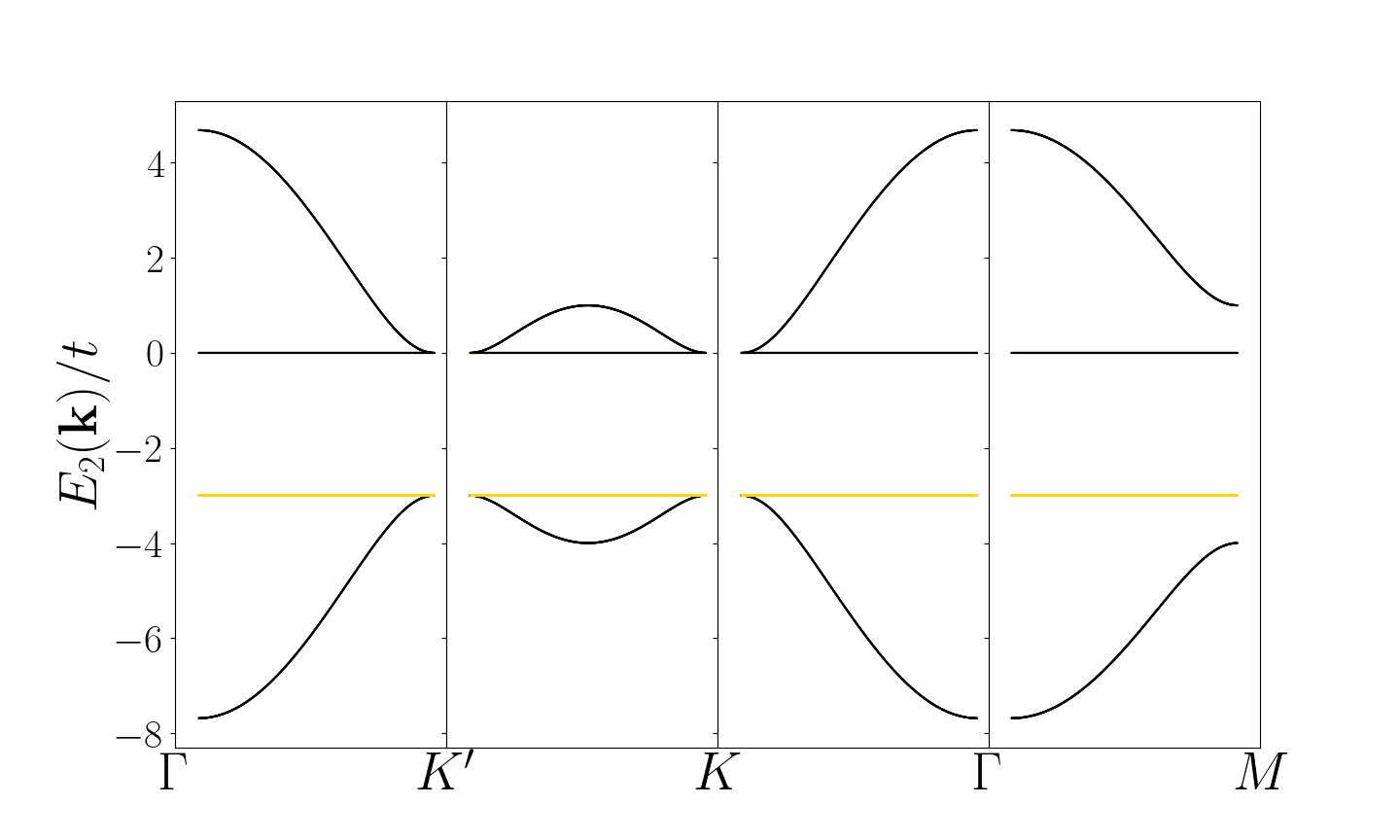}} &
\subfloat[]{\includegraphics[width=\overlap\textwidth]{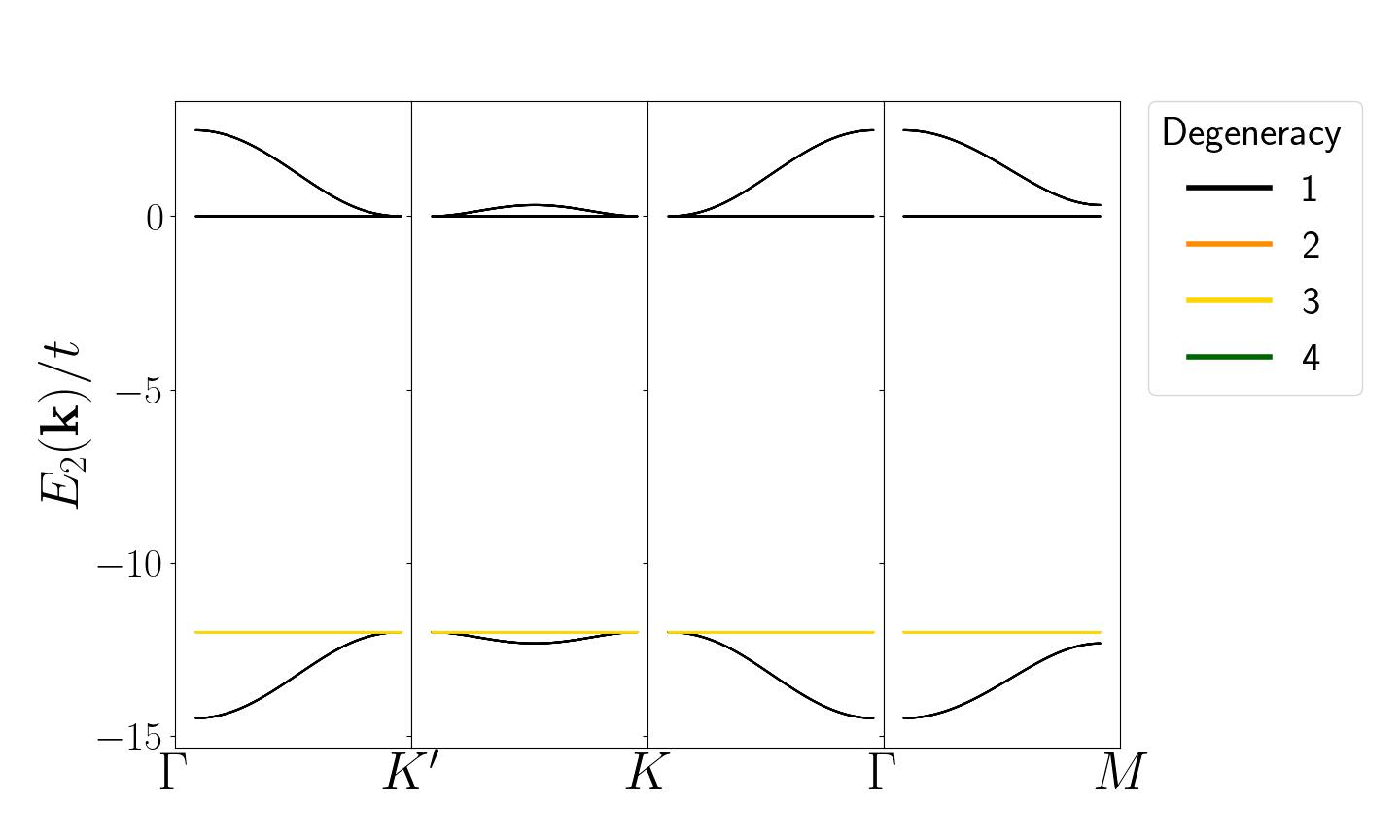}}
\end{tabularx}}
\caption{A comparison of the spectrum at half-filling (two electrons per unit cell) for a graphene tight-binding model with band HK interactions $\Bar{H}_{g}$ (top row), and orbital HK interactions $H_{g}$ (bottom row). 
(a) shows the single particle spectrum for the non-interacting Hamiltonian $H^{0}_{g}$.  
(b) shows the spectrum in the two particle sector $\Bar{E}_{2}(\mathbf{k})$ for the graphene band HK model with interactions strength $U_{1}$ equal to half the bandwidth ($U_{1}=3t$).
The ground state is degenerate when the green line is below the black.
(c) shows the spectrum in the two particle sector in the band basis HK model with interactions larger than the bandwidth $U_1=2W$ ($U_{1}=12t$). 
(d)  shows the (repeated) single particle spectrum for the non-interacting Hamiltonian $H^{0}_{g}$ with $t=1$. 
(e) shows the spectrum in the two particle sector $E_{2}(\mathbf{k})$ for the graphene orbital HK model with interactions strength $U_{1}$ equal to half the bandwidth ($U_{1}=3t$). 
(f) shows $E_{2}(\mathbf{k})$ for the orbital HK model with $U_{1}=2W$. 
The parameters of the non-interacting Hamiltonian are everywhere $t=1, \mu_{0}=U_1/2$.
We can see from the spectrum in the two particle sector $E_{2}(\mathbf{k})$ (b) and (c) that the ground state of the band model is degenerate across the region in the Brillouin zone where $U_{1}<2|tg(\mathbf{k})|$. 
For the orbital HK model, we can see from (e) and (f) that, for all $U_{1}>0$, the ground state is only degenerate at the $K$ and $K'$ points. }
\label{fig:GrapheneSpectra}

\end{figure*}

Since we can solve the two site Hubbard model analytically, we can immediately write down the energy eigenvalues at half-filling for a given $\mathbf{k}$ point:
\begin{align}
\label{eq:twositeHenergies}
    E^\pm_{2}(\mathbf{k},\mu_0)&=\frac{1}{2}\left(U_{1}\pm\sqrt{U^2_{1}+16|tg(\mathbf{k})|^2}\right)-2\mu_{0},\nonumber \\
    E^0_{2}(\mathbf{k},\mu_0)&=-2\mu_{0}, \\
    E^U_{2}(\mathbf{k},\mu_0)&=U_{1}-2\mu_{0}, \nonumber
\end{align}
with the ground state energy in the two-particle sector given by $E^-_2(\mathbf{k},\mu_0)$ in the first line.
In the left three panels of \Cref{fig:GrapheneSpectra} we plot the spectrum of $H_g(\mathbf{k})$ in the two-particle sector at each $\mathbf{k}$ point, $E^a_{2}(\mathbf{k},\mu_0)$, along the High Symmetry Points (HSPs) of the graphene Brillouin Zone for the non-interacting ($U_1=0$), intermediate interaction ($U_1\sim t$), and strongly interacting $(U_1\gg t)$ regimes.

To find the ground state of the entire $N$ particle system, we then need to find the number of particles at each $\mathbf{k}$ point, $n_\mathbf{k}$, that minimizes the total energy.
In general, this requires solving the corresponding optimization problem.

For particle-hole symmetric systems at half-filling, however, the solution is straightforward. 
Since adding a particle at half-filling necessitates double occupying an orbital state, it results in a higher energy. 
Since the system is particle-hole symmetric, this energy penalty cannot be made up by creating a hole elsewhere. 
This means that at half-filling the ground state of the particle-hole symmetric, $N$ particle system half fills the unit cell at \textit{every} $\mathbf{k}$-point. 
In our graphene example, this means the ground state of the particle block with two electrons per unit cell at any $\mathbf{k}$ point is the lowest energy state.

We can see this explicitly from the graphene spectrum in \Cref{eq:twositeHenergies}. 
We first set the chemical potential to $\mu_0=-\tfrac{U_1}{2}$ .
This makes the spectrum manifestly particle-hole symmetric. 
We introduce the particle-hole symmetry operator $P$ which acts on creation operators as
\begin{equation}\label{eq:grapheneph}
Pc_{\mathbf{k}\mu\sigma}P^{-1} = c^\dag_{\mathbf{k}\mu'\sigma'}\tau_z^{\mu'\mu}\sigma_x^{\sigma'\sigma}.
\end{equation}
$P$ maps the $n$-particle sector of the Hilbert space to the $2-n$ particle sector ($n=0,1,2$). 
Using the canonical anticommutation relations, we have that when acting on each fixed-$\mathbf{k}$ block of the graphene Hamiltonian given in \Cref{eq:graphenekblockham}
\begin{equation}
P\mathcal{H}_{g}(\mathbf{k})P^{-1} = \mathcal{H}_{g}(\mathbf{k}).
\end{equation} 
Hence when $\mu_0=U_1/2$, the spectrum of \Cref{eq:grapheneH} is particle-hole symmetric. 
Explicitly, the ground states of the one, two, and three particle block Hamiltonians $H_{g}(\mathbf{k})$ are given by:
\begin{align}
    &E^-_{1}(\mathbf{k},U_1/2)=-U_1/2-|tg(\mathbf{k})|\\
    \label{eq:E2graphenewithmu}
    &E^-_{2}(\mathbf{k},U_1/2)=-\frac{1}{2}\left(U-\sqrt{U^2_{1}+16|tg(\mathbf{k})|^2}\right)\\
    &E^-_{3}(\mathbf{k},U_1/2)=-U_1/2-|tg(\mathbf{k})|\text{} .
\end{align}
From this we see that for any $\mathbf{k}$, $U$, $t$ and $g$, $E^-_{2}(\mathbf{k},U_1/2)<E^-_{1}(\mathbf{k},U_1/2)=E^-_{3}(\mathbf{k},U_1/2)$. 
This means it is never energetically favorable to create a one and three-particle state from two two-particle states. 
Hence, the minimum energy $N$ particle state at half-filling half-fills \textit{every} $\mathbf{k}$ point. 

More generally, for the graphene Hamiltonian with orbital interactions $H_{g}$, so long as $\mu_0$ is in the range:
\begin{equation}
|tg(\mathbf{k})|-\sqrt{(\tfrac{U_1}{2})^2+4|tg(\mathbf{k})|^2}<\big|\mu_0-U_1/2\big|
\end{equation}
the ground state of $H_{g}$ will be at half-filling at every $\mathbf{k}$. 
In this range, the ground state will be the tensor product of the ground states of the half-filled states at every $\mathbf{k}$ point. Since only $\mu_0=U_1/2$ is particle-hole symmetric, this illustrates that the ground state is robust to particle-hole symmetry breaking, within a range.
Away from half-filling, this is only true for sufficiently large values of the interaction; in general, the ground state can have different numbers of electrons at each $\mathbf{k}$ point.  
We also note that since the many body ground state is a product state in momentum space:
\begin{equation}
    \ket{GS}=\frac{1}{\sqrt{N}}\bigotimes_{\mathbf{k}}\ket{\mathbf{k}},
\end{equation}
the representation of the ground state in localized real space orbitals is given by
\begin{equation}
\ket{GS}=\frac{1}{\sqrt{N}}\bigotimes_{\mathbf{k}}\left(\sum_{\mathbf{R}}e^{i\mathbf{k}\cdot\mathbf{R}}\ket{\mathbf{R}}\right)
\end{equation}
Hence the ground state has a high degree of long-range entanglement in real space. This follows from the long range of the HK interaction.

Having found the ground state, we can now also see why the ground state becomes degenerate at the $K$ and $K'$ points. 
At these points, $g(\mathbf{K})=g(\mathbf{K'})=0$ and so both $E^{-}_2(K,U_1/2)=E^{0}_2(K,U_1/2)=-U_1$ and $E^{-}_2(K',U_1/2)=E^{0}_2(K',U_1/2)=-U_1$ become degenerate.

From this perspective, we can also see that adding band HK interactions is a limiting case of orbital HK interactions. 
If we define creation operators for eigenstates of graphene in energy band $m$, $\bar{c}^\dagger_{\mathbf{k}m\sigma}$, such that the non-interacting Hamiltonian is diagonal:
\begin{equation}
\Bar{H}^{0}_g=\sum_{\mathbf{k}m\sigma}\xi_{m}(\mathbf{k})\Bar{c}^\dagger_{\mathbf{k}\sigma m}\Bar{c}_{\mathbf{k}m\sigma}
\end{equation}
and add HK interactions in this basis:
\begin{equation}
\label{eq:bandgrapheneH}
\Bar{H}_{g}=\sum_{\mathbf{k}m\sigma}\xi_m(\mathbf{k})\Bar{c}^\dagger_{\mathbf{k}m\sigma}\Bar{c}_{\mathbf{k}m\sigma}+U_1\sum_{\mathbf{k}}\Bar{n}_{\mathbf{k}m\uparrow}\Bar{n}_{\mathbf{k}m\downarrow}-\mu_{0}\sum_{\mathbf{k}m\sigma}\Bar{n}_{\mathbf{k}m\sigma},
\end{equation}
then following the same procedure as above but in the band basis leads to the two-particle matrix elements $\bar{\mathcal{H}}_{g,2}$:

\begin{equation}
\resizebox{0.95\linewidth}{!}{
$\Bar{\mathcal{H}}_{g,2}=\begin{pmatrix}
\xi_1+\xi_2&0&0&0&0&0\\
0&U_{1}+2\xi_1&0&0&0&0\\
0&0&\xi_1+\xi_2&0&0&0\\
0&0&0&\xi_1+\xi_2&0&0\\
0&0&0&0&U_1+2\xi_2&0\\
0&0&0&0&0&\xi_1+\xi_2
\end{pmatrix}$}-2\mu_{0}\mathbb{I}_{6}
\end{equation}
which corresponds to the two-site Hubbard model with zero hopping and a site-dependent chemical potential $-\xi_m$:
\begin{equation}
H_{H}=\sum^{1}_{m=0}\sum_{\sigma}(\xi_m-\mu_{0})\Bar{n}_{m\sigma}+U_{1}\sum^{1}_{m=0}\Bar{n}_{m\uparrow}\Bar{n}_{m\downarrow}
\end{equation}
Unlike the two-site Hubbard model with non-zero hopping, this Hamiltonian has a four-fold degenerate ground state whenever $\xi_{1}+\xi_{2}<U_{1}$, as in the orbital graphene model at the $K$ and $K'$ points.
This is why band HK models can have thermodynamically large ground state degeneracies, whereas orbital HK models do not.

In \Cref{fig:GrapheneSpectra} we illustrate this difference by comparing the two particle spectra for models of graphene with the band and with orbital HK interaction in the non-interacting [(a) and (d)], weakly interacting [(b) and (e)] and strongly interacting [(c) and (f)] limits. 
We see that the ground state energy of the orbital HK model is smooth as a function of $\mathbf{k}$, whereas the band HK ground state energy at intermediate interaction strength has a kink at the momenta where the interaction energy becomes comparable to the single-particle energy. 
We can also see from the comparison of the band and orbital HK spectra that the non-zero hopping in orbital models changes the shape of the non-interacting bands, whereas in band HK models the bands simply shift in overall energy.

We can understand the nature of the ground state in the two models by calculating the real-time retarded Green function matrix in the zero temperature limit, $G^{+}_{i,j}$, obtained via analytic continuation of the Matsubara Green function.
In the Lehmann representation, the matrix elements are:
\begin{align}
\label{eq:gfmatrix}
    &G^{+}_{i,j}(\mathbf{k},\omega)\equiv\mathcal{Z}_{g}^{-1}\sum_{mn}\frac{\bra{n}c_{\mathbf{k}i}\ket{m}\bra{m}c^\dagger_{\mathbf{k}j}\ket{n}}{\omega-(E_{m}-E_{g,n})+i\eta}\\
&+\frac{\bra{n}c^\dagger_{\mathbf{k}i}\ket{m}\bra{m}c_{\mathbf{k}j}\ket{n}}{\omega-(E_{g,n}-E_{m})+i\eta} ,\nonumber
\end{align}
where $m$ labels the energy eigenstates, $\eta\rightarrow 0^+$ is an infinitesimal positive number, and we have allowed for the possibility of a degenerate ground state by letting $n$ label the $n$-th linearly independent ground state, $E_{g,n}$ the corresponding ground state energy, and $\mathcal{Z}_{g}$ the partition function over the degenerate ground states.

From the properties of the retarded real-time Green function matrix $G^{+}$ we can read off a number of important properties that help to characetize the nature of the ground state.
Most importantly, the presence of a band of \textit{zero} eigenvalues of this matrix at every $\mathbf{k}$ point in the Brillouin zone, known as a Luttinger surface, signifies a divergence of the single-particle self-energy.
This indicates that the effect of interactions is nonperturbatively large.  
In turn, this implies that the system cannot be adiabatically connected to a trivial band insulator or Fermi liquid in the non-interacting limit. 
This is the defining feature of a Mott insulator \cite{dave2013absence}.

\begin{figure*}[ht]
\FPeval{\overlap}{0.53}
\FPeval{\scalevalue}{round(0.5*(1.12)/\overlap,2)}
\scalebox{\scalevalue}{
\setlength{\tabcolsep}{0pt} 
\def\tabularxcolumn#1{m{#1}}
\hskip-1.0cm\begin{tabularx}{\textwidth}{@{}XXX@{}}
\subfloat[]{\includegraphics[width=\overlap\textwidth]{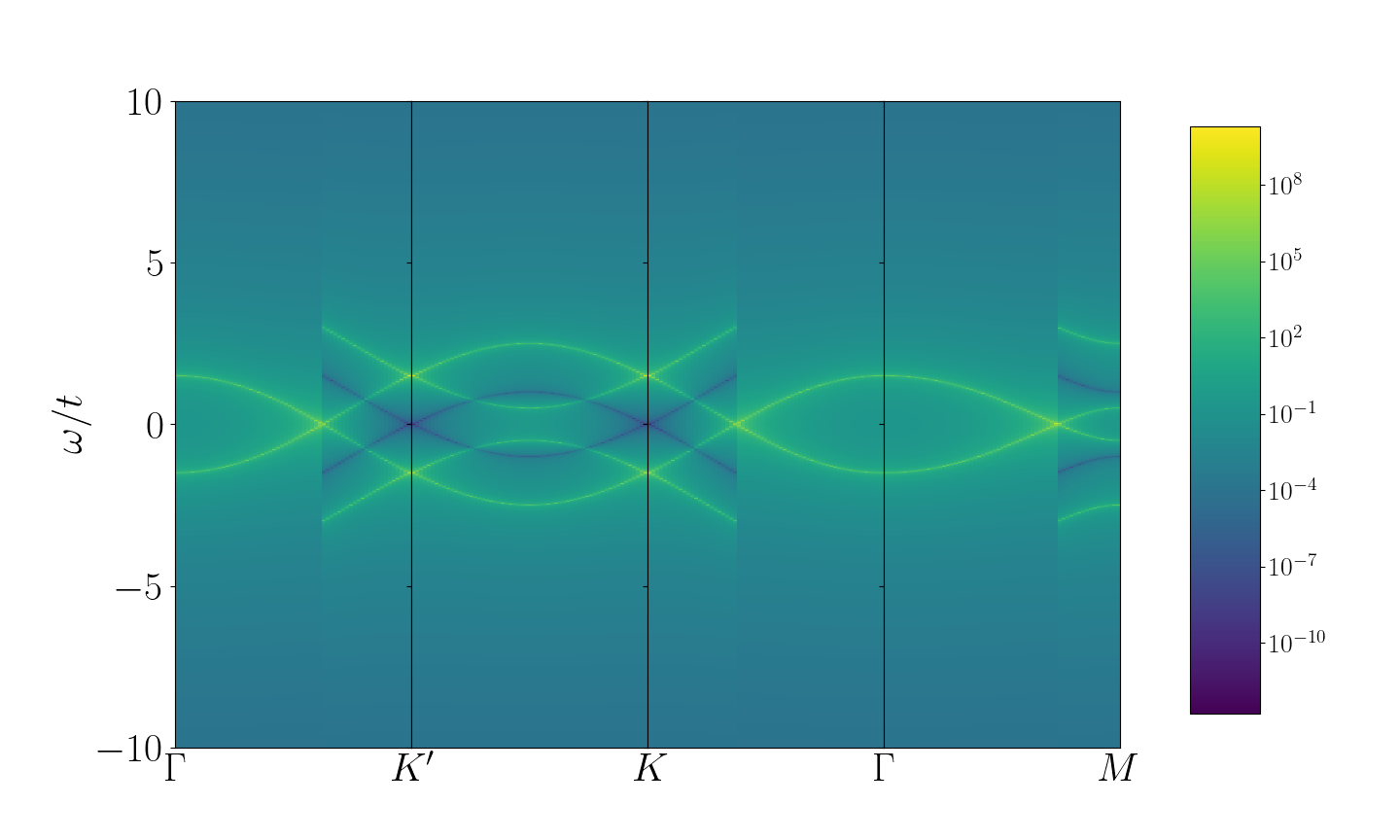}} &
\subfloat[]{\includegraphics[width=\overlap\textwidth]{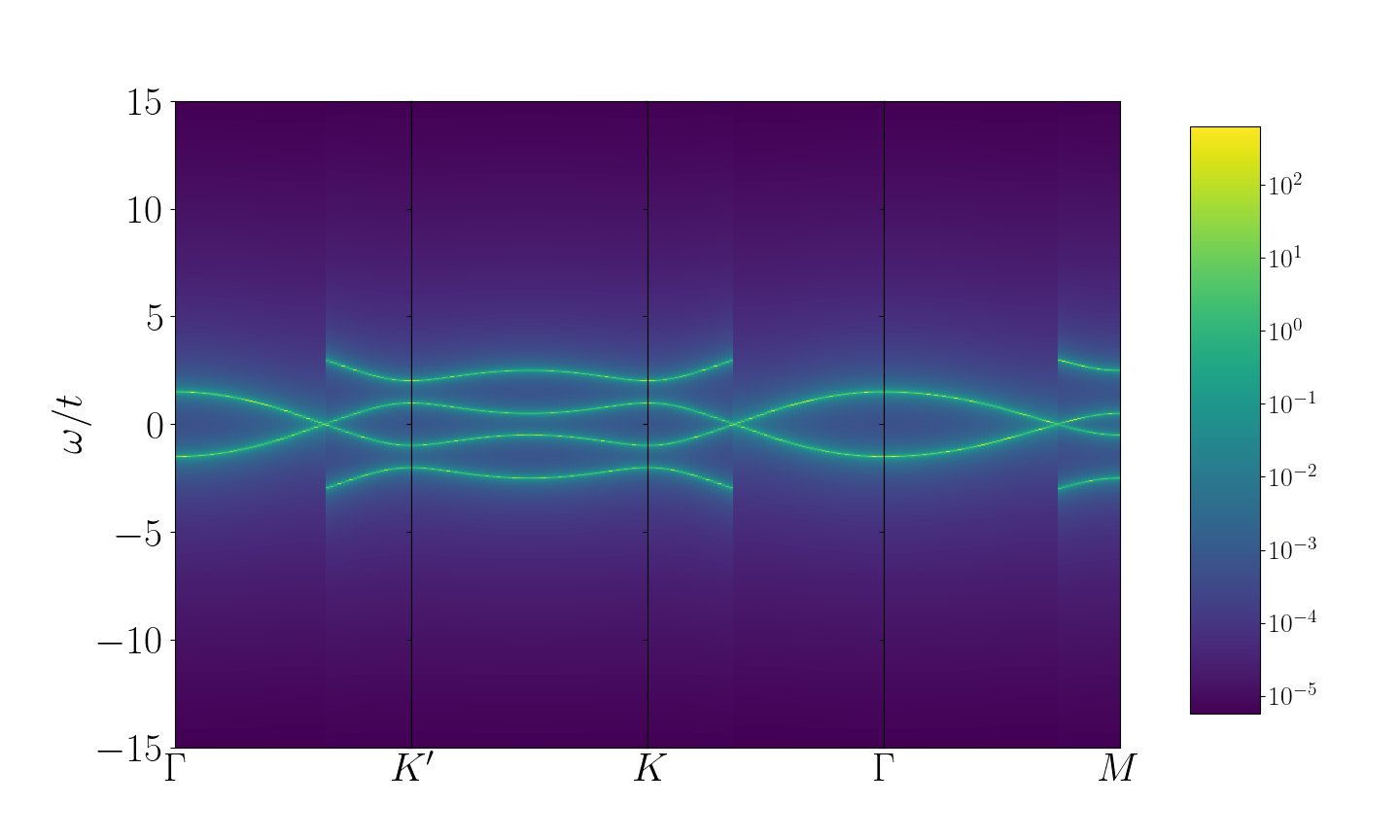}}
 \\[-3ex]
\subfloat[]{\includegraphics[width=\overlap\textwidth]{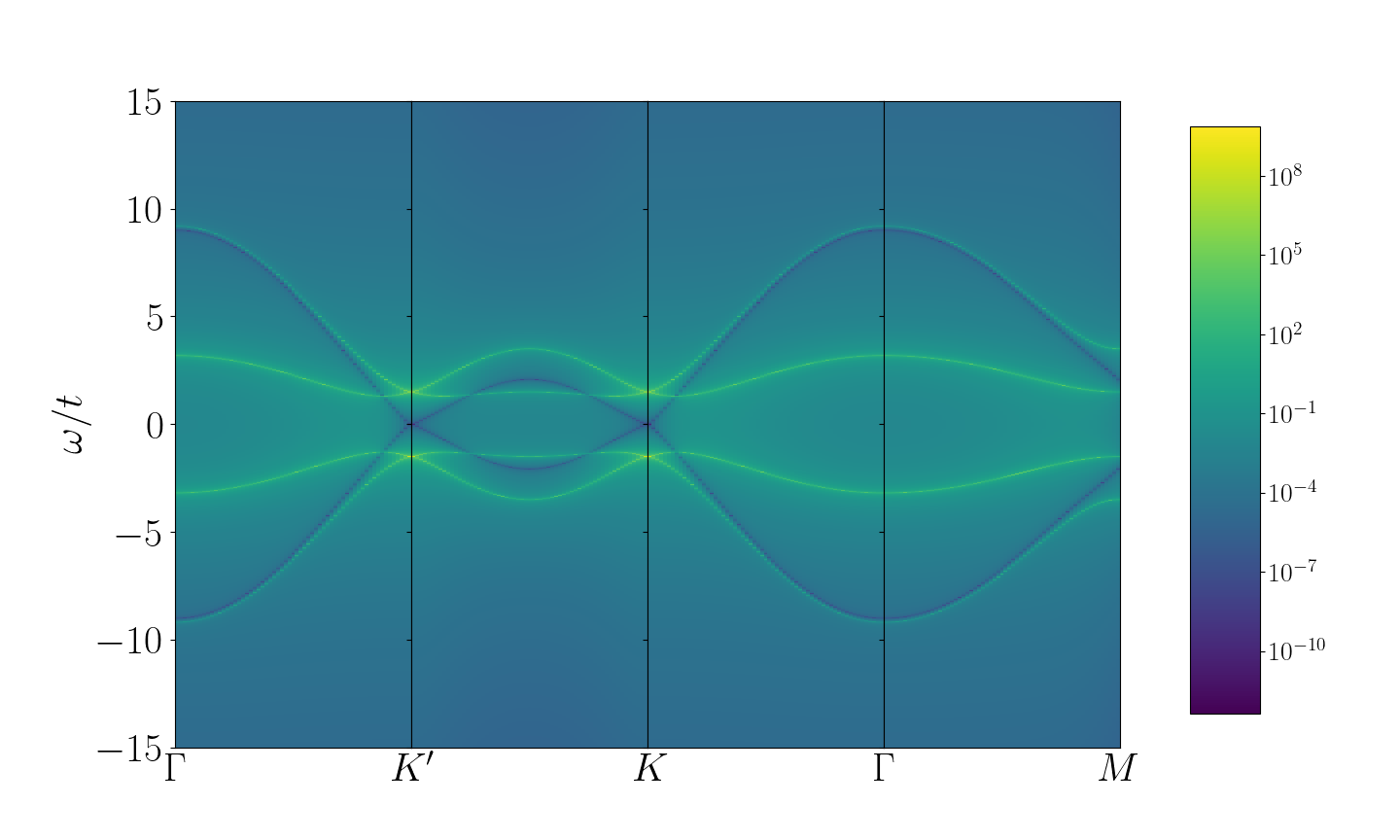}} &
\subfloat[]{\includegraphics[width=\overlap\textwidth]{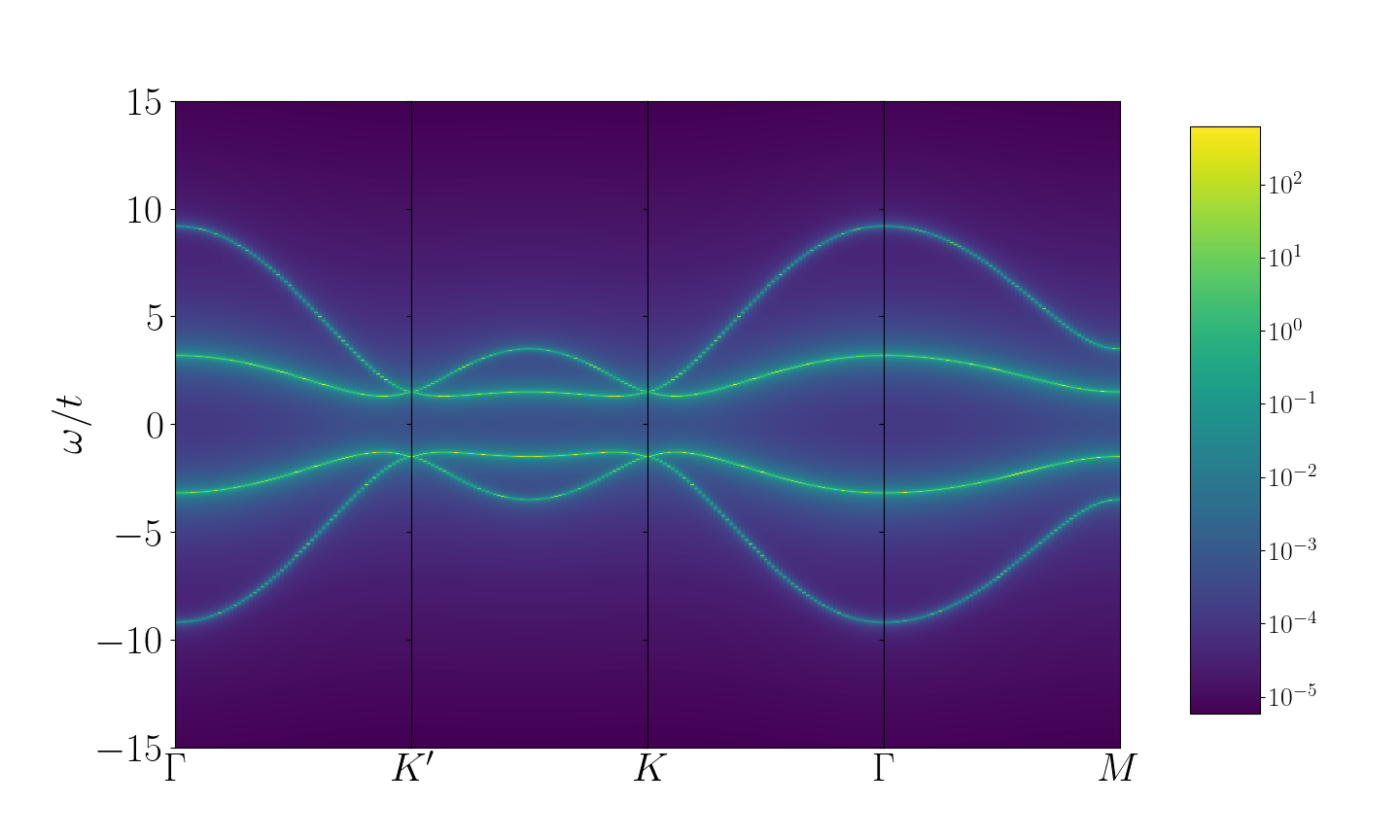}}
\end{tabularx}}
\caption{A comparison of the real time retarded Green function $G^{+}$ for band (top) and orbital (bottom) HK models of Graphene. 
In the left two panels (a) and (c), we show the absolute value of the determinant $|\det(G^{+})|$, and in the right two panels (b) and (d) the spectral function $-\frac{1}{\pi}\Im\Tr(G^{+}(\mathbf{k},\omega))$. 
(a) shows $|\det(G^{+})|$ for the intermediate interaction strength regime $U_{1}=W/2$ ($U_{1}=3t$) for the band HK model. 
Since the band of zeros is not present across the whole Brillouin zone, the band model in the intermediate regime is not a Mott insulator, but a general non-Fermi liquid.
(b)  shows the spectral function, $-\frac{1}{\pi}\Im\Tr(G^{+}(\mathbf{k},\omega))$, in the intermediate interaction strength regime for the band HK model. 
(c) shows $|\det (G^{+})|$ for the orbital HK model with $U_{1}=W/2$. 
The band of zeros is present across the whole Brillouin Zone.
(d) shows the spectral function for the orbital HK model in the intermediate interaction strength regime $U_{1}=W/2$ . 
The parameters of the non-interacting Hamiltonian are everywhere $t=1, \mu_{0}=U_1/2$.
The band of zeros in the determinant of the orbital model demonstrates that it is a Mott insulator even for interaction strengths lower than the bandwidth $U_{1}<W$.}
\label{fig:GrGF}

\end{figure*}

Conversely, \textit{poles} of $G^{+}$ correspond to single-particle charge excitations. 
As pointed out in Ref.~\cite{setty2023symmetry}, both the degeneracies of the poles of $G^{+}$, and the degeneracies of the zeroes of $G^{+}$ are constrained by crystal symmetries at the high symmetry points; the zero eigenvectors of both $G^{+}$ (zeros) and of $[G^{+}]^{-1}$ (poles) transform in irreducible representations of the space group. 
Here, we also observe the fact that zeroes and poles can be created and eliminated together.
This is an important feature of the discontinuities in the single-particle Green function of the graphene band-HK model (\Cref{fig:GrGF}).

We have plotted the absolute value of the determinant $|\det(G^{+})|$ and the spectral function $\frac{1}{\pi}\Im\Tr(G^{+}(\mathbf{k},\omega))$ in \Cref{fig:GrGF} for our graphene model with the band (top row) and the orbital (bottom row) HK interaction. 
We see that for interaction $U_1$ equal to half the single particle bandwidth $W$, the band HK interaction does not open a charge gap in the ground state, as evidenced by poles (bright regions) in both the spectral function and determinant crossing zero frequency. 
On the other hand, the orbital HK model leads to a ground state with a charge gap at every $\mathbf{k}$. 
Additionally both the band and orbital HK models have Green functions with zero eigenvalues, depicted as dark regions in the determinant. 
In the orbital model, there is a band of Green function zeros throughout the entire Brillouin Zone, and the zero surface traverses the charge gap. 
This indicates that graphene with the orbital HK interaction is in a Mott insulating state with a Luttinger surface, and so cannot be adiabatically connected to a non-interacting band insulator~\cite{dave2013absence}. 
We thus see that graphene with the orbital HK interaction is a Mott insulator with a fourfold degenerate ground state. 
For the band HK model, the Green function only contains zeros at $\mathbf{k}$ points for which the eigenvalues of the non-interacting Hamiltonian are less than half the interaction energy $\xi(\mathbf{k})<U_1/2$. 
In this regime, the retarded, real time Green function matrix is diagonal with elements given by:
\begin{align}
    G^{+}_{\pm,\sigma;\pm,\sigma}(\omega,\mathbf{k})&=\frac{1}{2}\bigg[\frac{1}{\omega-(\pm\xi(\mathbf{k})+U_1/2)+i\eta}\\ \nonumber+&\frac{1}{\omega-(\pm\xi(\mathbf{k})-U_1/2)+i\eta}\bigg],
\end{align}
Where here we have used $\pm$ as the band index $m$, and $\xi(\mathbf{k})$ is defined as the positive non-interacting Hamiltonian eigenvalue. Zeros occur when the two contributions to $G_{\pm,\sigma;\pm,\sigma}(\omega,\mathbf{k})$ cancel, which occurs when $\omega=\pm\xi(\mathbf{k})$.

Once the single particle energy $|\xi(\mathbf{k})|$ exceeds the interaction energy $|U_{1}|$, the ground state of the band HK model changes discontinuously as a function of $\mathbf{k}$ from having one electron in the lower and upper single particle bands, to having both electrons in the lower single particle band. In this regime, the diagonal elements of the Green function matrix become:
\begin{align}
    G^{+}_{\pm,\sigma;\pm,\sigma}=\frac{1}{\omega -(\pm\xi(\mathbf{k})-U_1/2)+i\eta}
\end{align}
This leads to the discontinuous disappearance of the upper and lower poles, along with the zeros in the Green function at $\xi(\mathbf{k})=U_1/2$. 
Notice that the zeros and poles are created and annihilated in pairs at the discontinuity. 

We mention one more important general property of the ground state for \textit{any} HK model which preserves crystal and time reversal symmetries. On general grounds, the ground state of such a system, if it is non-degenerate, has no magnetic order. This follows from the fact that the ground state preserves the time-reversal symmetry of the Hamiltonian. 
Although in principle it is possible that there is spontaneous symmetry breaking, this only happens in the thermodynamic limit. 
Since the HK model is diagonal in momentum space we can consider the Hamiltonian formed from the $\mathbf{k}$ and $-\mathbf{k}$ blocks:
\begin{equation}
    H_{\Tilde{\mathbf{k}}}=H_{\mathbf{k}}\oplus H_{-\mathbf{k}}.
\end{equation}
This Hamiltonian is invariant under time-reversal, since time-reversal merely permutes the $\mathbf{k}$ and $-\mathbf{k}$ blocks. 
Since this is a finite-rank Hamiltonian, it cannot spontaneously break a symmetry and hence it's ground state manifold must also be time-reversal symmetric. 
The overall Hamiltonian for a HK model can be written as a direct sum of the $H_{\Tilde{\mathbf{k}}}$ Hamiltonians:
\begin{equation}
H=\bigoplus_{\mathbf{k}}H_{\mathbf{k}}=\bigoplus_{\mathbf{\tilde{k}\geq 0}}H_{\mathbf{\Tilde{k}}},
\end{equation}
and so the ground state is the tensor product of the necessarily time-reversal symmetric ground states of $H_{\Tilde{\mathbf{k}}}$, and so is itself time-reversal symmetric. This forbids magnetic ordering.  Hence, typical non-thermodynamically degenerate ground states of orbital-HK models have a single-particle charge gap with no magnetic order; this is what we will define as a \textit{spin liquid} ground state. The only way to have a magnetically ordered ground state in an HK model is then to have a thermodynamically large ground state degeneracy which is unstable to magnetic ordering, as in band-HK models.

\subsubsection{Band and orbital Graphene models in a magnetic field}
One of the consequences of the thermodynamically large degeneracy in the band HK model of graphene \Cref{eq:bandgrapheneH} is a ferromagnetic instability.
This comes from the fact that wherever the interaction energy is greater than the hopping energy, $U_1/2>\xi(\mathbf{k})$, the ground state at $\mathbf{k}$ is an equal weight mixture of all spin combinations. 
The addition of an infinitesimal magnetic field lowers the energy of the spins aligned with the field at every $\mathbf{k}$ point, and so forces the available spins in the ground state of the whole $N$ particle system to align.
Typically, the resulting ferromagnetic state is adiabatically connected to a noninteracting magnetic band insulator~\cite{zhao2023failure,yang2021exactly}. 
In orbital HK models, however, the tendency towards the spins aligning is stabilized by the non-degeneracy of the ground state. 
In our particular case of graphene, we see from \Cref{eq:graphenekblockham} that the ground state subspace at every $\mathbf{k}$ is supported on states $2$--$5$ from Table~\ref{table:HKKMbasis}. 
The energies of these states are unaffected by a Zeeman field. 

\begin{figure*}[ht]
\FPeval{\overlap}{0.33}
\FPeval{\scalevalue}{round(0.33*(1.05)/\overlap,2)}
\scalebox{\scalevalue}{
\setlength{\tabcolsep}{0pt} 
\def\tabularxcolumn#1{m{#1}}
\hskip-0.75cm\begin{tabularx}{\textwidth}{@{}XXX@{}}
\subfloat[]{\includegraphics[width=\overlap\textwidth]{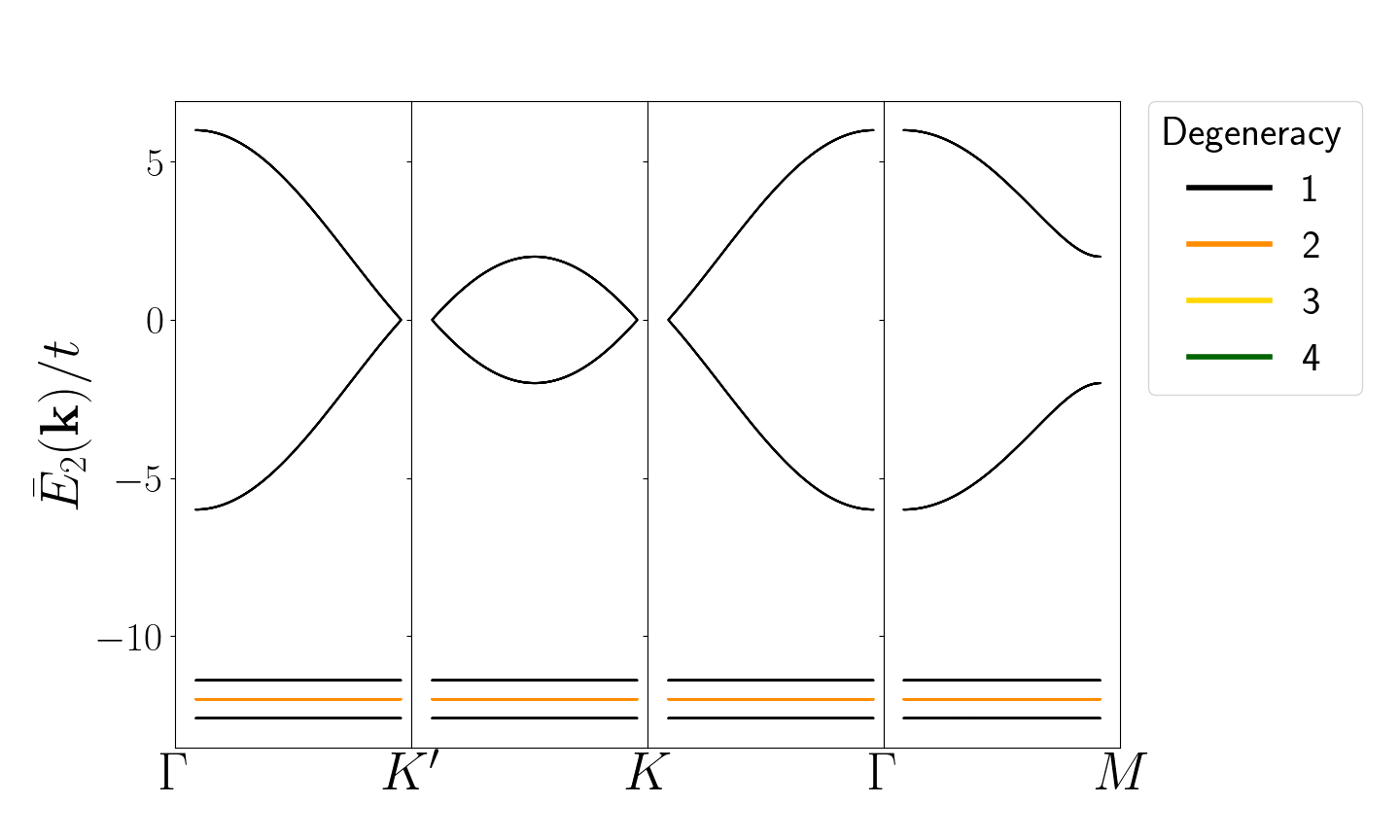}} &
\subfloat[]{\includegraphics[width=\overlap\textwidth]{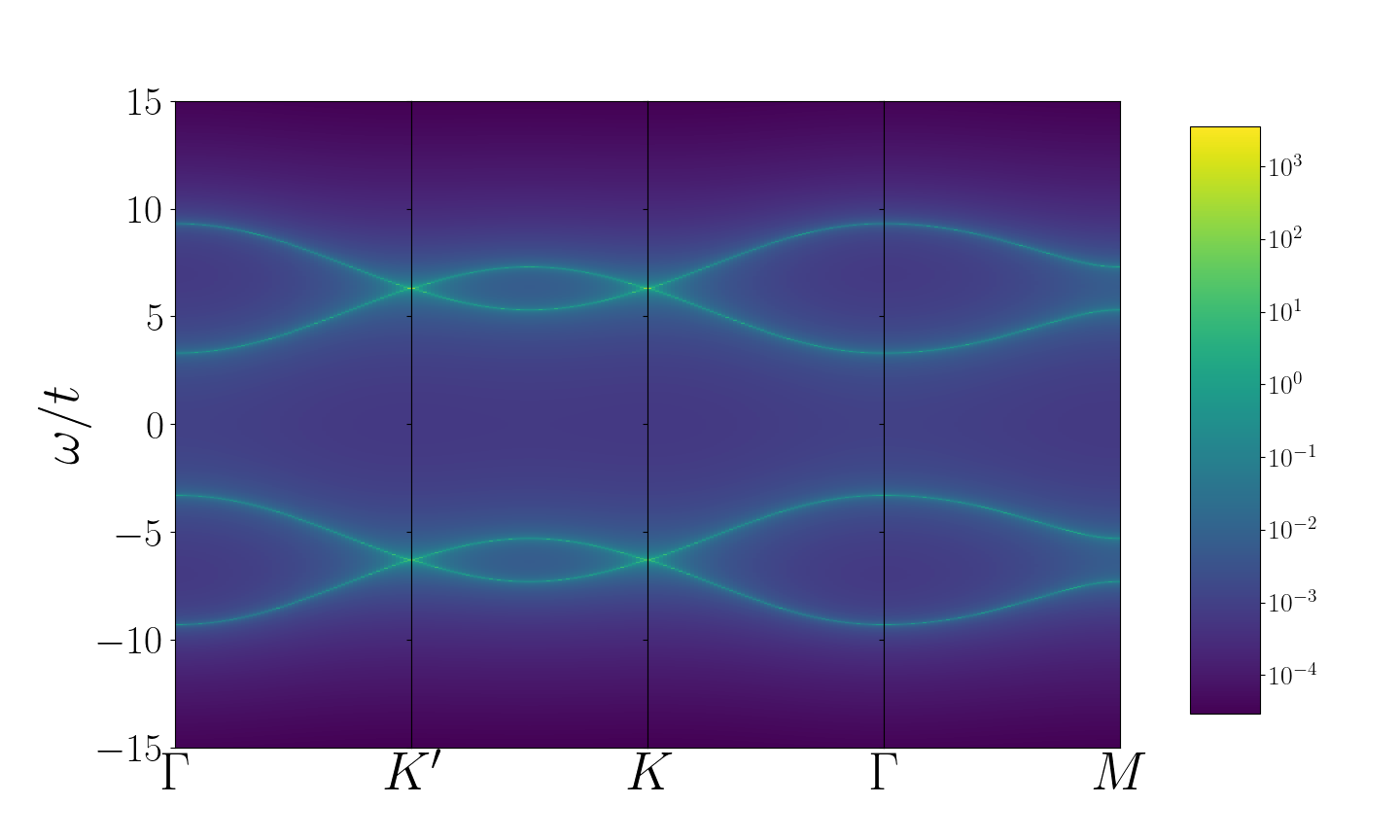}} &
\subfloat[]{\includegraphics[width=\overlap\textwidth]{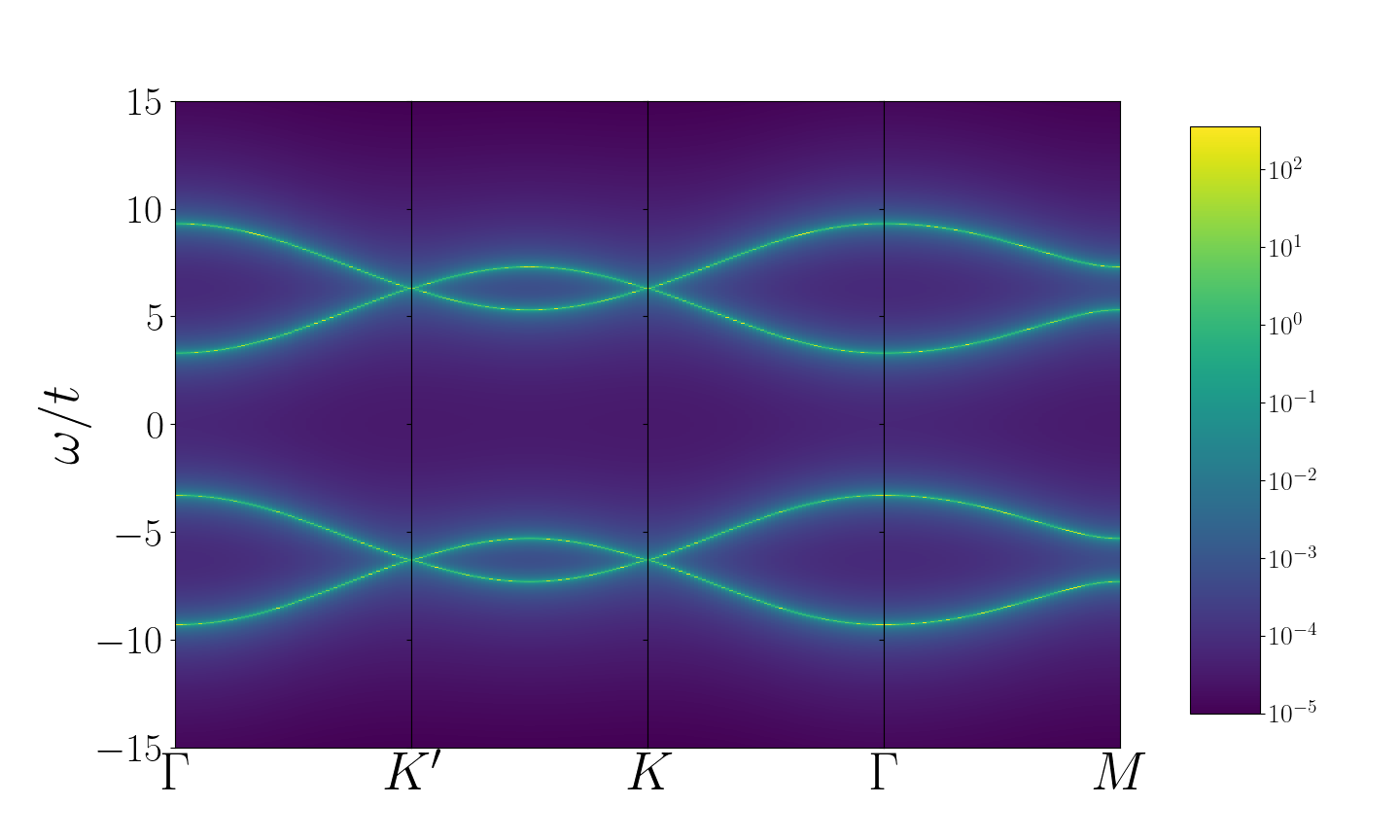}} \\[-3ex]
\subfloat[]{\includegraphics[width=\overlap\textwidth]{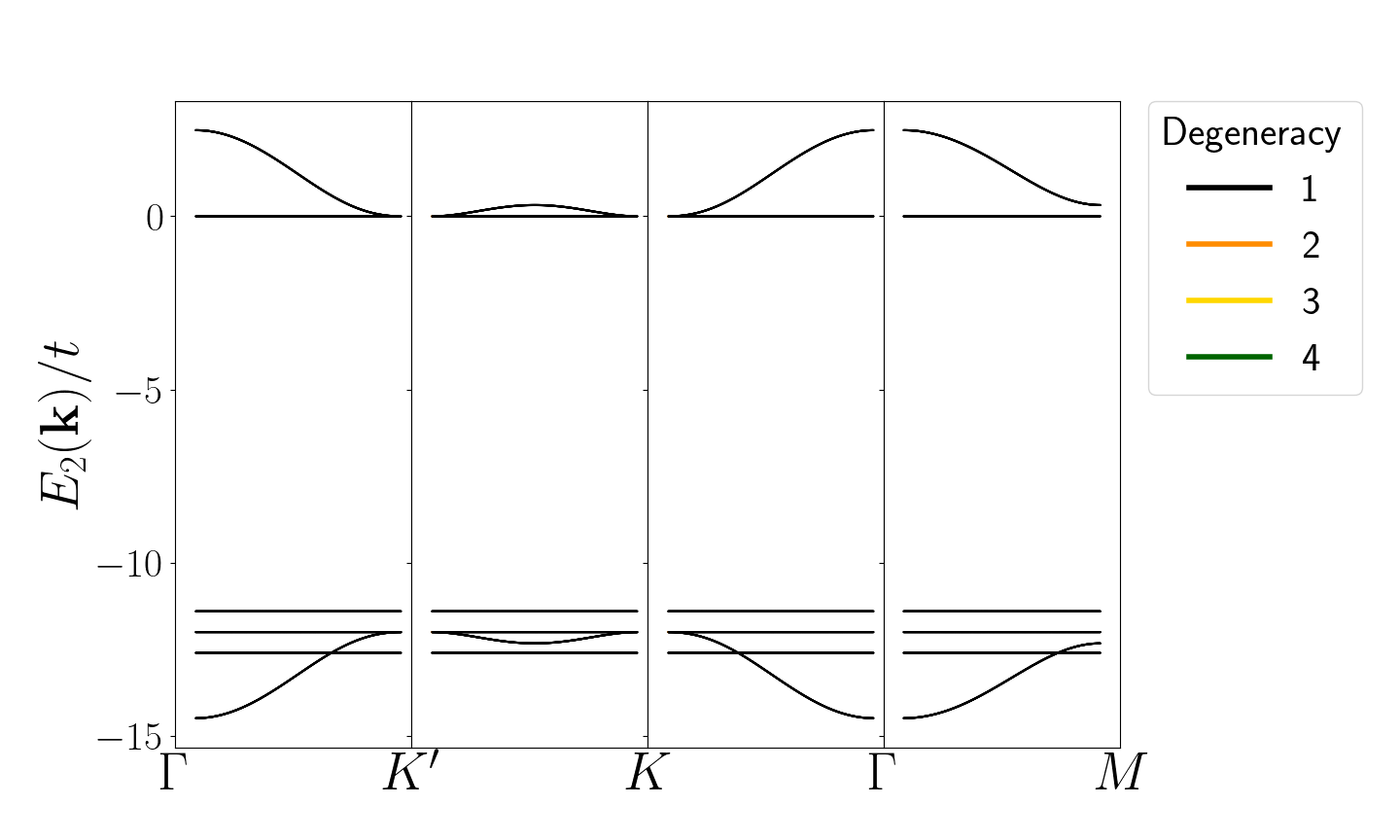}} &
\subfloat[]{\includegraphics[width=\overlap\textwidth]{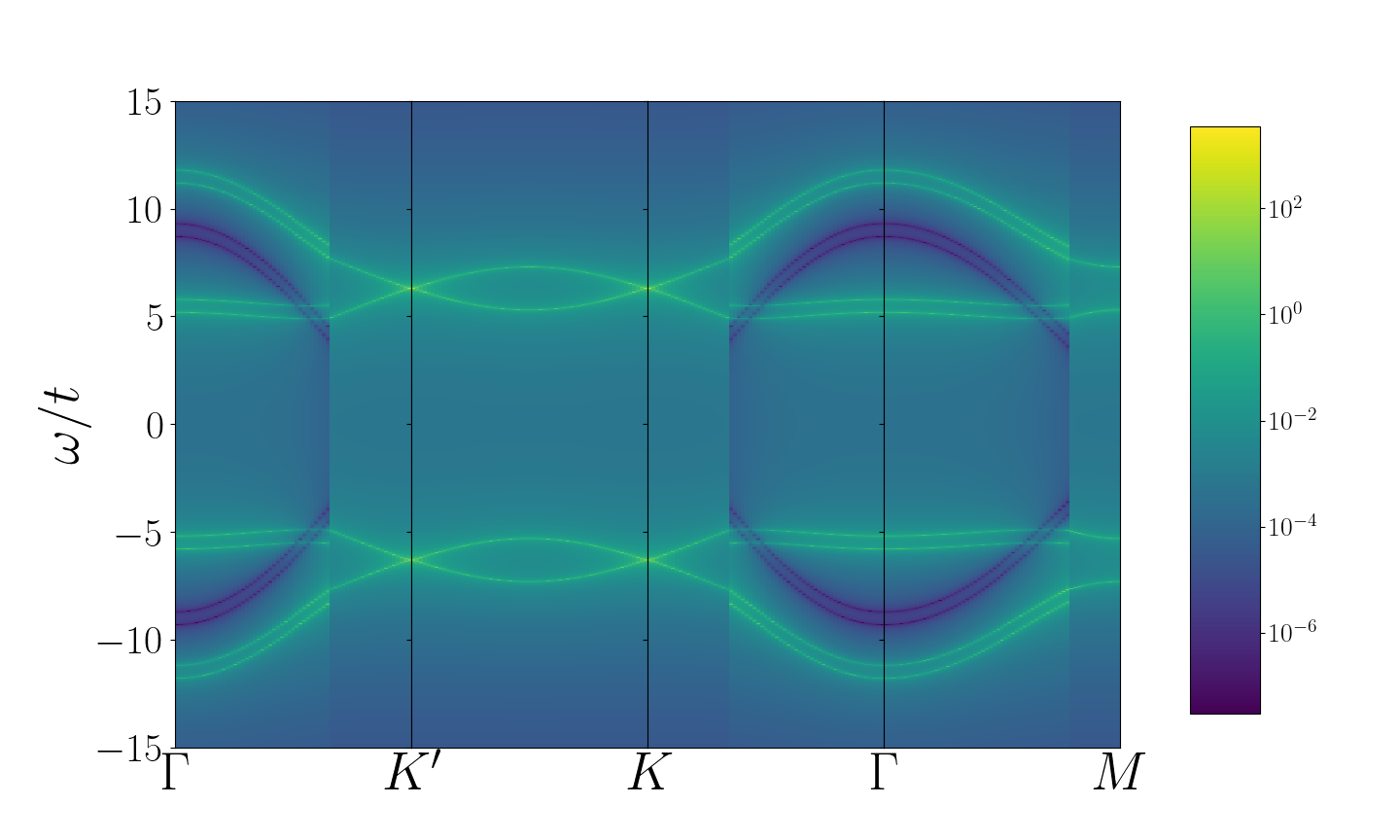}} &
\subfloat[]{\includegraphics[width=\overlap\textwidth]{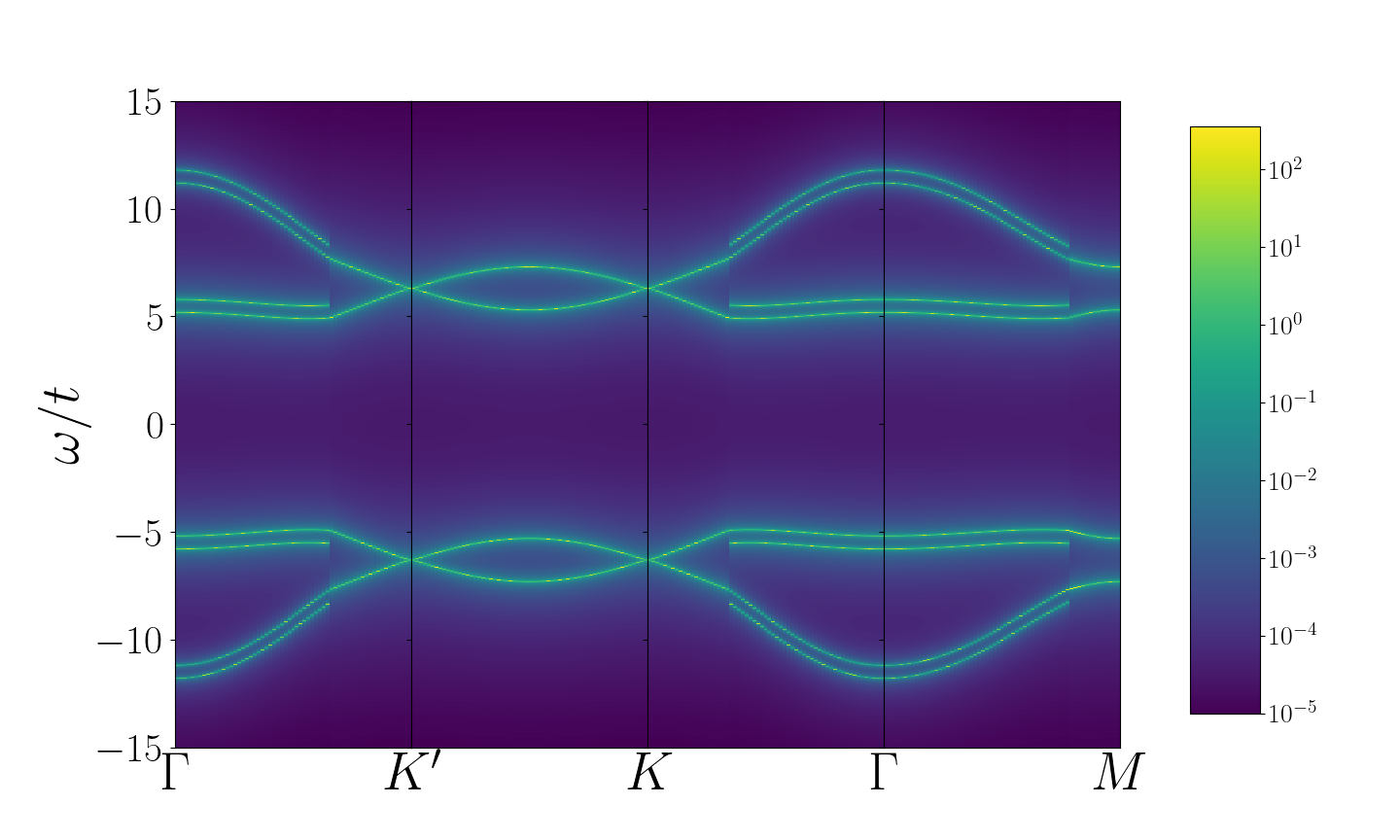}}
\end{tabularx}}
\caption{Energies and Green functions for band (top row)  and orbital (bottom row) graphene HK models with Zeeman interactions $\Bar{H}'_{g}$ and $H'_{g}$ in the strongly interacting limit $U_{1}=2W$ with a small Zeeman field $B=W/20$.
(a) shows the energy spectrum in the two particle sector for the band model $\Bar{H}'_{g}$. 
(b) shows $|\det(G^{+})|$ for the band model. 
There are no Green function zeros, indicating that the ferromagnetic state is adiabatically connected to a trivial insulator.
(c) shows the spectral function for the band model. 
(d) shows the energy spectrum in the two particle sector for the orbital model $H'_{g}$ with the same parameters as the band model.
(e) shows $|\det(G^{+})|$ for the orbital model. 
The presence of Green function zeros indicates that the orbital model remains a non-Fermi liquid.
(f) shows the spectral function of the orbital model.
Both models have parameter values in the strongly interacting regime: $t=1$, $U_{1}=12$, $B=0.3$, $\mu_{0}=6$.
We see that the band model becomes a trivial ferromagnetic insulator, whereas the orbital model retains some of the Green function zeros and hence remains a Mott insulator.}
\label{fig:HKZeemanGraphene}
\end{figure*}

To understand the response of the system to a Zeeman field, we calculate the real time retarded Green function matrix elements $G^{+}_{i,j}$ for the band and orbital HK models with a Zeeman field. 
For the band model, we consider the Hamiltonian:
\begin{align}
\label{eq:bandgraphenezeemanH}
\Bar{H}'_{g}&=\sum_{\mathbf{k}m\sigma}\xi_m(\mathbf{k})\Bar{c}^\dagger_{\mathbf{k}m\sigma}\Bar{c}_{\mathbf{k}m\sigma}+U_{1}\sum_{\mathbf{k}}\Bar{n}_{\mathbf{k}m\uparrow}\Bar{n}_{\mathbf{k}m\downarrow}+B\sigma c^\dagger_{\mathbf{k}\mu\sigma}c_{\mathbf{k}\mu\sigma} \nonumber \\
&-\mu_{0}\sum_{\mathbf{k}\mu\sigma}n_{\mathbf{k}\mu\sigma}\text{ }.
\end{align}
This band HK model with a Zeeman term has already been analyzed in Ref.~\cite{2022PhilipSpinHallHK} and we provide the spectrum in the two particle sector and the Green function in \Cref{fig:HKZeemanGraphene} for ease of comparison with the orbital model.

The Hamiltonian for the orbital HK model of graphene with a Zeeman field is given by
\begin{align}
\label{eq:grapheneorbzeemanH}
H'_{g}&=t\sum_{\mathbf{k}\sigma}g(\mathbf{k})c^\dagger_{\mathbf{k}B\sigma}c_{\mathbf{k}A\sigma}+h.c.+U_{1}\sum_{\mathbf{k}\mu}n_{\mathbf{k}\mu\uparrow}n_{\mathbf{k}\mu\downarrow} \nonumber \\
&+\sum_{\mathbf{k}\mu\sigma}B\sigma c^\dagger_{\mathbf{k}\mu\sigma}c_{\mathbf{k}\mu\sigma}-\mu_0\sum_{\mathbf{k}\mu\sigma}n_{\mathbf{k}\mu\sigma}.
\end{align}
Note that the Zeeman energy is invariant under the particle-hole symmetry operation $P$ given in \Cref{eq:grapheneph}, and hence the ground state at half filling still has two particles at every $\mathbf{k}$. 
In \Cref{fig:HKZeemanGraphene}(d) we show the spectrum of the Hamiltonian in the two particle sector. 
We see that the primary effect of the Zeeman field is to spin-split the threefold degenerate $E^0(\mathbf{k},\mu_0)$ band. 
The exact two-particle energies as a function of $B$ are given by
\begin{align}
\label{eq:twositeZeemanenergies}
    E^\pm_{2}(\mathbf{k},\mu_0,B)&=\frac{1}{2}\left(U_{1}\pm\sqrt{U^2_{1}+16|tg(\mathbf{k})|^2}\right)-2\mu_{0},\nonumber \\
    E^{0\uparrow}_{2}(\mathbf{k},\mu_0,B)&=-2B-2\mu_{0}, \nonumber \\
    E^{00}_{2}(\mathbf{k},\mu_0,B)&=-2\mu_{0}, \\
    E^{0\downarrow}_{2}(\mathbf{k},\mu_0,B)&=2B-2\mu_{0}, \\
    E^U_{2}(\mathbf{k},\mu_0,B)&=U_{1}-2\mu_{0}, \nonumber
\end{align}
The eigenstates with energies $E^{\pm}_{2}, E^{00}_2$, and $E^U_{2}$ are supported on states $2$--$5$ in Table~\ref{table:HKKMbasis}. 
The ferromagnetic state $\ket{A\uparrow;B\uparrow}$ aligned with the field has energy $E^{0\uparrow}_{2}(\mathbf{k},\mu_0,B)$, while the anti-aligned state $\ket{A\downarrow;B\downarrow}$ has energy $E^{0\downarrow}_{2}(\mathbf{k},\mu_0,B)$. 
The ground state at every $\mathbf{k}$ will either have energy $E^-_{2}(\mathbf{k},\mu_0,B)$ or $E^{0\uparrow}_{2}(\mathbf{k},\mu_0,B)$.
The absolute value of the resulting determinant and the spectral function are plotted in the large $U_1$ ($U_1=2W$) and small $B$ ($B=W/20$) limits in \Cref{fig:HKZeemanGraphene}. 
In the band model with Zeeman interactions, shown in \Cref{fig:HKZeemanGraphene}(b), the Green function zeros disappear. 
This means that the ferromagnetic state is adiabatically connected to a trivial band insulator, obtained by filling the spin up states in each single particle band. 
In the orbital model, however, the Green function zeros remain wherever the ground state has anti-aligned spins, i.e. for those $\mathbf{k}$ at which $E^\pm_{2}(\mathbf{k},\mu_0,B)<E^{0\uparrow}_{2}(\mathbf{k},\mu_0,B)$. 
This shows that the ground state of the orbital model in the large $U_1$ limit is not adiabatically connected to a trivial insulator when an infinitesimal magnetic field is added, and retains it's non-Fermi liquid behavior. 
Ultimately, this is a consequence of the difference in ground state degeneracies between these two models.

We can further analyze the response of the orbital-HK model to a Zeeman field by computing the magnetic susceptibility in the ground state.  
For small $B$, the region where $E^{0\uparrow}_{2}(\mathbf{k},\mu_0,B) < E^-_{2}(\mathbf{k},\mu_0,B)$ will be small, centered in two small pockets surrounding the $K$ and $K'$ points where $g(\mathbf{k})$ vanishes. 
The net magnetic moment per unit volume $m(B)$ in the ground state will then be $2\hbar N_{FM}/V$, where $N_{FM}$ is the number of $\mathbf{k}$ states surrounding the $K$ point where $E^{0\uparrow}_{2}(\mathbf{k},\mu_0,B) < E^-_{2}(\mathbf{k},\mu_0,B)$ (so that the ground state is ferromagnetic in this region), $V$ is the volume of the system, and the factor of $2\hbar=4\hbar/2$ comes from the number of valleys ($2$) times the magnetic moment per state ($2\hbar/2$). 
To determine $N_{FM}/V$ for small $B$, we can Taylor expand $E^-_{2}(\mathbf{k},\mu_0,B)$ about the $K$ point and solve for the critical $\mathbf{k_*}$ such that 
\begin{equation}
-B = \frac{1}{2}\left(U_{1}-\sqrt{U^2_{1}+16|tg(\delta\mathbf{k})|^2}\right) \approx -\frac{4v_F^2}{U_1^2}|\mathbf{\delta k}|^2
\end{equation}
where we have introduced the Fermi velocity $v_F = 3at/2$, and the deviation $\delta\mathbf{k} = \mathbf{k_*}-K$. $N_{FM}$ is thus given by the number of states inside a circle of radius $U_1\sqrt{B}/4v_F$. 
Letting $\Omega=3\sqrt{3}/2 a^2$ denote the volume of the unit cell, we find
\begin{align}
N_{FM}/V &= \frac{\Omega}{4\pi^2}\pi |\delta\mathbf{k}|^2 \\
&= \frac{BU^2}{8\sqrt{3}\pi t^2}
\end{align}
yielding a ground state magnetic moment of
\begin{equation}
m(B) = \frac{BU^2\hbar}{4\sqrt{3}\pi t^2}.
\end{equation}
We thus find that the magnetic susceptibility of the ground state is given by
\begin{equation}
\chi = \left.\frac{\partial m}{\partial B}\right|_{B=0} = \frac{\hbar U^2}{4\sqrt{3}\pi t^2},
\end{equation}
which is finite for all values of $U_1$ and $t\neq 0$. 
Other thermodynamic quantities can be calculated in a similar way, as was done for the band-HK model in Ref.~\cite{zhao2021determination} and for the Hall conductivity in Ref.~\cite{zhao2023failure}. A calculation of the magnetic instability from the partition function of a band-HK model was carried out in Ref.~\cite{zhao2023proof}.

\subsection{Extending to spinful models: HK Kane-Mele}
\label{subsec:HKKM}
Having demonstrated that we can lift the thermodynamically large degeneracy in band HK models, we now illustrate the second difference between band and orbital HK models: that orbital models  can be extended to spin-\textit{dependent}, non-interacting Hamiltonians without any assumption of spin conservation. 
The Kane-Mele model has a spin-dependent non-interacting Hamiltonian, and the interacting Kane-Mele-Hubbard model has been extensively studied \cite{2011WenKMH,zheng2011particle,hohenadler2012quantum}.

Previously, \cite{2022PhilipSpinHallHK} considered adding band HK interactions to the Kane-Mele model. 
It is possible to write down a band Kane-Mele-HK model because the spin projection $S_{z}$ commutes with the non-interacting Hamiltonian. 
Here we consider the \textit{orbital} Kane-Mele-HK model. 
In general, the orbital model does not require that the non-interacting Hamiltonian commute with $S_{z}$. 
This allows us to define and solve HK models for arbitrary spin dependent Hamiltonians. 
This will allow us to solve the orbital model for space group $P4/ncc1^\prime$ (\#130) and space group $P4_2/mbc1^\prime$ (\# 135) in \Cref{sec:DDSL}.

 To demonstrate this, we will first obtain the orbital HK Kane-Mele model from our interacting graphene Hamiltonian $H_{g}$. 
We add a spin-dependent hopping so that our total Hamiltonian is now:
\begin{align}
&H_{KM}=t\sum_{\mathbf{k}\sigma}g(\mathbf{k})c^\dagger_{\mathbf{k}B\sigma}c_{\mathbf{k}A\sigma}+i\lambda\sum_{\mathbf{k}\mu\sigma}\sigma\mu g_{1}(\mathbf{k})c^\dagger_{\mathbf{k}\mu\sigma}c_{\mathbf{k}\mu\sigma}+h.c.\nonumber \\
&+U_{1}\sum_{\mathbf{k}\mu}n_{\mathbf{k}\mu\uparrow}n_{\mathbf{k}\mu\downarrow}-\mu_{0}\sum_{\mathbf{k}\mu\sigma}n_{\mathbf{k}\mu\sigma} ,
\end{align}
where the new hopping: 
\begin{align}
g_1(\mathbf{k})&=\sum_{j}e^{i\mathbf{k}\cdot\mathbf{b}_j}
\end{align}
is to next nearest neighbor sites, as shown in \Cref{fig:graphene}. 
The block Hamiltonian at a given $\mathbf{k}$ point, $H_{KM}(\mathbf{k})$ is then given by:
\begin{align}
    &H_{KM}(\mathbf{k})=t\sum_{\sigma}g(\mathbf{k})c^\dagger_{\mathbf{k}B\sigma}c_{\mathbf{k}A\sigma}+h.c.\\
    &+2\lambda\sum_{\mu\sigma}\sigma\mu\tilde{g}_{1} (\mathbf{k})c^\dagger_{\mathbf{k}\mu\sigma}c_{\mathbf{k}\mu\sigma}+U_{1}\sum_{\mu}n_{\mathbf{k}\mu\uparrow}n_{\mathbf{k}\mu\downarrow} ,\nonumber
\end{align}
where, for convenience, we have defined:
\begin{equation}
\tilde{g}_{1}=\sum_{j}\sin(\mathbf{k}\cdot\mathbf{b}_j)\text{ }.
\end{equation}
 
\begin{figure*}
\FPeval{\overlap}{0.35}
\FPeval{\scalevalue}{round(0.33*(1.075)/\overlap,2)}
\scalebox{\scalevalue}{
\setlength{\tabcolsep}{0pt} 
\def\tabularxcolumn#1{m{#1}}
\hskip-0.75cm\begin{tabularx}{\textwidth}{@{}XXX@{}}
\subfloat[]{\includegraphics[width=\overlap\textwidth]{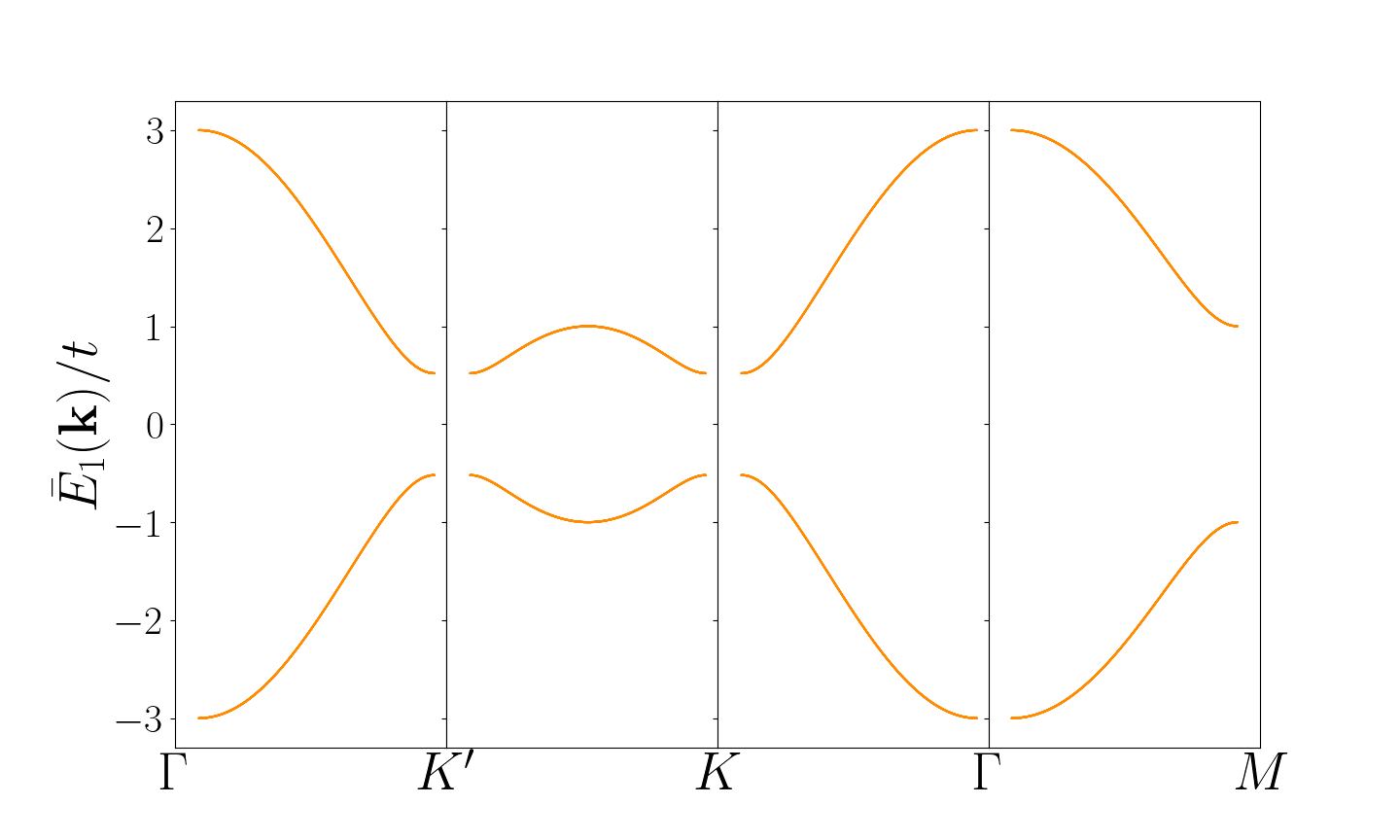}} &
\subfloat[]{\includegraphics[width=\overlap\textwidth]{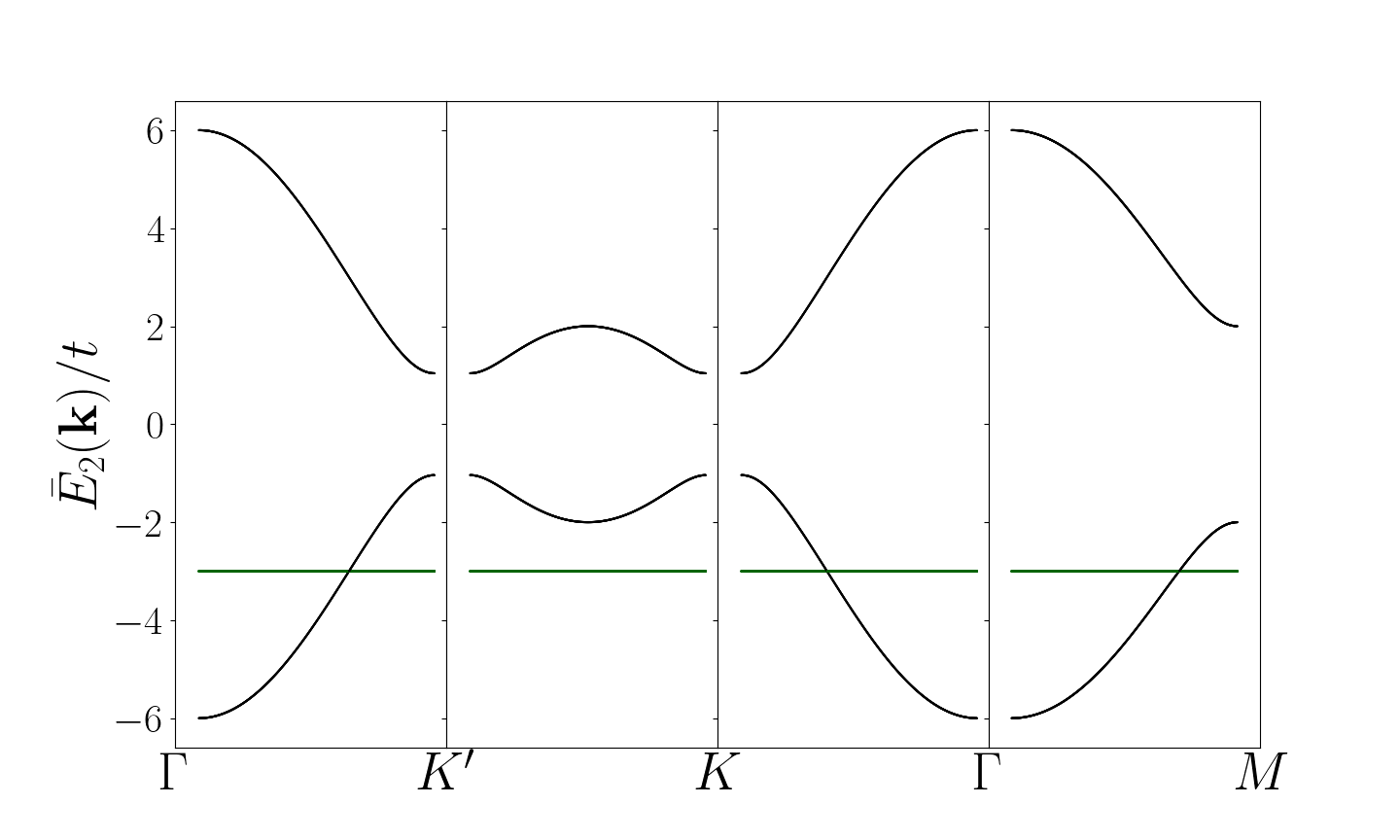}} &
\subfloat[]{\includegraphics[width=\overlap\textwidth]{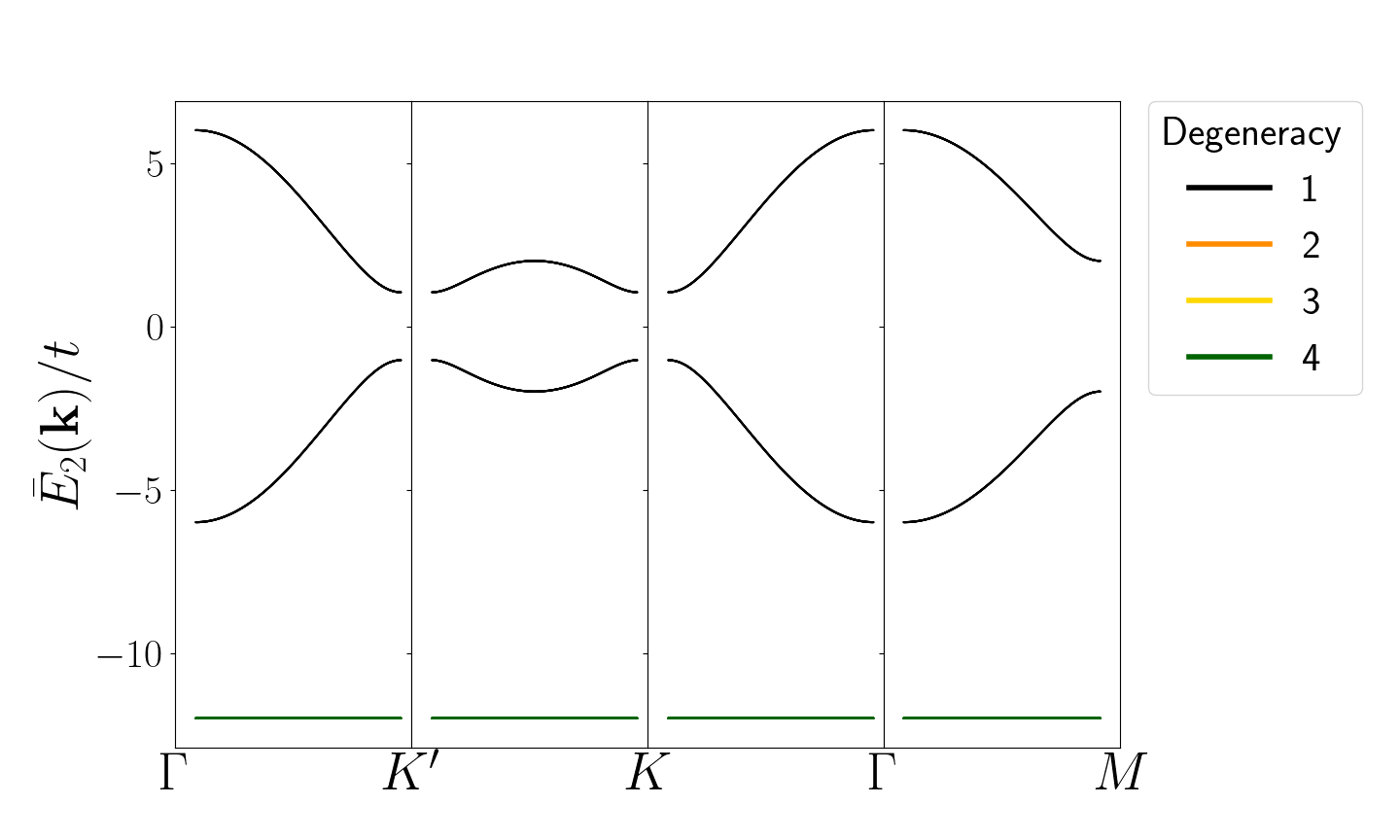}} \\[-3ex]
\subfloat[]{\includegraphics[width=\overlap\textwidth]{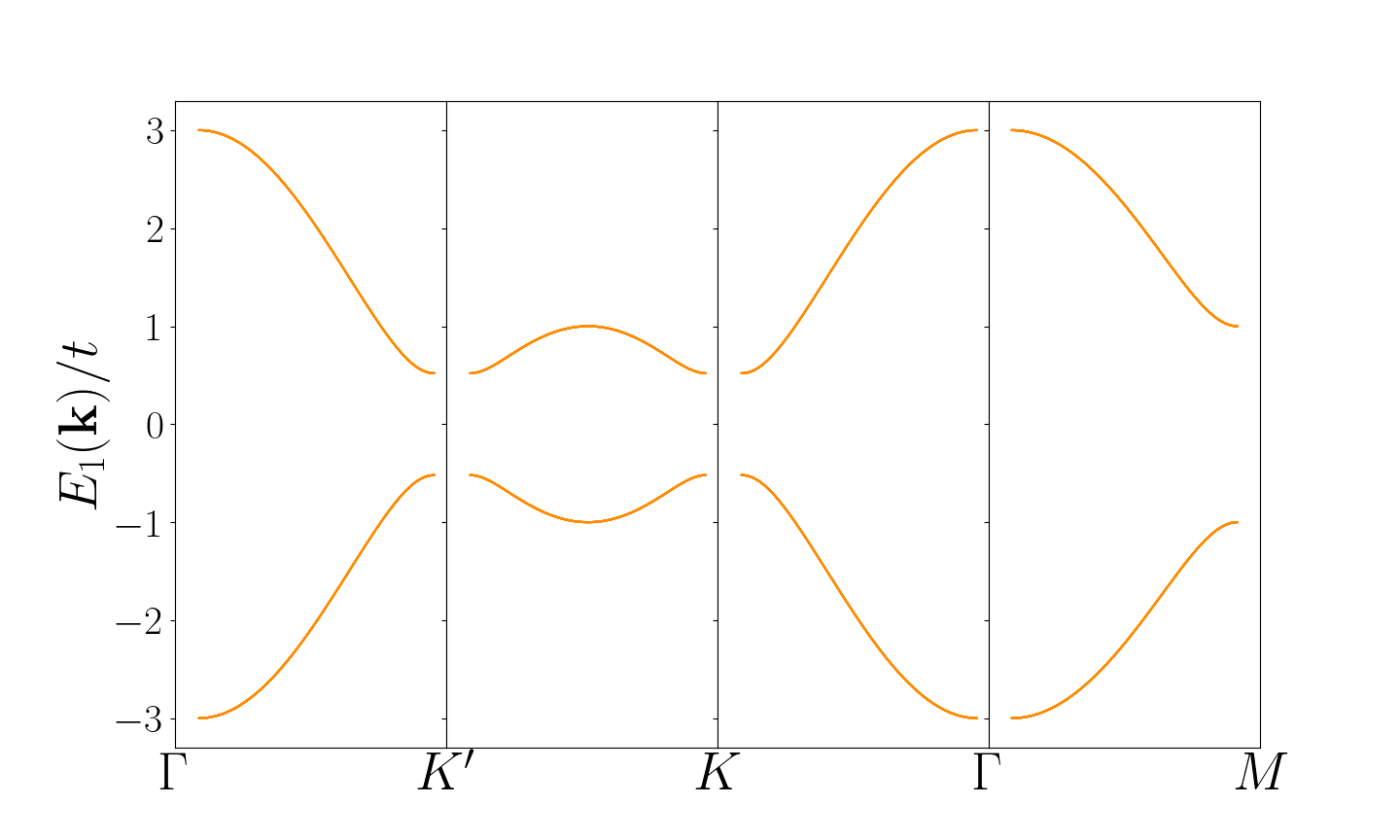}} &
\subfloat[]{\includegraphics[width=\overlap\textwidth]{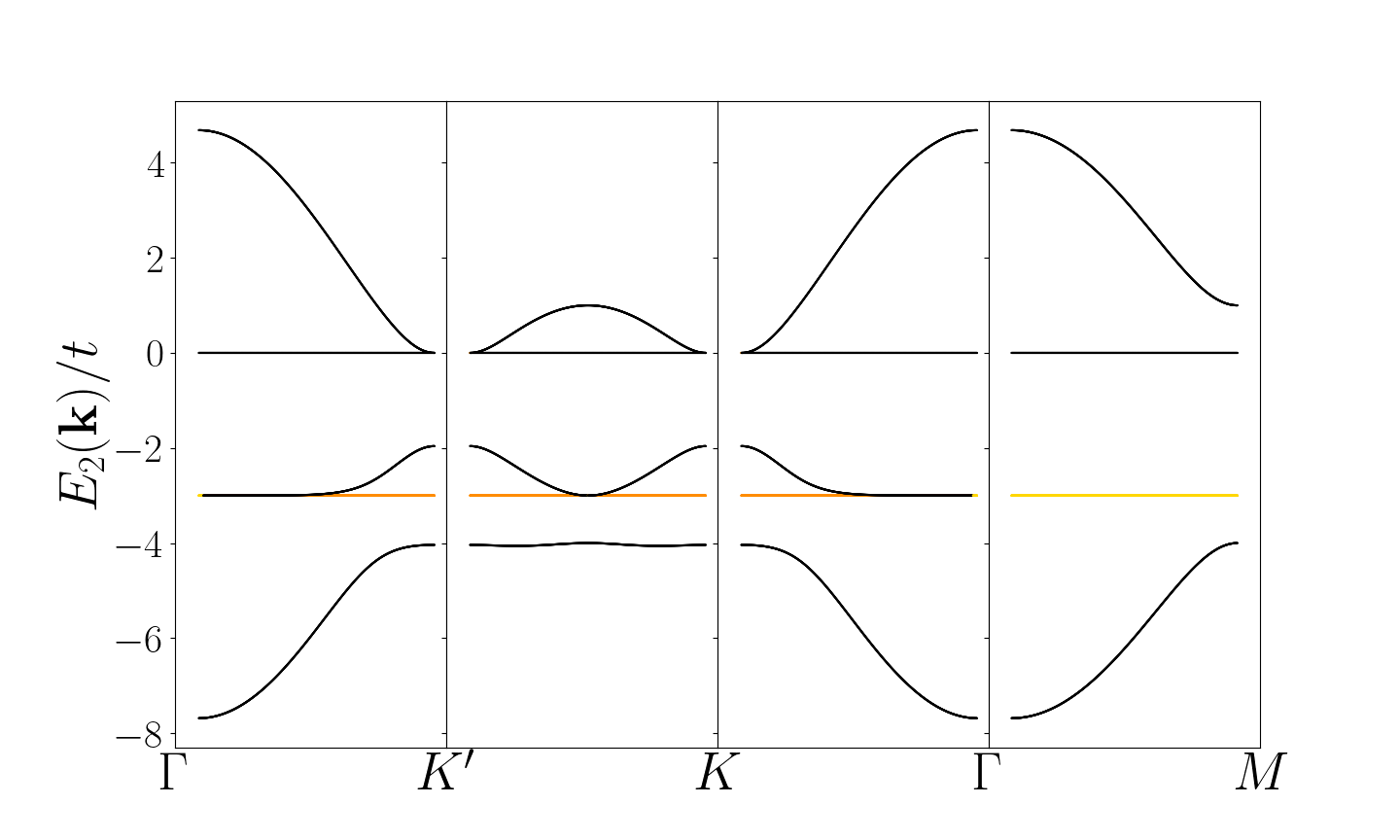}} &
\subfloat[]{\includegraphics[width=\overlap\textwidth]{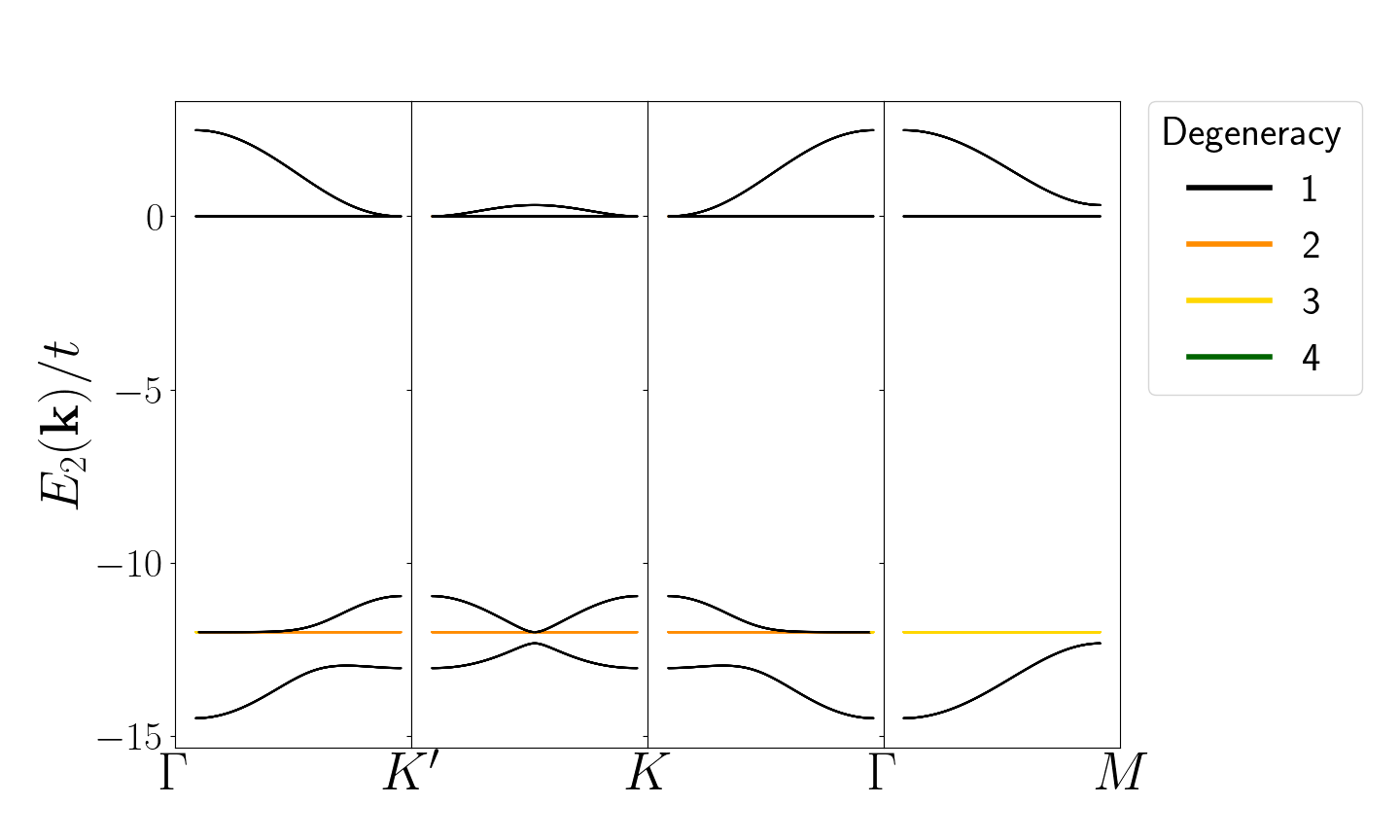}}
\end{tabularx}}
\caption{A comparison of the spectra in the two particle sector for band (top row) and orbital (bottom row) HK Kane-Mele models. 
(a) shows the single particle spectrum for the non-interacting Kane-Mele model.
(b) shows the two particle spectrum $\Bar{E}_{KM,2}(\mathbf{k}$) for the band HK Kane-Mele model with intermediate interaction strength $\Bar{H}_{KM}$ for $U_{1}=W/2$ ($U_{1}=3t$).
(c) shows the two particle spectrum $\Bar{E}_{KM,2}(\mathbf{k})$ for the band HK Kane-Mele model $\Bar{H}_{KM}$ with large interactions $U_{1}=2W$ ($U_{1}=12t$).
(d)  shows the single particle spectrum (repeated) for the non-interacting Kane-Mele model.
(e) shows the two particle spectrum $E_{KM,2}(\mathbf{k})$ for the orbital HK Kane-Mele model $H_{KM}$ with intermediate interaction strength $U_{1}=W/2$ ($U_{1}=3t$).
(f) shows the two particle spectrum $E_{KM,2}(\mathbf{k})$ for the orbital HK Kane-Mele model $H_{KM}$ with large interactions $U_{1}=2W$ ($U_{1}=12t$). 
Note that the ground state remains everywhere non-degenerate in the orbital model. 
The parameters of the non-interacting Hamiltonian are everywhere $t=1, \lambda=0.1$.
The topological gap in the non-interacting Hamiltonian means that the ground state of the orbital model is everywhere non-degenerate, whereas the ground state of the band HK model is only nondegenerate when the interaction energy is less than the topological gap. }
\label{fig:HKKMSpectra}
\end{figure*}

We can  solve the model by solving $H_{KM}(\mathbf{k})$ at every $\mathbf{k}$-point using the same procedure we used for our graphene tight binding model. 
Note that the spin-orbit coupling Hamiltonian is invariant under the particle-hole symmetry operation $P$ given in \Cref{eq:grapheneph}, and hence the ground state at half filling still has two particles at every $\mathbf{k}$.
In the same basis as before (\Cref{table:HKKMbasis}), the matrix elements of the Hamiltonian at half-filling $\mathcal{H}_{KM,2}(\mathbf{k})$ are given by:
\begin{equation}
\mathcal{H}_{KM,2}(\mathbf{k})=
\begin{pmatrix}
0&0&0&0&0&0\\
0&U_{1}&-tg^*&tg^*&0&0\\
0&-tg&-4\lambda \tilde{g}_1&0&-tg^*&0\\
0&tg&0&4\lambda \tilde{g}_1&tg^*&0\\
0&0&-tg&tg&U_1&0\\
0&0&0&0&0&0
\end{pmatrix}\text{ }.
\end{equation}
A comparison of the resulting spectrum in the two-particle sector for the band and orbital HK models is given in \Cref{fig:HKKMSpectra}. 
Since the non-interacting spectrum is gapped at the $K$ and $K'$ points, the resulting ground state is everywhere gapped and non-degenerate for all $U_{1}>0$. 

\begin{figure*}
\FPeval{\overlap}{0.53}
\FPeval{\scalevalue}{round(0.5*(1.12)/\overlap,2)}
\scalebox{\scalevalue}{
\setlength{\tabcolsep}{0pt} 
\def\tabularxcolumn#1{m{#1}}
\hskip-1.0cm\begin{tabularx}{\textwidth}{@{}XXX@{}}
\subfloat[]{\includegraphics[width=\overlap\textwidth]{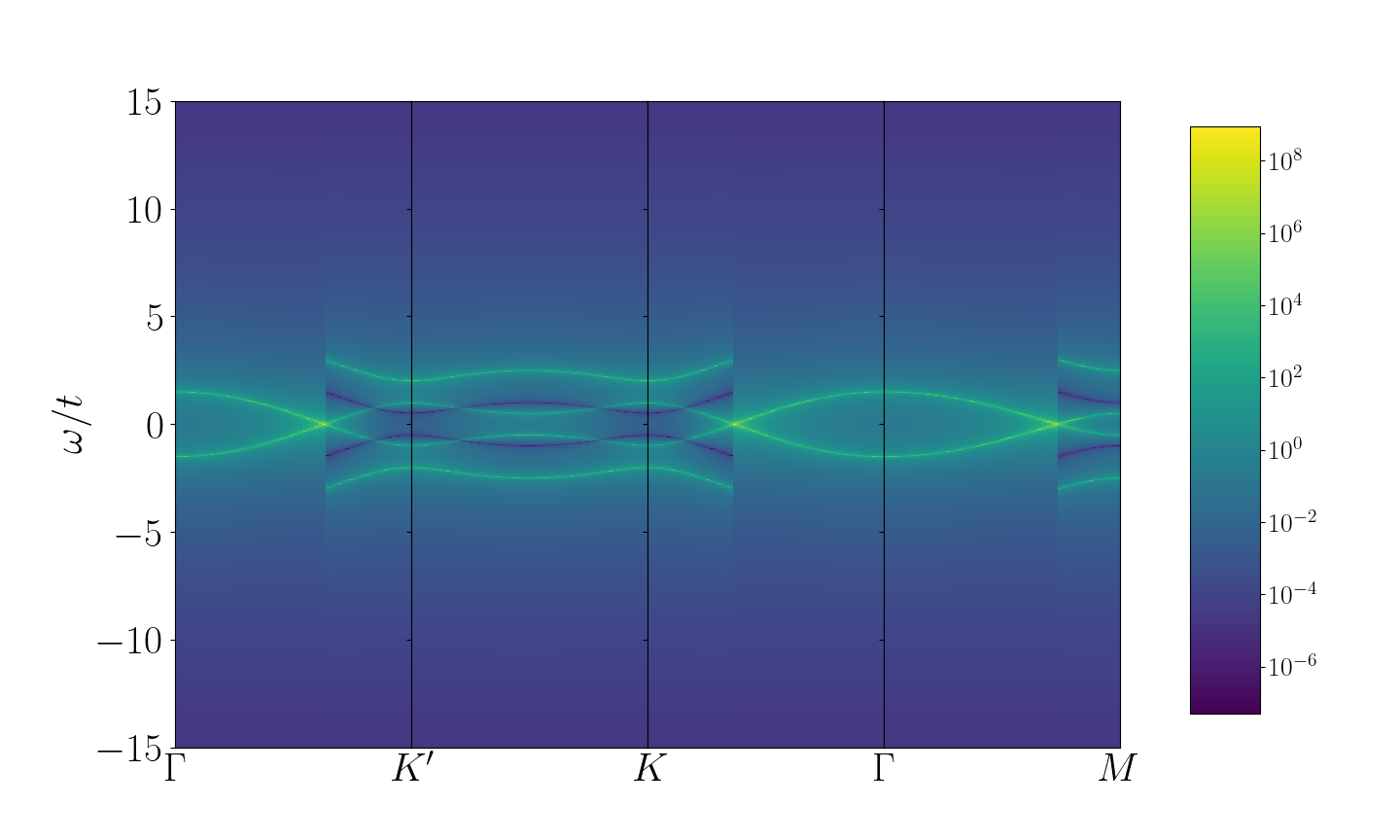}} &
\subfloat[]{\includegraphics[width=\overlap\textwidth]{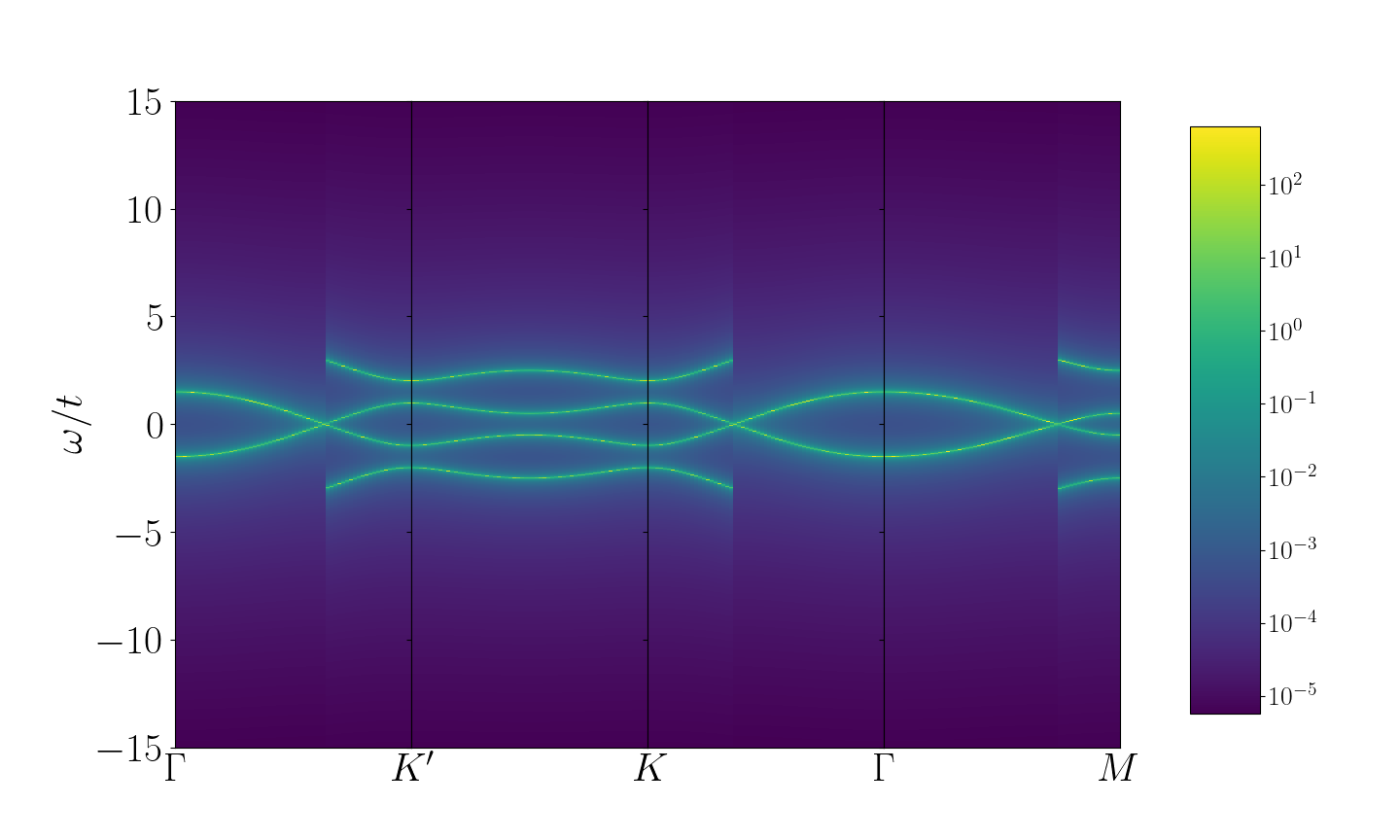}}
 \\[-3ex]
\subfloat[]{\includegraphics[width=\overlap\textwidth]{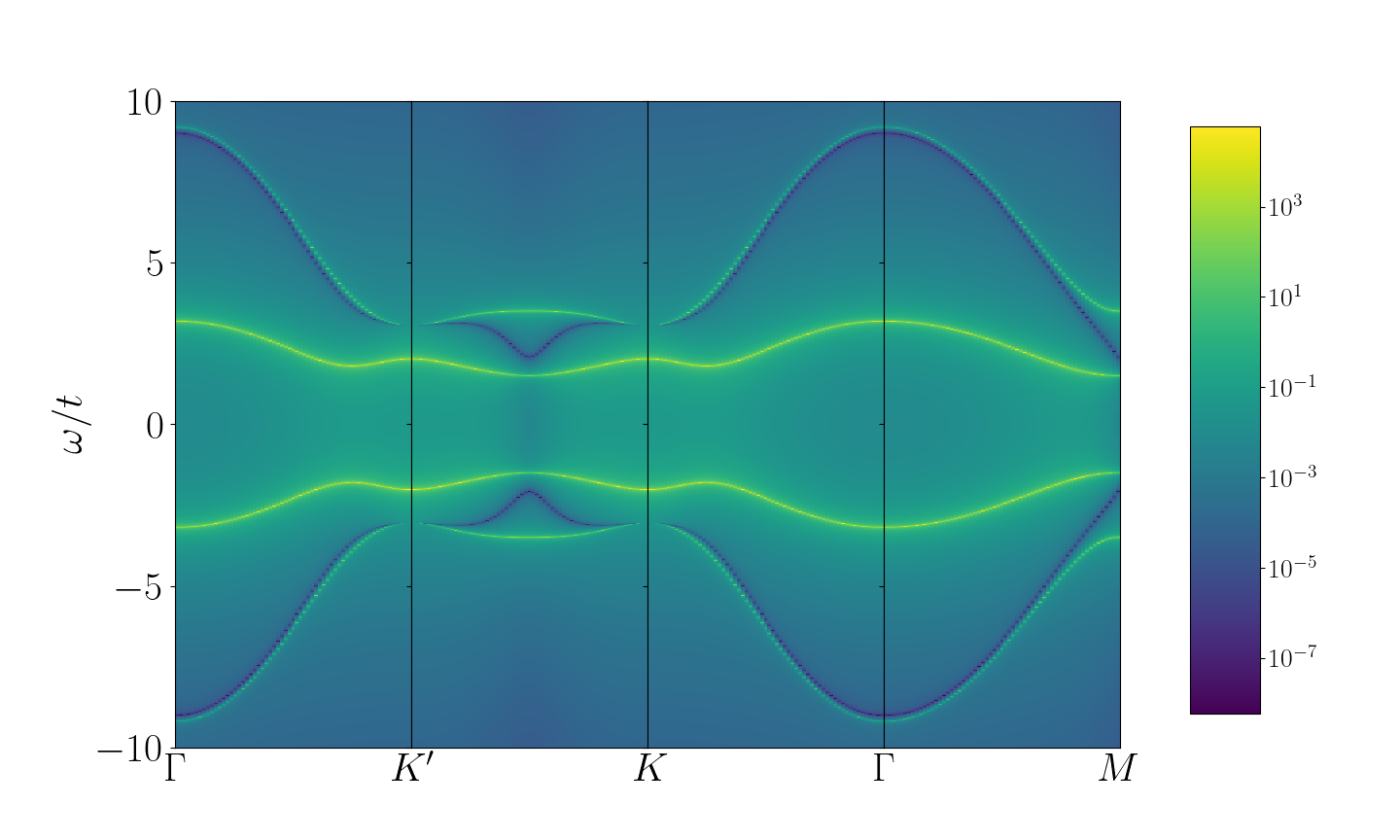}} &
\subfloat[]{\includegraphics[width=\overlap\textwidth]{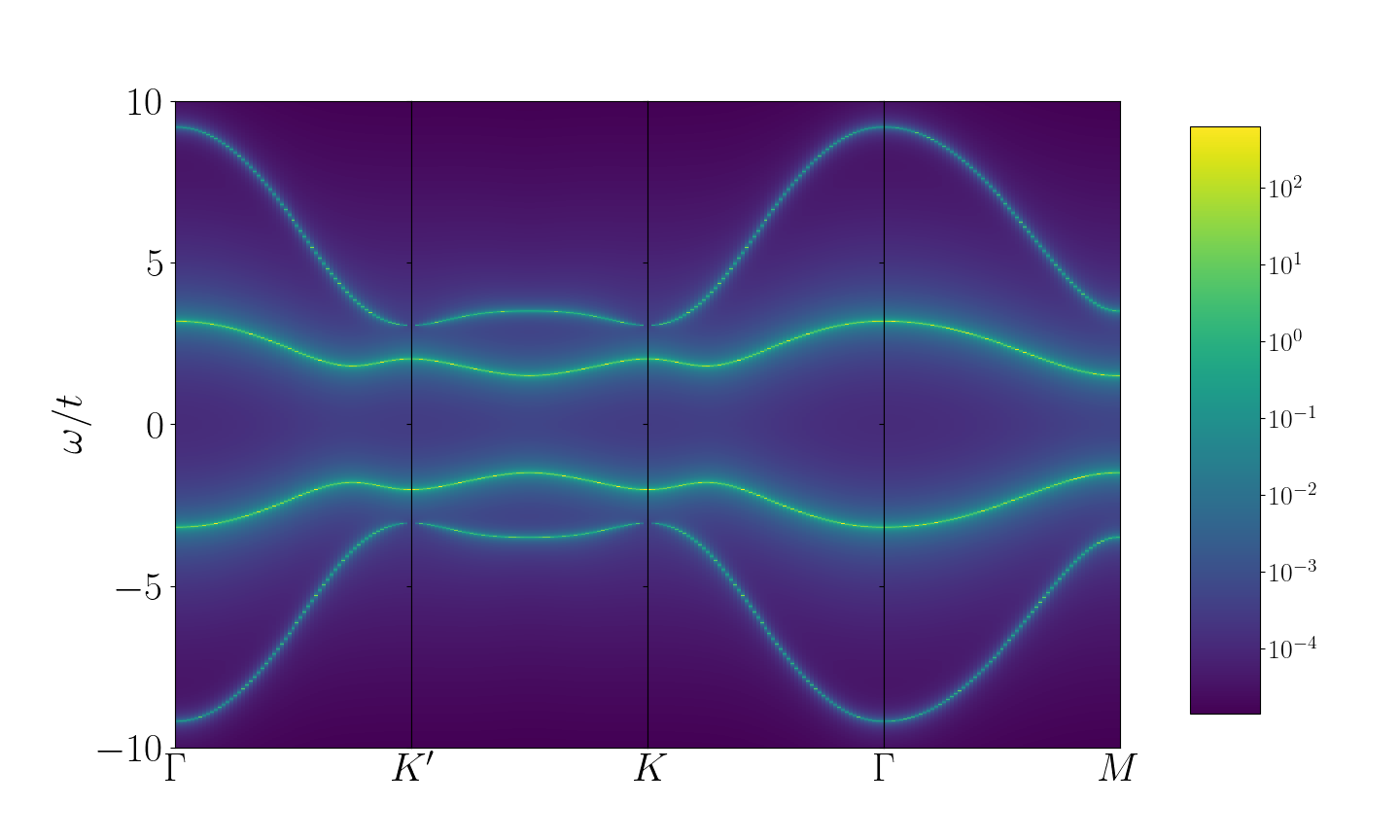}}
\end{tabularx}}
\caption{A comparison of the real time retarded Green function $G^{+}$ for band (top) and orbital (bottom) HK-KM models. 
In the left two panels (a) and (c), we show the absolute value of the determinant $|\det(G^{+})|$, and in the right two panels (b) and (d)  the spectral function $-\frac{1}{\pi}\Im\Tr(G^{+}(\mathbf{k},\omega))$. 
(a) shows $|\det(G^{+})|$ for the intermediate interaction strength regime $U_{1}=W/2$ ($U_{1}=3t$) for the band HK model. 
Since the band of zeros is not present across the whole Brillouin zone, the band model in the intermediate regime is not a Mott insulator, but a general non-Fermi liquid.
(b) shows  the spectral function, $-\frac{1}{\pi}\Im\Tr(G^{+}(\mathbf{k},\omega))$, in the intermediate interaction strength regime for the band HK model. 
(c) shows $|\det (G^{+})|$ for the orbital HK model with $U_{1}=W/2$. 
The band of zeros is present across the whole Brillouin Zone.
(d) shows  the spectral function for the orbital HK model in the intermediate interaction strength regime $U_{1}=W/2$ . 
The parameters of the non-interacting Hamiltonian are everywhere $t=1,\lambda=0.1, \mu_{0}=U_1/2$, and we have used a logarithmic scale for the heat map.
The gap in the spectral function of the orbital model demonstrates that it has an insulating ground state even for interaction strengths lower than the bandwidth $U_{1}<W$.}
\label{fig:KMGF}

\end{figure*}

In the \Cref{fig:KMGF} we provide the determinant $|\det(G^{+}(\mathbf{k},\omega))|$ and spectral function $-\frac{1}{\pi}\Im\Tr(G^{+}(\mathbf{k},\omega))$ for the band HK Kane-Mele [(a) and (b)] and the orbital HK Kane-Mele [(c) and (d)] models. 
We see that even for intermediate interaction strength $U=W/2$, the orbital model has a charge gap in the ground state.

\begin{figure*}
\FPeval{\overlap}{0.33}
\FPeval{\scalevalue}{round(0.33*(1.05)/\overlap,2)}
\scalebox{\scalevalue}{
\setlength{\tabcolsep}{0pt} 
\def\tabularxcolumn#1{m{#1}}
\hskip-0.75cm\begin{tabularx}{\textwidth}{@{}XXX@{}}
\subfloat[]{\includegraphics[width=\overlap\textwidth]{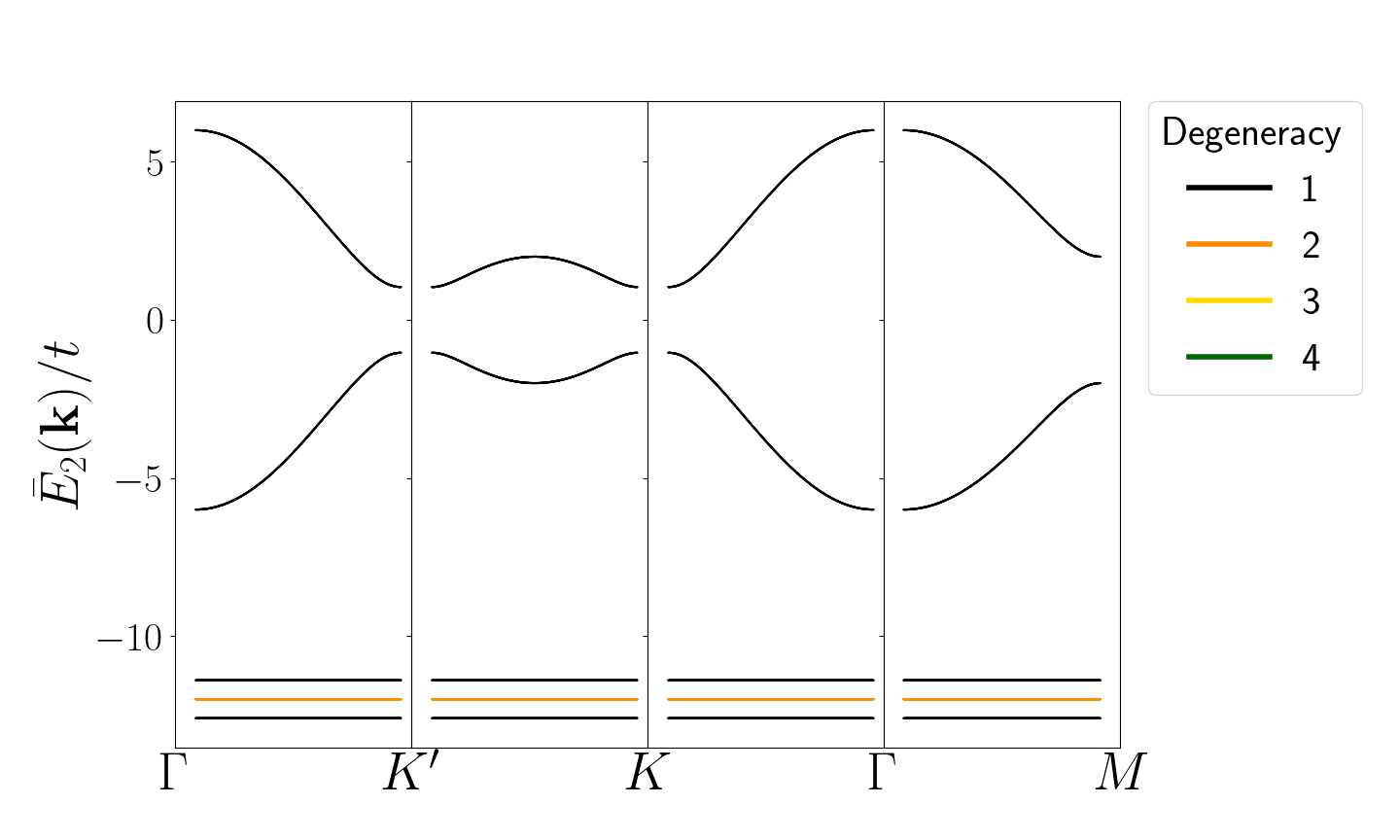}} &
\subfloat[]{\includegraphics[width=\overlap\textwidth]{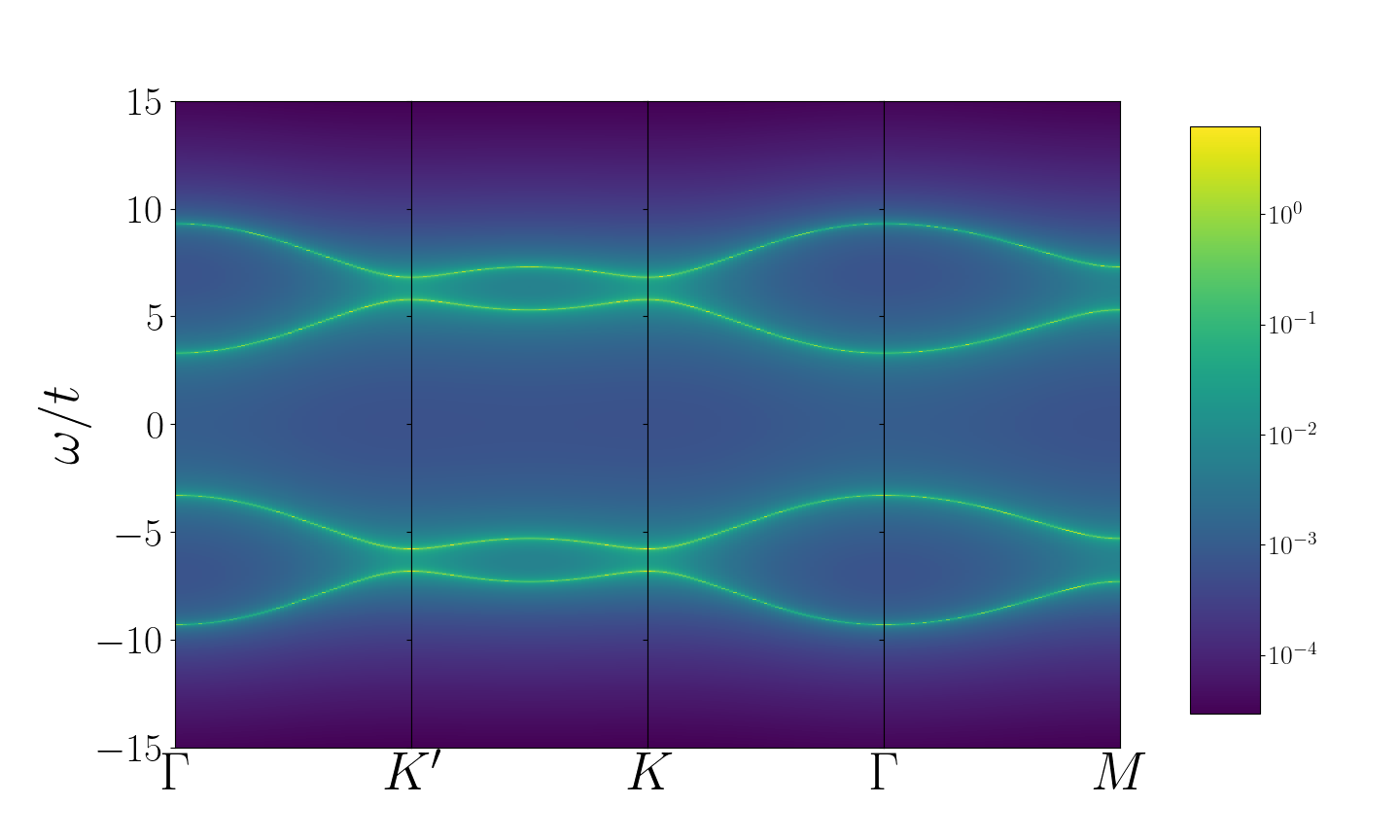}} &
\subfloat[]{\includegraphics[width=\overlap\textwidth]{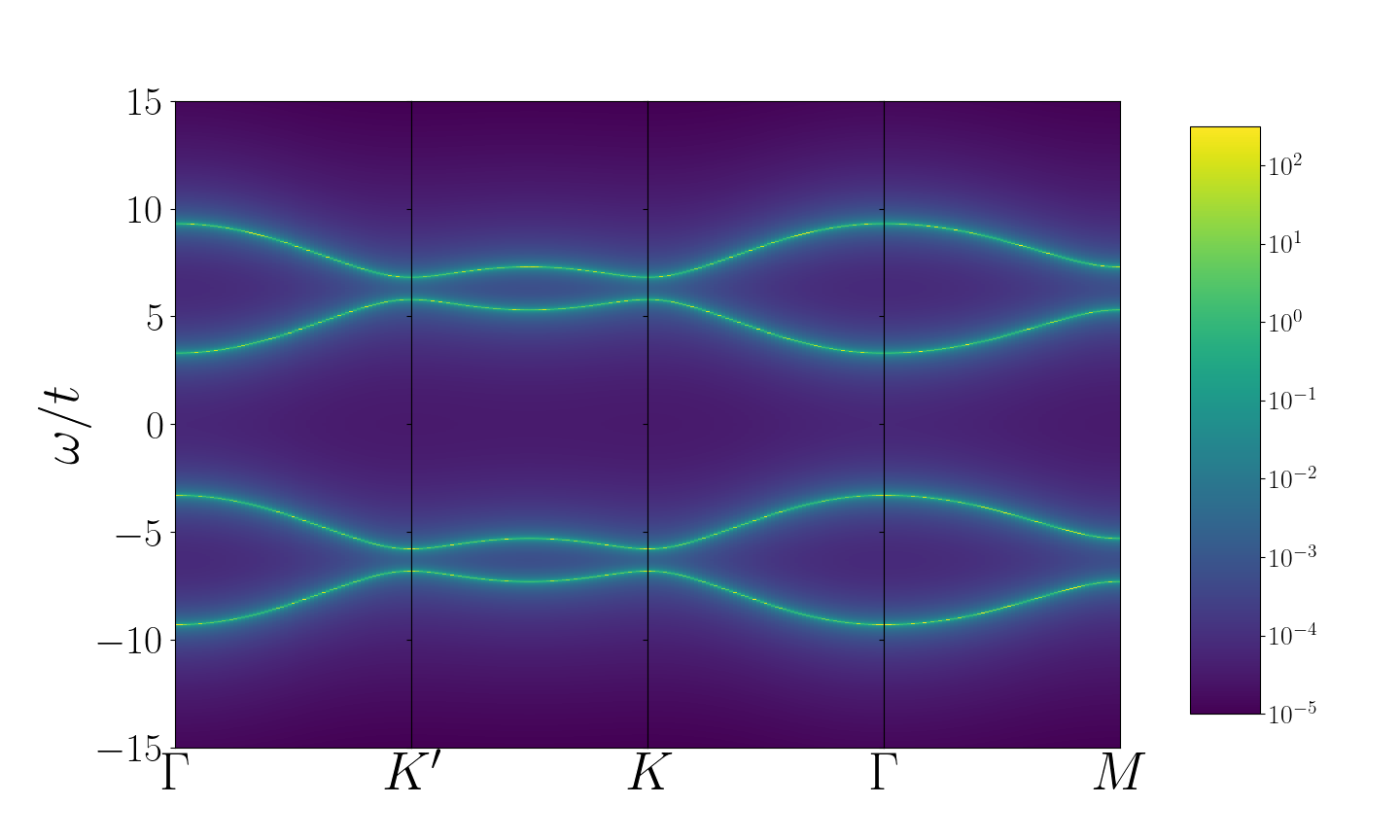}} \\[-3ex]
\subfloat[]{\includegraphics[width=\overlap\textwidth]{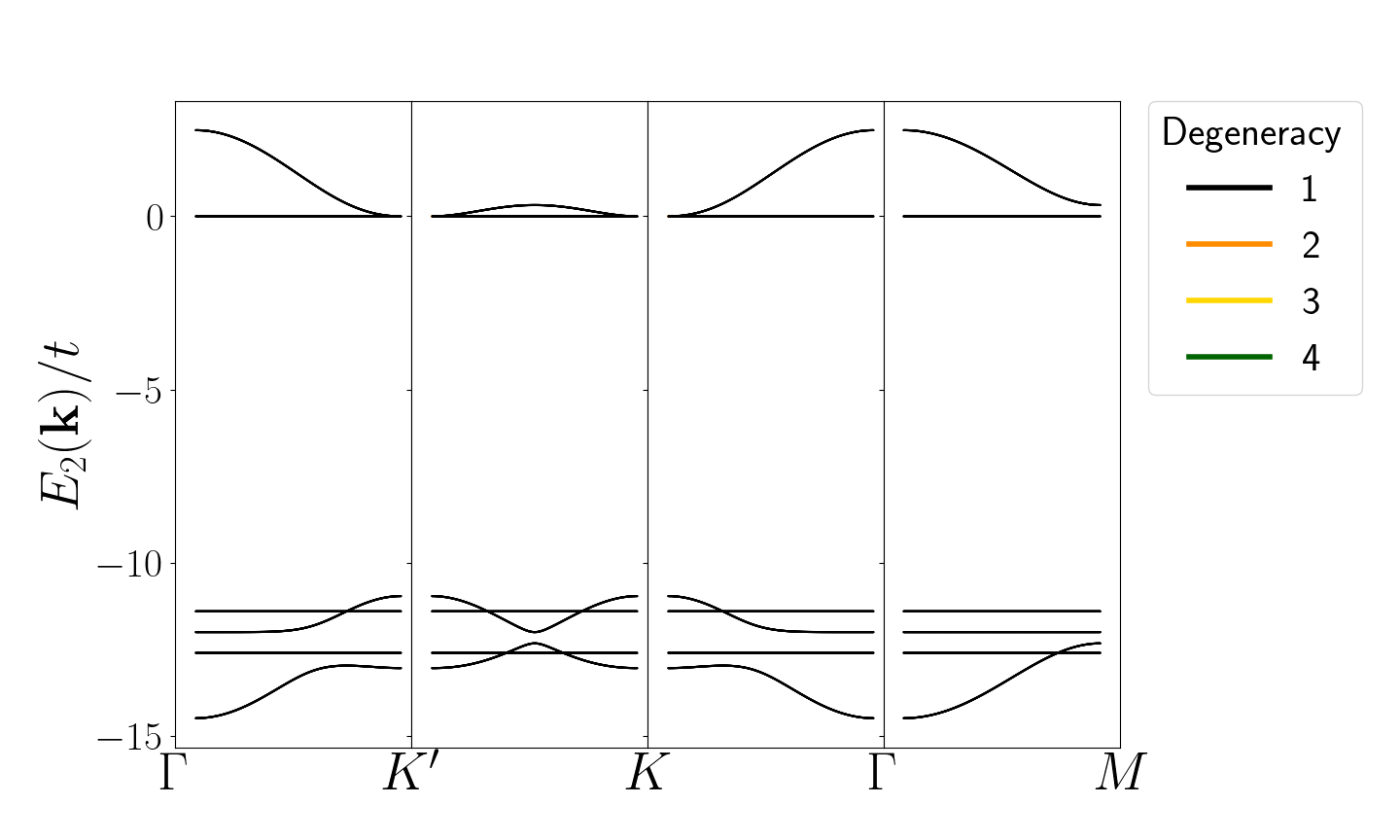}} &
\subfloat[]{\includegraphics[width=\overlap\textwidth]{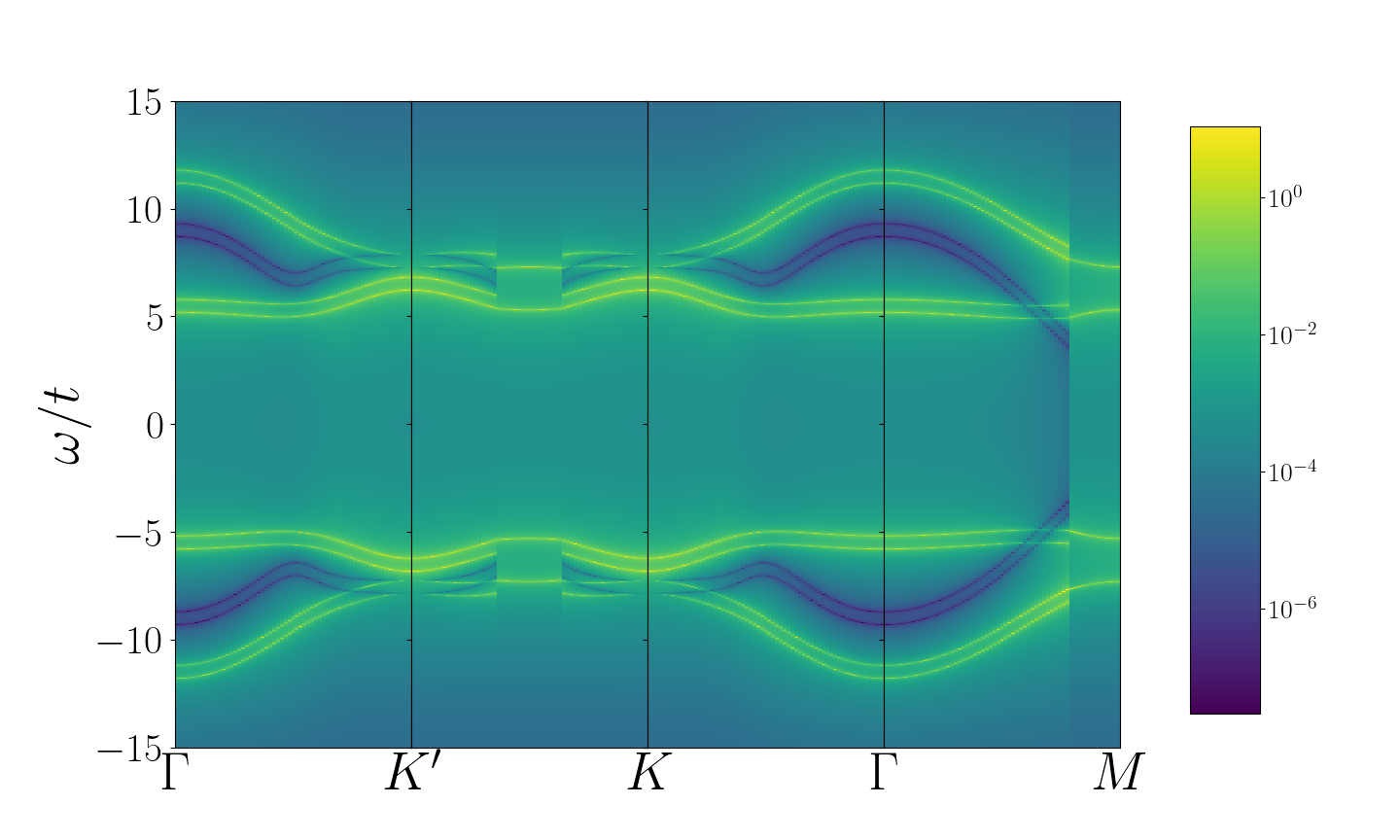}} &
\subfloat[]{\includegraphics[width=\overlap\textwidth]{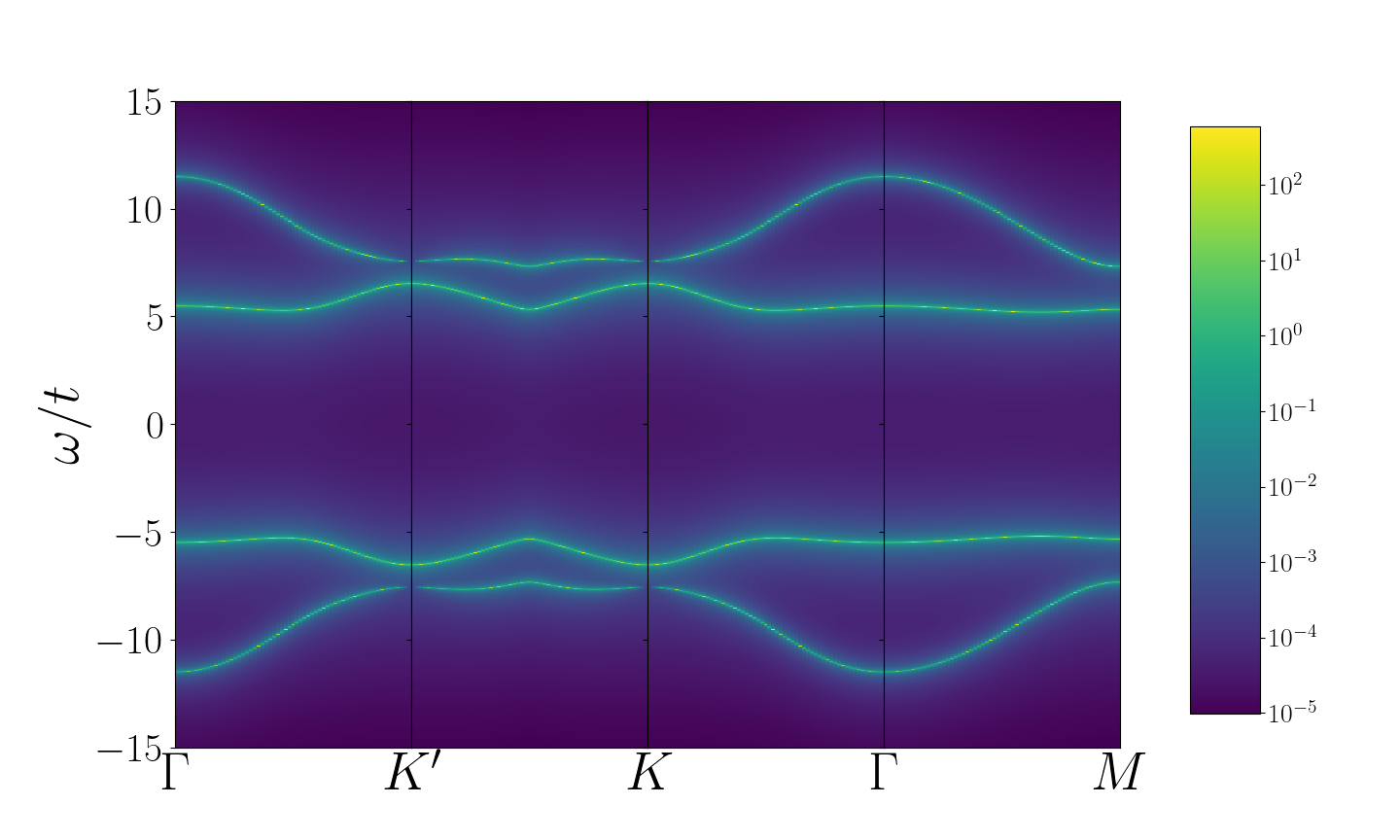}}
\end{tabularx}}
\caption{The Green function for Band and Orbital HK-KM models with a small Zeeman field in the large $U_1$ limit.
(a) shows the spectrum in the two particle sector for the band HK-KM model with Zeeman interactions.
(b) shows $|\det(G^+(\mathbf{k},\omega))$ for the band model. 
There is no band of Green function zeros which confirms that it is adiabatically connected to a trivial insulator.
(c) shows  the spectral function for the band model. 
(d) shows the spectrum in the two particle sector for the orbital HK-KM model with Zeeman interactions at the same parameters as the band model.
(e) shows $|\det(G^+(\mathbf{k},\omega))$ for the orbital model, in which we see that a band of zeros exists in the spectral gap along the $\Gamma-M$ line. 
(f) shows  the spectral function for the orbital model. 
The parameter values for both models are: $t=1$, $U_{1}=12t$, $\lambda=0.1$, $B=0.3t$, $\mu_{0}=U_1/2=6t$. 
As in the graphene model, we can see that an arbitrarily small magnetic field converts the band HK model into a ferromagnetic insulator, whereas the orbital HK model remains a non-fermi liquid with a finite magnetization.}
\label{fig:KMZeemanGFs}
\end{figure*}

As in the graphene model, we can understand the tendency towards ferromagnetism of the band and orbital Kane-Mele models by considering adding a small Zeeman term to the Hamiltonian. 
For the model with band HK interactions we consider the Hamiltonian
\begin{align}
&\Bar{H}_{KMZ}=\sum_{\mathbf{k}m\sigma}\xi_{nm\sigma}(\mathbf{k})\Bar{n}_{\mathbf{k}m\sigma}+U_{1}\sum_{\mathbf{k}m}n_{\mathbf{k}m\uparrow}\Bar{n}_{\mathbf{k}m\downarrow} +B\sum_{\mathbf{k}m\sigma}\sigma \Bar{n}_{\mathbf{k}m\sigma} \nonumber \\
&-\mu_{0}\sum_{\mathbf{k}m\sigma}\Bar{n}_{\mathbf{k}m\sigma} ,
\end{align}
and for the model with orbital HK interactions we have the Hamiltonian
\begin{align}
&H_{KMZ}=t\sum_{\mathbf{k}\sigma}g(\mathbf{k})c^\dagger_{\mathbf{k}B\sigma}c_{\mathbf{k}A\sigma}+i\lambda\sum_{\mathbf{k}\mu\sigma}\sigma\mu g_{1}(\mathbf{k})c^\dagger_{\mathbf{k}\mu\sigma}c_{\mathbf{k}\mu\sigma}\nonumber \\
&+h.c.+U_{1}\sum_{\mathbf{k}\mu}n_{\mathbf{k}\mu\uparrow}n_{\mathbf{k}\mu\downarrow}+B\sum_{\mathbf{k}\mu\sigma}\sigma n_{\mathbf{k}\mu\sigma}\mu_{0}-\sum_{\mathbf{k}\mu\sigma}n_{\mathbf{k}\mu\sigma}.
\end{align}
Note that the orbital model Hamiltonian $H_{KMZ}$ is particle-hole symmetric for $\mu_0=U_1/2$, with the particle-hole symmetry operator given by \Cref{eq:grapheneph}. 
The resulting determinant and trace of the Green function matrices $G^{+}$ are plotted in \Cref{fig:KMZeemanGFs} for each case in the strongly interacting limit $U>W$. 
As in the graphene model, we see that the orbital Kane-Mele model has a finite magnetic susceptibility as indicted by \Cref{fig:KMZeemanGFs}(d). 
Furthermore, the ground state retains its non-Fermi liquid behavior in the presence of a small magnetic field---indicated by the midgap Green function zeros near $M$---whereas the band HK interaction leads to a ferromagnetic state that is adiabatically connected to a band insulator. 

We have demonstrated that orbital HK models do not have thermodynamically large ground state degeneracies, and as a consequence are stable to small external magnetic fields. 
Furthermore, we have seen how orbital HK models can be extended to spin-dependent non-interacting Hamiltonians. 
These results provide the necessary tools to study the applicability of the LSM theorems to the HK models. 
Since we have lifted the degeneracy in band HK models, we can now ask whether an orbital HK model can have a topological degeneracy at a given filling. 
Since we can define orbital models even for systems that do not conserve spin, we can apply our method to filling constraints in the presence of spin-orbit coupling, as considered for short range entangled systems in Ref.~\cite{watanabe2015filling}.

In the next section, we will use these observations to introduce and solve interacting models for spin-orbit coupled Hamiltonians invariant under space group $P4_2/mbc1^\prime$ (\# 135) and space group $P4/ncc1^\prime$ (\#130). 
Crucially, the models we will consider do not conserve any component of the spin. 
In space group $P4_2/mbc1^\prime$ (\# 135) it is known that non-interacting systems with fewer than $\nu=8$ electrons per unit cell cannot be insulators; however, the tightest LSM theorem currently known allows for an interacting featureless insulator at $\nu=4$ electrons per unit cell. 
We will show that adding orbital HK interactions leads to a spin liquid ground state at $\nu=4$ and that adding a more general class of HK interactions leads to a non-degenerate, symmetric gapped phase at four electrons per unit cell. 
By considering space group $P4/ncc1^\prime$ (\#130), (and space group $P12_{1}/c1^\prime$ (\#14) in the appendix \cref{sec:SG14}) we will then show that generalized HK interaction are not subject to the same filling constraints as short-range entangled systems. 
This raises the question of how to understand and identify topology in a long range interacting system.

\section{The Double-Dirac Spin Liquid in Space Group $P4_2/mbc1^\prime$ (\# 135)}
\label{sec:DDSL}
\subsection{Model and Symmetries}\label{subsec:ddslmodel}

\begin{figure*}
\FPeval{\overlap}{0.5}
\FPeval{\lq}{1.1*\overlap}
\FPeval{\scalevalue}{round(0.5*(1)/\overlap,2)}

\scalebox{\scalevalue}{
\setlength{\tabcolsep}{0pt} 
\def\tabularxcolumn#1{m{#1}}
\hskip-1.0cm\begin{tabularx}{\columnwidth}{@{}XXX@{}}
\hskip-3.0cm\subfloat[]{\includegraphics[scale=1.3]{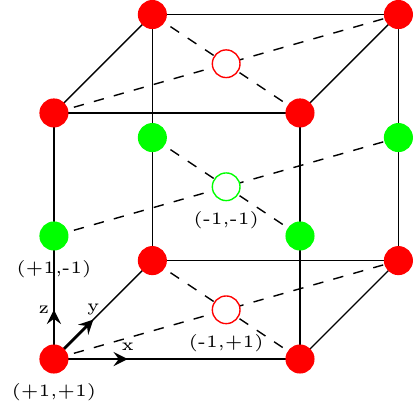}} &
\subfloat[]{\includegraphics[width=\overlap\textwidth]{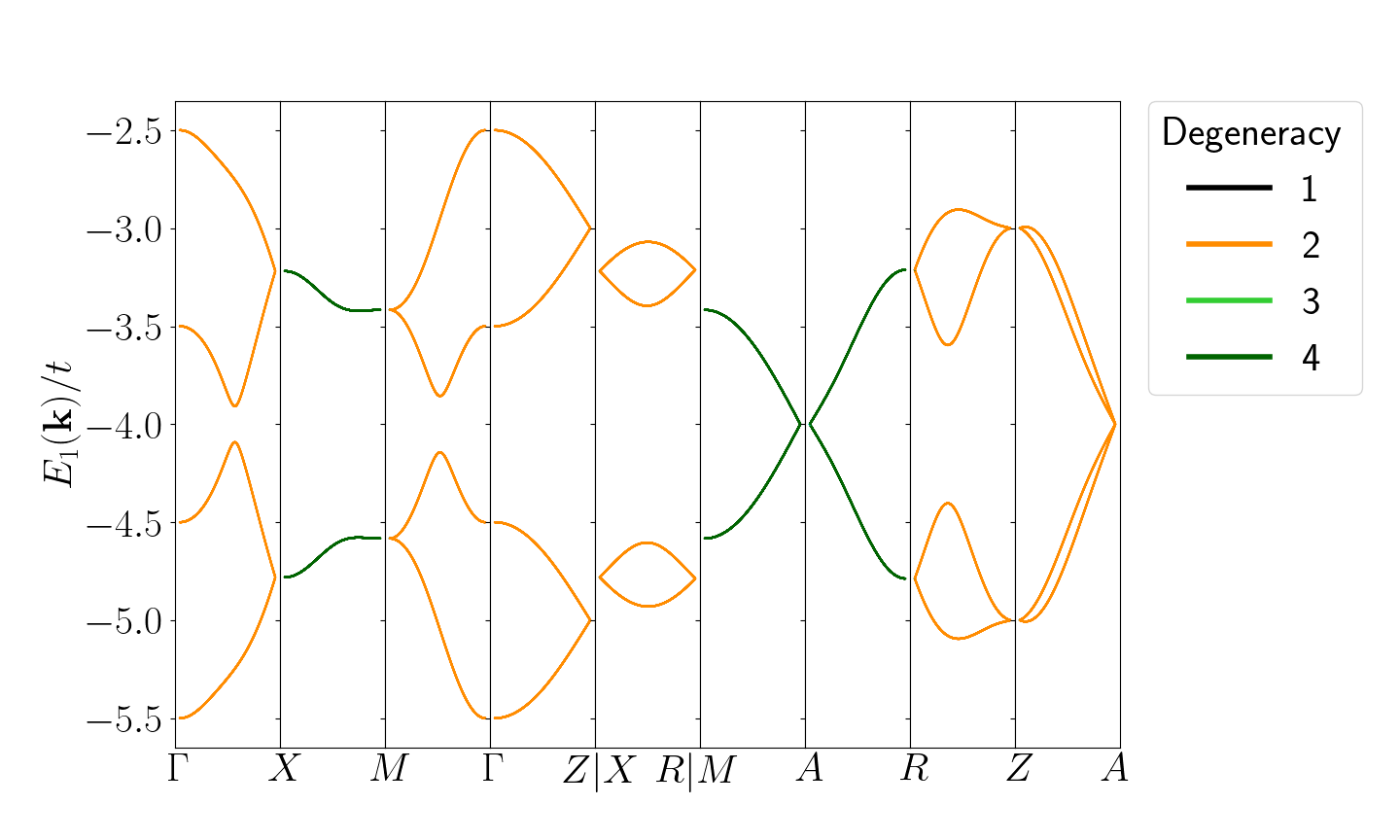}}
 \\[-3ex]
\hskip-3.0cm\subfloat[]{\includegraphics[scale=1.3]{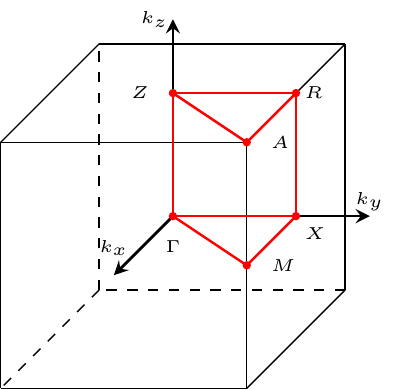}} &
\subfloat[]{\includegraphics[width=\overlap\textwidth]{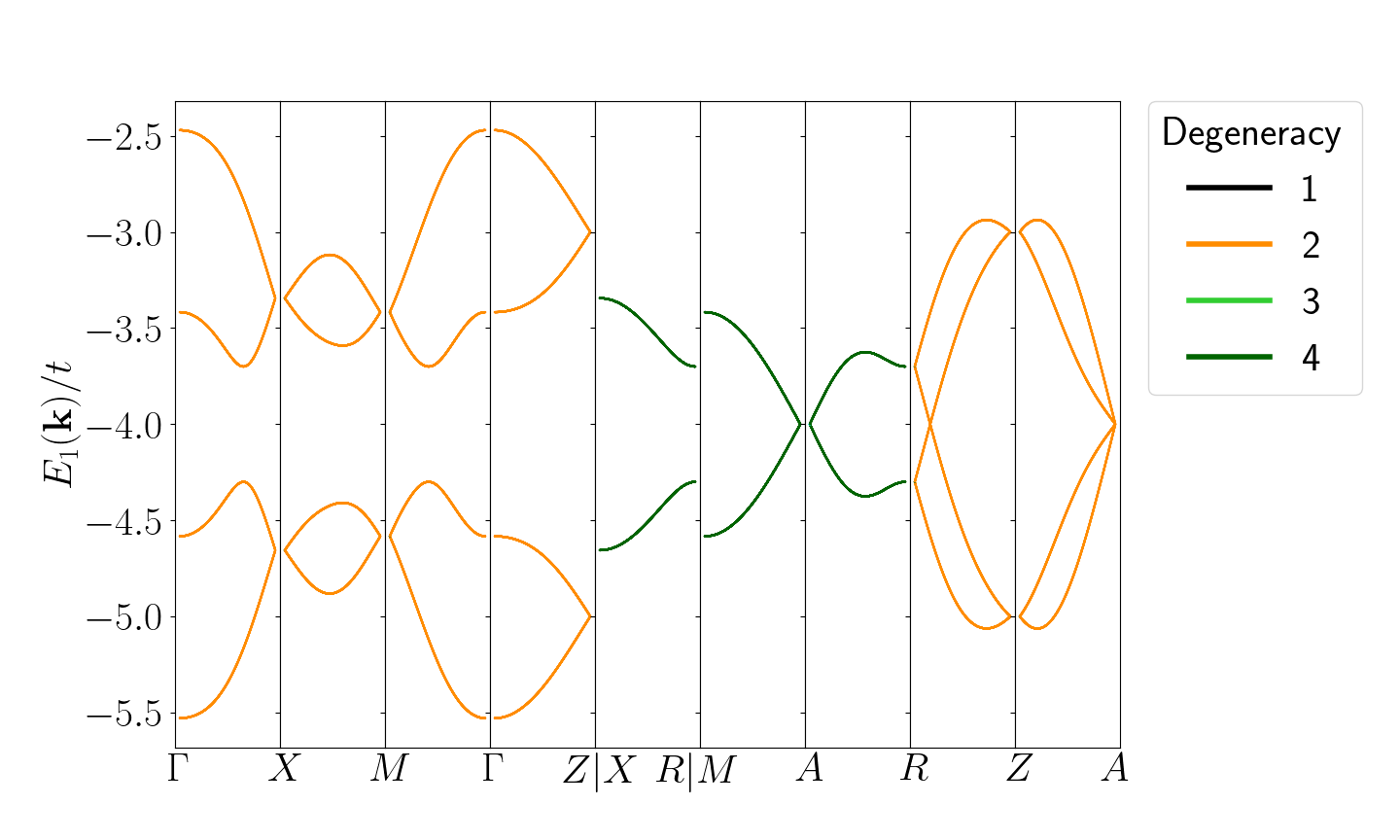}}
\end{tabularx}}

\caption{The parent lattice, Brillouin Zone, and non-interacting one-particle energy spectra. (a) shows the sublattice degrees of freedom. 
The color indicates the value of $\mu$: red is $\mu=+1$ ($A$ sublattice), green is $\mu=-1$ ($B$ sublattice) and whether they are filled ($+1$ for the $A$ sublattice) or empty ($-1$ for the $B$ sublattice) indicates the value of $\tau$. (b) shows the single particle spectrum for the non-interacting Hamiltonian $H^{0}_{135}$ in \Cref{eq:135 Hamiltonian}. 
The bands are colored according to their degeneracy, given on the right. 
We use the same non-interacting parameter values for all space group \sg135 models: $t_{xy}=1,t_z=0.5,t'_{1}=t'_2=0.3,\lambda_1=0.5,\lambda_2=0.1,\lambda_3=0.15$ throughout the text. (c) shows the corresponding tetragonal Brillouin Zone of space groups $P4/ncc1^\prime$ (\#130) and $P4_2/mbc1^\prime$ (\# 135).
(d) shows the single particle spectrum for the non-interacting Hamiltonian for space group $P4/ncc1^\prime$ (\#130) $H^{0}_{130}$. 
We use the same non-interacting parameter values for all $P4/ncc1^\prime$ (\#130) models throughout: $t_{xy}=1,t_{z}=0.5,\lambda_1=0.3,\lambda_2=0.3,\lambda_3=0.3$.}
\label{fig:SPandCubes}
\end{figure*}

We start with a non-interacting Hamiltonian $H^{0}_{135}$ constructed so as to be invariant under space group $P4_2/mbc1^\prime$ (\# 135). 
The same Hamiltonian was considered in Ref.~\cite{wieder2016double}\footnote{We are grateful to Benjamin Wieder for pointing out a misprint in Ref.~\cite{wieder2016double}: the $\mu^x\tau^y\sigma^y$ term should have a coefficient $\cos\tfrac{k_x}{2}\sin\tfrac{k_y}{2}$ but was printed as $\cos\tfrac{k_x}{2}\cos\tfrac{k_y}{2}$.}. 
The Hamiltonian is defined on a tetragonal Bravais lattice with four occupied sublattice sites per unit cell (see \Cref{fig:SPandCubes}).
The horizontal sublattice sites are indexed by $\tau=\pm 1$, and are located at $(0,0,0)$ and $(\tfrac{1}{2},\tfrac{1}{2},0)$ respectively (in reduced coordinates). 
The vertical sublattice sites are indexed by $\mu=\pm 1$ and are located at $(0,0,0)$ and $(0,0,\tfrac{1}{2})$ (in reduced coordinates). 
Together these four sites form the $4a$ Wyckoff position.

The single-particle Hamiltonian consists of a spin-independent hopping between these sublattice sites $H^{1}_{135}$ and a spin-dependent hopping $H^{2}_{135}$. 
In momentum space, we can write the corresponding Hamiltonians $H^{n}_{135}$ in the orbital basis of second quantized operators as
 \begin{equation}
 H^{n}_{135} = \sum_\mathbf{\mathbf{k},i,j,\sigma,\sigma'}c^\dagger_{\mathbf{k},i,\sigma}(\mathcal{H}^{n}_{135}(\mathbf{k}))_{i\sigma,j\sigma'} c_{\mathbf{k},j,\sigma'},
 \end{equation}
 with
\begin{align}
    \label{eq:135 Hamiltonian}
    &\mathcal{H}^{0}_{135}(\mathbf{k})=\mathcal{H}^{1}_{135}(\mathbf{k})+\mathcal{H}^{2}_{135}(\mathbf{k})\nonumber \\
    &\mathcal{H}^{1}_{135}(\mathbf{k})=t_{xy}\tau^x\cos\tfrac{k_x}{2}\cos\tfrac{k_y}{2}+t_{z}\mu^x\cos\tfrac{k_z}{2} \\
    &+t'_{1}\mu^z(\cos(k_x)-\cos(k_y))+t'_{2}\mu^y\tau^y\sin\tfrac{k_x}{2}\sin\tfrac{k_y}{2}\cos\tfrac{k_z}{2} \nonumber \\
    &\mathcal{H}^{2}_{135}(\mathbf{k})=\lambda'_{1}\mu^x\tau^y\left(\sigma^x\sin\tfrac{k_x}{2}\cos\tfrac{k_y}{2}+\sigma^y\cos\tfrac{k_x}{2}\sin\tfrac{k_y}{2}\right)\sin\tfrac{k_z}{2} \nonumber\\
    &+\lambda'_{2}\mu^y\tau^x\left(\sigma^x\cos\tfrac{k_x}{2}\sin\tfrac{k_y}{2}+\sigma^y\sin\tfrac{k_x}{2}\cos\tfrac{k_y}{2}\right)\sin\tfrac{k_z}{2} \nonumber \\
    &+\lambda'_3\mu^z\tau^y\sigma^z\cos\tfrac{k_x}{2}\cos\tfrac{k_y}{2}(\cos k_x-\cos k_y)\nonumber
\end{align}
To this Hamiltonian we add the orbital HK interaction $H^{1}_{HK}$:
\begin{equation}\label{eq:sg135orbitalhk}
H^{1}_{HK}=U_{1}\sum_{\mathbf{k}\mu\tau}n_{\mathbf{k}\mu\tau\uparrow}n_{\mathbf{k}\mu\tau\downarrow}
\end{equation}

In this basis of single-particle operators, the representations of the generators of space group $P4_2/mbc1^\prime$ (\# 135) [as well as those for space group $P4/ncc1^\prime$ (\#130), for later convenience] are given in \Cref{table:generatorsGamma}.
We can see from the single particle spectrum (\Cref{fig:SPandCubes}) that it hosts an eightfold degenerate double-Dirac fermion at the $A$-point as the only feature near the Fermi level at half filling.

\begin{table}[ht]
\caption{Representation of the symmetry generators for space group $P4/ncc1^\prime$ (\#130) and space group $P4_2/mbc1^\prime$ (\# 135) in the spin and sublattice basis. $\mathcal{K}$ denotes complex conjugation.} 
\centering 
\renewcommand*{\arraystretch}{1.4}
\begin{tabular}{c c c c | c c c c} 
\hline\hline 
\multicolumn{4}{c}{$P4/ncc1^\prime$ (\#130)} &\multicolumn{4}{c}{$P4_2/mbc1^\prime$ (\# 135)}\\[0.5ex]
\hline 
$\{g|\mathbf{t}\}$ & & $\rho\left(\{g|\mathbf{t}\}\right)$ & & $\{g|\mathbf{t}\}$ & & $\rho\left(\{g|\mathbf{t}\}\right)$ & \\ 
$\{C_{4z}|000\}$ & &  $e^{i\pi\sigma^z/4}$ & &  $\{C_{4z}|00\tfrac{1}{2}\}$ & & $\mu^xe^{i\pi\sigma^z/4}$ &  \\
$\{C_{2x}|\tfrac{1}{2}\tfrac{1}{2}0\}$ & &  $i\tau^x\sigma^x$ & &  $\{C_{2x}|\tfrac{1}{2}\tfrac{1}{2}0\}$ & &  $i\tau^x\sigma^x$ & \\
$\{I|\tfrac{1}{2}\tfrac{1}{2}\tfrac{1}{2}\}$ & & $\mu^x\tau^x$ & & $\{I|000\}$ & 
 & $1$ &  \\
TR & & $i\sigma^y \mathcal{K}$ & & TR &  & $i\sigma^y \mathcal{K}$ & \\ [1ex] 
\hline 
\end{tabular}
\label{table:generatorsGamma} 
\end{table}
.

\subsection{Excitation Spectrum and Ground State Degeneracy}\label{subsec:ddslprops}
\begin{figure*}
\FPeval{\overlap}{0.53}
\FPeval{\scalevalue}{round(0.5*(1.12)/\overlap,2)}
\scalebox{\scalevalue}{
\setlength{\tabcolsep}{0pt} 
\def\tabularxcolumn#1{m{#1}}
\hskip-1.0cm\begin{tabularx}{\textwidth}{@{}XXX@{}}
\subfloat[]{\includegraphics[width=\overlap\textwidth]{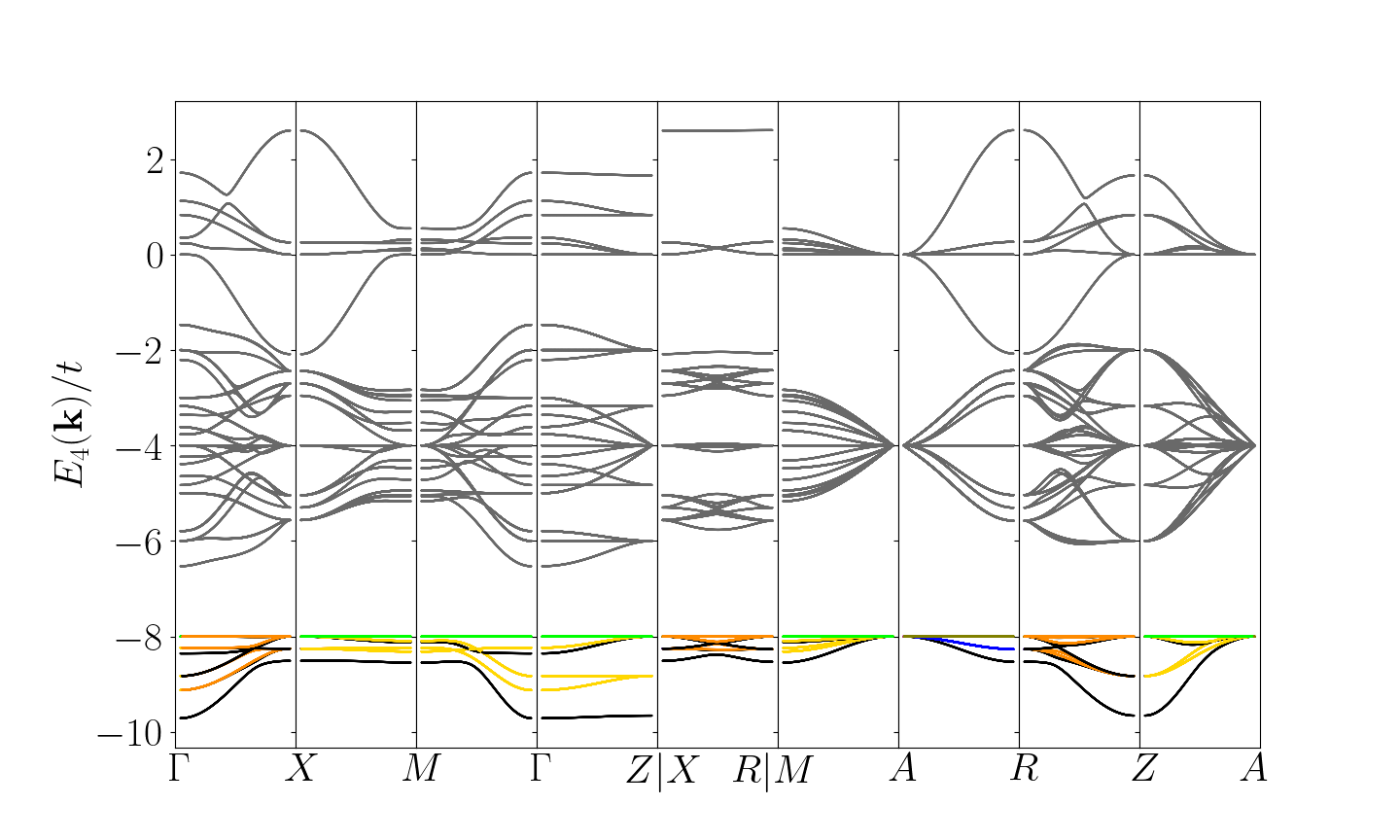}} &
\subfloat[]{\includegraphics[width=\overlap\textwidth]{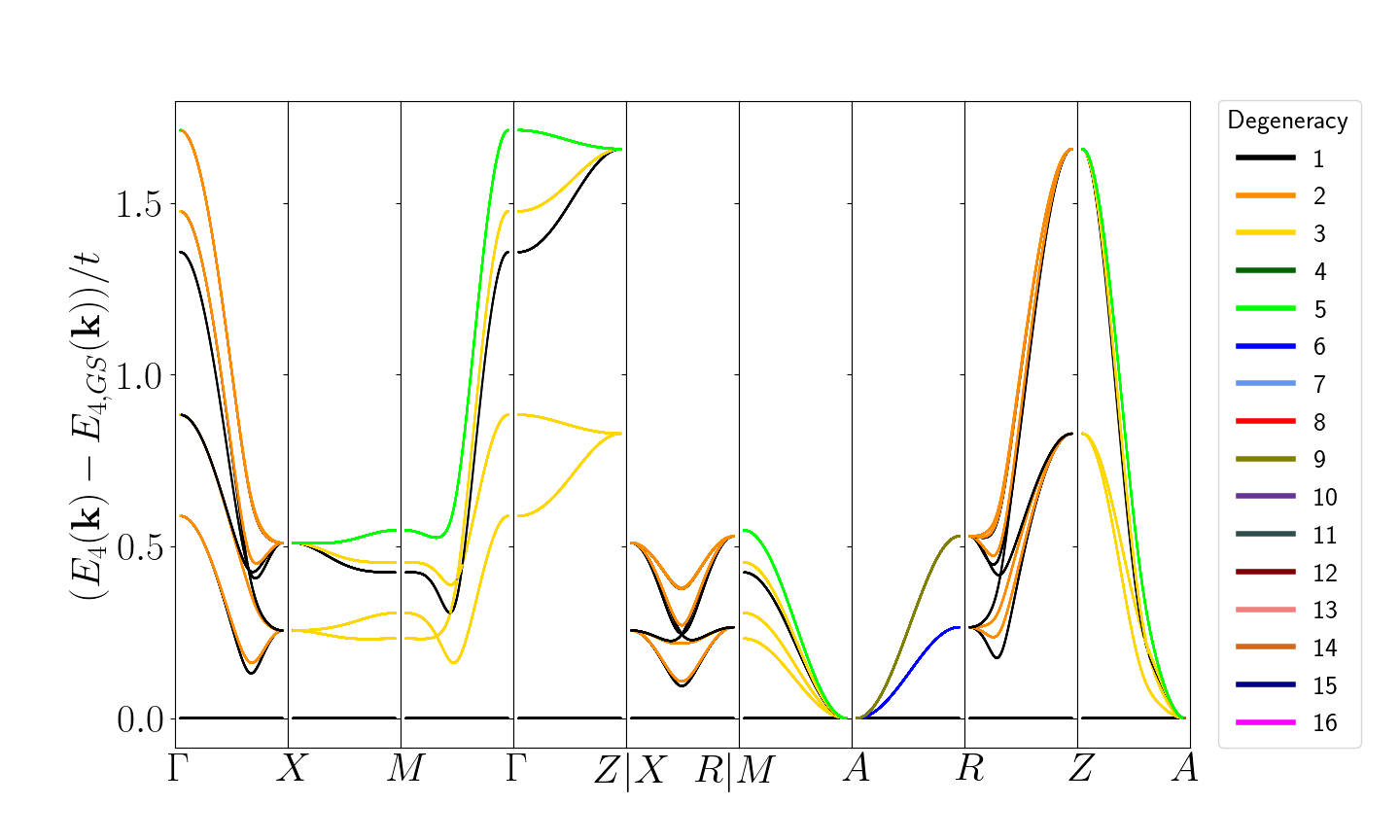}}
 \\[-3ex]
\subfloat[]{\includegraphics[width=\overlap\textwidth]{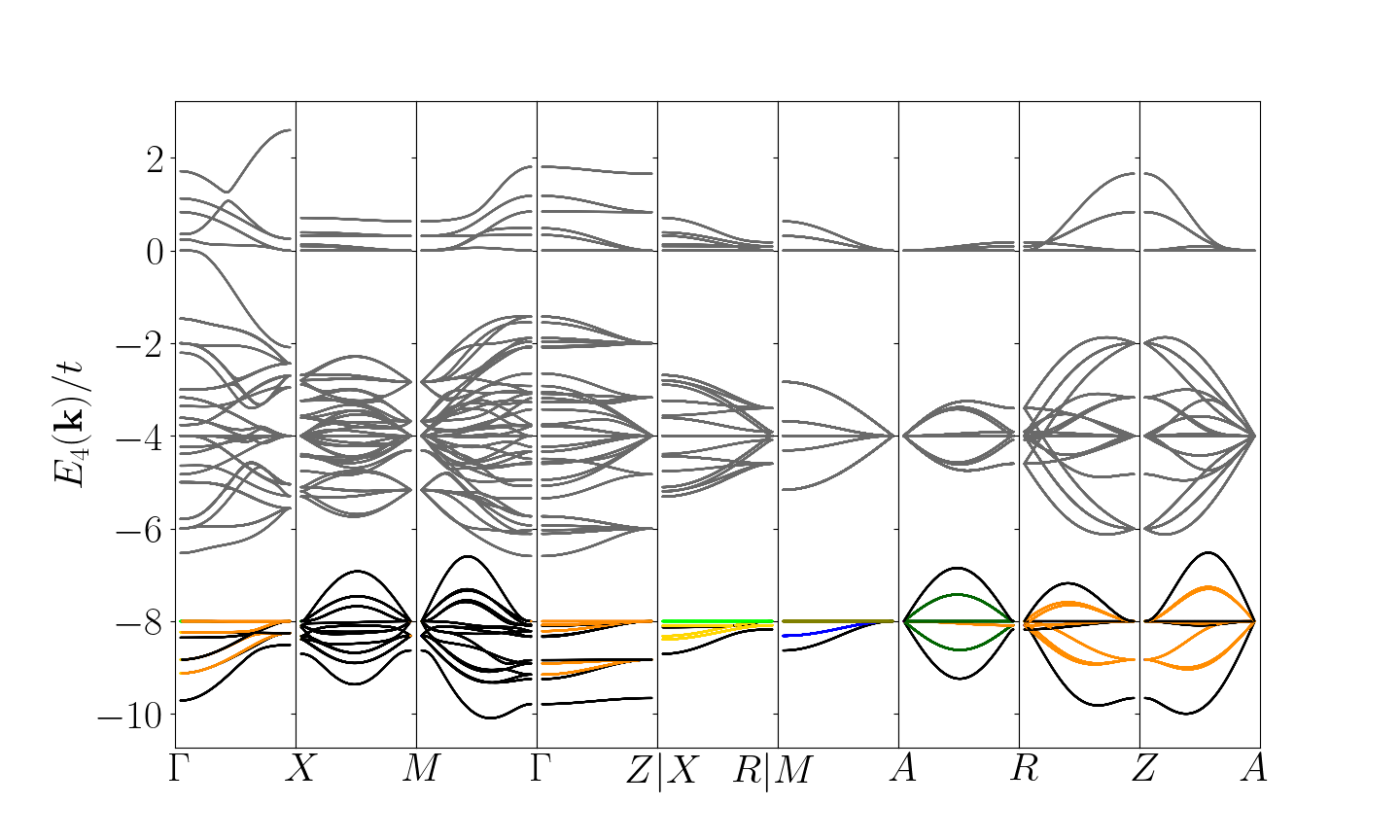}} &
\subfloat[]{\includegraphics[width=\overlap\textwidth]{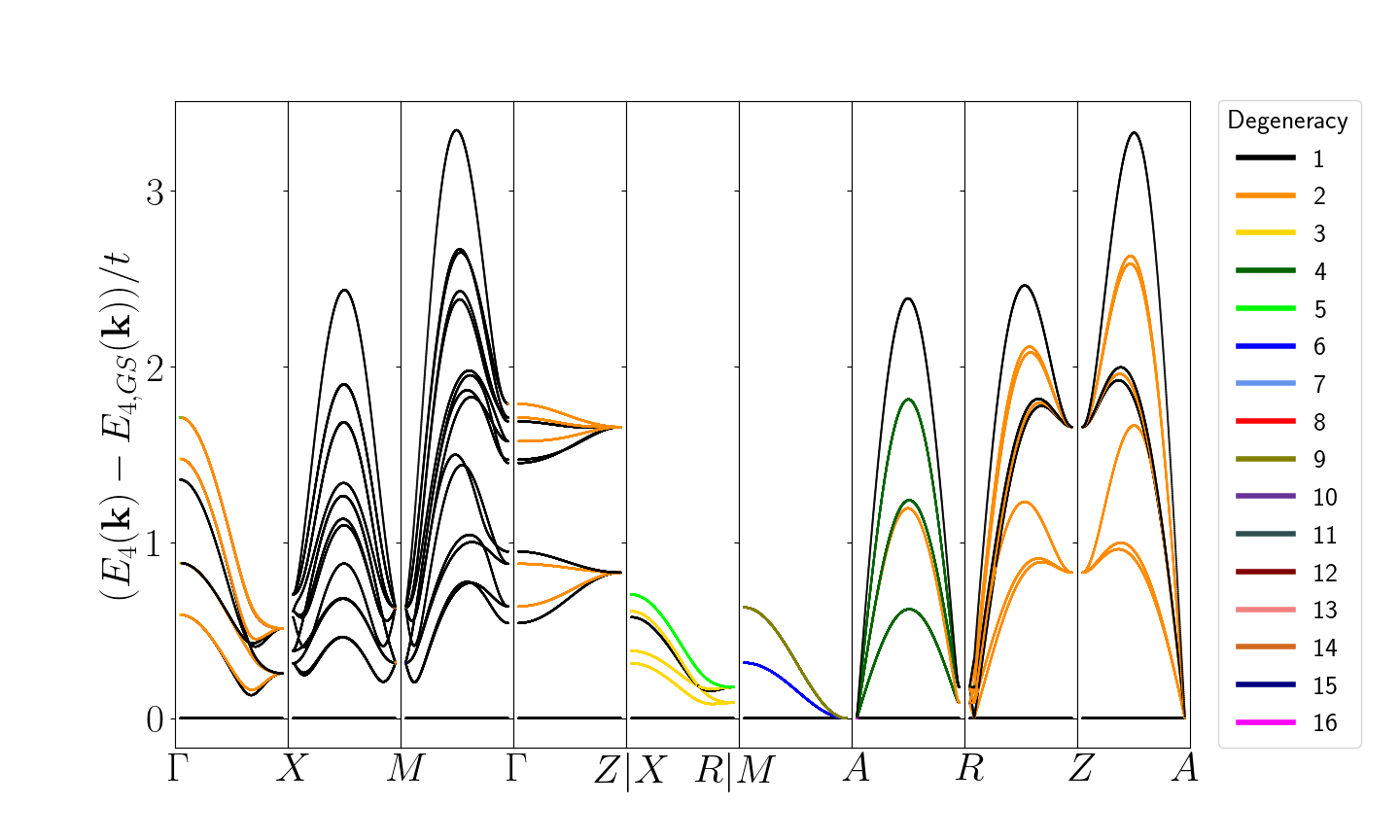}}
\end{tabularx}}
\caption{Spectra along the High Symmetry Paths (HSP) for space group $P4_2/mbc1^\prime$ (\# 135) (top row) and space group $P4/ncc1^\prime$ (\#130) (bottom row) with $HK$ interactions $H^{1}_{HK}$ in the half filled (four electrons per unit cell) sector. 
The low energy states have been colored according to their degeneracy, given in the rightmost panel. 
The degeneracy of the higher energy states has been grayed out for clarity. 
(a) shows the spectrum $E_{4}(\mathbf{k})$ of the orbital HK model in space group $P4_2/mbc1^\prime$ (\# 135) with HK interaction $H^{1}_{HK}$ in the four particle sector.
(b) shows the low energy part of the excitation spectrum $E_{4}(\mathbf{k})-E_{4, GS}(\mathbf{k})$. 
Here we show only the sixteen states below the gap.
(c) shows the spectrum $E_4(\mathbf{k})$ of the orbital HK model in space group $P4/ncc1^\prime$ (\#130) with HK interactions $H^{1}_{HK}$ in the four particle sector. 
(d) shows the low energy part of the excitation spectrum $E_{4}(\mathbf{k})-E_{4, GS}(\mathbf{k})$ for the model in space group $P4/ncc1^\prime$ (\#130). 
Note that the ground state is non-degenerate at the $A$-point and along the line $R-Z$.
The parameter values for space group $P4_2/mbc1^\prime$ (\# 135) are $t_{xy}=1,t_{z}=0.5,t'_{1}=0.3, t'_2=0.3, \lambda'_1=0.5,\lambda'_2=0.1,\lambda'_3=0.15, U_{1}=4, \mu_0=2$.
The parameter values for space group $P4/ncc1^\prime$ (\#130) are $t_{xy}=1,t_z=0.5,\lambda_1=0.3,\lambda_2=0.3,\lambda_3=0.3, U_1=4, \mu_0=2$.}
\label{fig:130135Spectra}

\end{figure*}

We can solve this model at half-filling (here four electrons per unit cell) using the same procedure that we used to solve the interacting graphene model \Cref{eq:grapheneH}. 
However, we note that, unlike the graphene models, the Hamiltonian $H^{1}_{135}=H_{135}^{0}+H_{HK}^{1}$ is not in general particle-hole symmetric. 
Specializing to $\mu_0=U_1/2$, we find that the Hamiltonian is particle-hole symmetric only when $t_2'=0$. 
Nevertheless, as we show in Appendix~\ref{sec:phsymm}, particle-hole symmetry breaking is weak provided $t_2\ll U_1$, and the ground state at half-filling still consists of four-particle states at every $\mathbf{k}$. 
To find the ground state in this regime, we first construct the seventy possible four-particle states at every $\mathbf{k}$-point and use them to form the matrix $\mathcal{H}_{135}(\mathbf{k})$ of the Hamiltonian in the four particle sector. 
We then diagonalize the resulting matrix to obtain the spectrum in the four particle sector and the corresponding eigenstates at each $\mathbf{k}$ point. 
The excitation spectrum is shown in the top row of \Cref{fig:130135Spectra}. \Cref{fig:130135Spectra}(a) shows the spectrum of the Hamiltonian $\mathcal{H}_{135}(\mathbf{k})$ in the four-particle sector at low energies, while \Cref{fig:130135Spectra}(b) shows the same spectrum with the ground state energy subtracted off. 
Since the low-energy excitations of $\mathcal{H}_{135}(\mathbf{k})$ are in the four particle sector, \Cref{fig:130135Spectra}(b) can be interpreted as the spectrum of low-energy excitations of the model at half filling.
We can see from the spectrum that the ground state is degenerate only at the $A$ point. 
This is occurs for the same reason that it occurred in our interacting graphene model: the non-interacting Hamiltonian vanishes at this point, and so the equivalent four-site Hubbard model at the $A$-point has zero hopping and is hence degenerate.

\begin{figure*}
\FPeval{\overlap}{0.53}
\FPeval{\scalevalue}{round(0.5*(1.12)/\overlap,2)}
\scalebox{\scalevalue}{
\setlength{\tabcolsep}{0pt} 
\def\tabularxcolumn#1{m{#1}}
\hskip-1.0cm\begin{tabularx}{\textwidth}{@{}XXX@{}}
\subfloat[]{\includegraphics[width=\overlap\textwidth]{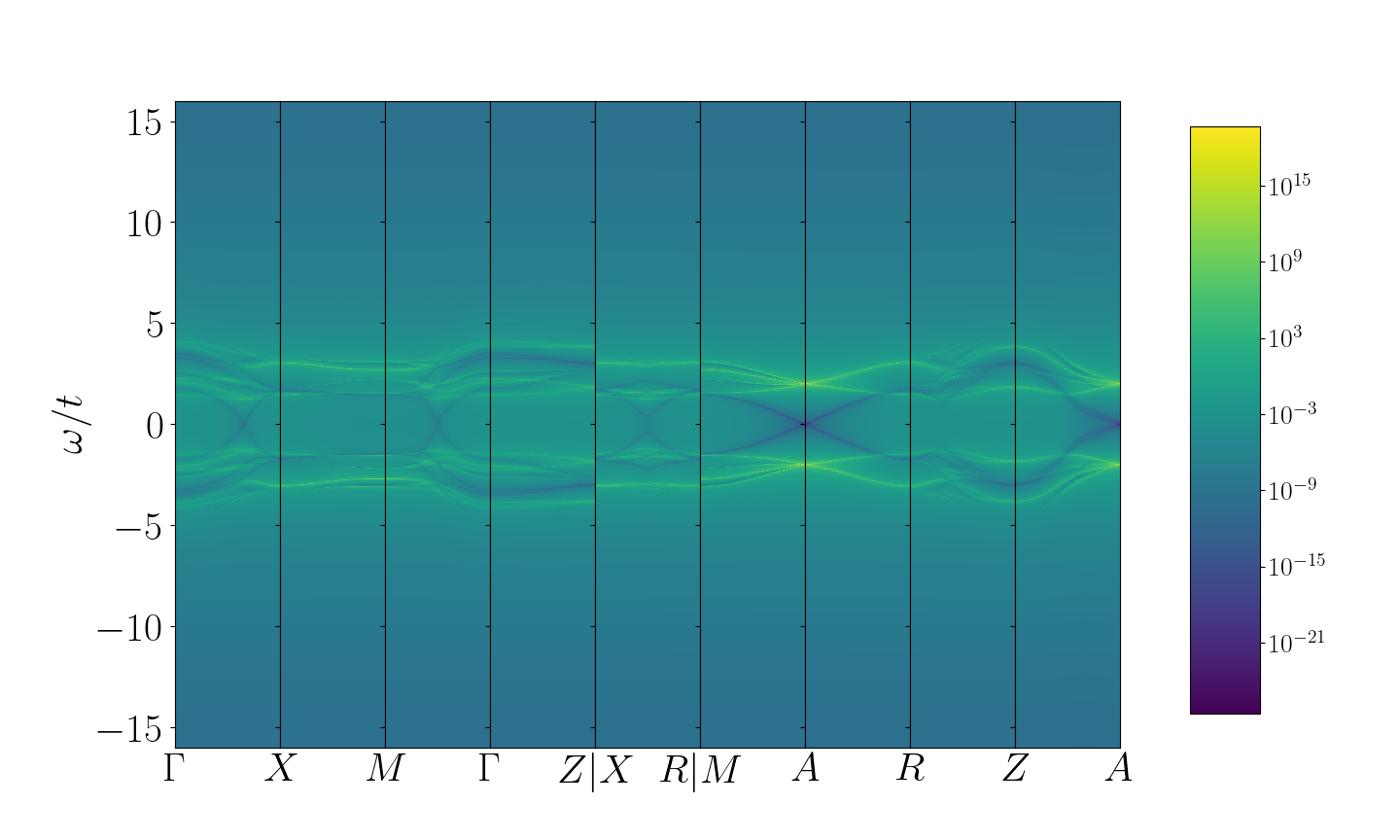}} &
\subfloat[]{\includegraphics[width=\overlap\textwidth]{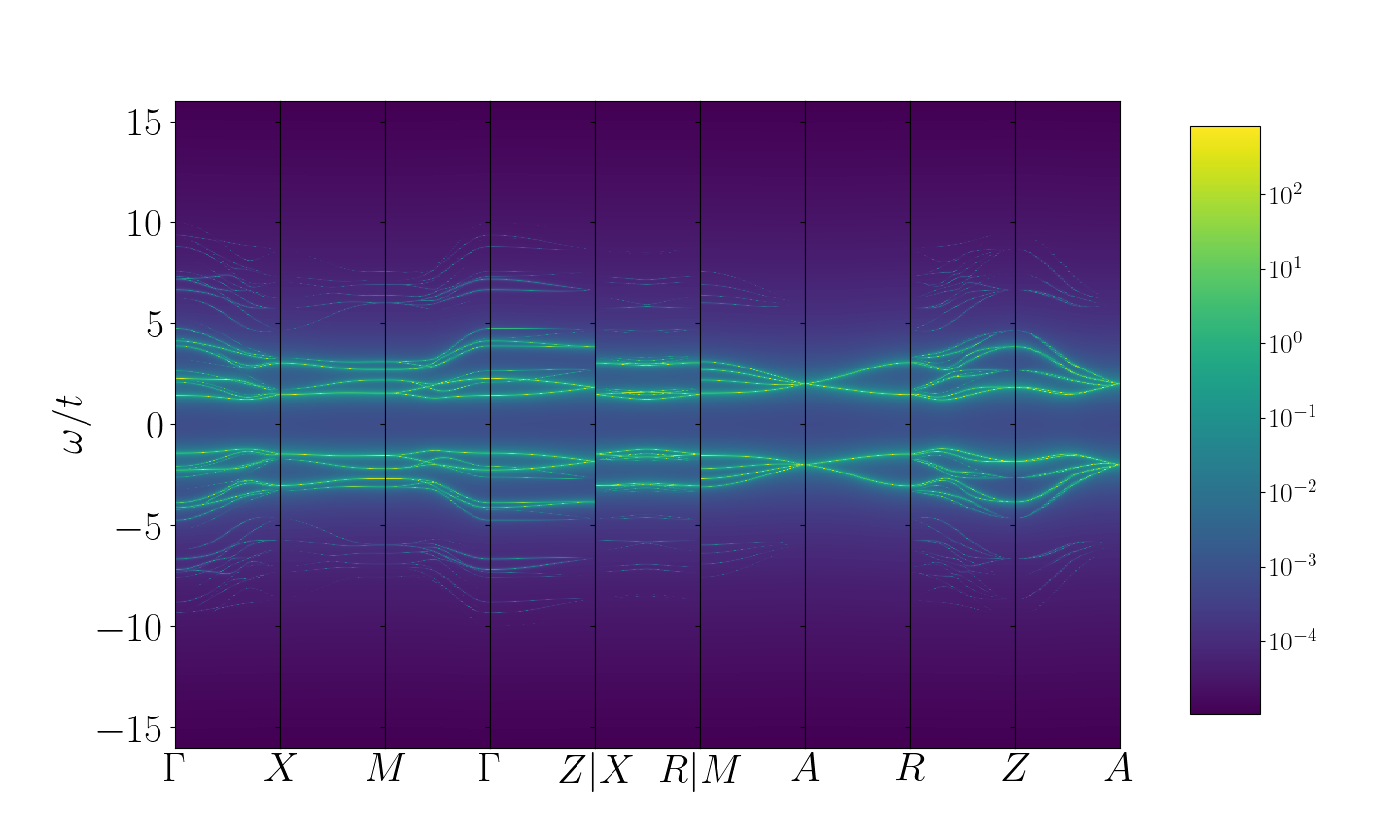}}
\end{tabularx}}
\caption{The absolute value of the determinant of the real time retarded Green function matrix $G^{+}$ at zero temperature (left) and the spectral function (right) for the orbital HK model invariant under space group $P4_2/mbc1^\prime$ (\# 135), $H^{1}_{135}=H^{0}_{135}+H^{1}_{HK}$.
(a) shows $|\det(G^{+})|$ for the orbital HK model $H^{1}_{135}=H^{0}_{135}+H^{1}_{HK}$. 
From the band of zeros in between the poles we can infer that $H^{1}_{135}=H^{0}_{135}+H^{1}_{HK}$ has a Mott insulating ground state. 
We also find that, at the $A$-point, the eigenvalues of the Green function matrix $G^{+}$ are eightfold degenerate.
(b) shows the spectral function $-\frac{1}{\pi}\Im\Tr{G}$ for $H^{0}_{135}+H^{1}_{HK}$. 
We have used the same non-interacting parameter values as previously: 
$t_{xy}=1,t_z=0.5,t'_{1}=t'_2=0.3,\lambda_1=0.5,\lambda_2=0.1,\lambda_3=0.15, \mu_0=2$, and the interaction strength is $U_{1}=4$.
}
\label{fig:135GFs}
\end{figure*}

We then calculate the zero temperature retarded real time Green function matrix $G^{+}_{i,j}$ using \Cref{eq:gfmatrix} and show the results in \Cref{fig:135GFs}.  
The surface of zeros in between the poles of $|\det(G^{+})|$ in \Cref{fig:135GFs}(a) confirms that the ground state is a Mott insulator, and the gap in the poles confirms that there is a charge gap. 
The determinant and spectral function are approximately symmetric around $\omega=0$ because of the approximate particle-hole symmetry of the model. 
As pointed out in Ref.~\cite{setty2023symmetry}, the eigenstates of the Green function $G^{+}_{i,j}$ transform in irreducible representations of the space group, just like eigenstates of the single-particle Hamiltonian. 
In particular, the degeneracies of the poles and zeros in $G^{+}_{i,j}$ match the allowed degeneracies of bands in the single particle spectrum. 
We have verified that the poles and zeros at the $A$ point in \Cref{fig:135GFs} for instance are eightfold degenerate, and the poles and zeros along the $M-A$ line are each fourfold degenerate. 
Additionally, unlike in band HK models, we see that there are regions in $\mathbf{k}$ and $\omega$ in which poles and zeros of the Green function coexist. 
At negative frequency (i.e. in the lower Hubbard bands) along the $M-A$ line for instance we see from \Cref{fig:135GFs}(a) that four bands of fourfold degenerate poles merge with two bands of fourfold degenerate zeros at the $A$ point to yield an eightfold degenerate pole at $A$. 
This annihilation of zeros and poles in the lower Hubbard bands is not observed in band HK models in the Mott insulating regime; it suggests that although the lower Hubbard band eigenstates share the same degeneracies as the single particle spectrum, they cannot be adiabatically connected to free fermion excitations. 
This establishes that the ground state has a charge gap across the Brillouin zone, and the Luttinger surface implies that it is a Mott insulating ground state.

We define a spin liquid ground state as one which has a charge gap and no magnetic order.
The poles of the green function in \Cref{fig:135GFs}(a) establish that there exists a charge gap throughout the Brillouin Zone.
As shown in \Cref{subsec:graphene} since neither time-reversal nor the crystal symmetries are spontaneously broken, the ground state also has no magnetic order. Hence, the ground state satisfies our definition of a spin liquid.

\subsection{Effective Hamiltonian}
\label{subsec:Schrieffer}
Th HK interaction $H^1_{HK}$ increases the energy of states which contain a doubly occupied orbital. 
Hence, the low-energy spectrum at half-filling is dominated by the sixteen states which have no interacting pairs, although in general other states have a small but non-zero overlap with the ground state. 
In particular, the ground state at the $A$ point contains only singly occupied orbitals.
Since the low-energy excitation spectrum shown in Fig.~\ref{fig:130135Spectra}(b) consists of excitations in the four-particle subspace near the $A$ point, this means that we expect charge degrees of freedom to be frozen, and the low-energy excitations to consist of spin-flip excitations with momentum near the $A$ point, $\mathbf{k}\approx (\pi,\pi,\pi)$. 
To make this precise and to derive a quantitative expression for the low-energy excitations, we can use a Schrieffer-Wolff transformation to derive a low-energy effective spin model for our system~\cite{bravyi2011schrieffer}.

We take the HK Hamiltonian $H^{1}_{HK}$ in \Cref{eq:sg135orbitalhk} as the initial Hamiltonian $H_0$ and the non-interacting Hamiltonian $H^{0}_{135}$ as the perturbation $V$.  
We note first that $H^{1}_{HK}$ is block-diagonal, with each block corresponding to a sector of the Hilbert space with a fixed number of interacting pairs.
This means that, to second order in the ratio of the bandwidth $W$ to the interaction energy $U$, the matrix elements of the Schrieffer-Wolff Hamiltonian are:
\begin{align}
&\bra{m}H_{SW}\ket{n}=\bra{m}H_{0}\ket{n}+\bra{m}V\ket{n}+\nonumber \\
&\frac{1}{2}\sum_{l}\bra{m}V\ket{l}\bra{l}V\ket{n}\left[\frac{1}{E_{m}-E_{l}}+\frac{1}{E_{n}-E_{l}}\right],
\end{align}
where $m,n$ index the eigenstates of the low-energy subspace of $H_{0}=H^{1}_{HK}$ (at half-filling the sixteen states with no HK interacting pairs) and $l$ runs over all eigenstates not in the low-energy subspace.

\begin{figure*}
\FPeval{\overlap}{0.50}
\FPeval{\scalevalue}{round(0.5*(1.12)/\overlap,2)}
\scalebox{\scalevalue}{
\setlength{\tabcolsep}{0pt} 
\def\tabularxcolumn#1{m{#1}}
\hskip-1.0cm\begin{tabularx}{\textwidth}{@{}XXX@{}}
\subfloat[]{\includegraphics[width=\overlap\textwidth]{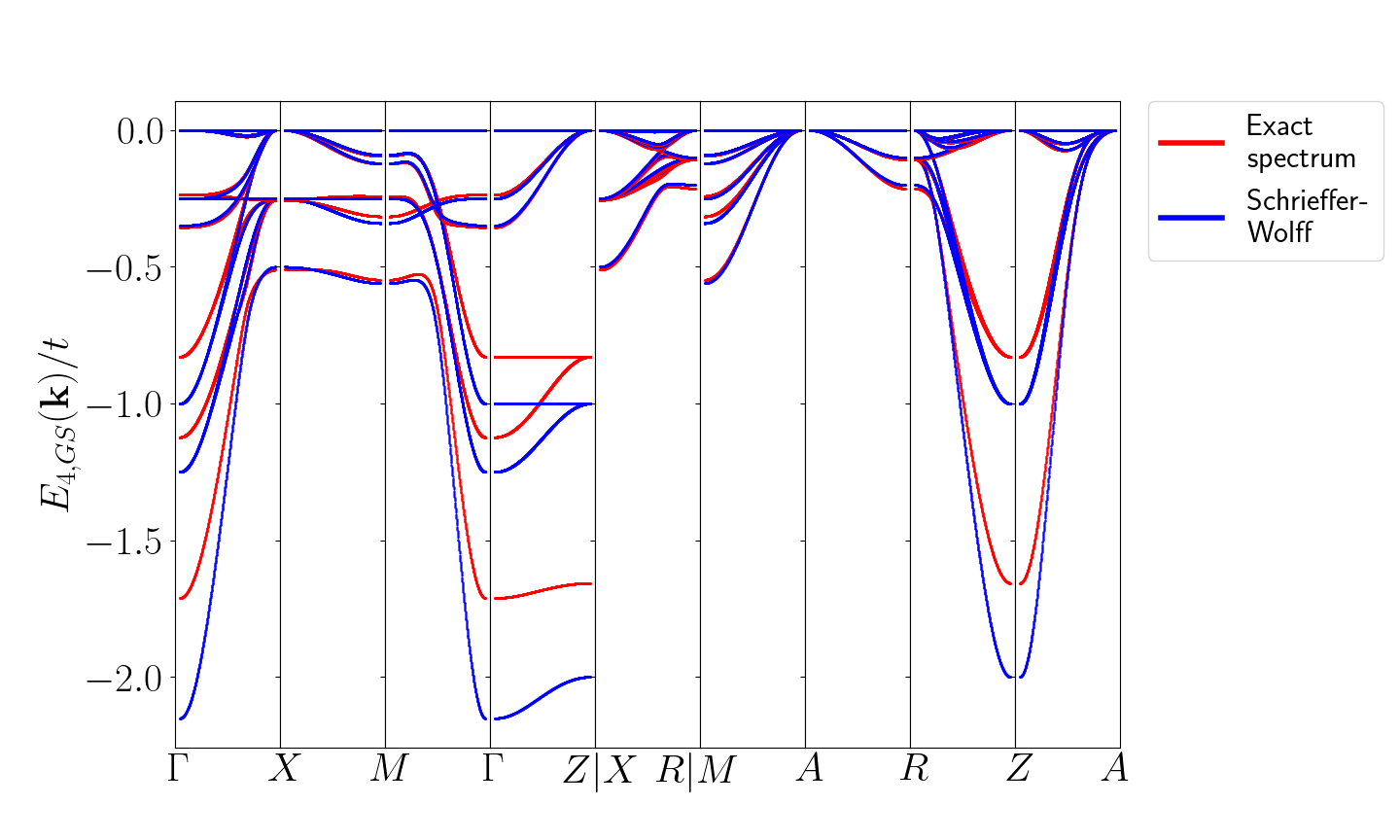}} &
\subfloat[]{\includegraphics[width=\overlap\textwidth]{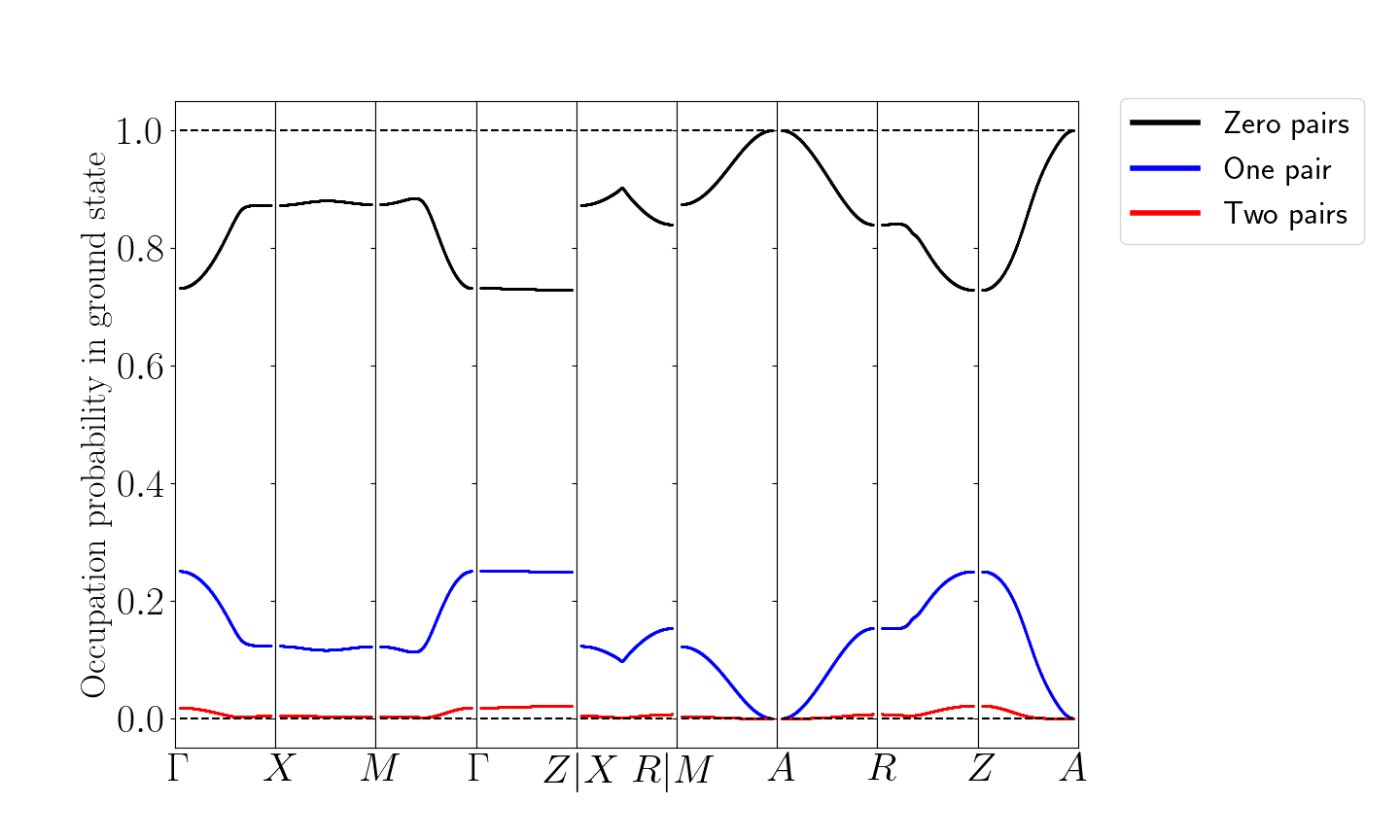}}
\end{tabularx}}
\caption{A comparison of the spectrum of the effective Schrieffer-Wolff Hamiltonian with the spectrum of the full Hamiltonian (left) and the probability of doubly occupied an orbital in the ground state (right) at each $\mathbf{k}$ point. 
(a) The spectrum of the Schrieffer-Wolff Hamiltonian \Cref{eq:Schrieffer-Wolff}, in blue, plotted with the full Hamiltonian in red. 
(b) The probability of having zero (black), one (blue) and two (red) HK interacting pairs in the ground state. 
We see that when the probability of having a doubly occupied orbital is low, the effective Hamiltonian is a good approximation to the full Hamiltonian.
}
\label{fig:SWcomparison}
\end{figure*}

Second, at half-filling every creation operator acting on a state with no HK interacting pairs creates one. 
This means that the first order contribution to the matrix elements $\bra{m}V\ket{n}$ is zero, and $E_{i}-E_{k}=U_{1}$ for all $i,k$ in the second-order contribution. 
This allows us to simplify the matrix elements to:
\begin{equation}
\label{eq:Schrieffer-Wolff}
\bra{i}H_{SW}\ket{j}=\bra{i}H_{0}\ket{j}+\frac{1}{U_{1}}\bra{i}V^{2}\ket{j}\text{ }.
\end{equation}
A comparison of the spectrum obtained from this Hamiltonian and the low-energy subspace of the full Hamiltonian is shown in \Cref{fig:SWcomparison}. 
Fig.~\ref{fig:SWcomparison}(a) shows both the exact and Schrieffer-Wolff excitation spectrum, while Fig.~\ref{fig:SWcomparison}(b) shows the probability of finding zero, one, or two doubly-occupied orbitals in the ground state.
As we would expect, the effective and full Hamiltonians agree closely where there is only a small amplitude for having a doubly occupied state. 
In particular, the Schrieffer-Wolff spectrum is a good approximation to the low-energy excitation spectrum near the $A$ point. 

We can take the physical interpretation further. 
Since the HK interaction has `frozen out' the charge degrees of freedom, the Schrieffer-Wolff Hamiltonian is a pure spin Hamiltonian. 
This means that we can use the orthogonality of spin-operators under the trace to find it's form in terms of spin operators.
We give the full spin-spin Hamiltonian and the details of it's derivation in \Cref{sec:Schrieffer-Wolff Appendix}. 
Since the HK interaction is infinitely long-ranged in position space, we find that our spin Hamiltonian contains interactions between spins at all distance scales. 

\section{More General Interactions}
\label{sec:General Interactions}
 We now show that it is possible to lift the degeneracy at the $A$-point by considering more general HK interactions that respect the space group symmetries. 
 First, we note that the ground state space at the $A$ point is spanned by the sixteen states with no HK interacting pairs. 
 A basis for this subspace can be written as
 \begin{equation}
 \{\ket{AA\sigma_1;AB\sigma_2;BA\sigma_3;BB\sigma_4}\},
 \end{equation} 
where the first letter in each triplet gives the $\mu_z$ orbital eigenvalue, the second letter gives the $\tau_z$ orbital eigenvalue (i.e., $A$ corresponds to $+1$ and $B$ corresponds to $-1$ as in our analysis of graphene), $\sigma_1$--$\sigma_4$ are spin indices, and we have left the $\mathbf{k}=(\pi,\pi,\pi)$ dependence implicit.
These sixteen states transform in a (reducible) representation of the space group, which we can write as a direct sum of irreducible representations. 
To do so, we first examine how each state transforms under Bravais lattice translations. 
Since each of the four particles has the same momentum $\mathbf{k}=(\pi,\pi,\pi)$, we find that under a Bravais lattice translation by lattice vector $\mathbf{t}$--represented by a unitary operator $U_\mathbf{t}$--the states transform as
\begin{align}
U_\mathbf{t}&\ket{AA\sigma_1;AB\sigma_2;BA\sigma_3;BB\sigma_4}= \nonumber \\ 
& =e^{-4i\mathbf{k}\cdot\mathbf{t}}\ket{AA\sigma_1;AB\sigma_2;BA\sigma_3;BB\sigma_4} \nonumber \\
& = \ket{AA\sigma_1;AB\sigma_2;BA\sigma_3;BB\sigma_4}.
\end{align}
In other words, the sixteen low-energy states at $A$ in the four-particle sector have zero crystal momentum (modulo a reciprocal lattice vector). 
This means that the states will transform in representations of the space group induced from the $\Gamma$ point in the BZ. 
Furthermore, since these states contain an even number of fermions, they have integer angular momentum. 
We can thus decompose our sixteen state basis into a direct sum of single-valued space group representations induced from the $\Gamma$ point. 
We refer the reader to the Bilbao Crystallographic Server (BCS) for the character tables and representation matrices for the space group representations~\cite{aroyo2006bilbaoa,aroyo2006bilbao,aroyo2011crystallography,elcoro2017double}. 
Using the Schur orthogonality relations coupled with the character tables on the BCS, we find that the representations of the symmetries of space group $P4_2/mbc1^\prime$ (\# 135) and time reversal on these sixteen four-particle states at the $A$-point decompose into eight one-dimensional and four two-dimensional irreducible representations.
These are tabulated in \Cref{table:Ainvariantsubspaces}.

\begin{table}[ht]
\centering 
\renewcommand*{\arraystretch}{1}
\begin{tabular}{c c c} 
\hline\hline 
Representation & Multiplicity & Dimension\\[0.5ex]
\hline 
$\Gamma^{+}_{1}$&4&1\\
$\Gamma^{+}_{2}$&2&1\\
$\Gamma^{+}_{4}$&2&1\\
$\Gamma^{+}_{5}$&4&2\\
\hline 
\end{tabular}
\caption{Decomposition into space group irreducible representations of the sixteen degenerate states four particle states at the $A$ point for the orbital HK model $H^0_{135}+H^1_{HK}$ in space group $P4_2/mbc1'$ (\# 135). 
We follow the labeling convention used on the Bilbao Crystallographic Server~\cite{aroyo2006bilbaoa,aroyo2006bilbao,aroyo2011crystallography,elcoro2017double}.}
\label{table:Ainvariantsubspaces} 
\end{table}

\begin{table*}[ht]
\centering 
\renewcommand*{\arraystretch}{1}
\begin{tabular}{c c c} 
\hline\hline 
Index & \multicolumn{2}{c}{Basis state} \\[0.5ex]
\hline 
\rule{0pt}{1\normalbaselineskip}
1&\multicolumn{2}{c}{$\tfrac{1}{\sqrt{2}}(\ket{AA\uparrow;AB\downarrow;BA\uparrow;BB\downarrow}+\ket{AA\downarrow;AB\uparrow;BA\downarrow;BB\uparrow})$}\\[2ex]
2&\multicolumn{2}{c}{$\tfrac{i}{\sqrt{2}}(\ket{AA\uparrow;AB\downarrow;BA\uparrow;BB\downarrow}-\ket{AA\downarrow;AB\uparrow;BA\downarrow;BB\uparrow})$}\\[2ex]
3&\multicolumn{2}{c}{$\tfrac{1}{\sqrt{2}}(\ket{AA\uparrow;AB\uparrow;BA\downarrow;BB\downarrow}+\ket{AA\downarrow;AB\downarrow;BA\uparrow;BB\uparrow})$}\\[2ex]
4&\multicolumn{2}{c}{$\tfrac{1}{\sqrt{2}}(\ket{AA\uparrow;AB\downarrow;BA\downarrow;BB\uparrow}+\ket{AA\downarrow;AB\uparrow;BA\uparrow;BB\downarrow})$}\\[1ex]
\hline 
\end{tabular}
\caption{The states which transform under the four copies of the trivial representation in the ground state subspace of $H^0_{135}+H^1_{HK}$ in space group $P4_2/mbc1^\prime$ (\# 135) at the $A$-point. 
The first two states can become disentangled upon breaking time-reversal symmetry, whereas to disentangle the third and fourth states requires breaking one of the spatial symmetries.}
\label{table:Trivialrepstates} 
\end{table*}

In Table~\ref{table:Trivialrepstates} we show the four states that transform in copies of the trivial representation. 
We see that each is a cat state superposition of spin- and orbital-ordered states at fixed crystal momentum $\mathbf{k}=(\pi,\pi,\pi)$. 
This suggests that the long-range interactions in the HK model can stabilize long-range entanglement in the ground state. 
In particular, we note that states $1$ and $2$ in Table~\ref{table:Trivialrepstates} can become disentangled by spontaneously breaking time-reversal symmetry, while states $3$ and $4$ can become disentangled by spontaneously breaking fourfold rotation symmetry. 

\subsection{Splitting the Degeneracy}\label{subsec:splitting}
Our symmetry analysis in Table~\ref{table:Ainvariantsubspaces} shows that the sixteen-fold-degenerate ground state subspace decomposes into eight one-dimensional and four two-dimensional irreducible representations of the space group. 
This allows for the possibility of perturbing our model to lift the degeneracy of the ground state; if the resulting perturbed model has a ground state that transforms in a one-dimensional representation at the $A$ point, then we can realize an HK model with a nondegenerate ground state and no symmetry breaking. 
This would give us a candidate state realizing a filling $\nu=4$ gapped, symmetric nondegenerate ground state as allowed for by the LSM theorem of Ref.~\cite{watanabe2015filling} that cannot be adiabatically connected to a band insulator.

We now show that we can form this insulating phase by generalizing our initial HK interaction $H^{1}_{HK}$ to a wider class of symmetry preserving interactions. 
First, we consider adding terms of the form:
\begin{equation}\label{eq:hkNterms}
H^{2}_{HK}=\sum_{i,j}n_{\mathbf{k},i}N_{i,j} n_{\mathbf{k},j},
\end{equation}
Where for convenience we have abused notation to absorb the spin into the orbital index so that $i,j$ are a shorthand for a set of $(\mu,\tau,\sigma)$ indices and $N$ is a symmetric matrix in the basis of $(\mu,\tau,\sigma)$ indices. This corresponds to our $H^{1}_{HK}$ for $N=\sigma^x$, and a special case of this term was considered for a band-HK model in Ref.~\cite{yanghk2019}.

\begin{figure*}
\FPeval{\overlap}{0.53}
\FPeval{\scalevalue}{round(0.5*(1.12)/\overlap,2)}
\scalebox{\scalevalue}{
\setlength{\tabcolsep}{0pt} 
\def\tabularxcolumn#1{m{#1}}
\hskip-1.0cm\begin{tabularx}{\textwidth}{@{}XXX@{}}
\subfloat[]{\includegraphics[width=\overlap\textwidth]{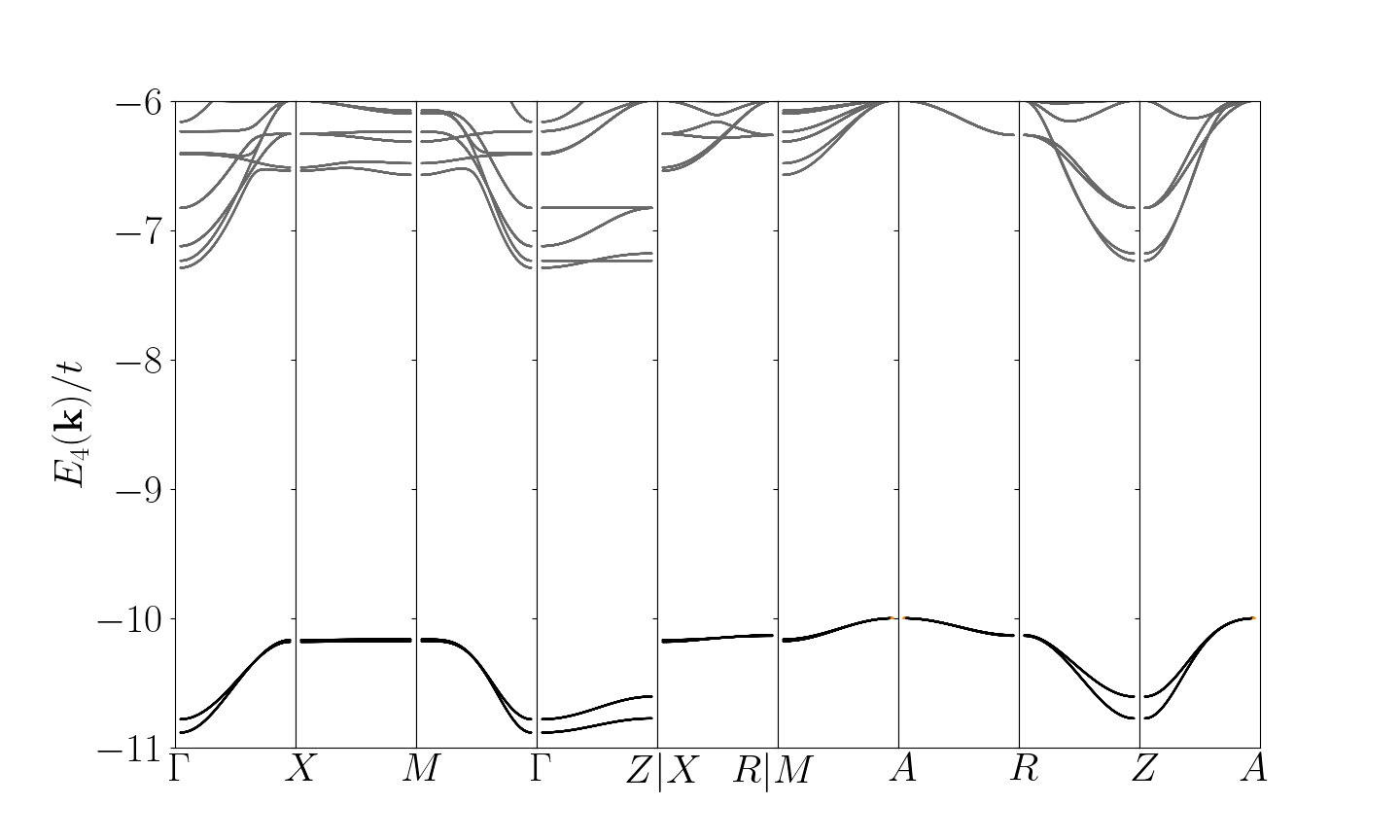}} &
\subfloat[]{\includegraphics[width=\overlap\textwidth]{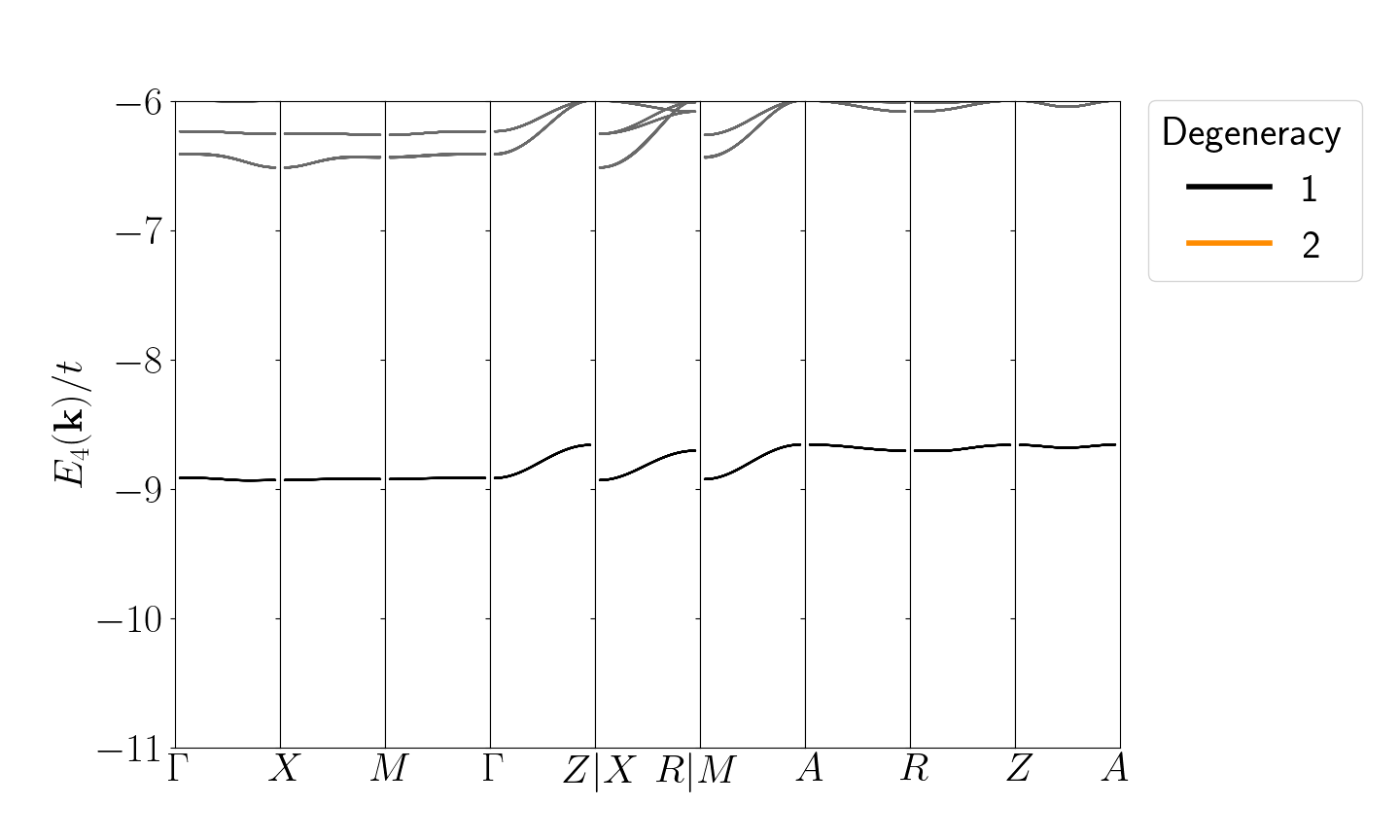}}
 \\[-3ex]
\subfloat[]{\includegraphics[width=\overlap\textwidth]{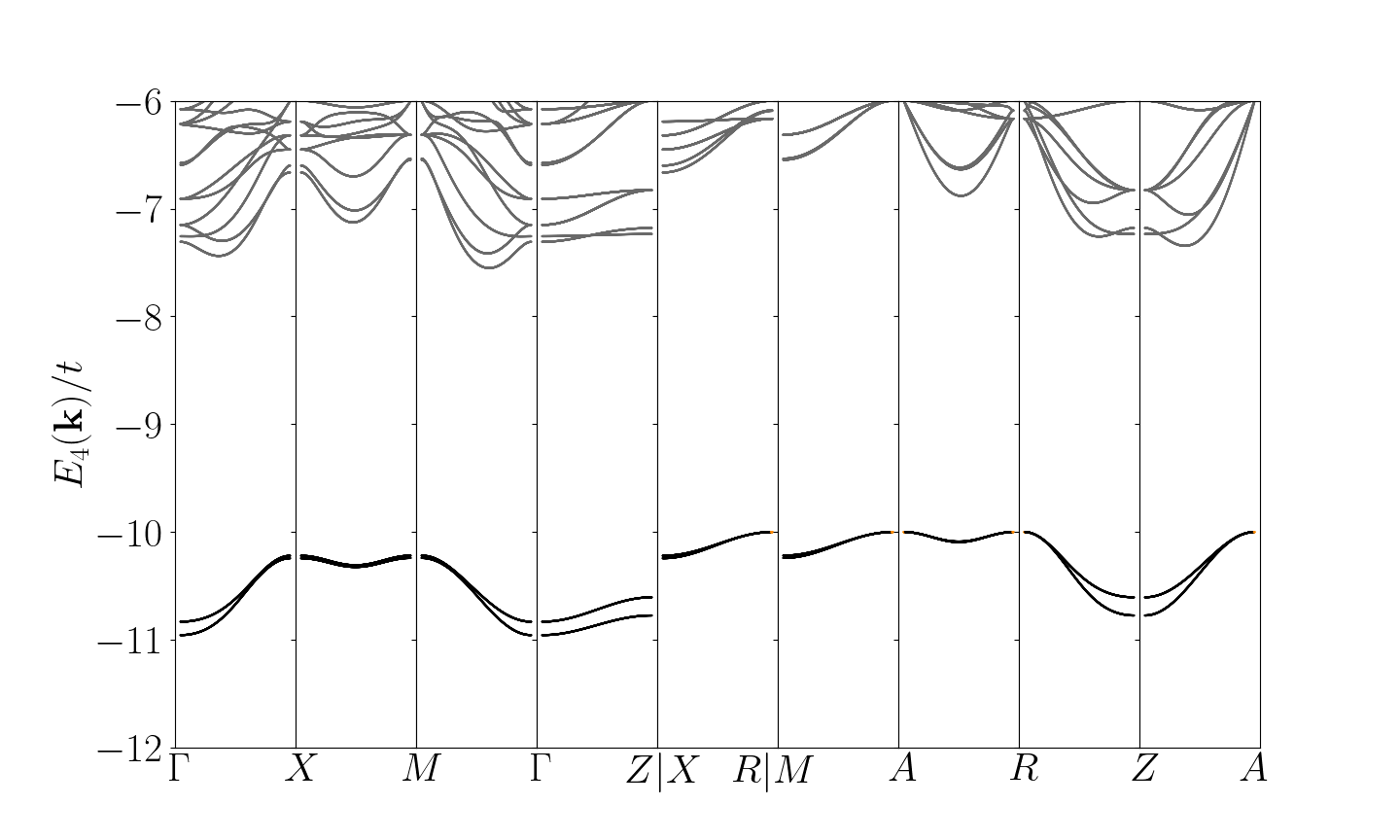}} &
\subfloat[]{\includegraphics[width=\overlap\textwidth]{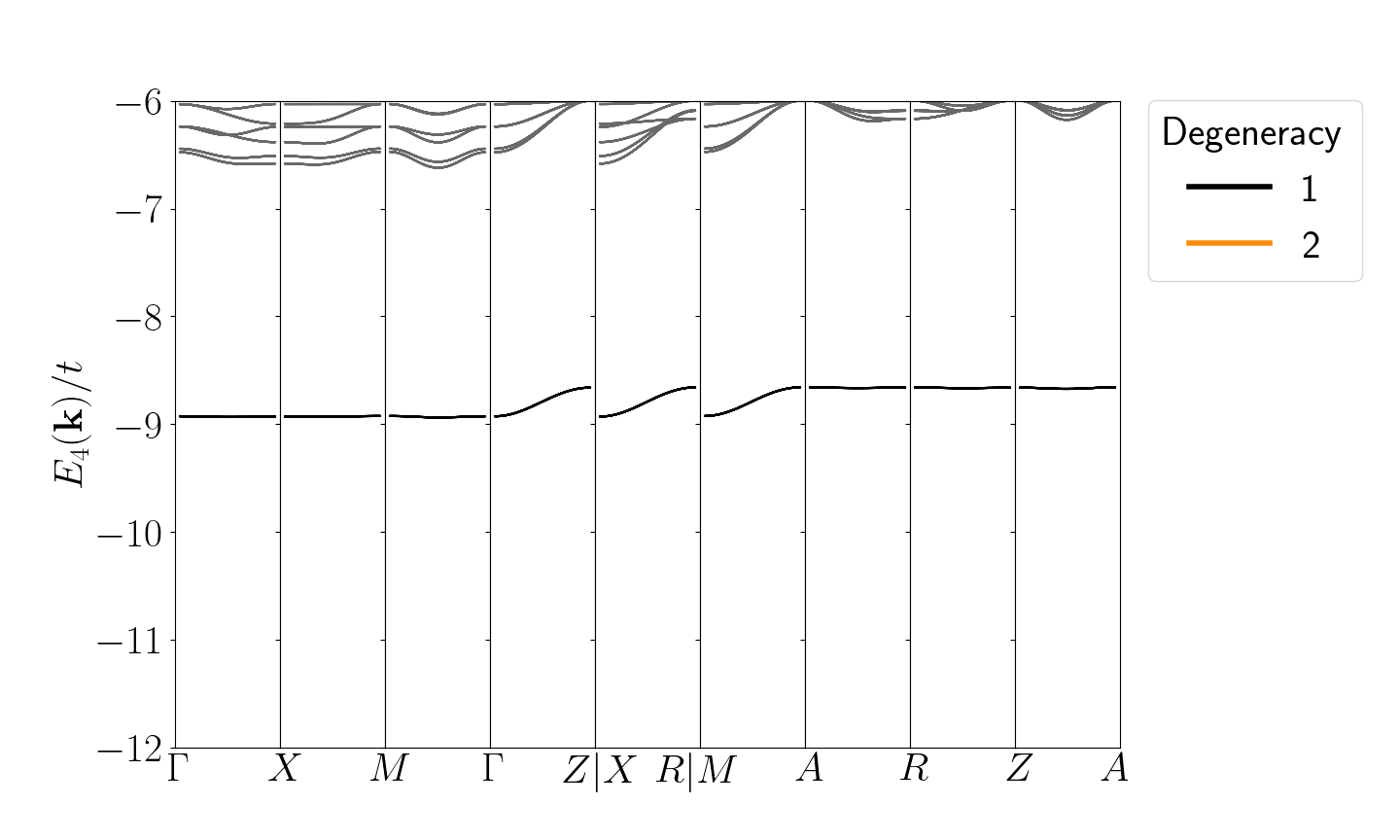}}
\end{tabularx}}
\caption{Energy spectra in the four particle sector, $E_{4}(\mathbf{k})$, for models symmetric under space group $P4_2/mbc1^\prime$ (\# 135) (top row) and space group $P4/ncc1^\prime$ (\#130) (bottom row) with generalized $HK$ interactions.
The low energy states have been colored according to their degeneracy, given in the rightmost panel. 
The degeneracy of the higher energy states has been grayed out for clarity.
(a) shows the spectrum in the four particle sector $E_{4}(\mathbf{k})$ for $H^{0}_{135}+H^{1}_{HK}+H^{2}_{HK}$. 
The ground state is two-fold degenerate at the $A$-point. 
(b) shows the spectrum in the four particle sector for $H^{0}_{135}+H^{1}_{HK}+H^{2}_{HK}+H^{3}_{HK}$. 
Adding $H^{3}_{HK}$ removes the degeneracy at the $A$-point, leading to a ground state which is everywhere gapped and non-degenerate.
(c) shows the spectrum in the four particle sector $E_{4}(\mathbf{k})$ for $H^{0}_{130}+H^{1}_{HK}+H^{2}_{HK}$. 
The ground state is again two-fold degenerate at the $A$-point.
(d) shows the spectrum in the four particle sector for $H^{0}_{135}+H^{1}_{HK}+H^{2}_{HK}+H^{3}_{HK}$. 
Adding $H^{3}_{HK}$ removes the degeneracy at the $A$-point in space group $P4/ncc1^\prime$ (\#130) as well, again leading to a ground state which is everywhere gapped and non-degenerate. 
This violates the filling bound in Ref.~\cite{watanabe2015filling}.
The parameter values for space group $P4_2/mbc1^\prime$ (\# 135) are $t_{xy}=1,t_{z}=0.5,t'_{1}=0.3, t'_2=0.3, \lambda'_1=0.5,\lambda'_2=0.1,\lambda'_3=0.15, U_{1}=4, U_{2}=2, U_{3}=2, \mu_0=(U_{1}+2U_{2})/2=4$.
The parameter values for space group $P4/ncc1^\prime$ (\#130) are $t_{xy}=1,t_z=0.5,\lambda_1=0.3,\lambda_2=0.3,\lambda_3=0.3, U_1=4, U_{2}=2, U_{3}=2, \mu_0=(U_{1}+2U_{2})/2=4$. }
\label{fig:130135allUSpectra}
\end{figure*}

Of the sixty-four possible independent $N$ matrices, only the eight listed in \Cref{table:HKNterms} preserve the symmetries of space group $P4_2/mbc1^\prime$ (\# 135). 
Adding to our initial Hamiltonian:
\begin{equation}\label{eq:hhk2135}
H^{2}_{HK}=U_{\mu^x}\sum_{\mu\tau\sigma}n_{\mu\tau\sigma}n_{-\mu\tau\sigma}+
    U_{\tau^x}\sum_{\mu\tau\sigma}n_{\mu\tau\sigma}n_{\mu-\tau\sigma}
\end{equation}
allows us to lift the sixteen fold degeneracy at the $A$-point into a two-fold degeneracy. 
Note that $H_{HK}^{2}$ is particle-hole symmetric up to an overall shift of the chemical potential. 
 Thus, as discussed in Sec.~\ref{sec:DDSL}, the spectrum of $H_{135}^{0}+H_{HK}^{1}+H_{HK}^{2}$ is approximately particle-hole symmetric at half filling for $t_2'\ll U_1+U_{\mu^x}+U_{\tau^x}$. 
 This means that the ground state still consists of four-particle states at every $\mathbf{k}$. 
 We verify this numerically in Appendix~\ref{sec:phsymm}. 
 We show the spectrum in the four-particle subspace in \Cref{fig:130135allUSpectra}a. 
 The ground state is given by the lowest energy eigenstate at each $\mathbf{k}$. 
 The ground state is twofold degenerate at $A$ and nondegenerate for all other $\mathbf{k}$.
 
We can then consider a further generalization to HK type interactions $H^{3}_{HK}$ of the form:
\begin{equation}
\sum_{i,j,k,l}\left[c^{\dagger}_{\mathbf{k}i}F^{1}_{ij}c_{\mathbf{k}j}\right]\left[c^{\dagger}_{\mathbf{k}l}F^{2}_{lm}c_{\mathbf{k}m}\right],
\end{equation}
Where $i,j,k,l$ run over all eight possible $(\mu,\tau,\sigma)$ indices. 
We again exhaustively search over all terms to find those that preserve the symmetries.
In this case, there is an additional constraint due to the fact that space group $P4_2/mbc1^\prime$ (\# 135) is nonsymmorphic \cite{wieder2018wallpaper}.
There are only three such terms, provided with the details of the derivation in \Cref{sec:symmetry}. 
We add one of them, $H^{3}_{HK}$, to our Hamiltonian:
\begin{equation}
H^{3}_{HK}=U_{\tau^x\tau^x}\left[\sum_{\mu\tau\sigma}c^\dagger_{\mu\tau\sigma}c_{\mu-\tau\sigma}\right]
\left[\sum_{\mu'\tau'\sigma'}c^\dagger_{\mu'\tau'\sigma'}c_{\mu'-\tau'\sigma'}\right],
\end{equation}
and note that $H^{3}_{HK}$ is explicitly particle-hole symmetric, as we discuss in Appendix~\ref{sec:phsymm}. 
The Hamiltonian $H^{3}_{135}=H^{0}_{135}+H^{1}_{HK}+H^{2}_{HK}+H^{3}_{HK}$ results in the spectrum shown in \Cref{fig:130135allUSpectra}b.
This ground state is everywhere gapped and non-degenerate, with four particles at every $\mathbf{k}$. 

\begin{figure*}
\FPeval{\overlap}{0.53}
\FPeval{\scalevalue}{round(0.5*(1.12)/\overlap,2)}
\scalebox{\scalevalue}{
\setlength{\tabcolsep}{0pt} 
\def\tabularxcolumn#1{m{#1}}
\hskip-1.0cm\begin{tabularx}{\textwidth}{@{}XXX@{}}
\subfloat[]{\includegraphics[width=\overlap\textwidth]{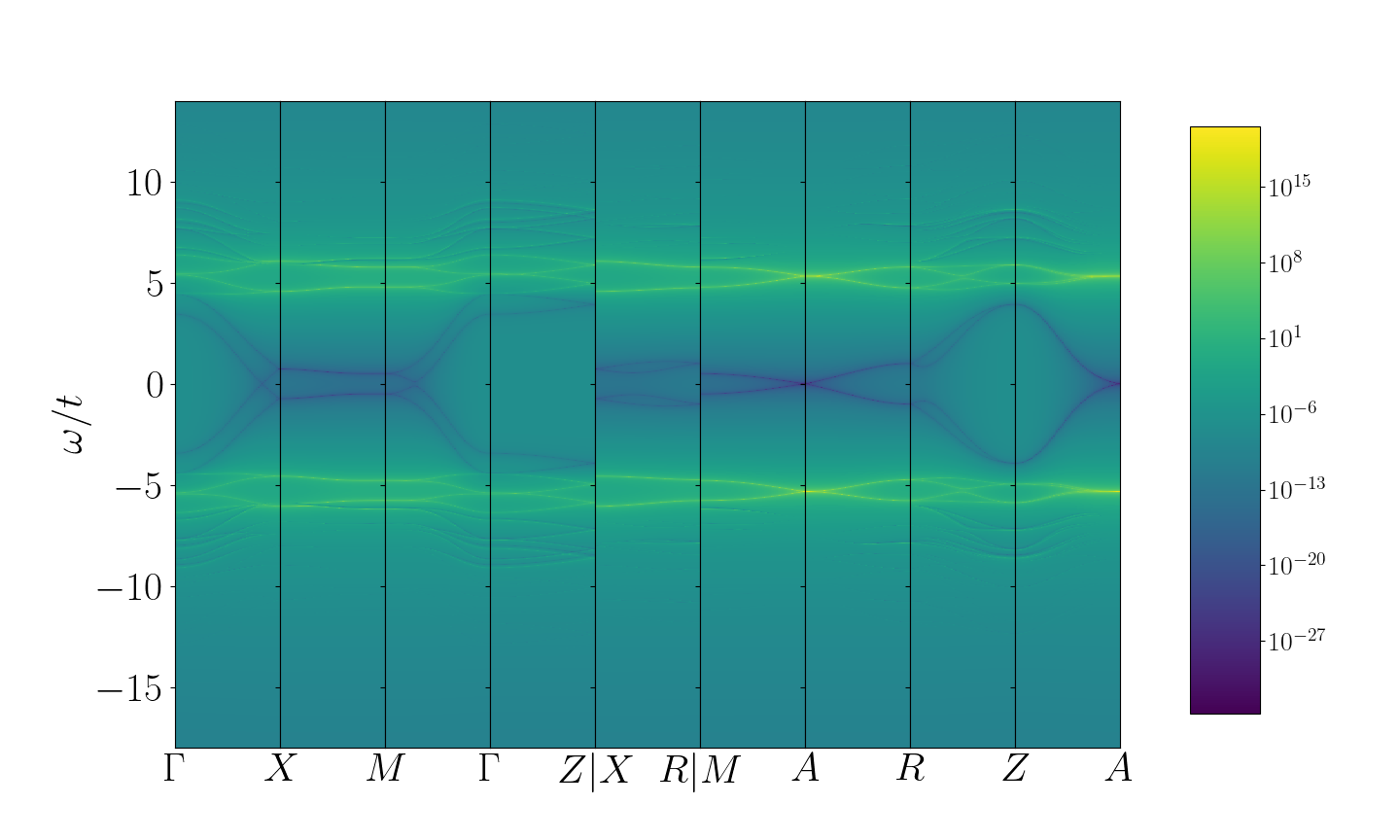}} &
\subfloat[]{\includegraphics[width=\overlap\textwidth]{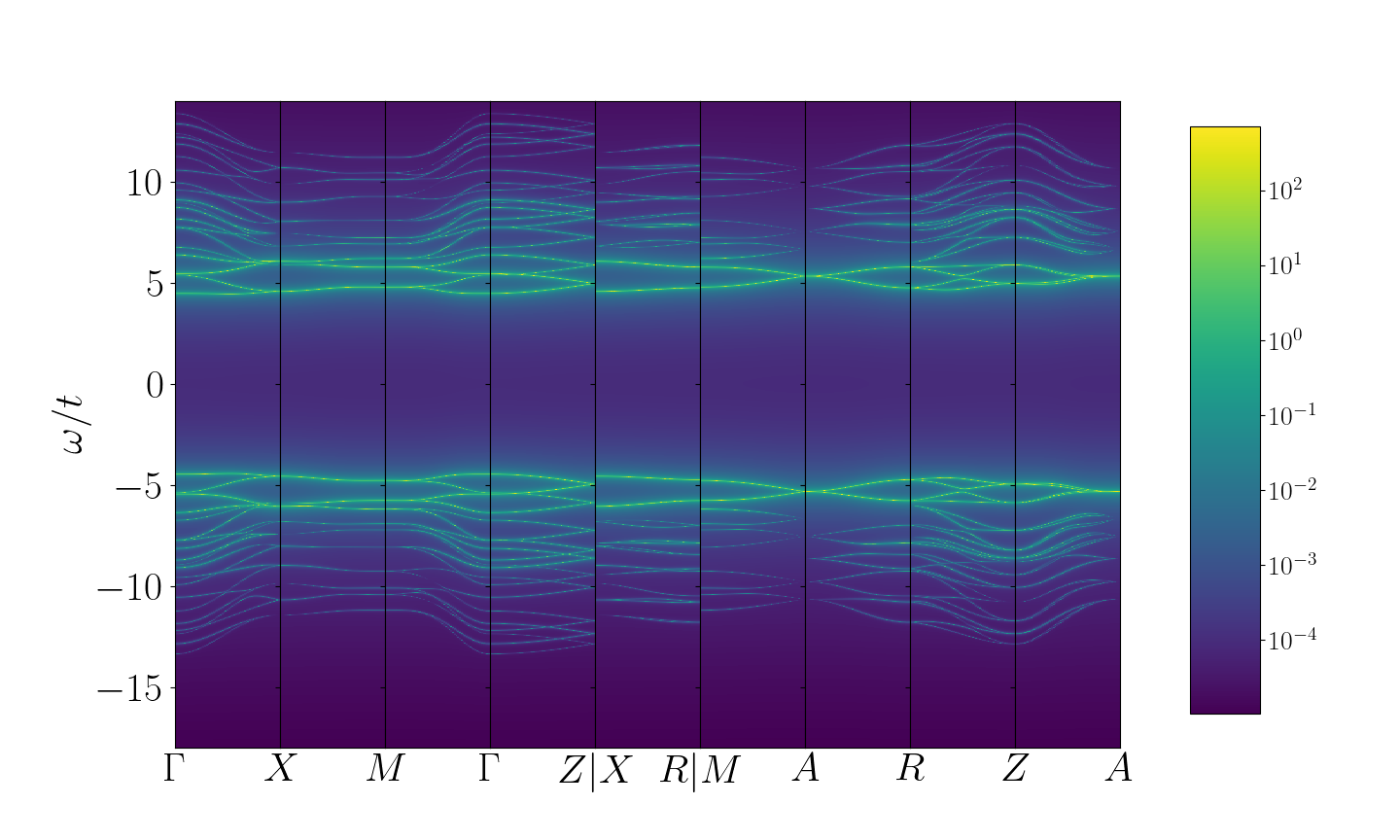}}
 \\[-3ex]
\subfloat[]{\includegraphics[width=\overlap\textwidth]{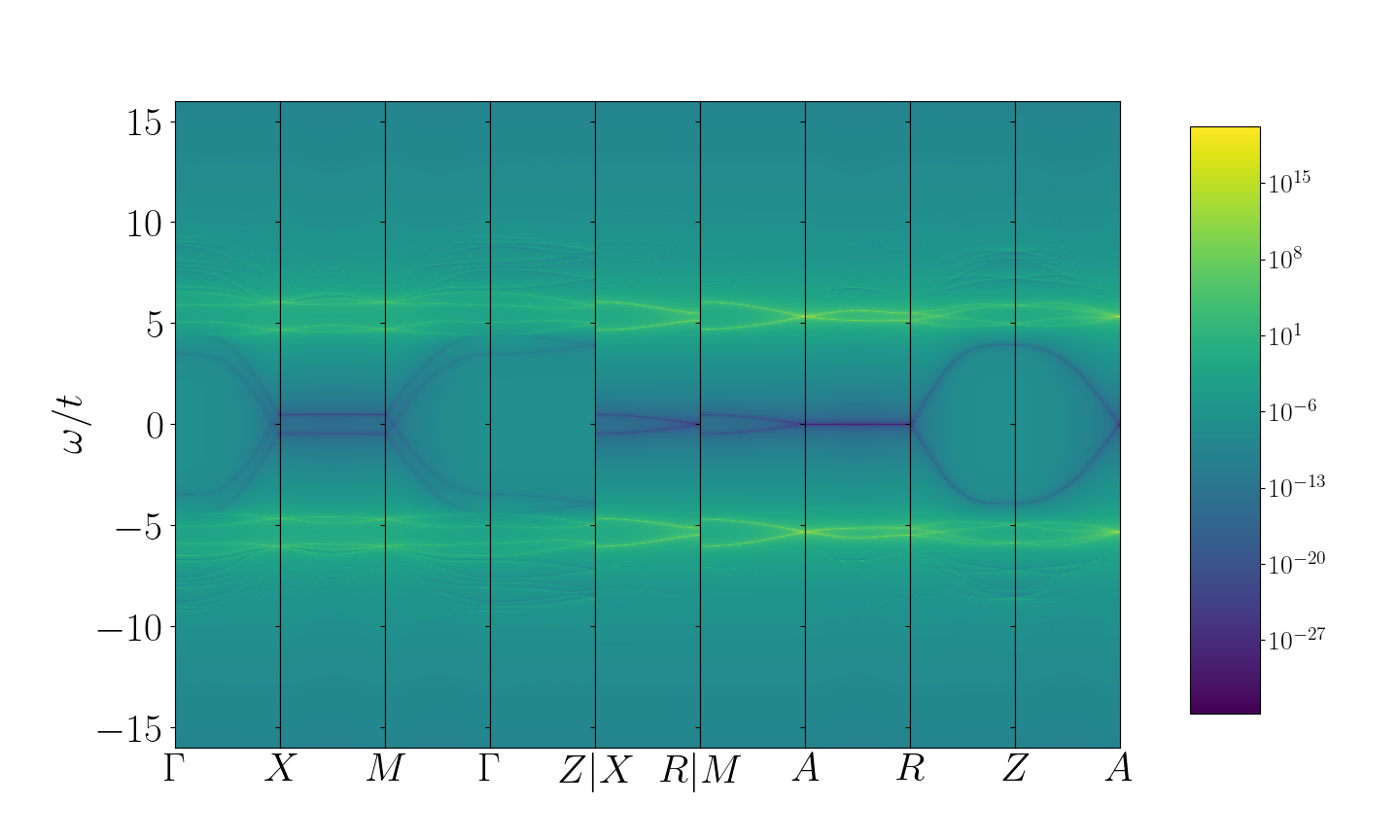}} &
\subfloat[]{\includegraphics[width=\overlap\textwidth]{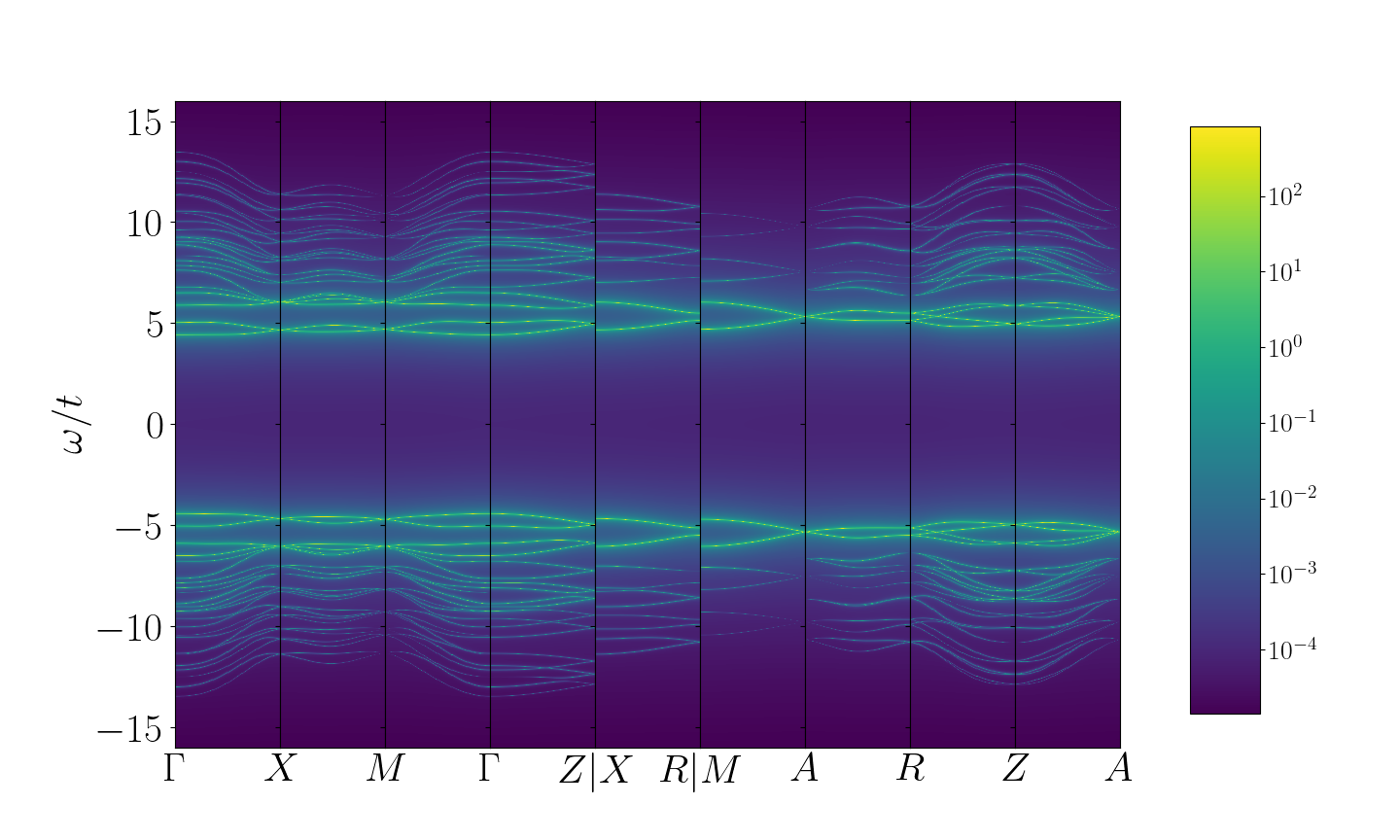}}
\end{tabularx}}
\caption{Green function determinant and spectral function for Hamiltonians invariant under space groups $P4_2/mbc1^\prime$ (\# 135) (top row) and $P4/ncc1^\prime$ (\#130) (bottom row) with $HK$ interactions $H^{1}_{HK}+H^{2}_{HK}+H^{3}_{HK}$ at half filling (four electrons per unit cell).
In the left two panels (a) and (c), we show the absolute value of the determinant $|\det(G^{+}(\mathbf{k},\omega))|$, and in the right two panels (b) and (d) the spectral function $-\frac{1}{\pi}\Im\Tr(G^{+}(\mathbf{k},\omega))$. The band of zeros in the determinant of the orbital model demonstrates that models in both space groups remain Mott insulators when the ground state becomes non-degenerate.
(a) shows $|\det(G^+(\mathbf{k},\omega))|$ for the space group $P4_2/mbc1^\prime$ (\# 135) Hamiltonian $H^{0}_{135}+H^{1}_{HK}+H^{2}_{HK}+H^{3}_{HK}$.
(b) shows the spectral function, $-\frac{1}{\pi}\Im\Tr(G^{+}(\mathbf{k},\omega))$, for $H^{0}_{135}+H^{1}_{HK}+H^{2}_{HK}+H^{3}_{HK}$.
(c) shows $|\det(G^{+}(\mathbf{k},\omega))|$ for the space group $P4/ncc1^\prime$ (\#130) Hamiltonian $H^{0}_{130}+H^{1}_{HK}+H^{2}_{HK}+H^{3}_{HK}$ . 
(d) shows  the spectral function, $-\frac{1}{\pi}\Im\Tr(G^{+}(\mathbf{k},\omega))$, for the space group $P4/ncc1^\prime$ (\#130) Hamiltonian $H^{0}_{130}+H^{1}_{HK}+H^{2}_{HK}+H^{3}_{HK}$.
The parameter values for space group $P4_2/mbc1^\prime$ (\# 135) are $t_{xy}=1,t_{z}=0.5,t'_{1}=0.3, t'_2=0.3, \lambda'_1=0.5,\lambda'_2=0.1,\lambda'_3=0.15, U_{1}=4, U_{2}=2, U_3=4 \mu_0=4$.
The parameter values for space group $P4/ncc1^\prime$ (\#130) are $t_{xy}=1,t_z=0.5,\lambda_1=0.3,\lambda_2=0.3,\lambda_3=0.3, U_1=4, U_2=2, U_3=2, \mu_0=4$. }
\label{fig:130135allUGF}

\end{figure*}

We plot the determinant of the Green function and the spectral function for the Hamiltonian $H^{3}_{135}=H^{0}_{135}+H^{1}_{HK}+H^{2}_{HK}+H^{3}_{HK}$ in Figs.\ref{fig:130135allUGF}(a) and (b). 
We see a clear gap between the poles corresponding to the lower and upper Hubbard band excitations. 
We have also confirmed numerically that this ground state transforms under the trivial representation at the $A$ point. 
This implies that there is no spontaneous symmetry breaking in this ground state. 
Nevertheless the determinant of the Green function has a band of zeros in the single particle gap. 
We also confirm numerically that the eigenvalues of the determinant of the Green function matrix are consistent with the symmetry allowed degeneracy in space group $P4_2/mbc1^\prime$ ($\#135$); in particular the zero eigenvalues at the $A$-point are eightfold degenerate. 
Thus, we arrive at a gapped, symmetric, nondegenerate Mott insulating ground state.

\section{Minimal Insulating Filling in HK Models}\label{sec:comparison}

We have seen from Figs.~\ref{fig:130135allUSpectra}(a), (b) and Figs.~\ref{fig:130135allUGF}(a) and (b) that the Hamiltonian $H^{0}_{135}+H^{1}_{HK}+H^{2}_{HK}+H^{3}_{HK}$ realizes a gapped, nondegenerate ground state at filling $\nu=4$ electrons per unit cell that does not spontaneously break any symmetries and is not adiabatically connected to a free fermion band insulator. 
Our ground state is thus a candidate for the featureless interacting insulator in space group $P4_2/mbc1^\prime$ (\# 135) allowed for by the LSM theorems of Ref.~\cite{watanabe2015filling}. 
However, unlike the family of Hamiltonians considered in Ref.~\cite{watanabe2015filling}, our HK model contains arbitrarily long-range interactions. 
This opens up the possibility that the ground state of $H^{0}_{135}+H^{1}_{HK}+H^{2}_{HK}+H^{3}_{HK}$ evades the LSM theorem by being a long-range entangled cat state. 
To see whether this is the case, we consider a similar HK model in space group $P4/ncc1^\prime$ (\# 130), which also realizes a double-Dirac semimetal in the noninteracting limit. 
The LSM theorem in space group $P4/ncc1^\prime$ (\# 130), however, forbids a featureless insulator at $\nu=4$. 
Nevertheless, we will see that we can construct an HK model in space group $P4/ncc1^\prime$ (\# 130) with a gapped, nondegenerate, symmetric ground state at filling $\nu=4$.

\subsection{A Comparison: Space Group $P4/ncc1^\prime$ (\#130)}
We now ask whether we can also turn a Hamiltonian invariant under space group $P4/ncc1^\prime$ (\#130) into a featureless insulator. 
We use the same non-interacting Hamiltonian that was considered in Ref.~\cite{wieder2016double}\footnote{We are grateful to Benjamin Wieder for pointing out a misprint in this paper. 
The first spin orbit coupling term with coefficient $\lambda_1$ should have a $\sigma^z$.}:
\begin{align}
    \label{eq:130noninteractingH}
    \mathcal{H}^{0}_{130}(\mathbf{k})&=\mathcal{H}^{1}_{130}(\mathbf{k})+\mathcal{H}^{2}_{130}(\mathbf{k})\\
    \mathcal{H}^{1}_{130}(\mathbf{k})&=t_{xy}\tau^x\cos\tfrac{k_x}{2}\cos\tfrac{k_y}{2}+t_{z}\mu^x\cos\tfrac{k_z}{2}\\
    \mathcal{H}^{2}_{130}(\mathbf{k})&=\lambda_1\mu^y\tau^z\sigma^z\cos\tfrac{k_z}{2}+\lambda_2\tau^z(\sigma^x\sin k_y-\sigma^y\sin k_x)\nonumber\\
    +&\lambda_3\mu^z\tau^x\left(\sigma^x\sin\tfrac{k_x}{2}\cos\tfrac{k_y}{2}+\sigma^y\cos\tfrac{k_x}{2}\sin\tfrac{k_y}{2}\right)
\end{align}
With the corresponding Hamiltonian given by:
\begin{equation}\label{eq:130ham0}
    H^{0}_{130}=\sum_{\mathbf{k}, i\sigma,j\sigma'}c^\dagger_{\mathbf{k}i\sigma}(\mathcal{H}^{0}_{130}(\mathbf{k}))_{i\sigma,j\sigma'}c_{\mathbf{k}j\sigma'}
\end{equation}

The single particle spectrum is given in the bottom right panel of \Cref{fig:SPandCubes}.
We then consider adding to \Cref{eq:130ham0} the original orbital HK interaction $H^{1}_{HK}$ from \Cref{eq:sg135orbitalhk}:
\begin{equation}\label{eq:sg130u1ham}
    H^{1}_{130}=H^{0}_{135}+H^1_{HK}.
\end{equation}
As we show in Appendix~\ref{sec:phsymm}, the ground state of \Cref{eq:sg130u1ham} at half filling $\nu=4$ consists of four electrons at every $\mathbf{k}$, and the low-energy excitation spectrum is given by excitations in the four particle sector. 
We show the spectrum in the four particle subspace in \Cref{fig:130135allUSpectra}[c-d]. 
As for space group $P4_2/mbc1^\prime$ (\# 135), we find a ground state that is only degenerate at the $A$ point. 

We then add the generalized HK interactions, $H^{2}_{HK},H^{3}_{HK}$ to \Cref{eq:sg130u1ham}, to obtain the Hamiltonian
\begin{align}
\label{eq:130fullinteractingH}
H_{130}=H^{0}_{130}+H^{1}_{HK}+H^2_{HK}+H^{3}_{HK},
\end{align}
As shown in Appendix~\ref{sec:phsymm}, the low energy physics of $H_{130}$ is dominated by excitations in the four-particle sector, and the spectrum of $H_{130}$ in the four particle sector is shown \Cref{fig:130135allUSpectra}[c-d]. 
The spectrum is everywhere gapped and non-degenerate, which means that our model in space group $P4/ncc1^\prime$ (\#130) with $\nu=4$ electrons per unit cell can also be made insulating without breaking any of the symmetries.

We additionally show the determinant of the Green function and the spectral function for our model in \Cref{fig:130135allUGF}(c) and (d). 
As for our model in space group $P4_2/mbc1^\prime$ (\# 135), the gap between the poles in the spectral function and the band of zeros in the determinant of the Green function confirms that the ground state is a Mott insulator.
.

A gapped, nondegenerate, symmetric, short-range entangled ground state in space group $P4/ncc1^\prime$ (\#130) at filling $\nu=4$ would explicitly violate the LSM theorem of Ref.~\cite{watanabe2015filling}. 
Since we have constructed a ground state that is gapped, nondegenerate, and symmetric at filling $\nu=4$, we thus must conclude that it is not short-range entangled. 
To understand this, we can revisit the proof of the LSM theorem for nonsymmorphic space groups. 
The derivation of the filling bounds relies on the local properties of the system being invariant under a change of the boundary conditions. 
Since HK interactions are infinitely long range, however, every point in the bulk is coupled to the boundary and so the theorem breaks down. 

This suggests that our models in space groups $P4_2/mbc1^\prime$ (\# 135) and $P4/ncc1^\prime$ (\#130) realize insulators at filling $\nu=4$ because the ground states are long-range entangled via the HK interaction. 
This long-range entanglement need not be topological, however. 
As pointed out in Ref.~\cite{watanabe2015filling}, cat states such as the symmetric linear combination of two opposite ferromagnetic states have just the sort of long-range entanglement needed to violate the LSM theorem.

This in turn raises three questions. 
First, we can ask what filling constraints, if any, carry over to HK models from the LSM theorem for systems invariant under time-reversal and crystal symmetries. 
For the case of a 1D, translationally invariant HK model it was shown in Ref.~\cite{lsmyin2022} that an insulating state must have an integer number of electrons per unit cell. Since it is simpler to calculate the filling than to determine whether the system is insulating, we are primarily concerned here with the converse statement: whether a given filling implies an insulating state.
For the case of time-reversal and crystal symmetric systems, we conjecture that the only remaining constraints beyond integer filling are those that arise from the on-site symmetries--i.e. from Kramers' theorem. 
To support this conjecture, we show in \Cref{sec:SG14} that the LSM bound is violated by an HK model in space group $P4_1/c1^\prime$ (\# 14) at filling $\nu=2$, which represents the smallest enhancement of the LSM theorem by nonsymmorphic symmetries. 
Second, since Hubbard models can generate short-range entangled phases that \textit{are} subject to the bound, we have found an explicit example in which the phase of the HK model (insulating for space group $P4/ncc1^\prime$ (\#130) at half-filling) is different to the phase of a Hubbard model required by the LSM theorem (i.e. metallic, magnetic, or topologically-ordered for space group $P4/ncc1^\prime$ (\#130) at half-filling). 
This raises the question of which topological properties of the Hubbard model can be inferred from HK models from the renormalization group arguments of Refs.~\cite{2022Philipuniversalityargument,zhao2023proof}. 
Third, since the HK model is local in momentum space, it would be interesting to explore whether any constraints analogous to LSM theorems arise from considering the momentum space (orbital) entanglement spectrum~\cite{li2008entanglement,chandran2011bulk}.

\section{Conclusion}\label{sec:conclusion}
In this work, we have initiated a systematic exploration of the ground state degeneracy and filling constraints in long-range HK type models. 
First, in Sec.~\ref{sec:orbital-hk} we introduced a general class of HK models with interactions written in the orbital basis. 
We showed that previously-considered band-basis HK models arise as a special case. 
We showed how orbital HK models can be solved via a mapping to a few-site Hubbard model at each $\mathbf{k}$. 
The main advantage of orbital-basis HK models is that they have ground states which are in general nondegenerate except possibly at isolated high-symmetry $\mathbf{k}$ points in the BZ. 
This is in contrast to band-basis HK models which have thermodynamically large ground state degeneracies (there is a degeneracy at every $\mathbf{k}$) in the strongly-interacting Mott insulator regime. 
Because of this, orbital-basis HK models are stable against symmetry-breaking order. 
We showed this explicitly through an analysis of a tight-binding model of graphene with a simple orbital HK interaction in Sec.~\ref{subsec:graphene}. 
Additionally, the orbital-basis allows for a natural treatment of HK models with spin-orbit coupling, such as the Kane-Mele model considered in Sec.~\ref{subsec:HKKM}. 

To fully demonstrate the utility of orbital HK models, we next turned our attention in Sec.~\ref{sec:DDSL} to orbital HK models with the symmetries of space group $P4_2/mbc1^\prime$ (\# 135), where LSM theorems seem to allow for a gapped, symmetric, short-range-entangled insulator at filling $\nu=4$ that is nevertheless not adiabatically connected to a band insulator. 
We presented a model of an interacting double-Dirac semimetal in space group $P4_2/mbc1^\prime$ at filling $\nu=4$ with orbital HK interactions, and showed that the ground state is a Mott insulator with a sixteenfold degeneracy at the $A$ point in the Brillouin zone. 
We identified the low-energy excitations above the ground state with spin waves, and showed that the system could be mapped via a Schrieffer-Wolff transformation to a long-range interacting spin model. 
Next, we showed through a careful symmetry analysis of possible HK interactions in Sec.~\ref{sec:General Interactions} that there exists an orbital HK model in space group $P4_2/mbc1^\prime$ at filling $\nu=4$ that has a gapped, symmetric, nondegenerate ground state. 
Furthermore, an analysis of the single-particle Green function demonstrates that the ground state is a Mott insulator that is not adiabatically connected to a band insulator. 
However, we show that the orbital HK interaction stabilizes long-range entanglement of seemingly nontopological origin: the symmetric ground state is, at each (star of) $\mathbf{k}$ a superposition of orbital- and spin-ordered states that transforms trivially under the symmetries of the space group. 
While for short-range Hamiltonians such cat states always come in degenerate pairs (i.e., the two ground states of the ferromagnetic Ising model with opposite spins), the HK interaction stabilizes a cat state as a nondegenerate ground state. 
We provide evidence for this picture by considering in Sec.~\ref{sec:comparison} an orbital HK model with symmetries of space group $P4/ncc1^\prime$ (\# 130) at filling $\nu=4$, where LSM theorems forbid the existence of any gapped, symmetric, short-range entangled, nondegenerate ground state. 
Nevertheless, our orbital HK model has a gapped, symmetric, nondegenerate ground state, which therefore must have long-range entanglement. 

Our work shows that the features of Mott physics that are captured by HK models are not due to any ground state degeneracy. 
Our orbital HK models with nondegenerate or order-one degenerate ground states are nevertheless Mott insulators---they feature a charge gap, do not spontaneously break any symmetry in the ground state, and have midgap zeros of their single-particle Green functions that point to violations of Luttinger's theorem. 
The lack of an extensive ground state degeneracy also shows that the Mott insulating ground state of our orbital HK models is stable to symmetry-breaking perturbations. 
This is in contrast to band-HK models, which become adiabatically connected to trivial insulators under the application of an infinitesimal symmetry breaking field.

Additionally, our work highlights several delicate issues that must be addressed in studies that use HK models as a probe of topological order. 
We have seen through our study of filling constraints that the ground states of HK-type Hamiltonians can violate LSM theorems. 
This arises in part because proofs of the LSM theorems for crystalline symmetries rest on the assumption of a short-range Hamiltonian. 
In both entanglement based and flux-insertion based proofs, it is crucial that the Hamiltonian in position space at a point deep in the bulk of the system is insensitive to perturbations at a distant point. 
This assumption is maximally violated for HK-like Hamiltonians, which contain position-space couplings of arbitrarily long range. 
In our examples of graphene and of Mott insulators in space groups $P4_2/mbc1^\prime$ (\# 135), $P4/ncc1^\prime$ (\# 130), and $P4_1/c1^\prime$ (\# 14), this resulted in a ground state that was nondegenerate but long-range entangled. 
Care must be taken to distinguish this ``trivial'' cat-state like long-range entanglement from more exotic topological order in any HK model.  Furthermore, our results suggest that even in HK models of topologically ordered phases, the long-range HK interaction may split the otherwise topologically protected ground state degeneracy. 
Thus, one cannot say by looking at the ground state degeneracy alone whether or not the ground state of an HK Hamiltonian has topological order. 

Our work opens up several intriguing directions for future research. 
First, although we focused on half-filled systems in this work, orbital HK models for quarter filled systems can be studied using the same techniques. 
Examinations of partially-filled quantum anomalous Hall and quantum spin Hall bands with the band-HK interaction have shown several signatures of nontrivial topology, but have been complicated by the presence of an extensive ground state degeneracy and ferromagnetic instability~\cite{2022PhilipSpinHallHK,mai2023topological}. 
Orbital-HK models for these topological Mott insulators can allow for an analysis of robust observables that do not depend on the ground state degeneracy, such as the Hall conductivity~\cite{zhao2023failure}. 
Second, recent work has renewed interest in the study of topological invariants of zeros and poles of the single-particle Green function in interacting Mott insulators~\cite{setty2023symmetry,wagner2023mott}. 
Our analysis of orbital HK models in graphene and in three-dimensional nonsymmorphic space groups has shown that generically there are both Green function zeros and Green function poles in the lower and upper Hubbard bands. 
While the degeneracy of zeros and poles are both restricted by space group symmetries, their coexistence raises interesting possibilities for computing topological properties of the Hubbard band and Luttinger surface eigenstates. 
Third, since the Schrieffer-Wolff approximation allows us to approximate the orbital HK Hamiltonian as a long-range spin model, it would be interesting to explore the connection between the HK ground state and spin liquids, or other spin-charge separated descriptions of Mott insulators. 
Finally, although we have presented evidence that our exotic ground states in space groups $P4_2/mbc1^\prime$ (\# 135), $P4_1/c1^\prime$ (\# 14), and $P4/ncc1^\prime$ (\#130) at filling $\nu=4$ have a trivial form of long-range entanglement, we have not ruled out the presence of hidden topological order. 
A systematic study of the ground state properties of non-degenerate orbital HK ground states beyond single-particle observables could be explored in future work.

\begin{acknowledgments}
We thank  J. Gliozzi, E. Huang, C.~L. 
Kane, P.~Phillips, and J. Zhao for useful discussion. 
We also thank B.~Wieder for pointing out corrections to Ref.~\cite{wieder2016double}, given in Footnotes [52] and [59]. 
This work was supported by the Alfred P Sloan Foundation, and the National Science Foundation under grant DMR-1945058.
\end{acknowledgments}

\onecolumngrid
\appendix

\section{Schrieffer-Wolff Hamiltonian}
\label{sec:Schrieffer-Wolff Appendix}
We found in \Cref{eq:Schrieffer-Wolff} that the matrix elements of the Schrieffer-Wolff approximation to the Hamiltonian $H_{135}^{0}+H_{HK}^{1}$ [\Cref{eq:135 Hamiltonian,eq:sg135orbitalhk}] are:
\begin{equation}
\bra{i}H_{SW}\ket{j}=\bra{i}H_{0}\ket{j}+\frac{1}{U_{1}}\bra{i}V^{2}\ket{j}
\end{equation}
Since the charge degrees of freedom are frozen out, we expect this Hamiltonian to consist solely of couplings between spin operators at the four sublattice sites of the form:
\begin{equation}
    S^{i}_{\mu_{1}\tau_{1}}S^{j}_{\mu_{2}\tau_{2}}
\end{equation}
where $i,j$ are Pauli matrix indices running over $0,x,y,z$, and the spin operator $S^{i}_{\mu_1\tau_1}$ is defined as:
\begin{equation}
    S^{i}_{\mu_1\tau_1}\equiv \sum_{k,l}c^\dagger_{\mu\tau\sigma_k}[\sigma^{i}]^{kl}c_{\mu\tau\sigma_l}
\end{equation}
To decompose this Hamiltonian into spin operators, we use the fact that the spin operators are orthonormal under the trace:
\begin{equation}
\Tr([S^{i}_{\alpha}S^{j}_{\beta}][S^{k}_{\gamma}S^{l}_{\epsilon}])=\lambda^{ij}_{\alpha\beta}\delta_{\alpha\gamma}\delta_{\beta\epsilon}\delta^{ik}\delta^{jl}
\end{equation}
Where $\alpha,\beta,\gamma,\delta$ are a shorthand for a set of $\mu$ and $\tau$ sublattice indices, and $\lambda^{ij}_{\alpha\beta}$ is a normalization constant that in general depends on the indices. 

We can compute the coefficient $\alpha^{ij}_{\mu_1\tau_1\mu_2\tau_2}$ of each spin matrix pair in the Hamiltonian as:
\begin{equation}
    \alpha^{ij}_{\alpha\beta}=\frac{1}{\lambda^{ij}_{\alpha\beta}}\Tr(H_{SW}S^{i}_{\mu_{1}\tau_{1}}S^{j}_{\mu_{2}\tau_{2}})
\end{equation}

We can now exhaustively iterate over all 256 possible spin matrices to find the decomposition of $H_{SW}$.

We split the Hamiltonian into three parts: an identity term, a non-spin orbit coupling term, and a spin-orbit coupling term.
To make it easier to look at, we also use the shorthand that the coefficient includes the $\mathbf{k}$ dependent hopping, for example $\lambda_{11}\equiv\lambda_{1}\sin\tfrac{k_x}{2}\cos\tfrac{k_y}{2}\sin\tfrac{k_z}{2}(\mathbf{k})$. 
We have:
\begin{equation}
H_{I}=\mathds{1}\left[4(t^2_{xy}+t^2_{z}+\lambda^2_{3})+10(t^2_{2}+\lambda^2_{11}+\lambda^2_{12}+\lambda^2_{21}+\lambda^2_{22})\right]
\end{equation}
    
\begin{align}
H_{1}&=(t^2_{xy}+t^2_{z}+2t^2_{2})\left(\sum_{\substack{\mu_1,\tau_1,\mu_2,\tau_2\\(\mu_1,\tau_1)\neq(\mu_2,\tau_2))}}S^{0}_{\mu_{1}\tau_{1}}S^{0}_{\mu_{2}\tau_{2}}\right)-\frac{1}{2}\sum_{i,\mu,\tau}\left[t^2_{xy}S^{i}_{\mu\tau}S^{i}_{\mu(-\tau)}
+t^2_{z}S^{i}_{\mu\tau}S^{i}_{(-\mu)\tau}\right]
\end{align}

\begin{align}
H_{SO}&=2\left(\lambda_{11}^2+\lambda_{12}^2+\lambda_{21}^2+\lambda_{22}^2+\frac{\lambda_{3}^2}{2}\right)\left(\sum_{\substack{\mu_1,\tau_1,\mu_2,\tau_2\\(\mu_1,\tau_1)\neq(\mu_2,\tau_2))}}S^{0}_{\mu_{1}\tau_{1}}S^{0}_{\mu_{2}\tau_{2}}\right) \\
&+\left(\lambda_{11}^2+\lambda_{12}^2+\lambda_{21}^2+\lambda_{22}^2\right)\left(\sum_{i}S^{i}_{AA}S^{i}_{BB}+S^{i}_{AB}S^{i}_{BA}\right)-2(\lambda^2_{11}+\lambda^{2}_{21})(S^{x}_{AA}S^{x}_{BB}+S^{x}_{AB}S^{x}_{BA}) \nonumber \\
&-2(\lambda^2_{21}+\lambda^{2}_{22})(S^{y}_{AA}S^{y}_{BB}+S^{y}_{AB}S^{y}_{BA})+\frac{\lambda^{2}_3}{2}\sum_{i,\mu,\tau}\left[S^{i}_{\mu\tau}S^{i}_{\mu(-\tau)}-2S^{z}_{\mu\tau}S^{z}_{\mu(-\tau)}\right] \nonumber\\
&+2t_{2}\left[
((\lambda_{11}+\lambda_{21})(S^{z}_{AA}S^{y}_{BB}-S^{y}_{AA}S^{z}_{BB})+(\lambda_{11}-\lambda_{21})(S^{z}_{AB}S^{y}_{BA}-S^{y}_{AB}S^{z}_{BA})
\right]\nonumber \\
&+t_{2}\bigg[
((\lambda_{11}+\lambda_{21})(S^{x}_{AA}S^{z}_{BB}-S^{z}_{AA}S^{x}_{BB}+S^{z}_{BB}S^{x}_{AA}-S^{x}_{BB}S^{z}_{AA})\nonumber \\
&+(\lambda_{11}-\lambda_{21})(S^{x}_{AB}S^{z}_{BA}-S^{z}_{AB}S^{x}_{BA}+S^{z}_{BA}S^{x}_{AB}-S^{x}_{BA}S^{z}_{AB})\bigg]\nonumber \\
&-\left[(\lambda_{11}\lambda_{21}+\lambda_{21}\lambda_{22})+(\lambda_{11}\lambda_{22}+\lambda_{12}\lambda_{21})\right]
(S^{x}_{AA}S^{y}_{BB}+S^{y}_{AA}S^{x}_{BB}+S^{x}_{BB}S^{y}_{AA}+S^{y}_{BB}S^{x}_{AA}) \nonumber\\
&+\left[(\lambda_{11}\lambda_{22}+\lambda_{12}\lambda_{21})-(\lambda_{11}\lambda_{12}+\lambda_{21}\lambda_{22})\right]
(S^{x}_{AB}S^{y}_{BA}+S^{y}_{AB}S^{x}_{BA}+S^{x}_{BA}S^{y}_{AB}+S^{y}_{BA}S^{x}_{AB})\nonumber\\
&+2\bigg[(\lambda_{22}\lambda_{12}-\lambda_{11}\lambda_{21})(S^{x}_{AA}S^{x}_{BB}-S^{y}_{AA}S^{y}_{BB}-S^{x}_{AB}S^{x}_{BA}+S^{y}_{AB}S^{y}_{BA})-\nonumber\\
&(\lambda_{22}\lambda_{12}+\lambda_{11}\lambda_{21})(S^{z}_{AA}S^{z}_{BB}+S^{z}_{AB}S^{z}_{BA})\bigg] \nonumber \\
&+t_{xy}\lambda_{3}\left[
    S_{AA}^{x}S_{AB}^{y}-S_{AA}^{y}S_{AB}^{x}
    +S_{AB}^{y}S_{AA}^{x}-S_{AA}^{y}S_{AA}^{x}
    +S_{BA}^{y}S_{BB}^{x}-S_{BA}^{y}S_{BB}^{x}
    +S_{BB}^{x}S_{AA}^{y}-S_{BB}^{y}S_{BA}^{x}
    \right]
\end{align}

\section{Generalized HK interaction terms preserving the symmetry}
\label{sec:symmetry}
In this section we explain how we calculated the symmetry preserving terms for the generalized HK interactions. 
Each term needs to satisfy two conditions:
\begin{enumerate}
    \item It needs to preserve the space group symmetries, including time reversal, whose single particle representations for space group $P4_2/mbc1^\prime$ (\# 135) and space group $P4/ncc1^\prime$ (\#130) were given in \Cref{table:generatorsGamma}.
    \item The Hamiltonian has to have the periodicity of the Brillouin Zone.
\end{enumerate}
To check the first condition, we first find the representation of the symmetry group elements $g$ on the four particle states $\rho(g)$. 
Then we test whether:
\begin{equation}
\mathcal{H}(g\mathbf{k})=\rho^\dagger(g)\mathcal{H}(\mathbf{k})\rho(g)
\label{eq:symmetrycondition}
\end{equation}
is satisfied for the space group generators, and whether:
\begin{equation}
\mathcal{H}(-\mathbf{k})=\mathcal{T}\mathcal{H}(\mathbf{k})\mathcal{T}^{-1}
\label{eq:trcondition}
\end{equation}
is satisfied for the time-reversal operator $\mathcal{T}$. 

The second condition is not automatically satisfied because both space group $P4/ncc1^\prime$ (\#130) and space group $P4_2/mbc1^\prime$ (\# 135) are nonsymmorphic space groups \cite{wieder2018wallpaper}. 
This requires the presence of electronic orbitals at points away from the origin of the unit cell. 
The matrix $V(\mathbf{G})$ maps the space of states at $\mathbf{k}$ to the space of states at another $\mathbf{k}$ point related by a reciprocal lattice vector $\mathbf{k}'=\mathbf{k}+\mathbf{G}$:
\begin{equation}
    c_{\mathbf{k}+\mathbf{G},i}=\sum_{\beta}V_{i,j}c_{\mathbf{k},j}
\end{equation}
Where $i,j$ are a shorthand for a set of $(\mu,\tau,\sigma)$ indices, and $V(\mathbf{G})$ is given by:
\begin{equation}
V(\mathbf{G})_{i,j}=\delta_{i,j}e^{i\mathbf{G}\cdot\mathbf{r}_{i}}
\end{equation}
with $\mathbf{r}_{i}$ being the position vector of the sublattice sites indexed by $i$ (for example, $(\tfrac{1}{2},\tfrac{1}{2},0)$ for $\mu=1,\tau=-1$ or $(0,0,\tfrac{1}{2})$ for $\mu=-1,\tau=1$). 
This transformation has two consequences. 

 First, the transformation on the second quantized operators implies a condition on the Hamiltonian at a given $\mathbf{k}$ $H_{\mathbf{k}}$ in order that it be invariant under translation by a reciprocal lattice vector.
\begin{equation}
    \label{eq:periodicitycondition}
    H_{\mathbf{k}}\overset{!}{=}H_{\mathbf{k}+\mathbf{G}}
\end{equation}
Second, the inequivalence of the space of states at different $\mathbf{k}$ points means that the action of a symmetry group element on a state at $\mathbf{k}$ will map it into a \textit{different} set of states at $\mathbf{k}'=g\mathbf{k}$. 
If the group element $g$ is not in the little group of $\mathbf{k}'$ (i.e. there is no reciprocal lattice vector $\mathbf{G}$ such that $\mathbf{k}'=\mathbf{k}+\mathbf{G}$) then we cannot find the action of symmetries on the original space of states at $\mathbf{k}$.

Carrying this out, we find \Cref{table:135V(G)matrices} for the $V(\mathbf{G})$ matrices for space group $P4_2/mbc1^\prime$ (\# 135) and space group $P4/ncc1^\prime$ (\#130).

\begin{table}[ht]
\caption{$V(\mathbf{G})$ matrices at the High Symmetry Points (HSPs) for space group $P4_2/mbc1^\prime$ (\# 135)} 
\centering 
\begin{tabular}{c c c c c c c} 
\hline\hline 
Generator & $\Gamma - (0,0,0)$ & $A - (\pi,\pi,\pi)$ & $M - (\pi,\pi,0)$ & $R - (0,\pi,\pi)$ & $X -(0,\pi,0)$ & $Z - (0,0,\pi)$\\[0.5ex] 
\hline 
$\{C_{4z}|000\}$ &$\mu^0\tau^0$ & $\mu^0\tau^z$ &$\mu^0\tau^z$ &$\mu^0\tau^0$ & $\mu^0\tau^0$ & $\mu^0\tau^0$\\
$\{C_{2x}|\tfrac{1}{2}\tfrac{1}{2}0\}$ & $\mu^0\tau^0$ & $\mu^z\tau^z$ & $\mu^0\tau^z$ & $\mu^z\tau^z$ & $\mu^0\tau^z$ & $\mu^0\tau^0$ \\
$\{I|000\}$ & $\mu^0\tau^0$ & $\mu^z\tau^0$ & $\mu^0\tau^0$ & $\mu^0\tau^0$ & $\mu^0\tau^0$ & $\mu^0\tau^0$ \\
TR & $\mu^0\tau^0$ & $\mu^z\tau^0$ & $\mu^0\tau^0$ & $\mu^z\tau^z$ & $\mu^0\tau^z$ & $\mu^z\tau^0$ \\ [1ex] 
\hline 
\end{tabular}
\label{table:135V(G)matrices} 
\end{table}
Which allow us to find the representations on the space of states at the high symmetry points.

We can now test whether the periodicity \Cref{eq:periodicitycondition} and symmetry conditions \Cref{eq:symmetrycondition}, \Cref{eq:trcondition} are satisfied for an arbitrary HK term.
For the HK number terms \Cref{eq:hkNterms}, for example, the periodicity condition \Cref{eq:periodicitycondition} is automatically satisfied since the induced transformation on $n_{\mathbf{k},i}$ is trivial:
\begin{align}
n_{\mathbf{k}+\mathbf{G},i}=c^\dagger_{\mathbf{k}+\mathbf{G},i}c_{\mathbf{k}+\mathbf{G},i}=\sum_{jk}V^\dagger_{ij}V_{ik}c^\dagger_{\mathbf{k},j}c_{\mathbf{k},k}=\delta_{ji}\delta_{ki}c^\dagger_{\mathbf{k},j}c_{\mathbf{k},k}=n_{\mathbf{k},i}
\end{align}
This means we only need to check whether the $N$ matrices in the generalized HK number term:
\begin{equation}
\sum_{\mathbf{k},i,j}n_{\mathbf{k},i}N_{ij}n_{\mathbf{k},j}
\end{equation}
satisfy the symmetry and time-reversal preserving invariance conditions for the generators in \Cref{table:generatorsGamma}. 
Of the sixty four possible terms, the eight that satisfy these conditions are given in \Cref{table:HKNterms}. 
\begin{table}[ht]
\caption{Symmetry preserving HK number terms in space groups $P4_2/mbc1^\prime$ (\#135) and $P4/ncc1^\prime$ (\# 130)} 
\centering 
\renewcommand*{\arraystretch}{1.4}
\begin{tabular}{c c c c} 
\hline\hline 
\multicolumn{2}{c}{Number} &\multicolumn{2}{c}{$N$}\\[0.5ex]
\hline 
$1$ & &$I$&\\
$2$ & &$\mu^0\tau^0\sigma^x$&\\
$3$ & &$\mu^0\tau^x\sigma^0$&\\
$4$ & &$\mu^0\tau^x\sigma^x$&\\
$5$ & &$\mu^x\tau^0\sigma^0$&\\
$6$ & &$\mu^x\tau^0\sigma^x$&\\
$7$ & &$\mu^x\tau^x\sigma^0$&\\
$8$ & &$\mu^x\tau^x\sigma^x$&\\
\hline 
\end{tabular}
\label{table:HKNterms} 
\end{table}
For the generalized HK terms of the form:
\begin{equation}
\sum_{i,j,k,l}\left[c^\dagger_{\mathbf{k},i}F^{1}_{ij}c_{\mathbf{k},j}\right]\left[c^\dagger_{\mathbf{k},i}F^{2}_{ij}c_{\mathbf{k},j}\right]
\end{equation}
the transformations of the creation and annihilation operators are not trivial. 
Each $F$ transforms as:
\begin{align}
c^\dagger_{\mathbf{k}+\mathbf{G},i}F_{ij}c_{\mathbf{k}+\mathbf{G},j}=V^\dagger_{in}F_{ij}V_{jm}c^\dagger_{\mathbf{k},n}c_{\mathbf{k},m}
\end{align}
And so the overall term transforms as:
\begin{equation}
\sum_{\substack{i,j,l,r}}\left[c'^\dagger_{\mathbf{k}+\mathbf{G},i}F^{1}_{ij}c'_{\mathbf{k}+\mathbf{G},j}\right]
\left[c'^\dagger_{\mathbf{k}+\mathbf{G},l}F^{2}_{lr}c'_{\mathbf{k}+\mathbf{G},r}\right]=
\sum_{\substack{i,j,l,r\\n,m,p,q}}\left[c^\dagger_{\mathbf{k},n}V^\dagger_{i,n}F^{1}_{ij}V_{j,m}c_{\mathbf{k},j}\right]
\left[c^\dagger_{\mathbf{k},l}V^\dagger_{l,p}F^{2}_{lr}V_{r,q}c_{\mathbf{k},r}\right]
\end{equation}
Since the $F^1,F^2$ matrices are not functions of $\mathbf{k}$, we require that:
\begin{equation}
\label{eq:Fcondition}
\sum_{\substack{i,j,l,r\\n,m,p,q}}\left[c^\dagger_{\mathbf{k},n}V^\dagger_{i,n}F^{1}_{ij}V_{j,m}c_{\mathbf{k},j}\right]
\left[c^\dagger_{\mathbf{k},l}V^\dagger_{l,p}F^{2}_{lr}V_{r,q}c_{\mathbf{k},r}\right]\overset{!}{=}
\sum_{i,j,k,l}\left[c^\dagger_{\mathbf{k},i}F^{1}_{ij}c_{\mathbf{k},j}\right]\left[c^\dagger_{\mathbf{k},i}F^{2}_{ij}c_{\mathbf{k},j}\right]
\end{equation}
We can test which of the $(64)^2$ possible $F^{1},F^{2}$ terms satisfy both the symmetry and periodicity requirements. 
There are only three such terms:
\begin{align}
&H^{3}_{HK}=U_{\tau^x\tau^x}\left[\sum_{\mu\tau\sigma}c^\dagger_{\mu\tau\sigma}c_{\mu-\tau\sigma}\right]
\left[\sum_{\mu'\tau'\sigma'}c^\dagger_{\mu'\tau'\sigma'}c_{\mu'-\tau'\sigma'}\right] \label{eq:135f1}\\
&H^{3'}_{HK}=U_{\mu^x\mu^x}\left[\sum_{\mu\tau\sigma}c^\dagger_{-\mu\tau\sigma}c_{\mu\tau\sigma}\right]
\left[\sum_{\mu'\tau'\sigma'}c^\dagger_{\mu'\tau'\sigma'}c_{-\mu'{\tau'}\sigma'}\right]\label{eq:135f2}\\
&H^{3''}_{HK}=U_{(\mu^x\tau^x)(\mu^x\tau^x)}\left[\sum_{\mu\tau\sigma}c^\dagger_{-\mu-\tau\sigma}c_{\mu\tau\sigma}\right]
\left[\sum_{\mu'\tau'\sigma'}c^\dagger_{\mu'\tau'\sigma'}c_{-\mu'-\tau'\sigma'}\right]\label{eq:135f3}
\end{align}

While we have focused here on HK terms with interaction strengths that are $\mathbf{k}$-independent, the procedure outlined here generalizes straightforwardly to allow for the classification of symmetric, momentum-dependent HK interactions as well.

To conclude, we also sketch the procedure necessary to check the invariance of multiparticle states under the space group symmetries at high symmetry points. 
At the high symmetry points, there will be at least one group element for which $g\mathbf{k}=k+\mathbf{G}$. 
In this case, we can use the transformation to map back to the original space of states.
For a representation $\rho(g)$ of the symmetries on the single particle states:
\begin{equation}
    \rho(g)c_{\mathbf{k},i}=\sum_{j}\rho(g)_{ij}c_{g\mathbf{k},j}=\sum_{j}\rho_{ij}c_{\mathbf{k}+\mathbf{G},j}=\sum_{j,k}\rho_{ij}V_{jk}(\mathbf{G})c_{\mathbf{k},k}
\end{equation}
The consequence of this is that to find the representation of the space group symmetries on the space of states at a high symmetry point $\mathbf{Q}$ whose little group is equal to the full space group, we can follow this simple procedure:
\begin{enumerate}
    \item Find the reciprocal lattice vector between the high symmetry point and it's image under the space group: $\mathbf{G}=\rho(g)\mathbf{Q}-\mathbf{Q}$.
    \item Find the matrix elements $V(\mathbf{G})_{ij}=\delta_{ij}e^{i\mathbf{G}\cdot\mathbf{r}_{i}}$.
    \item Multiply the single particle representation matrices by the $V(\mathbf{G})$ matrices to map the space of states in the image back to the original $\mathbf{k}$ point.
\end{enumerate}

\section{Violation of minimal filling bound: Space Group $P2_1/c1^\prime$ (\# 14)}
\label{sec:SG14}
We conjectured that for models with long-range HK interactions, the only constraint left over from the LSM theorem for time-reversal symmetric systems is Kramers degeneracy. 
Kramers degeneracy requires that an insulator have a minimum filling of $2n$, with $n$ an integer. 
This means that the lowest additional constraint would enforce an insulator at a filling of $4n$. 
The LSM theorem for space group $P2_1/c1^\prime$ (\# 14) given in Ref.~\cite{watanabe2015filling} requires a featureless insulator in this space group to have filling $\nu=4n$. 
We now show that even this minimal additional bound is violated in an orbital HK model.

Space group $P2_1/c1^\prime$ (\# 14) is generated by the twofold screw $\{C_{2x}|00\tfrac{1}{2}\}$, spatial inversion, and the Bravais lattice of translations. 
It can be viewed as an index-four subgroup of $P4_2/mbc1^\prime$ (\# 135). 
To construct a Hamiltonian invariant under space group $P2_1/c1^\prime$ (\# 14), we can consider a monoclinic parent lattice with two sublattice sites per unit cell indexed by $\tau=\pm 1$. 
The symmetry generators are given in \Cref{table:generatorsGammaSG14}.
\begin{table}[ht]
\caption{Representation matrices of the symmetry generators for space group $P2_1/c1^\prime$ (\# 14) in the spin and sublattice basis.} 
\centering 
\renewcommand*{\arraystretch}{1.4}
\begin{tabular}{c c c c} 
\hline\hline 
\multicolumn{4}{c}{SG14}\\[0.5ex]
\hline 
$\{g|\mathbf{t}\}$ & & $\rho\left(\{g|\mathbf{t}\}\right)$ & \\ 
$\{C_{2x}|\tfrac{1}{2}\tfrac{1}{2}0\}$ & &  $i\tau^x\sigma^x$ & \\
$\{I|000\}$ & & $I$ & \\
TR & & $i\tau^0\sigma^y K$ &\\ [1ex] 
\hline 
\end{tabular}
\label{table:generatorsGammaSG14} 
\end{table}

To obtain a Hamiltonian invariant under space group $P2_1/c1^\prime$ (\# 14), we can simply drop the $\mu$ index of our space group $P4_2/mbc1^\prime$ (\# 135) Hamiltonian \Cref{eq:135 Hamiltonian}, and retain only the tight-binding terms that are invariant under the space group symmetries and time reversal:
\begin{align}
    &\mathcal{H}^{14}_{0}(\mathbf{k})=\mathcal{H}^{14}_{1}(\mathbf{k})+\mathcal{H}^{14}_{2}(\mathbf{k})\nonumber \\
    &\mathcal{H}^{14}_1(\mathbf{k})=t_{xy}\tau^x\cos\tfrac{k_x}{2}\cos\tfrac{k_y}{2}\\
    &\mathcal{H}^{14}_{2}(\mathbf{k})=\lambda'_3\tau^y\sigma^z\cos\tfrac{k_x}{2}\cos\tfrac{k_y}{2}(\cos k_x-\cos k_y) \nonumber
    \label{eq:sg14ham}
\end{align}

\begin{figure*}
\FPeval{\overlap}{0.53}
\FPeval{\scalevalue}{round(0.5*(1.12)/\overlap,2)}
\scalebox{\scalevalue}{
\setlength{\tabcolsep}{0pt} 
\def\tabularxcolumn#1{m{#1}}
\hskip-1.0cm\begin{tabularx}{\textwidth}{@{}XXX@{}}
\subfloat[]{\includegraphics[width=\overlap\textwidth]{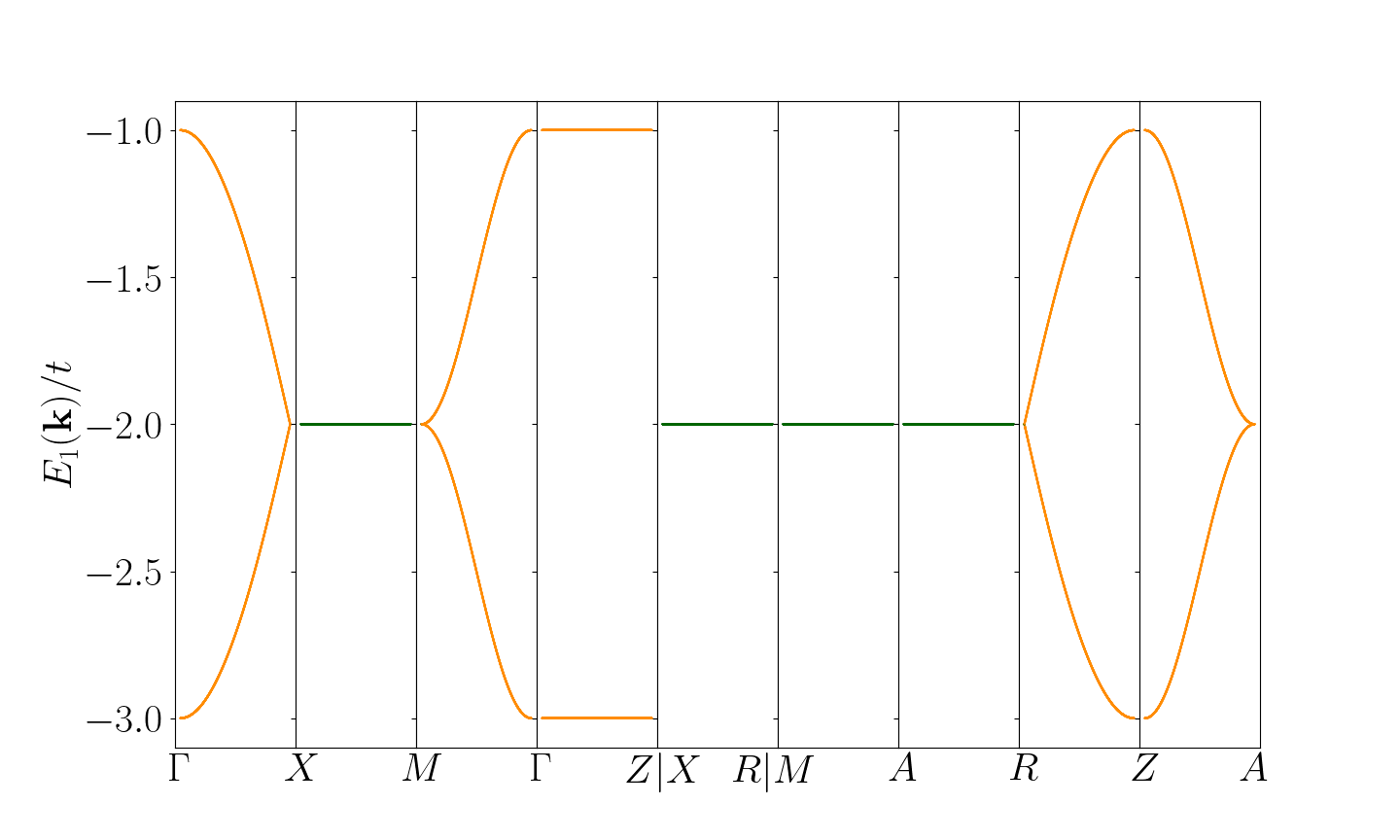}} &
\subfloat[]{\includegraphics[width=\overlap\textwidth]{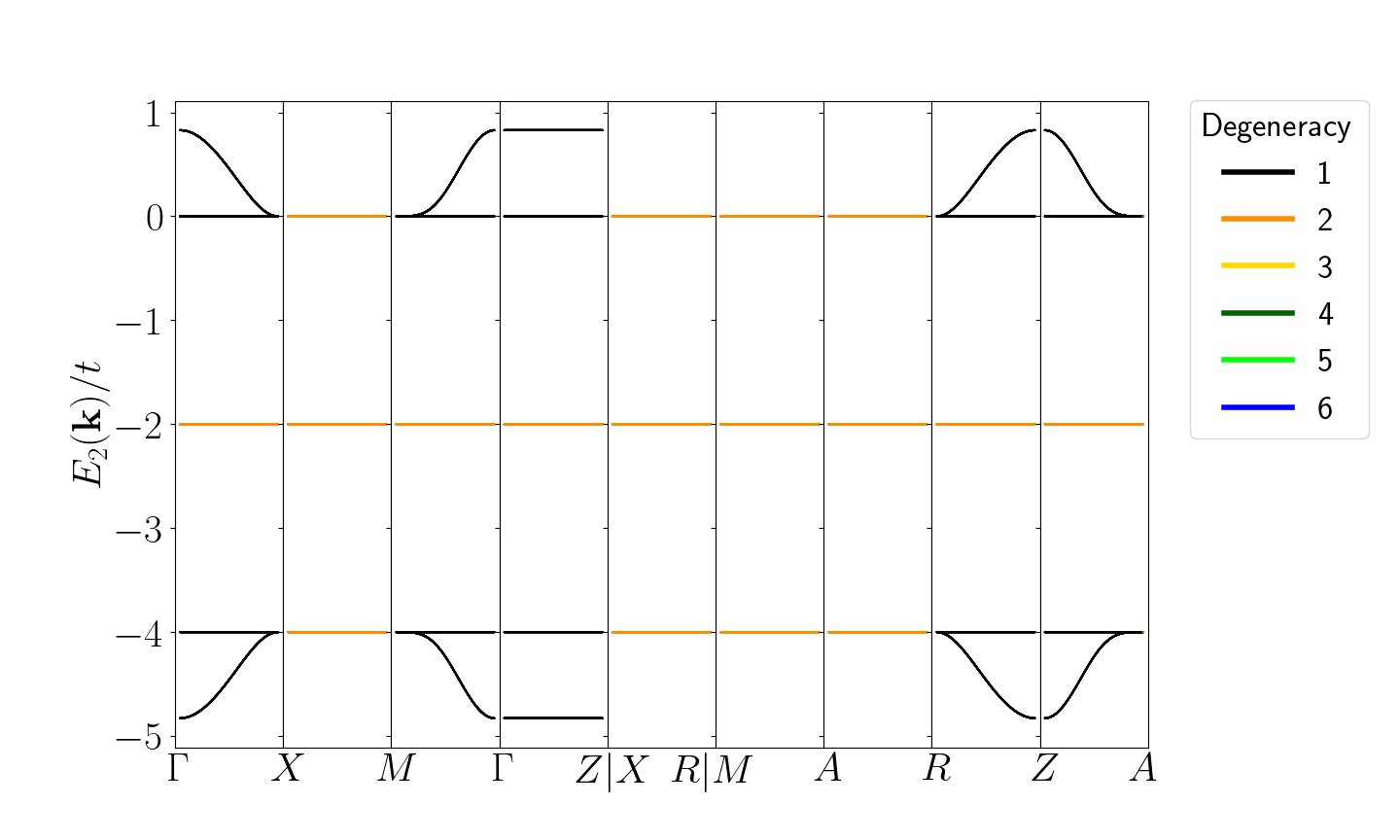}}
 \\[-3ex]
\subfloat[]{\includegraphics[width=\overlap\textwidth]{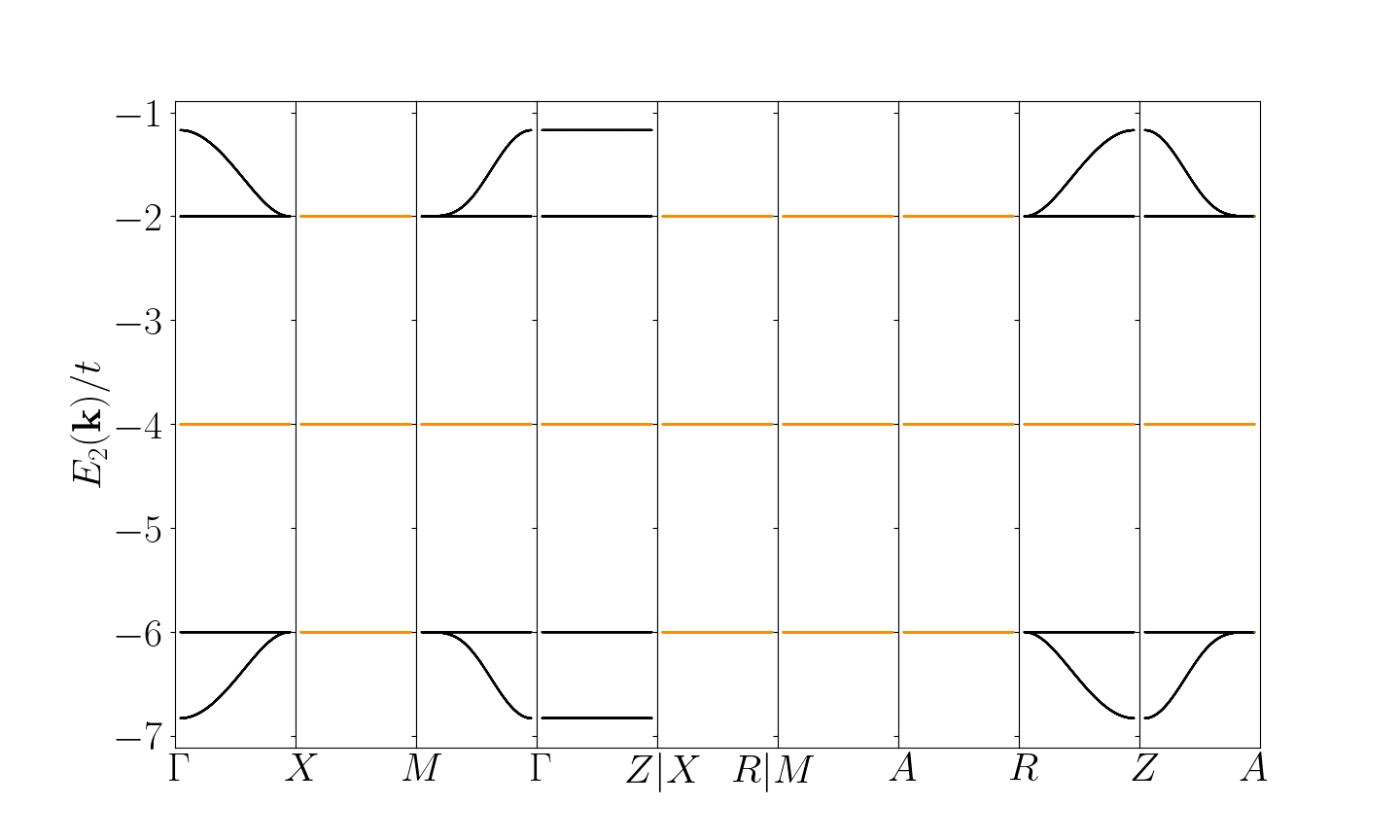}} &
\subfloat[]{\includegraphics[width=\overlap\textwidth]{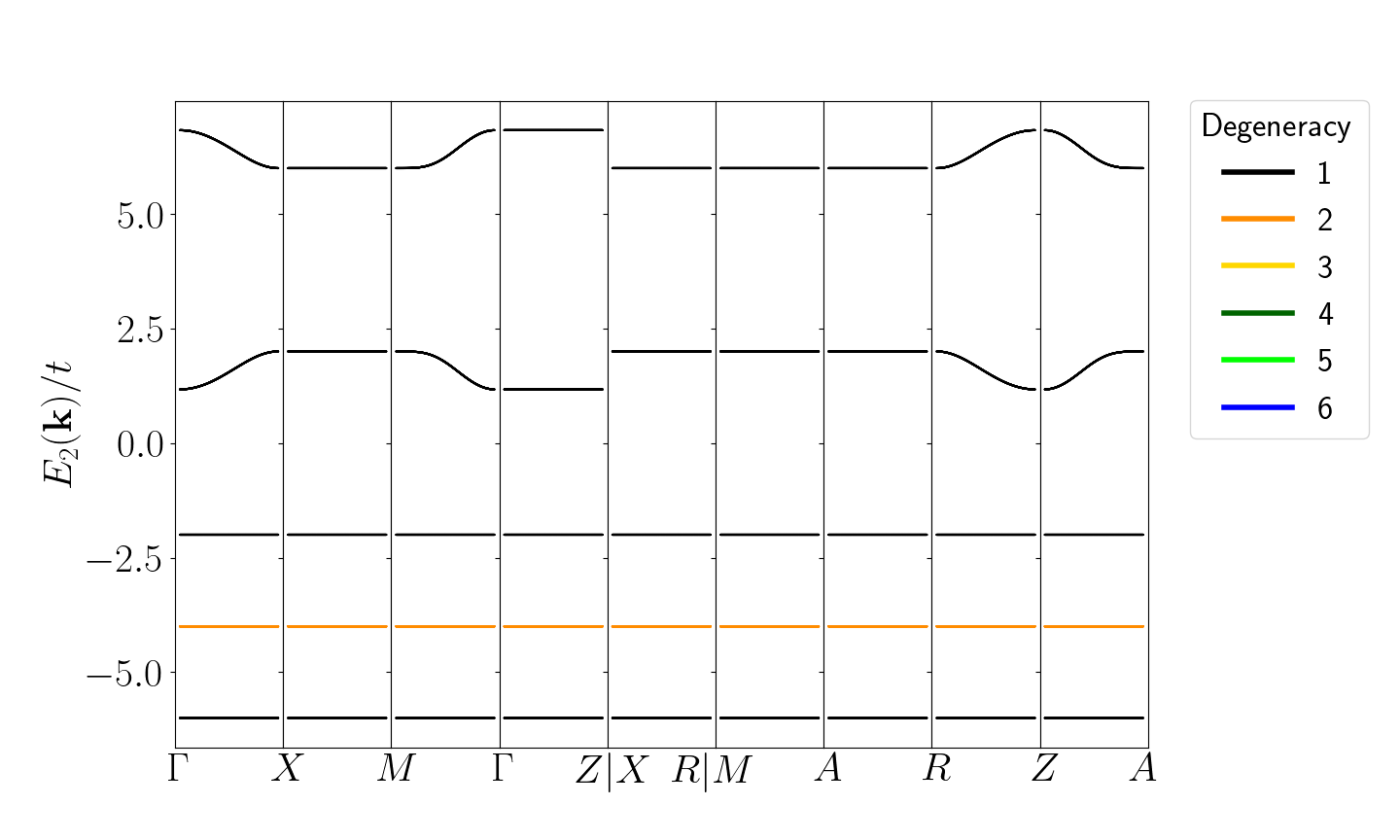}}
\end{tabularx}}
\caption{The spectrum in the half-filled (two particle) sector for the model in space group $P2_1/c1^\prime$ (\# 14) with generalized HK interactions. 
(a) shows the single particle spectrum. 
(b) shows the spectrum in the two particle sector $E_{2}(\mathbf{k})$ for the Hamiltonian with orbital HK interactions $H^{1}_{HK}$, $H=H^{14}_{0}+H^{1}_{HK}$ with $U_{1}=4$.
(c) shows the spectrum in the two particle sector $E_{2}(\mathbf{k})$ for the Hamiltonian with generalized HK interaction $H=H^{0}_{14}+H^{1}_{HK}+H^{2}_{HK}$ with $U_{1}=4,U_{2}=2$. 
(d) shows the spectrum in the two particle sector $E_{2}(\mathbf{k})$ for the sHamiltonian with generalized HK interaction $H=H^{14}_{0}+H^{1}_{HK}+H^{2}_{HK}+H^{3}_{HK}$ with $U_{1}=4,U_{2}=2, U_{3}=2$.
The non-interacting parameter values are: $t_{xy}=1, \lambda'_{3}=0.15, \mu_{0}=(U_{1}+2U_{2})/2$.}
\label{fig:SG14}
\end{figure*}

\begin{figure*}
\FPeval{\overlap}{0.53}
\FPeval{\scalevalue}{round(0.5*(1.12)/\overlap,2)}
\scalebox{\scalevalue}{
\setlength{\tabcolsep}{0pt} 
\def\tabularxcolumn#1{m{#1}}
\hskip-1.0cm\begin{tabularx}{\textwidth}{@{}XXX@{}}
\subfloat[]{\includegraphics[width=\overlap\textwidth]{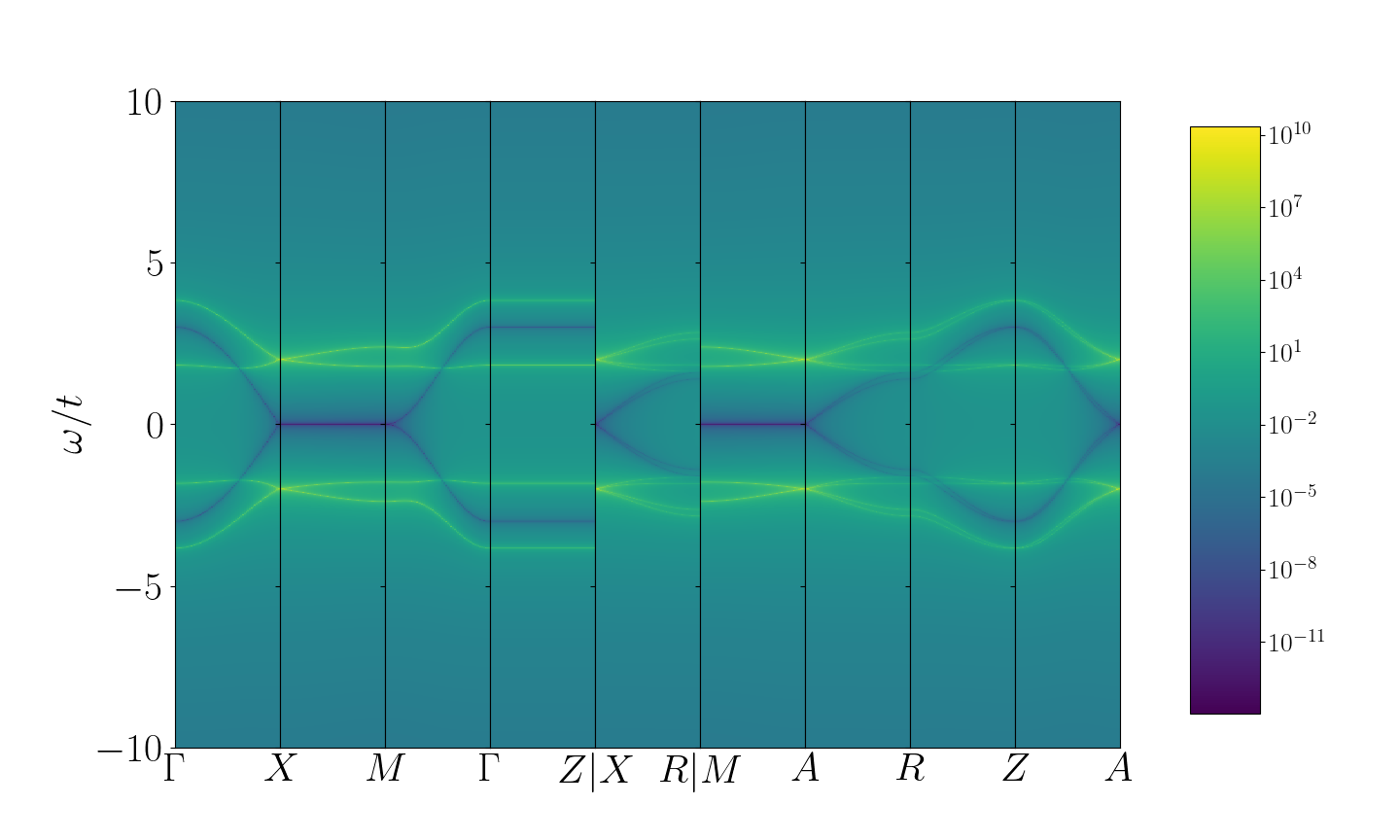}} &
\subfloat[]{\includegraphics[width=\overlap\textwidth]{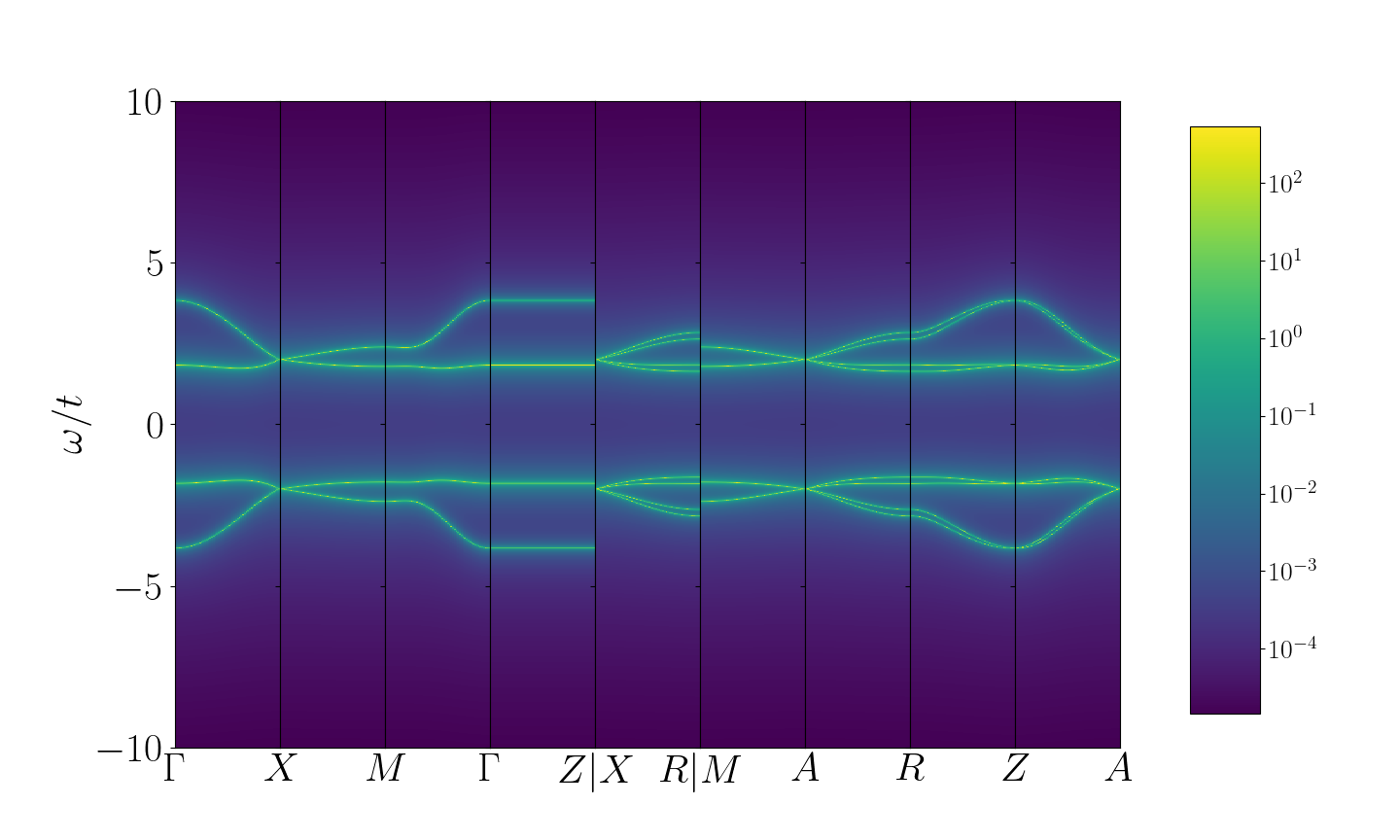}}
 \\[-3ex]
\subfloat[]{\includegraphics[width=\overlap\textwidth]{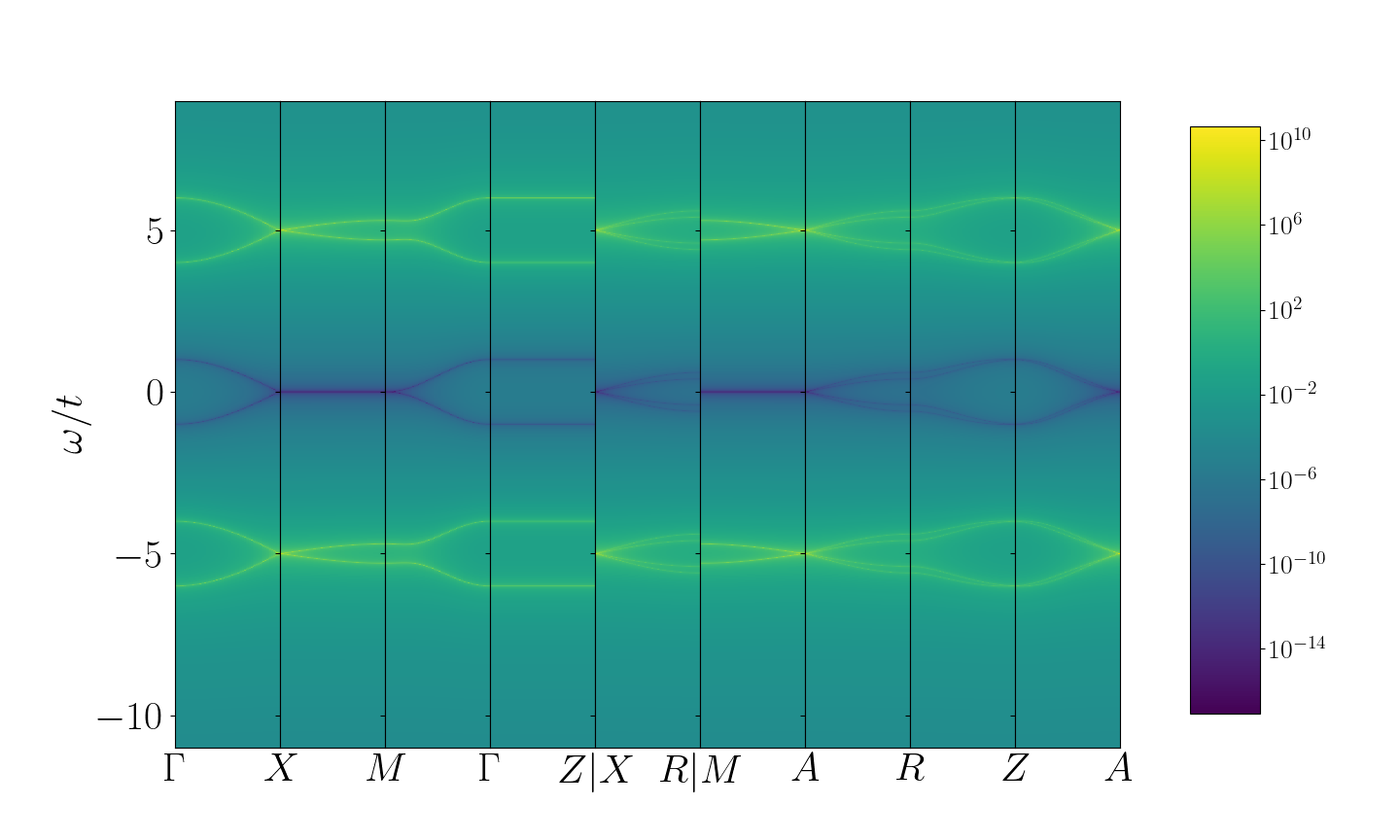}} &
\subfloat[]{\includegraphics[width=\overlap\textwidth]{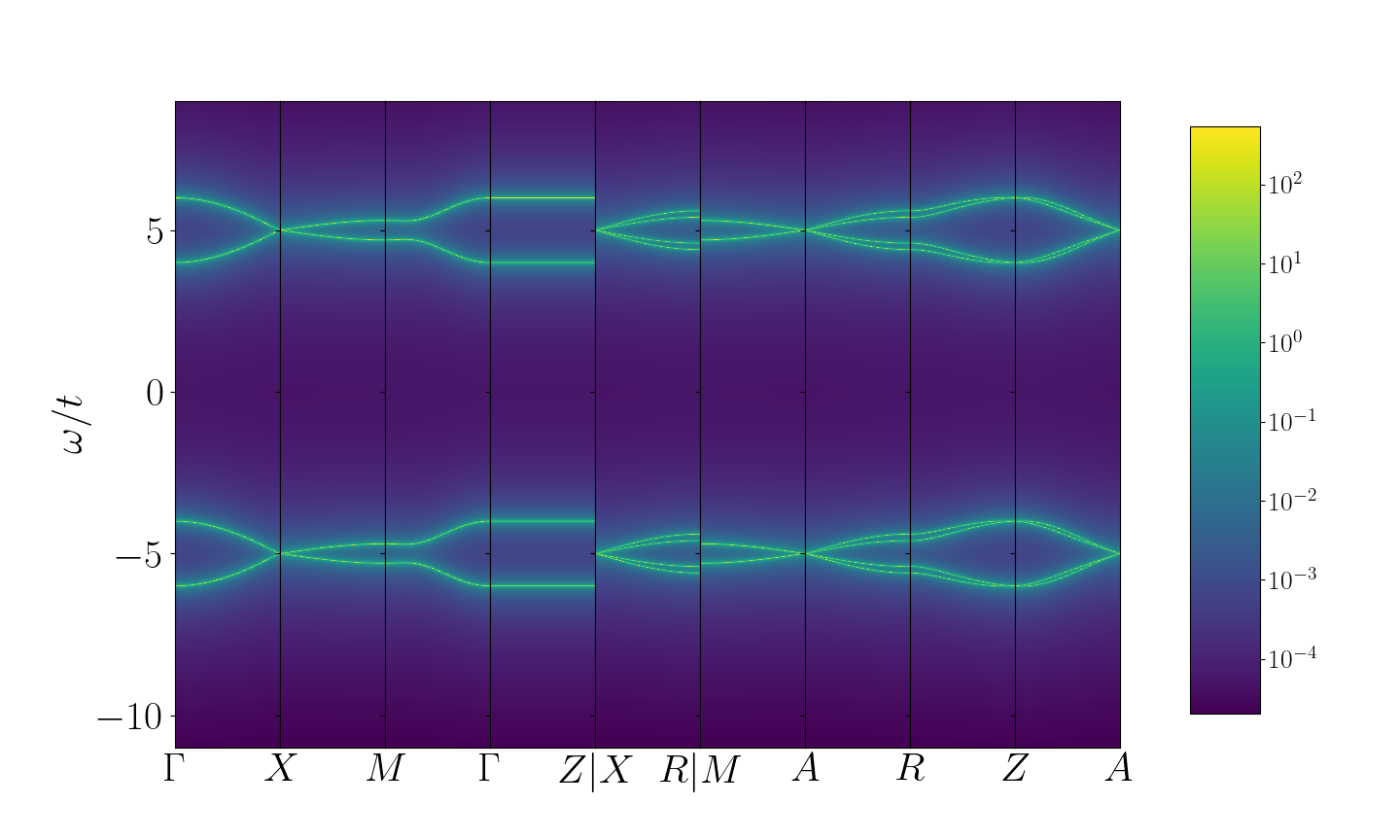}}
\end{tabularx}}
\caption{The absolute value of the determinant, $|\det(G^{+}(\mathbf{k},\omega))|$ and the spectral function $-\frac{1}{\pi}\Im\Tr(G^{+}(\mathbf{k},\omega))$ for orbital HK models in space group $P2_1/c1^\prime$ (\# 14). 
The top row shows the Green function for the Hamiltonian $H^{0}_{14}+\Tilde{H}^{1}_{HK}$ and the bottom row shows the Green function for the Hamiltonian $H^{0}_{14}+\Tilde{H}^{1}_{HK}+\Tilde{H}^{2}_{HK}+\Tilde{H}^{3}_{HK}$.
From the determinant we can see that SG14 with orbital HK interactions is a Mott insulator, just like space group $P4_2/mbc1^\prime$ (\# 135). 
(a) shows $|\det{G}|$ for the Hamiltonian $H^{0}_{14}+\Tilde{H}^{1}_{HK}$ with $U_{1}=4$.
The band of zeros confirms that it is a Mott insulator.
(b) shows the corresponding spectral function $-\frac{1}{\pi}\Im\Tr(G^{+}(\mathbf{k},\omega))$.
(c) shows $|\det(G^{+}(\mathbf{k},\omega))|$ for the model with all three HK interacting terms, $H^{0}_{14}+\Tilde{H}^{1}_{HK}+\Tilde{H}^{2}_{HK}+\Tilde{H}^{3}_{HK}$ with $U_{1}=4,U_{2}=2, U_3=2$.
As above, the band of zeros confirms that the system is a Mott insulator.
(d) shows the corresponding spectral function $-\frac{1}{\pi}\Im\Tr(G^{+}(\mathbf{k},\omega))$.
The non-interacting parameter values are: $t_{xy}=1, \lambda'_{3}=0.15, \mu_{0}=(U_{1}+2U_{2})/2$.
}
\label{fig:SG14GFs}
\end{figure*}

We can then add the same sequence of HK terms as before, dropping the $\mu$ sublattice degree of freedom. 
We first add the orbital HK term:
\noindent
\begin{align}
H^{1}_{HK}=U_{1}\sum_{\mathbf{k}\tau}n_{\mathbf{k}\tau\uparrow}n_{\mathbf{k}\tau\downarrow},
\end{align}
and find in \Cref{fig:SG14}(b) a ground state that has a two-fold degeneracy along the lines $X-M$, $X-R$, $M-A$ and $M-R$.
We can then add the generalized number HK term $
H^{2}_{HK}=U_{\tau^x}\sum_{\mathbf{k}\tau\sigma}n_{\mathbf{k}-\tau\sigma}n_{\mathbf{k}\tau\sigma}
$ 
and the final term $H^{3}_{HK}=U_{\tau^x\tau^x}\left[\sum_{\mathbf{k}\tau\sigma}c^\dagger_{\mathbf{k}\tau\sigma}c_{\mathbf{k}-\tau\sigma}\right]\left[\sum_{\mathbf{k}\tau'\sigma'}c^\dagger_{\mathbf{k}\tau'\sigma'}c_{\mathbf{k}-\tau'\sigma'}\right]$ to produce a ground state that is everywhere gapped and non-degenerate---again violating the LSM filling bound---as shown by the spectrum in \Cref{fig:SG14}d. 
We also provide the absolute value of the determinant and the spectral function of the real-time retarded Green function in \Cref{fig:SG14GFs}. 
Just like for space group $P4_2/mbc1^\prime$ (\# 135), the band of zeros in the GF confirms that the ground state of $H^{0}_{14}+H^{1}_{HK}$ and for $H^{0}_{14}+H^{1}_{HK}+H^{2}_{HK}+H^{3}_{HK}$  is a Mott insulator.

\section{Particle-hole symmetry of Generalized HK models}\label{sec:phsymm}

In this Appendix, we analyze the action of particle-hole symmetry on our HK models in space groups $P4_2/mbc1^\prime$ (\# 135), $P4/ncc1^\prime$ (\#130) and $P2_1/c1^\prime$ (\# 14). 

\begin{figure*}
\FPeval{\overlap}{0.33}
\FPeval{\scalevalue}{round(0.33*(1.05)/\overlap,2)}
\scalebox{\scalevalue}{
\setlength{\tabcolsep}{0pt} 
\def\tabularxcolumn#1{m{#1}}
\hskip-0.75cm\begin{tabularx}{\textwidth}{@{}XXX@{}}
\subfloat[]{\includegraphics[width=\overlap\textwidth]{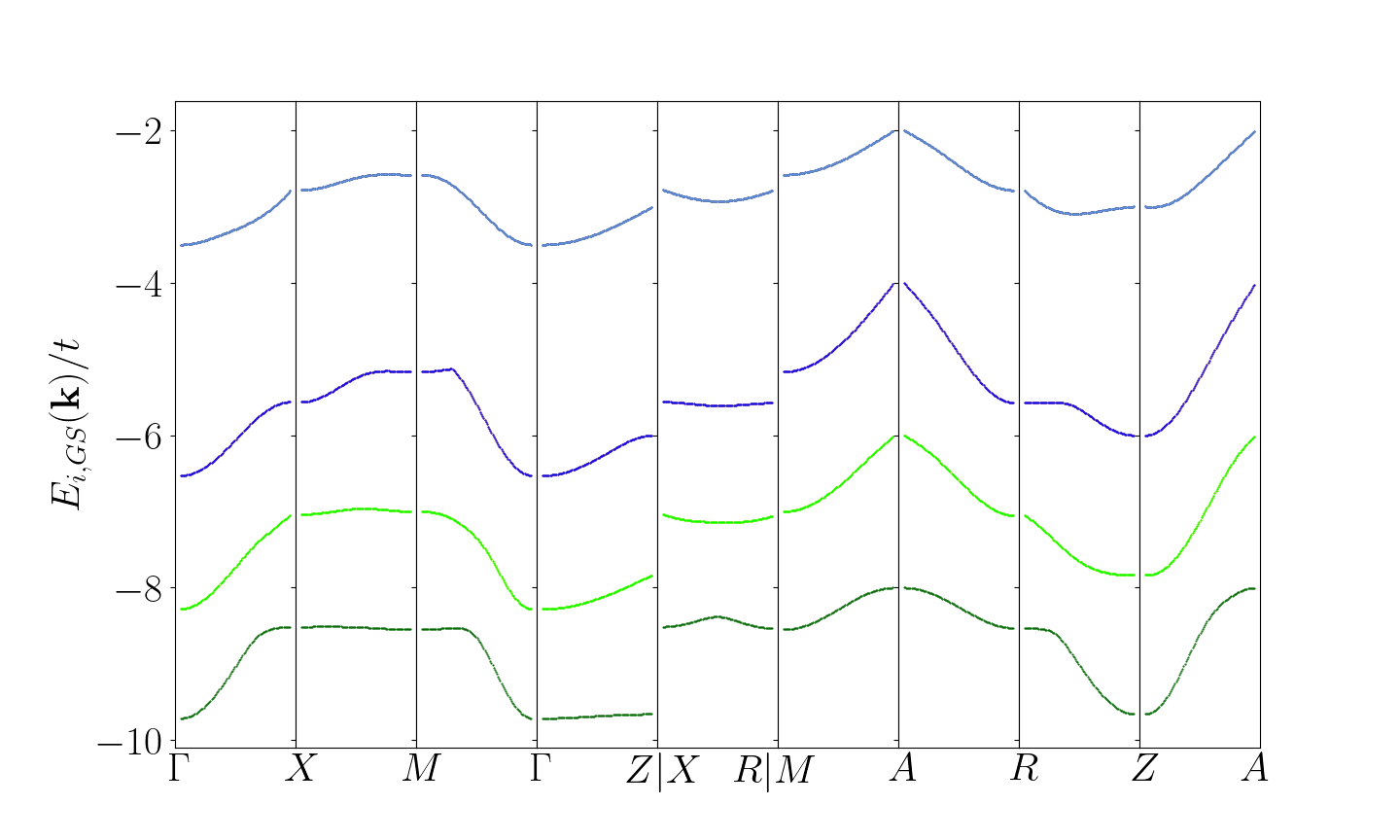}} &
\subfloat[]{\includegraphics[width=\overlap\textwidth]{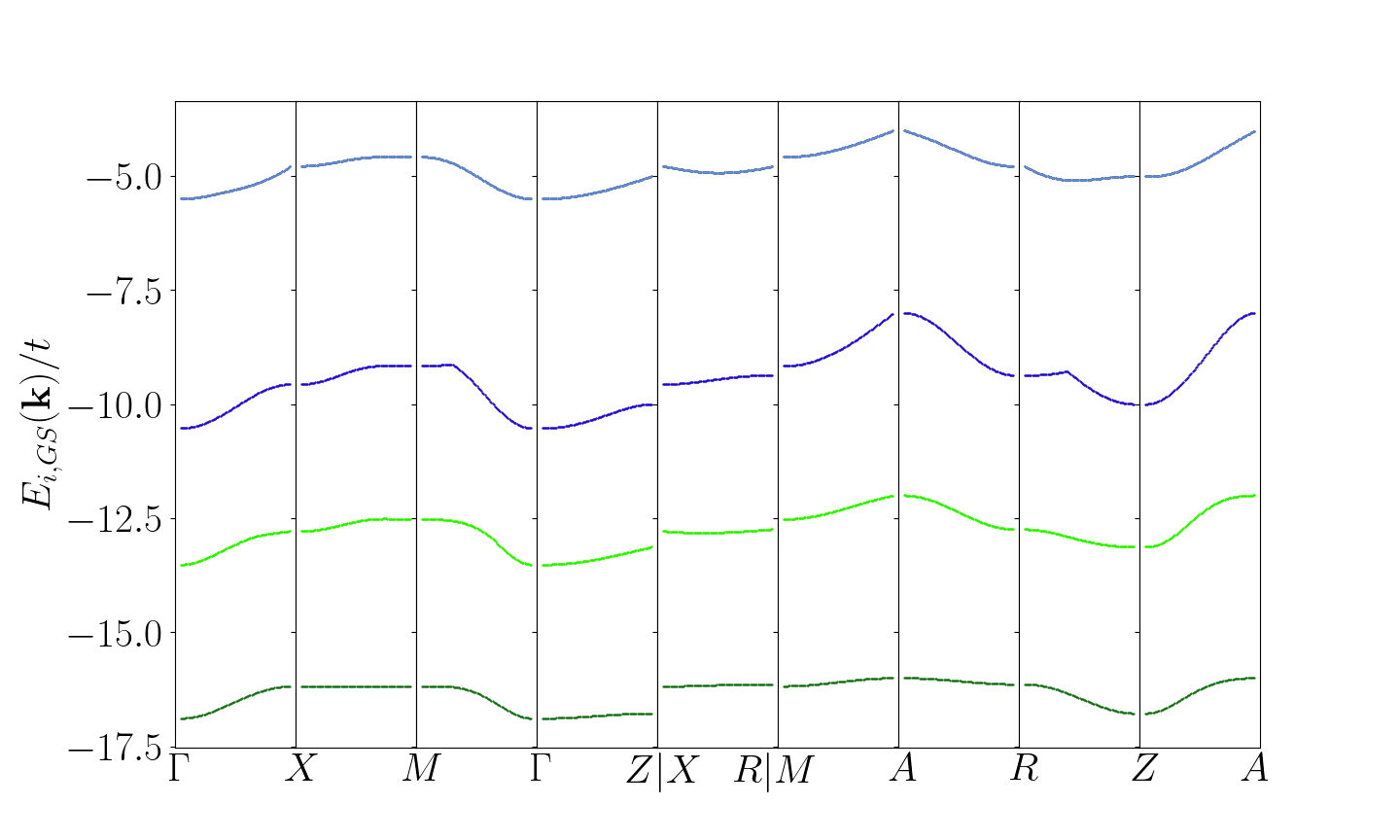}} &
\subfloat[]{\includegraphics[width=\overlap\textwidth]{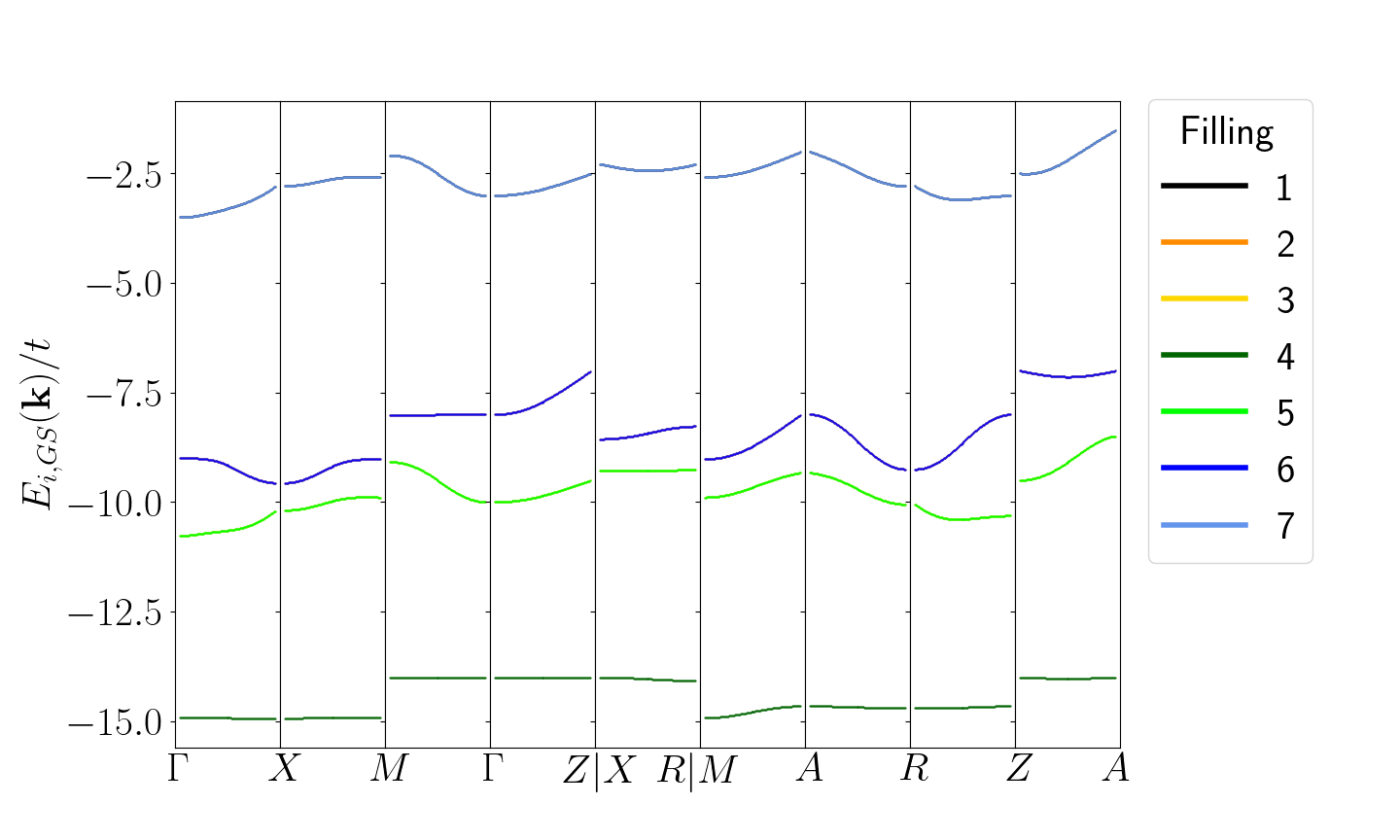}} \\[-3ex]
\subfloat[]{\includegraphics[width=\overlap\textwidth]{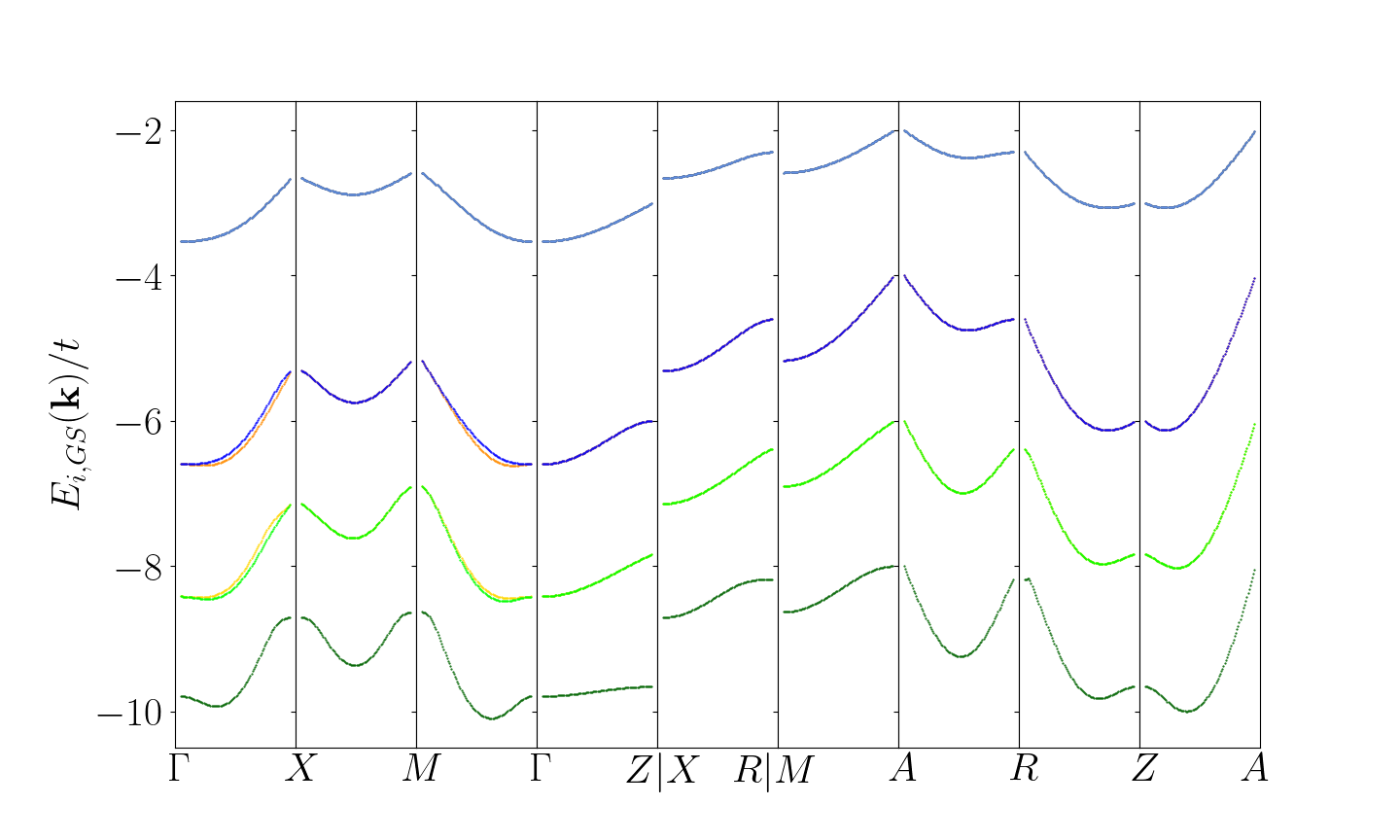}} &
\subfloat[]{\includegraphics[width=\overlap\textwidth]{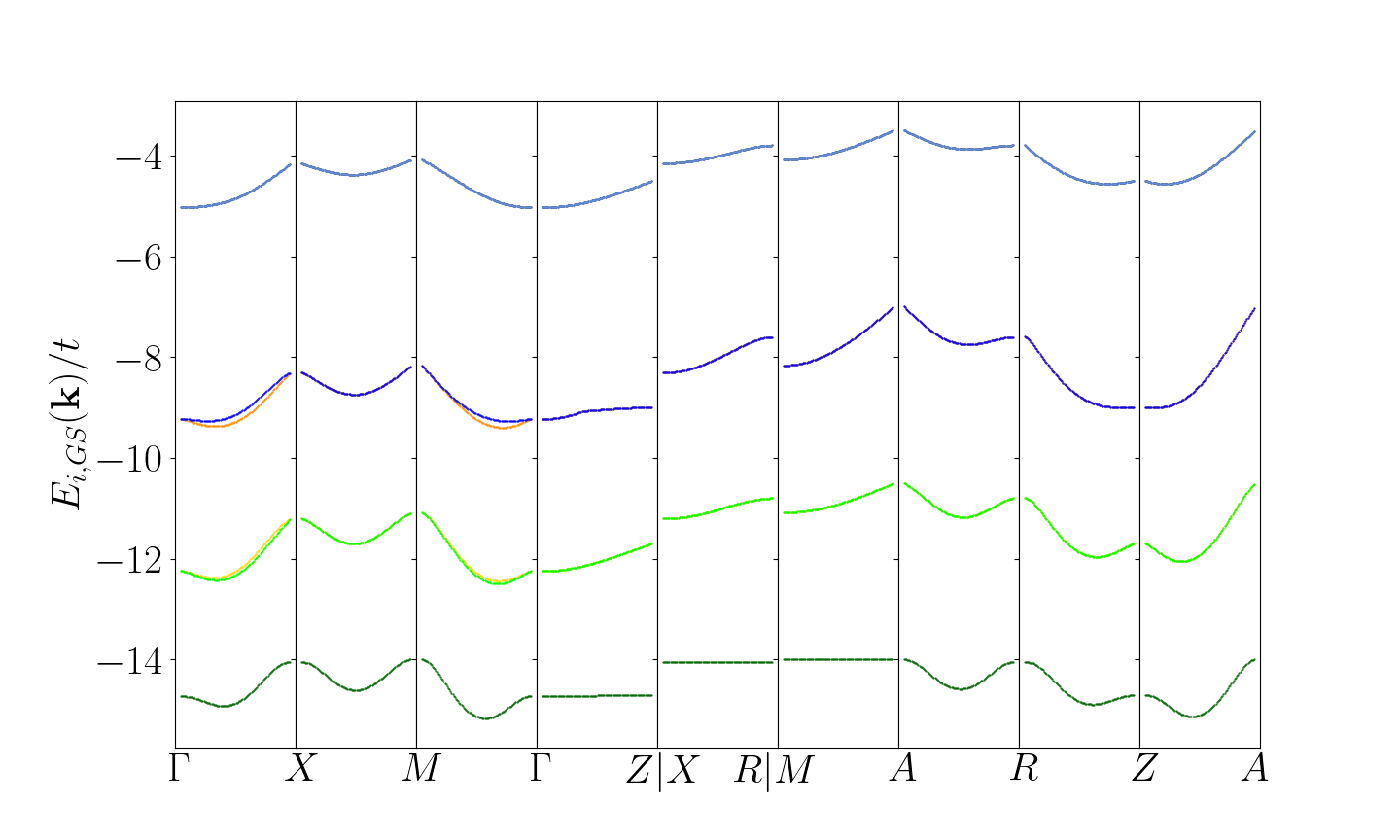}} &
\subfloat[]{\includegraphics[width=\overlap\textwidth]{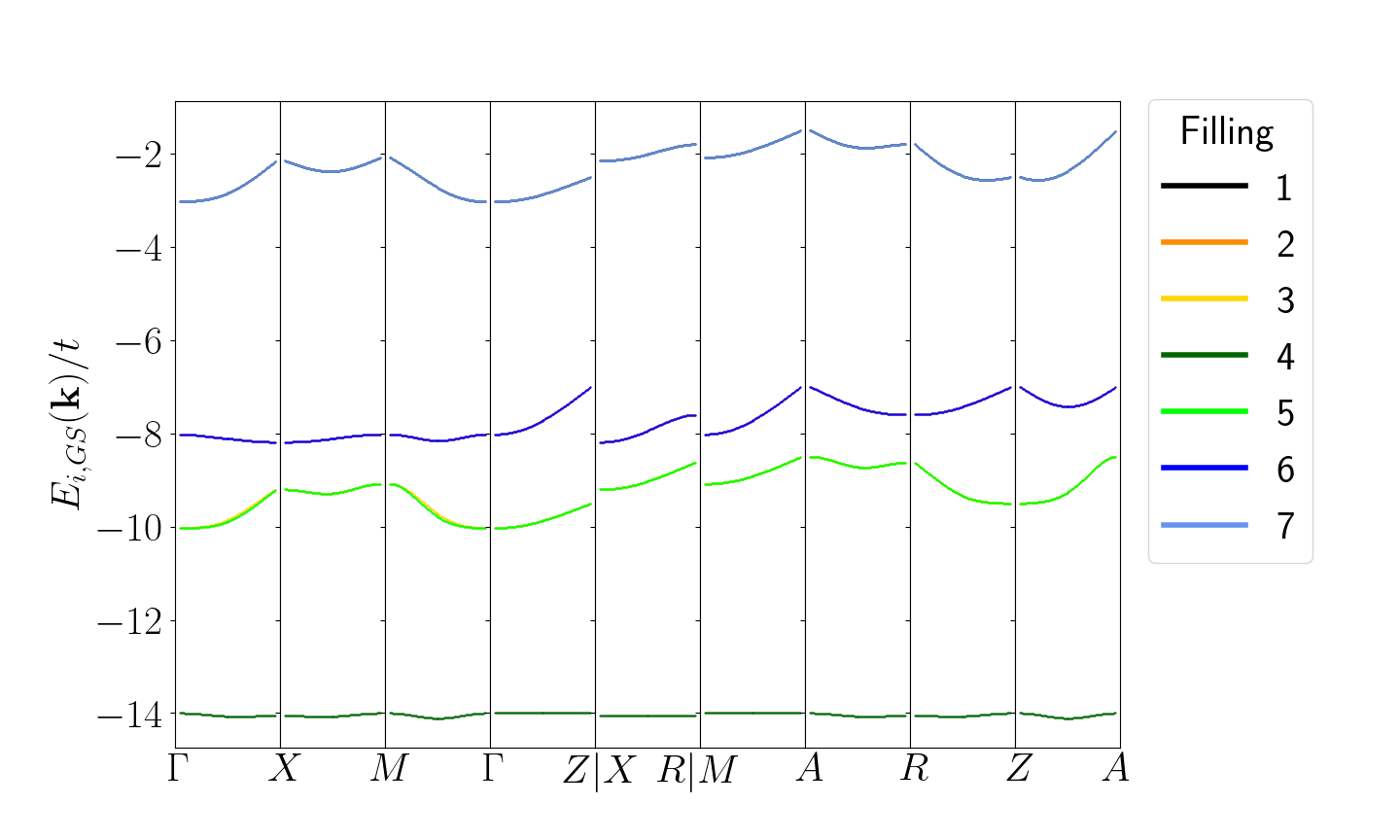}}
\end{tabularx}}
\caption{A comparison of the ground state energies $E_{i}(\mathbf{k})$ for different particle numbers at each $\mathbf{k}$ point for space group $P4_2/mbc1^\prime$ (\# 135) (top row) and space group $P4/ncc1^\prime$ (\#130) (bottom row). 
Where spectra overlap, only the larger filling is shown (i.e if $E_{2}(\mathbf{k})$ overlaps $E_{6}(\mathbf{k})$ we show only a blue line ).
From the fact that the spectra do not overlap for all $\mathbf{k}$ points away from half-filling, we can see that the spectrum is weakly particle-hole asymmetric. 
Nonetheless, for the parameter values in this paper, the lowest energy state of the $N$ particle Hamiltonian at half-filling is still the tensor product of the ground state of the four particle (i.e half-filled) Hamiltonian at every $\mathbf{k}$ point since the half-filled four particle energies (in green) are the lowest for every $\mathbf{k}$ point.
(a) shows the ground state energies $E_{i}(\mathbf{k})$ for different particle numbers for $H=H^{0}_{135}+H^{1}_{HK}$.
(b) shows the ground state energies $E_{i}(\mathbf{k})$ for different particle numbers for $H=H^{0}_{135}+H^{1}_{HK}+H^{2}_{HK}$.
(c) shows the ground state energies $E_{i}(\mathbf{k})$ for different particle numbers for $H=H^{0}_{135}+H^{1}_{HK}+H^{2}_{HK}+H^{3}_{HK}$.
(d) shows the ground state energies $E_{i}(\mathbf{k})$ for different particle numbers for $H=H^{0}_{130}+H^{1}_{HK}$.
(e) shows the ground state energies $E_{i}(\mathbf{k})$ for different particle numbers for $H=H^{0}_{130}+H^{1}_{HK}+H^{2}_{HK}$.
(f) shows the ground state energies $E_{i}(\mathbf{k})$ for different particle numbers for $H=H^{0}_{130}+H^{1}_{HK}+H^{2}_{HK}+H^{3}_{HK}$.}
\label{fig:ParticleHole}
\end{figure*}

\begin{figure*}
\FPeval{\overlap}{0.33}
\FPeval{\scalevalue}{round(0.33*(1.05)/\overlap,2)}
\scalebox{\scalevalue}{
\setlength{\tabcolsep}{0pt} 
\def\tabularxcolumn#1{m{#1}}
\hskip-0.75cm\begin{tabularx}{\textwidth}{@{}XXX@{}}
\subfloat[]{\includegraphics[width=\overlap\textwidth]{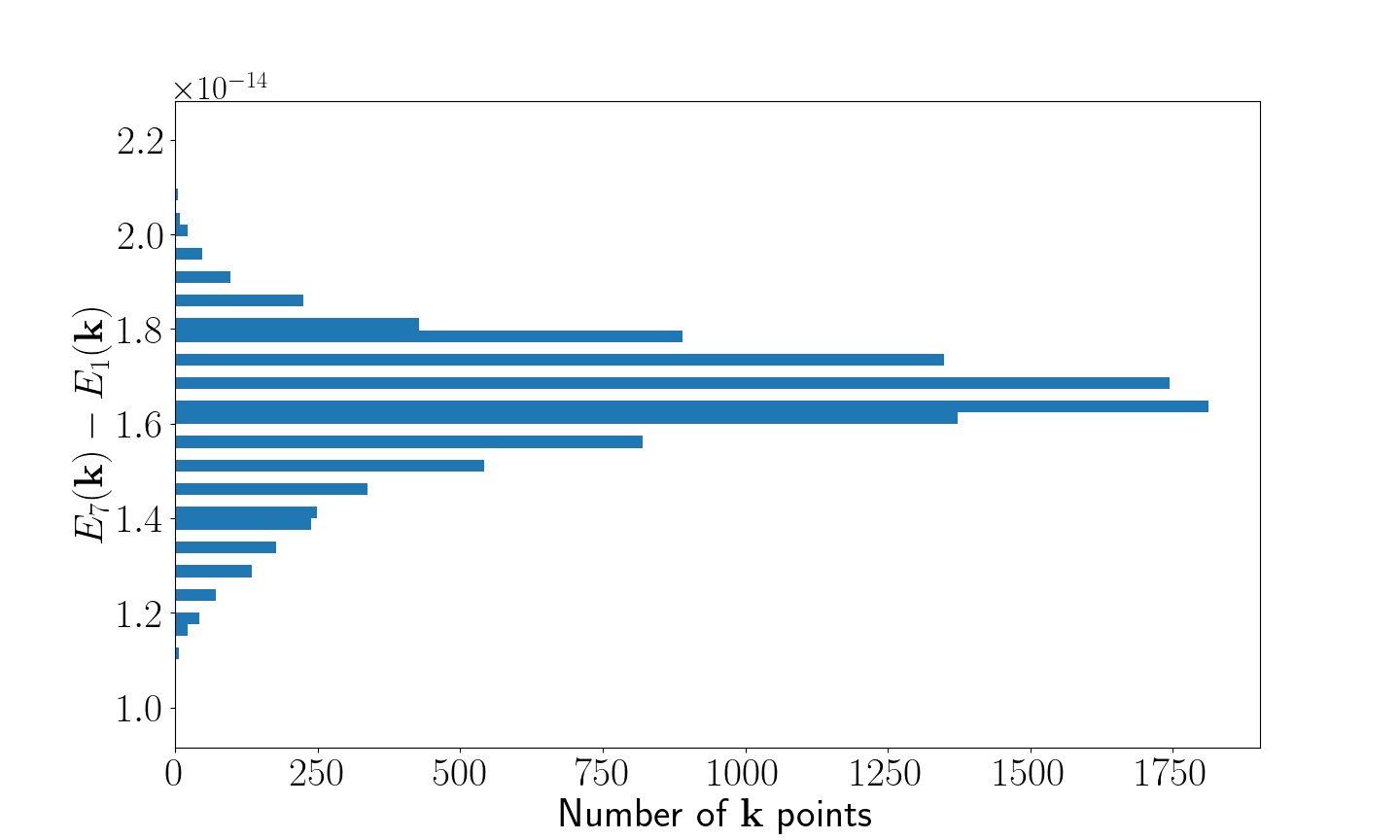}} &
\subfloat[]{\includegraphics[width=\overlap\textwidth]{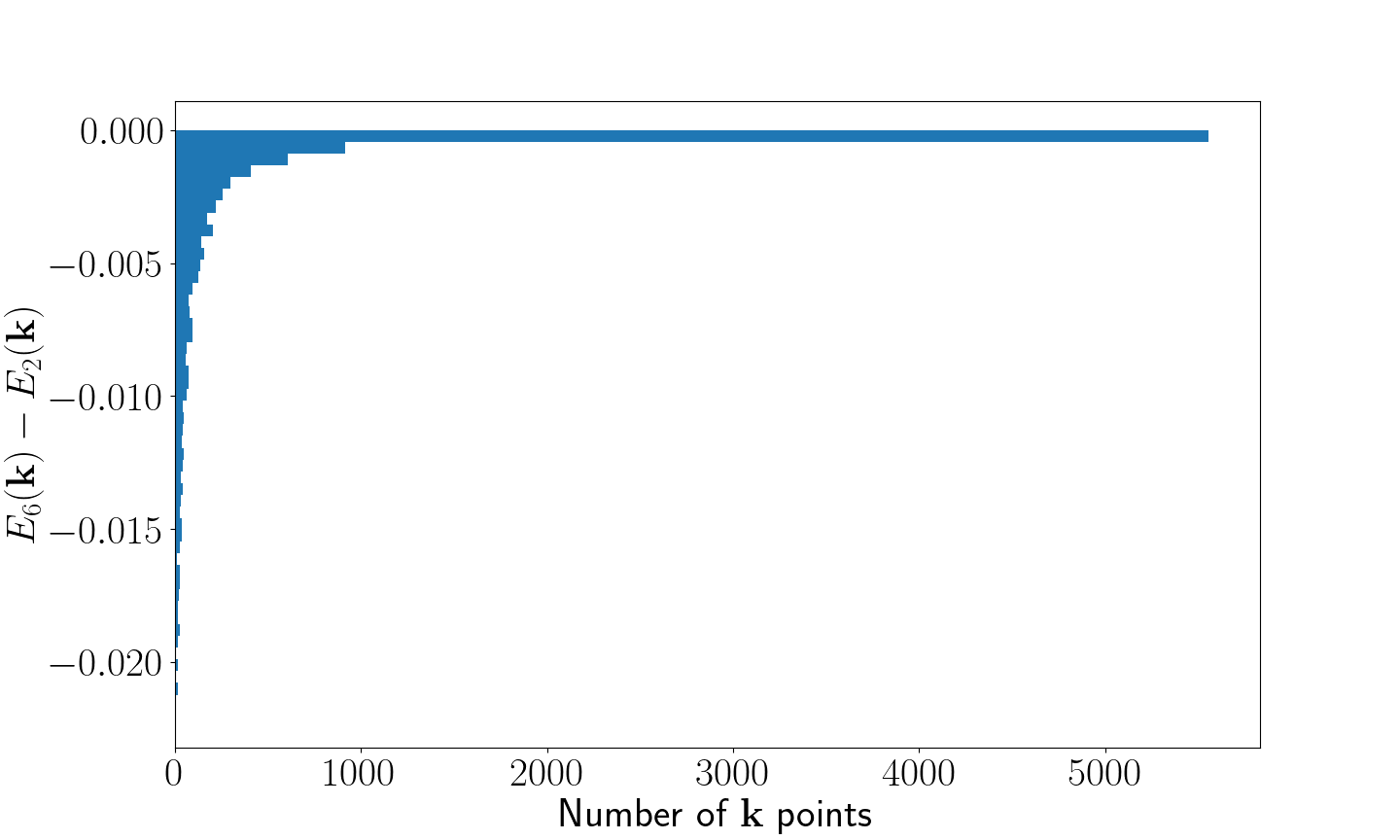}} &
\subfloat[]{\includegraphics[width=\overlap\textwidth]{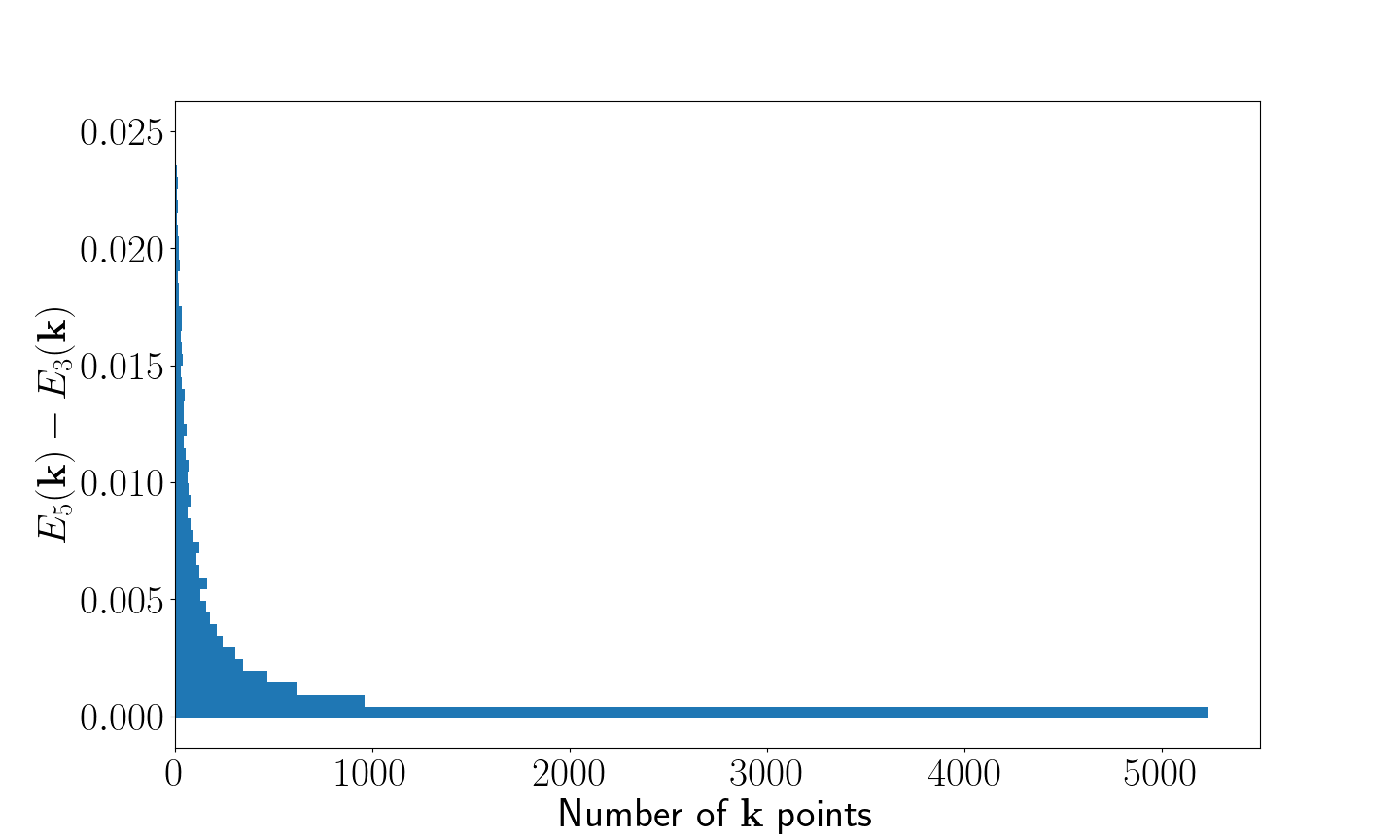}} \\[-3ex]
\subfloat[]{\includegraphics[width=\overlap\textwidth]{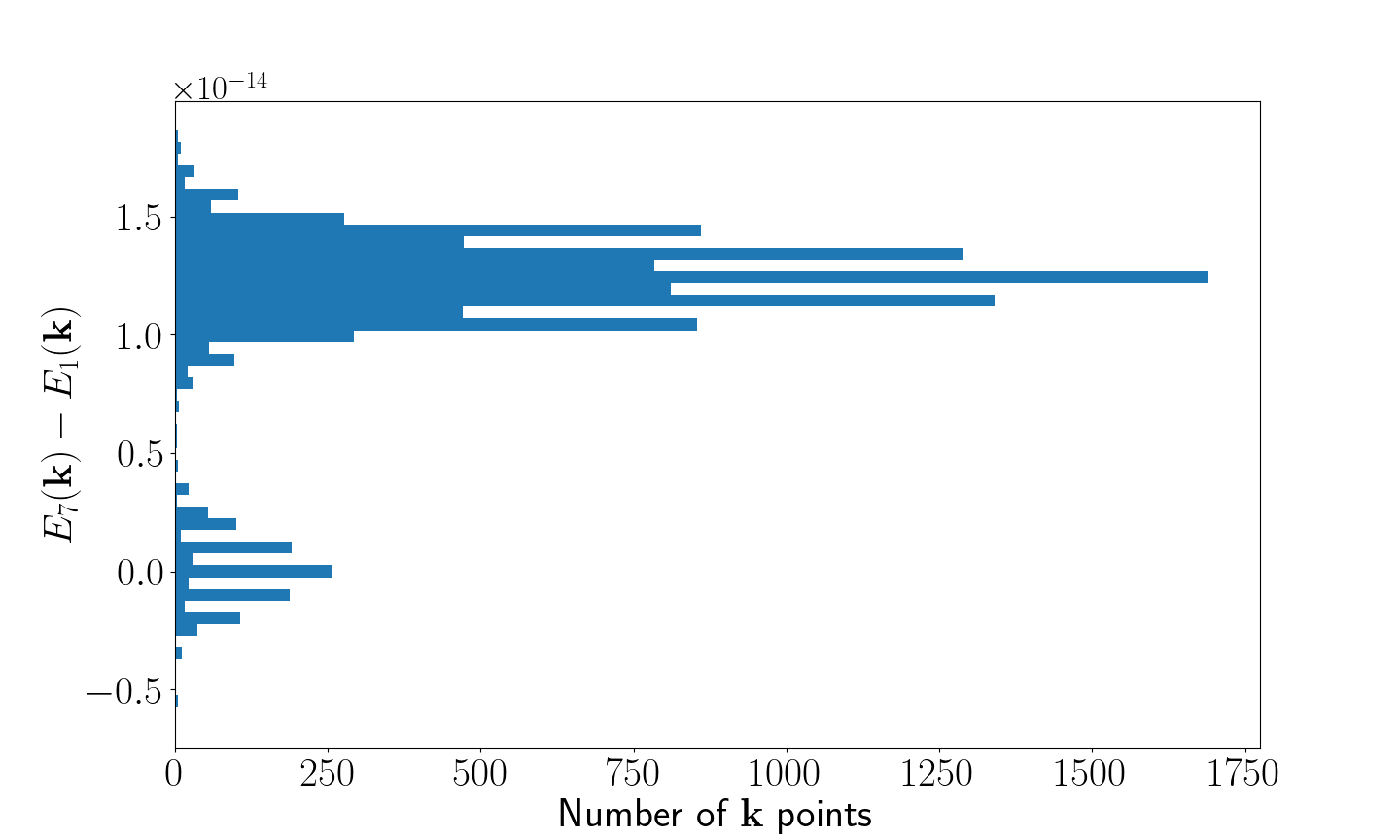}} &
\subfloat[]{\includegraphics[width=\overlap\textwidth]{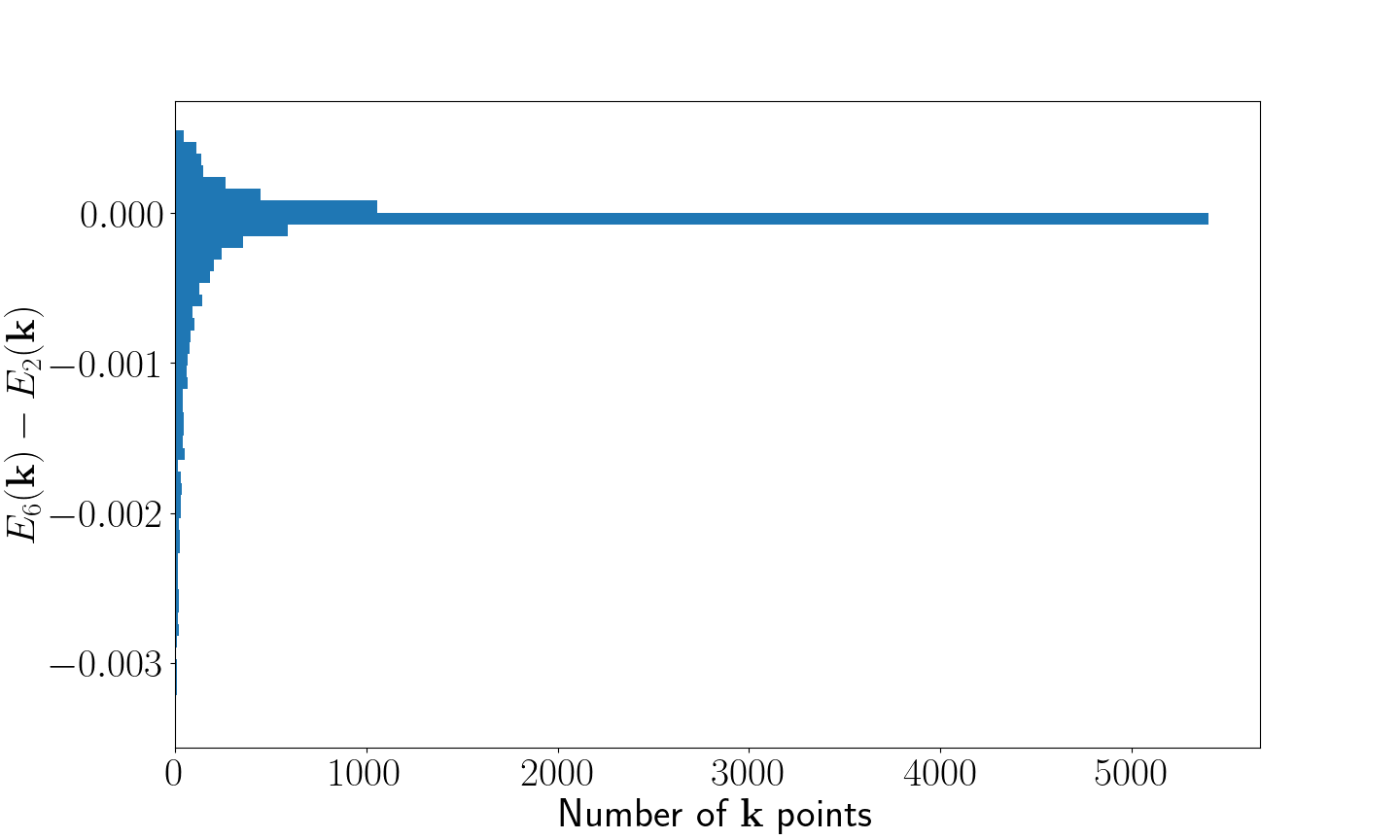}} &
\subfloat[]{\includegraphics[width=\overlap\textwidth]{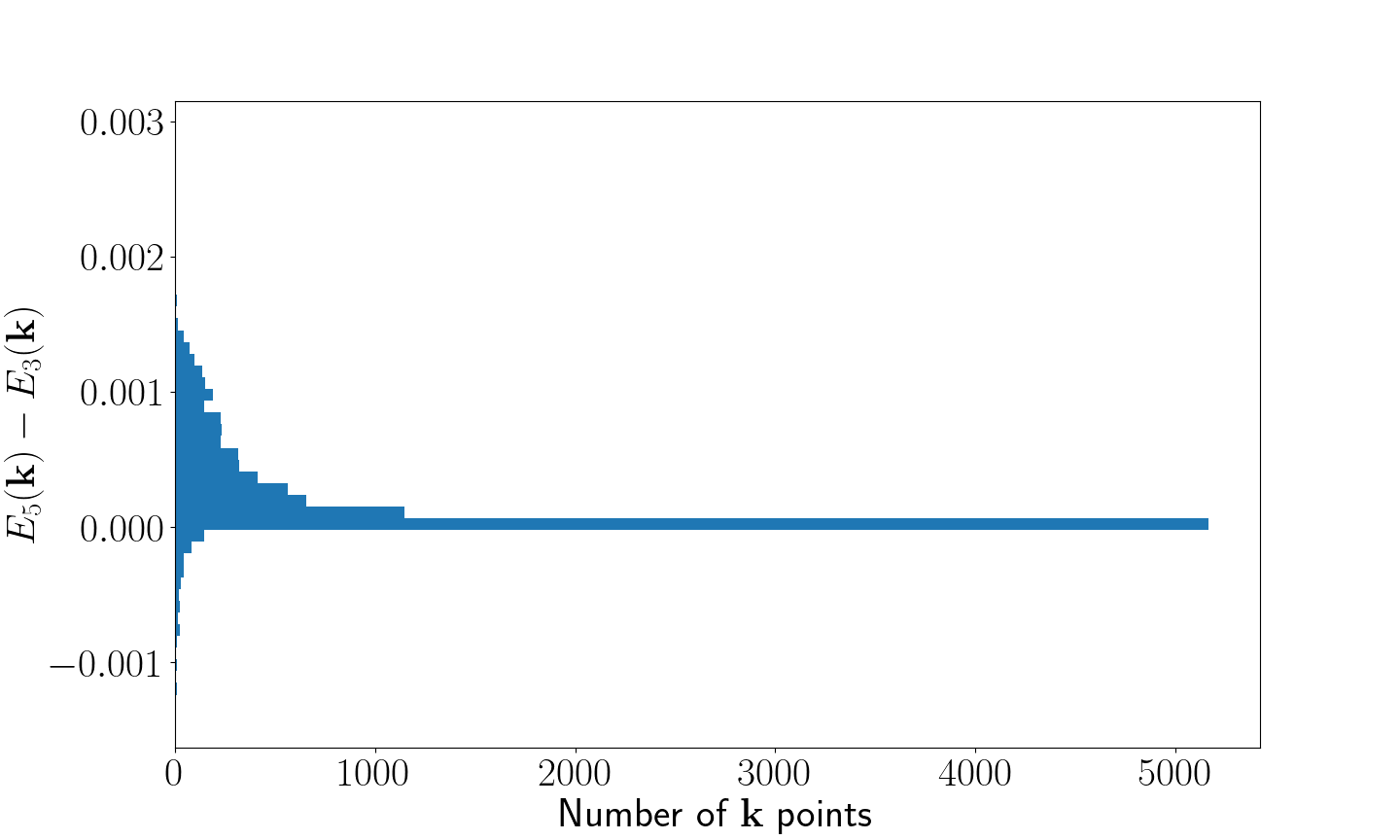}}\\[-3ex]
\subfloat[]{\includegraphics[width=\overlap\textwidth]{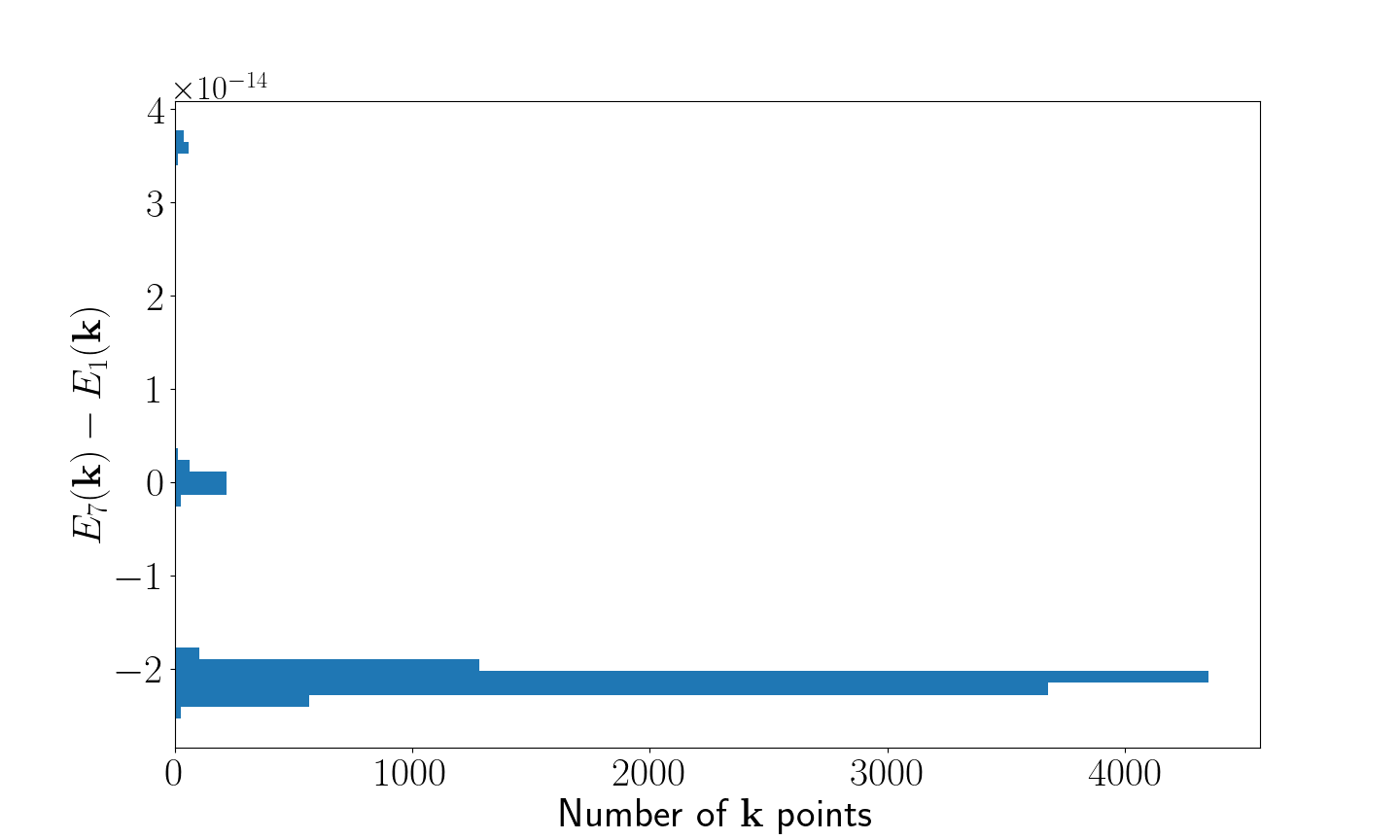}} &
\subfloat[]{\includegraphics[width=\overlap\textwidth]{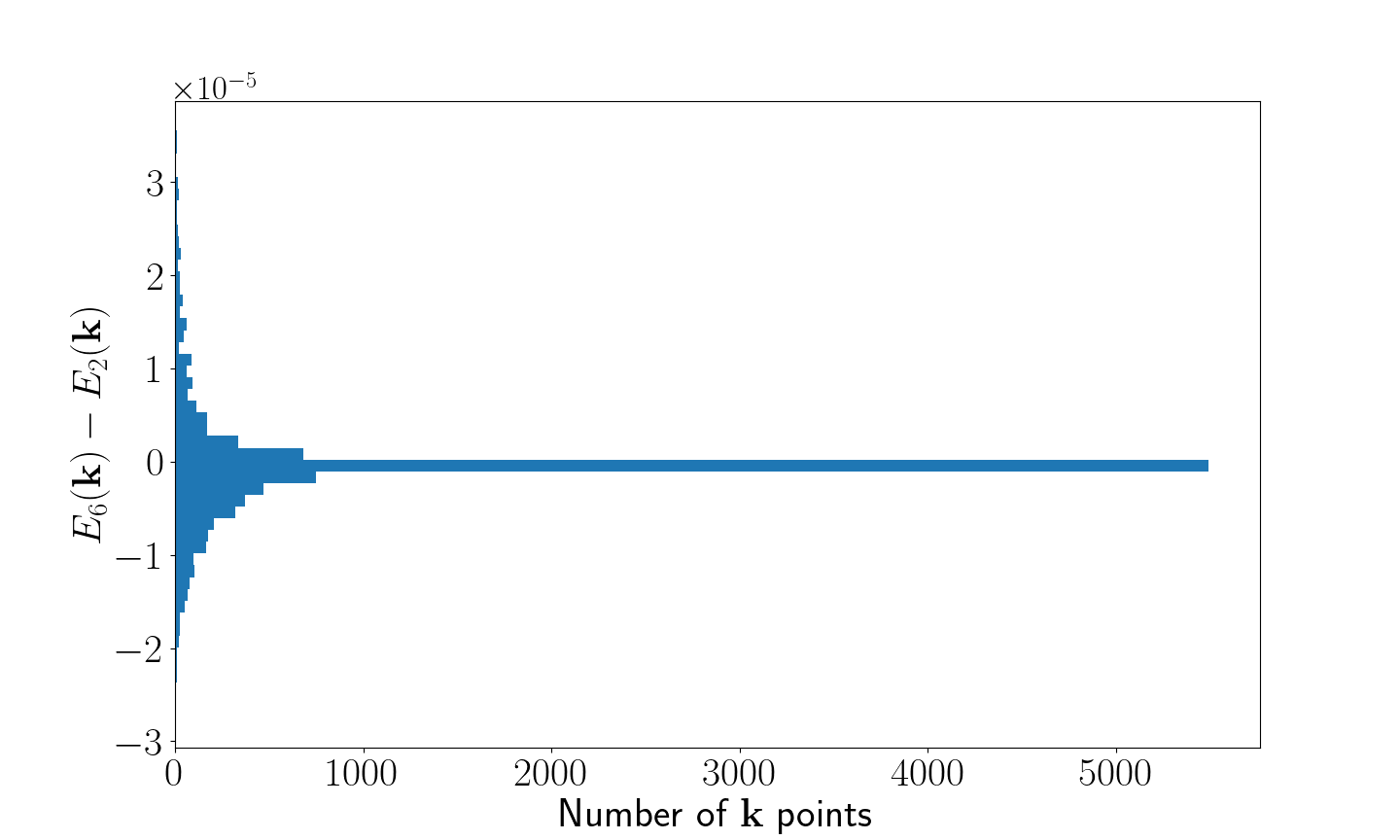}} &
\subfloat[]{\includegraphics[width=\overlap\textwidth]{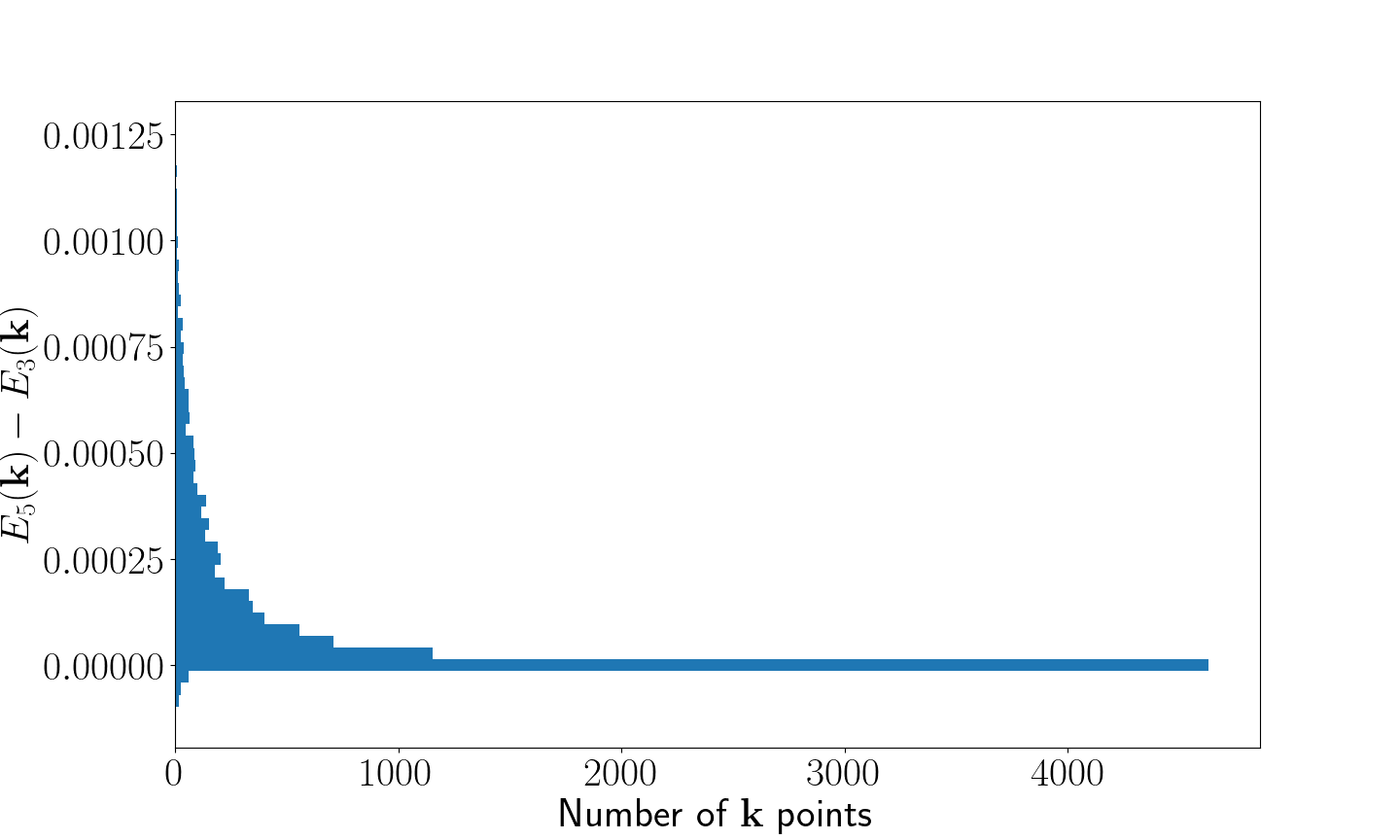}}
\end{tabularx}}
\caption{The difference between particle and hole doped ground state energies $E_{4+n}(\mathbf{k})-E_{4-n}(\mathbf{k})$ for the space group $P4_2/mbc1^\prime$ ($\# 135$) Hamiltonian $H^{0}_{135}$ with HK interactions $H^{1}_{HK}$ (top row), $H^{1}_{HK}+H^{2}_{HK}$ (middle row) and $H^{1}_{HK}+H^{2}_{HK}+H^{3}_{HK}$ (bottom row). 
The difference in energies is calculated for a cube of $10,000$ $\mathbf{k}$ points in the positive octant of the Brillouin Zone.
The particle hole symmetry breaking is of order $(t'_{2})^2/U$.
(a) shows the distribution of the difference between the seven and one particle ground state energies, $E_{7}(\mathbf{k})-E_{1}(\mathbf{k})$, for $H^{0}_{135}+H^{1}_{HK}$. 
As expected, there is no difference between the energies to within numerical error.
(b) shows the distribution of the difference between the six and two particle ground state energies, $E_{6}(\mathbf{k})-E_{2}(\mathbf{k})$, for $H^{0}_{135}+H^{1}_{HK}$. 
The maximum particle hole symmetry breaking is of order $(t'_{2})^2/U\sim 0.02$.
(c) shows the distribution of the difference between the five and three particle ground state energies, $E_{5}(\mathbf{k})-E_{3}(\mathbf{k})$, for $H^{0}_{135}+H^{1}_{HK}$. 
(d) shows the distribution of the difference between the seven and one particle ground state energies, $E_{7}(\mathbf{k})-E_{1}(\mathbf{k})$, across the positive octant of the Brillouin Zone for $H^{0}_{135}+H^{1}_{HK}+H^{2}_{HK}$. 
As expected, to within numerical error, there is no particle-hole symmetry breaking.
(e) shows the distribution of the difference between the six and two particle ground state energies, $E_{6}(\mathbf{k})-E_{2}(\mathbf{k})$, for $H^{0}_{135}+H^{1}_{HK}+H^{2}_{HK}$. 
Here, $(t'_{2})^2/(U_{1}+2U_{2})\sim 0.01$.
(f) shows the distribution of the difference between the five and three particle ground state energies, $E_{5}(\mathbf{k})-E_{3}(\mathbf{k})$, for $H^{0}_{135}+H^{1}_{HK}+H^{2}_{HK}$. 
(f) shows the distribution of the difference between the seven and one particle ground state energies, $E_{7}(\mathbf{k})-E_{1}(\mathbf{k})$, for $H^{0}_{135}+H^{1}_{HK}+H^{2}_{HK}+H^{3}_{HK}$. 
As expected, to within numerical error, there is no particle-hole symmetry breaking.
(g) shows the distribution of the difference between the six and two particle ground state energies, $E_{6}(\mathbf{k})-E_{2}(\mathbf{k})$, for $H^{0}_{135}+H^{1}_{HK}+H^{3}_{HK}+H^{2}_{HK}$. 
Here, $(t'_{2})^2/(U_{1}+2U_{2}+U_{3})\sim 0.01$.
(h) shows the distribution of the difference between the five and three particle ground state energies, $E_{5}(\mathbf{k})-E_{3}(\mathbf{k})$, for $H^{0}_{135}+H^{1}_{HK}+H^{2}_{HK}+H^{3}_{HK}$. 
}
\label{fig:ParticleHoleBrillouinZone135}
\end{figure*}

\begin{figure*}
\FPeval{\overlap}{0.33}
\FPeval{\scalevalue}{round(0.33*(1.05)/\overlap,2)}
\scalebox{\scalevalue}{
\setlength{\tabcolsep}{0pt} 
\def\tabularxcolumn#1{m{#1}}
\hskip-0.75cm\begin{tabularx}{\textwidth}{@{}XXX@{}}
\subfloat[]{\includegraphics[width=\overlap\textwidth]{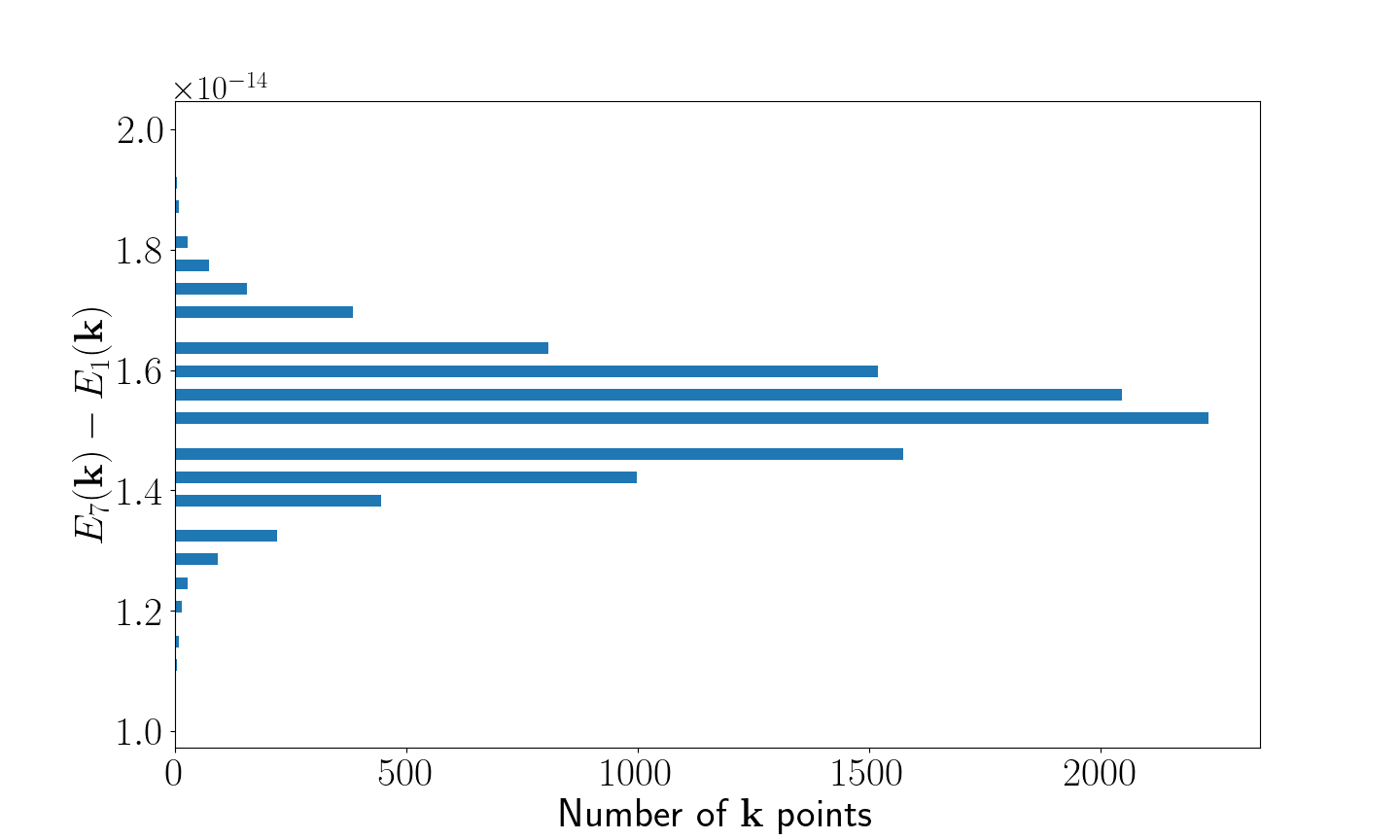}} &
\subfloat[]{\includegraphics[width=\overlap\textwidth]{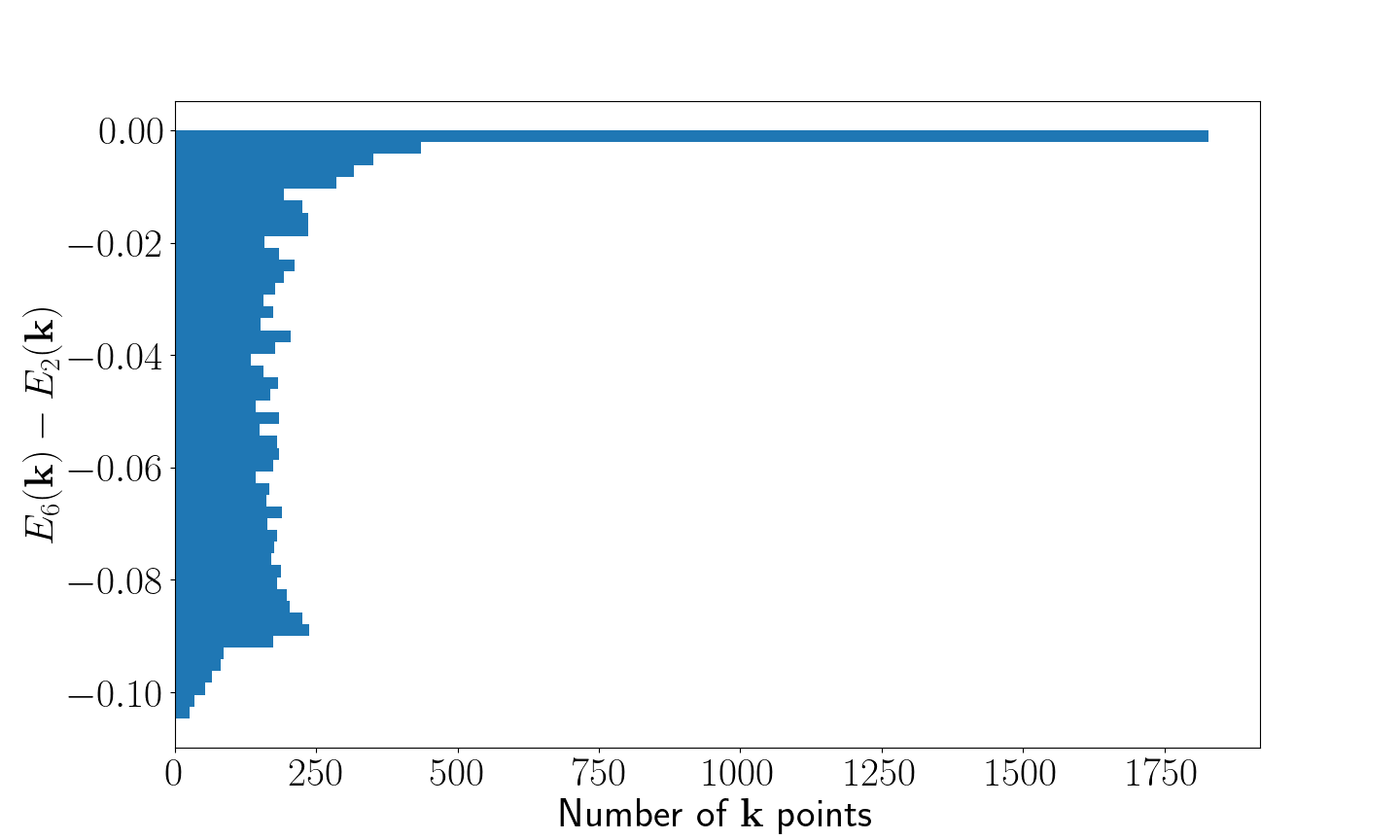}} &
\subfloat[]{\includegraphics[width=\overlap\textwidth]{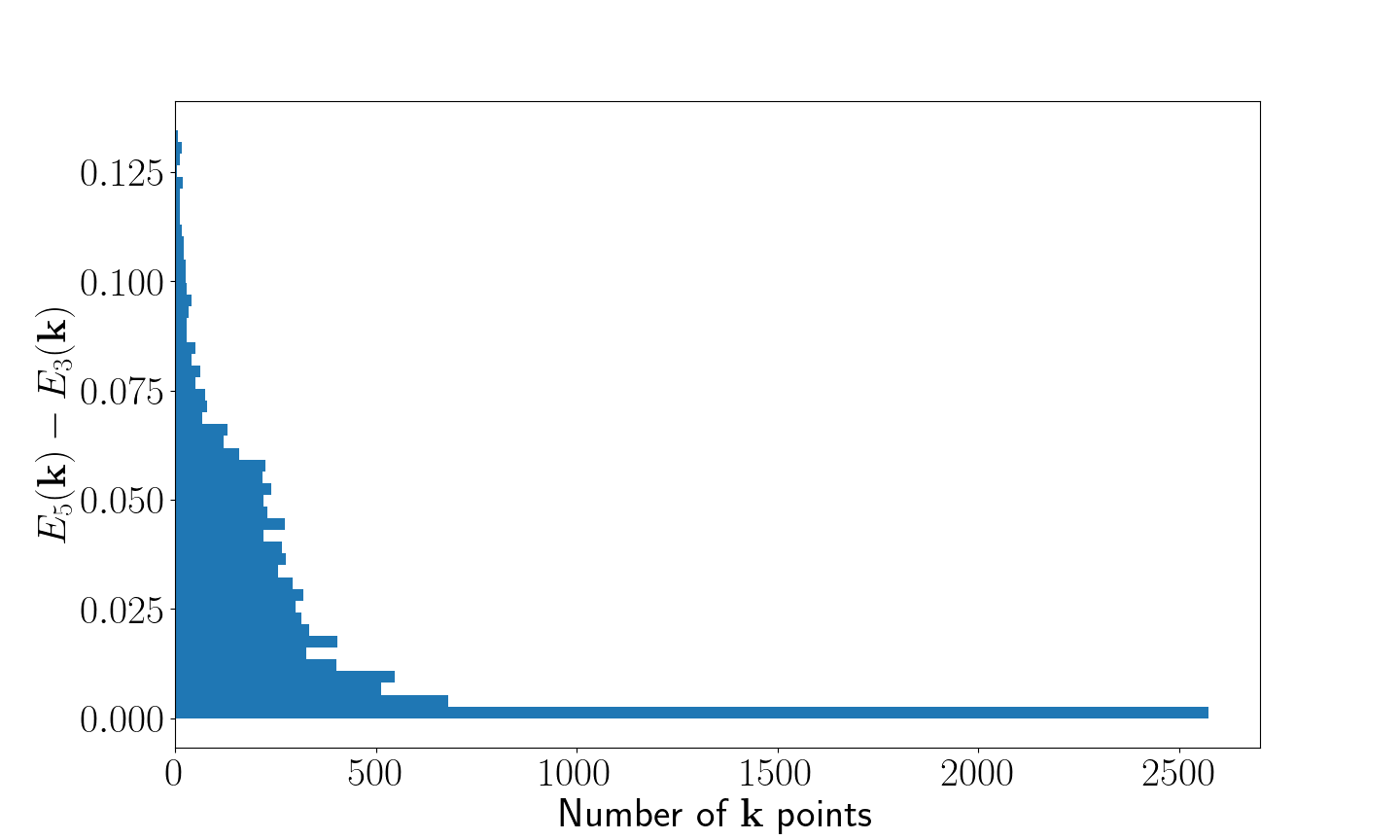}} \\[-3ex]
\subfloat[]{\includegraphics[width=\overlap\textwidth]{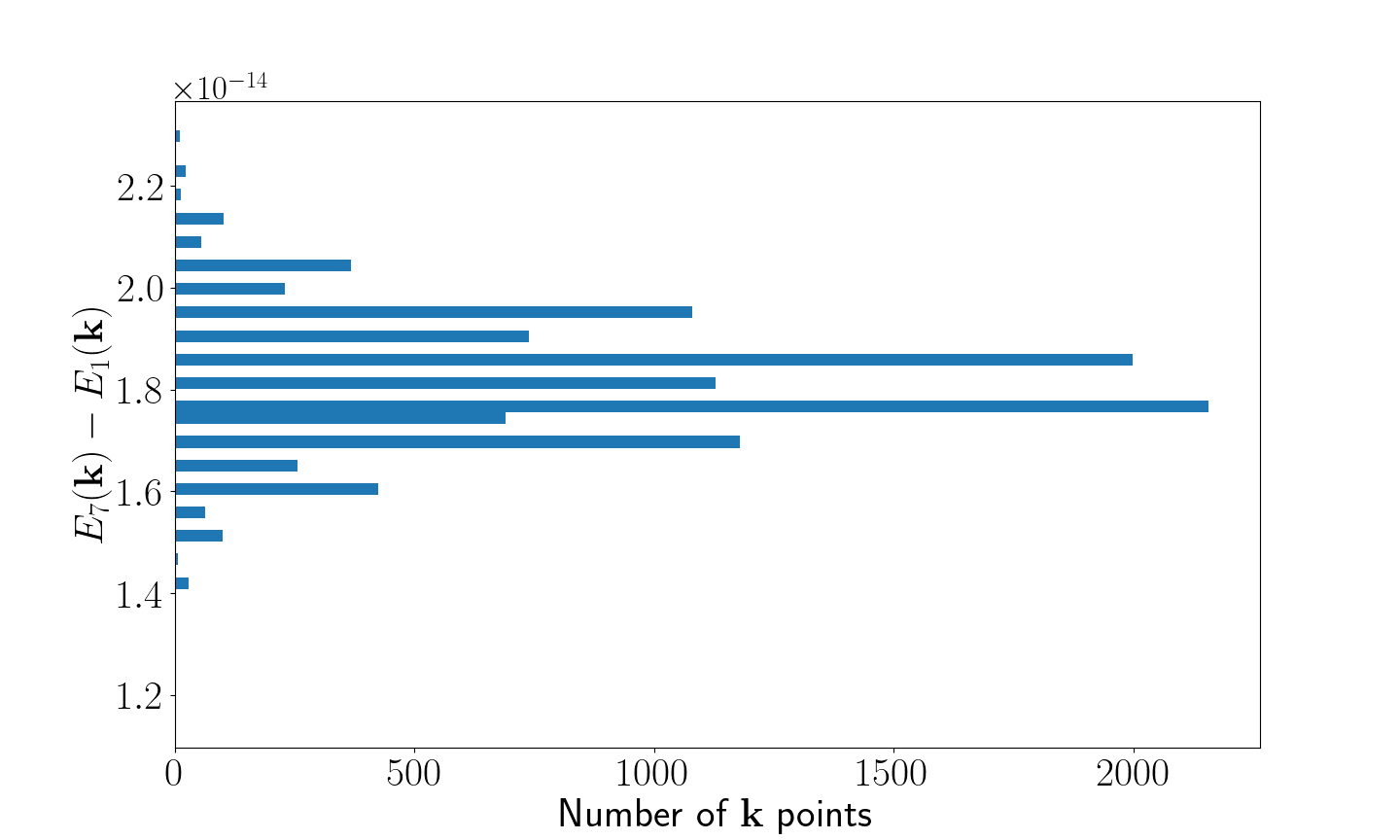}} &
\subfloat[]{\includegraphics[width=\overlap\textwidth]{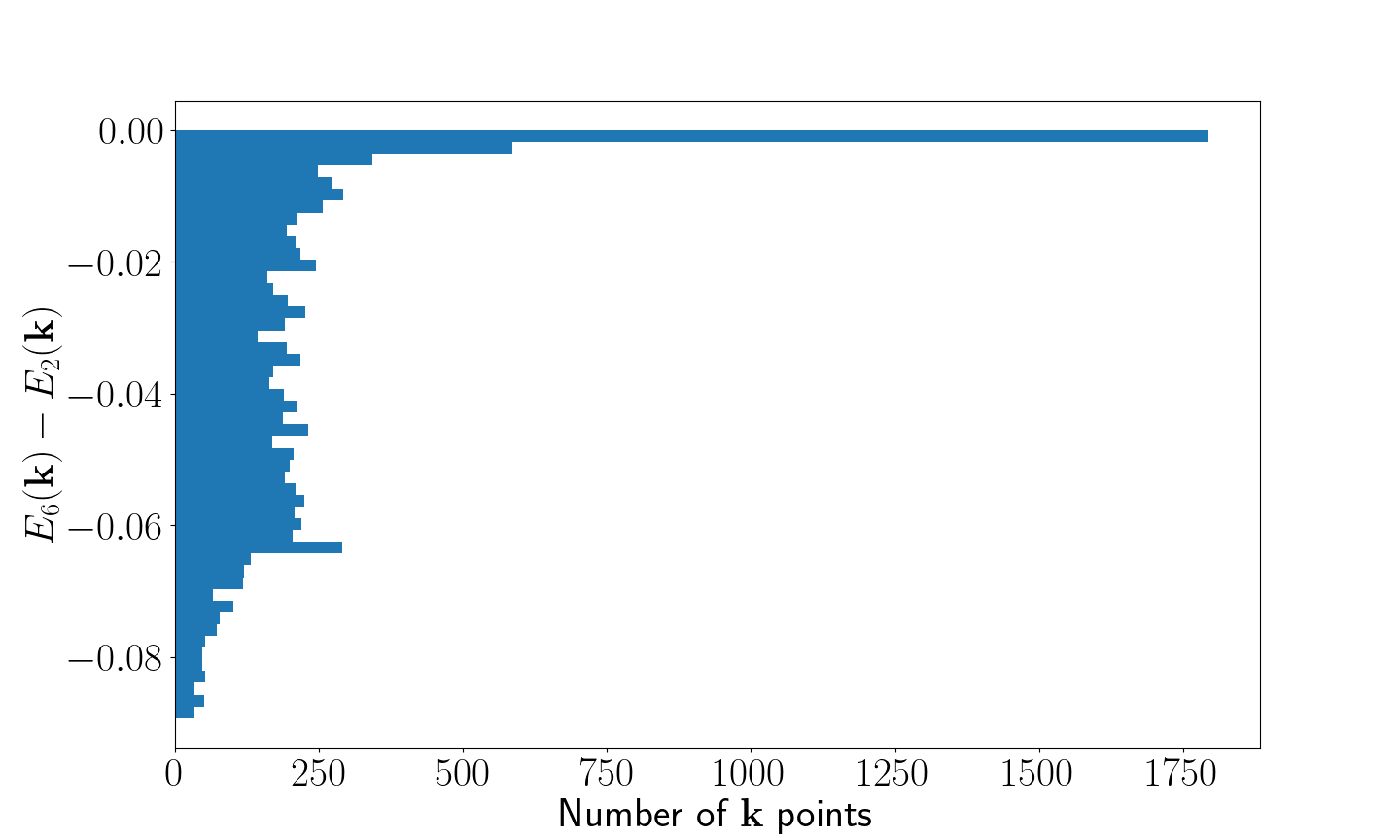}} &
\subfloat[]{\includegraphics[width=\overlap\textwidth]{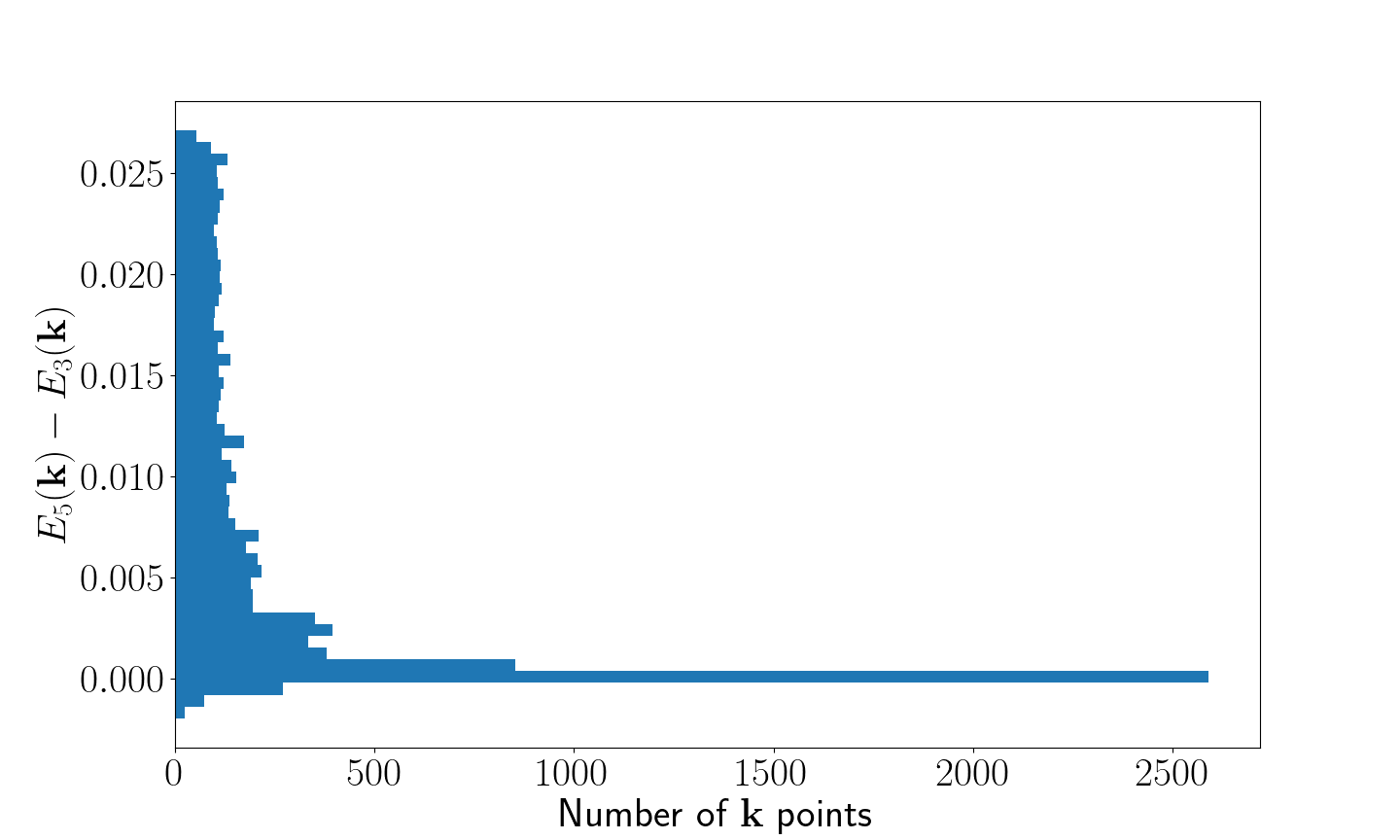}}\\[-3ex]
\subfloat[]{\includegraphics[width=\overlap\textwidth]{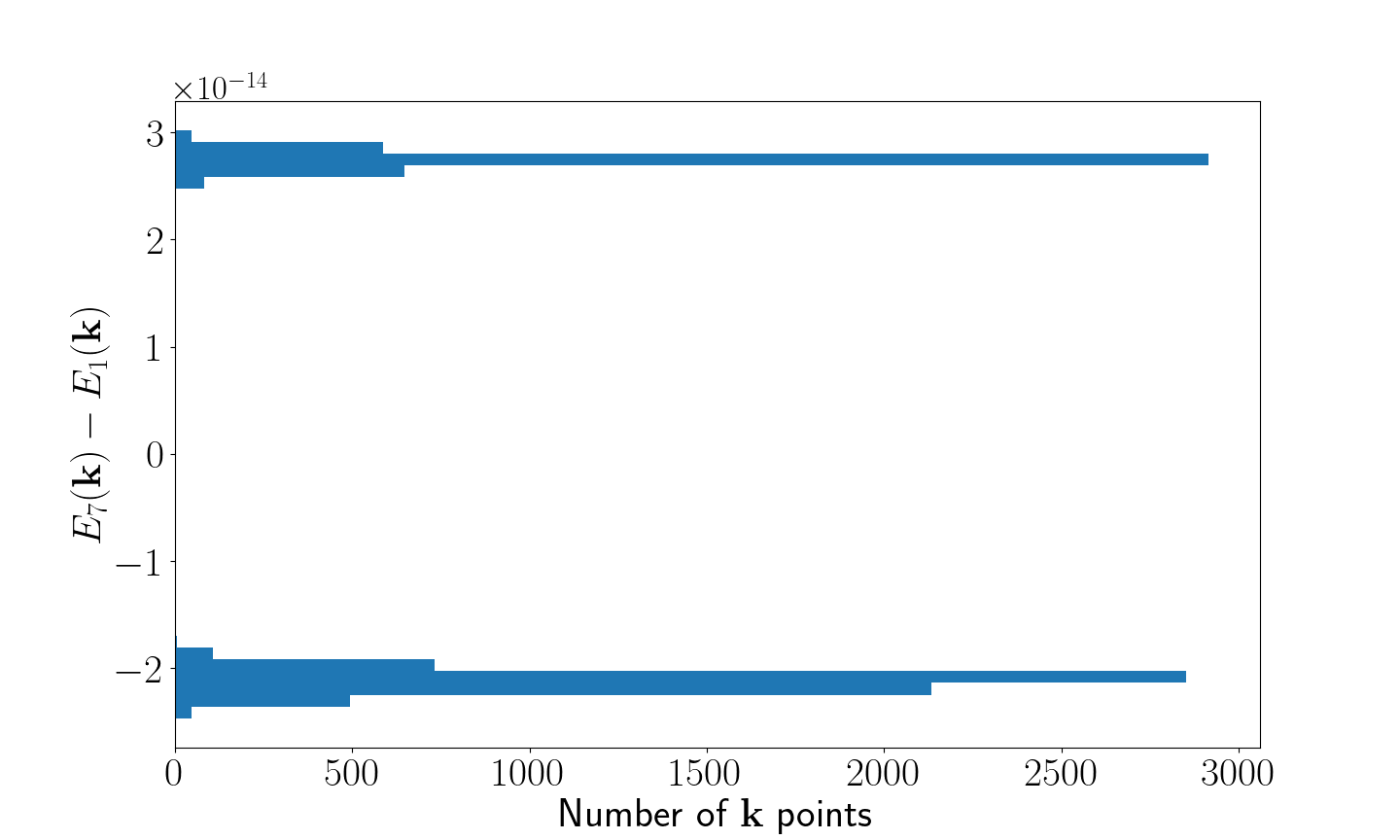}} &
\subfloat[]{\includegraphics[width=\overlap\textwidth]{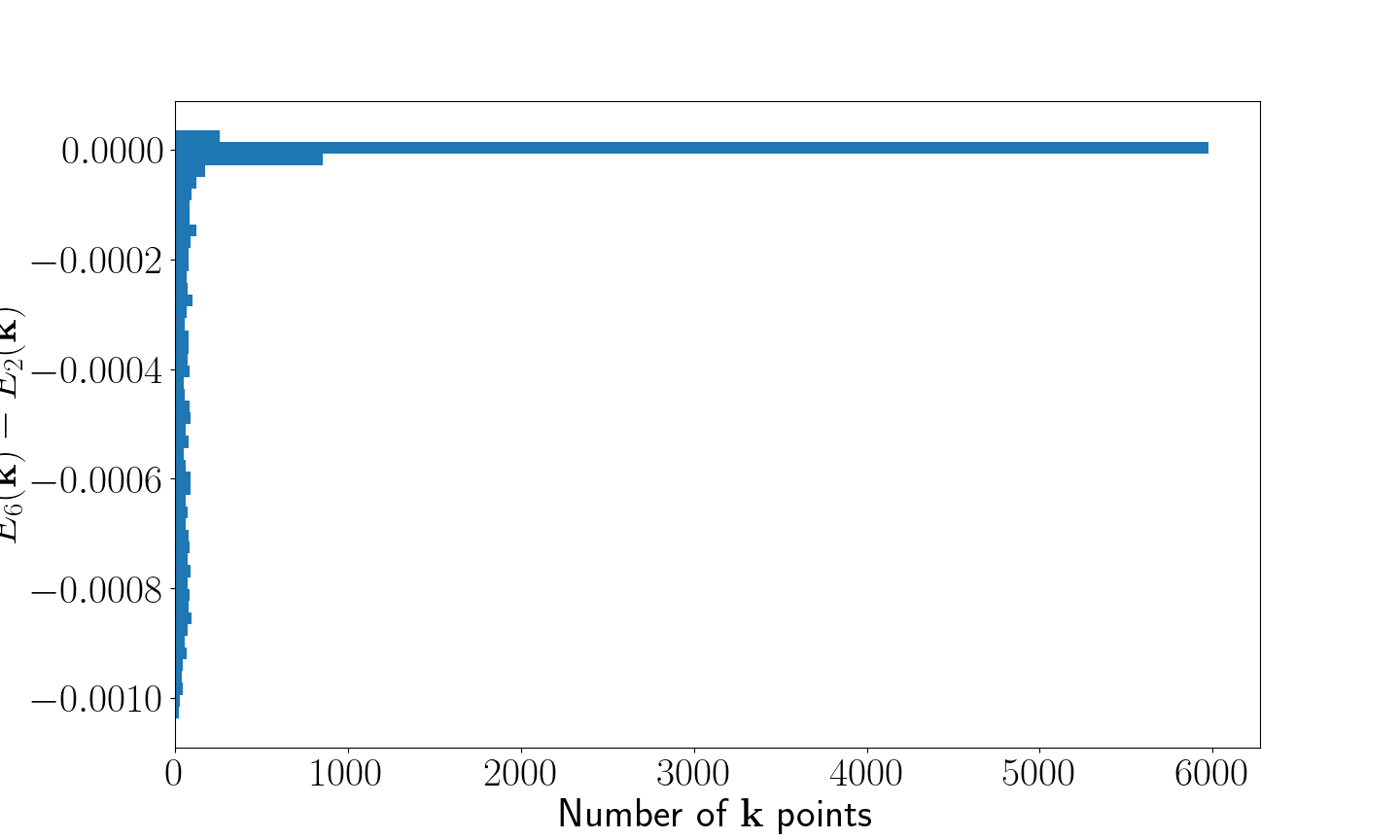}} &
\subfloat[]{\includegraphics[width=\overlap\textwidth]{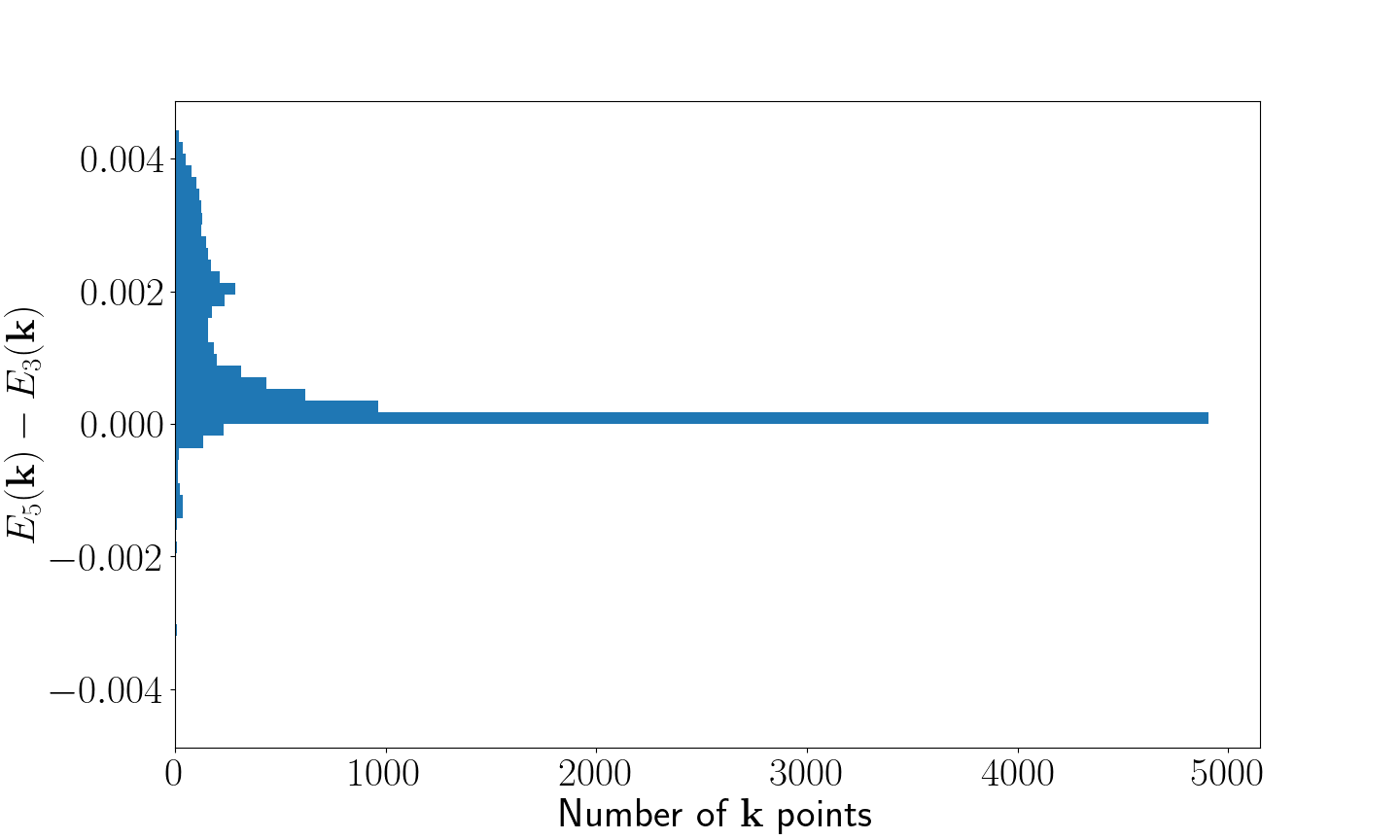}}
\end{tabularx}}
\caption{The difference between particle and hole doped ground state energies $E_{4+n}(\mathbf{k})-E_{4-n}(\mathbf{k})$ for the space group $P4/ncc1^\prime$ (\# 130) Hamiltonian $H^{0}_{130}$ with HK interactions $H^{1}_{HK}$ (top row), $H^{1}_{HK}+H^{2}_{HK}$ (middle row) and $H^{1}_{HK}+H^{2}_{HK}+H^{3}_{HK}$ (bottom row). 
The difference in energies is calculated for a cube of $10,000$ $\mathbf{k}$ points in the positive octant of the Brillouin Zone.
The particle hole symmetry breaking is of order $\lambda^2_1/U$.
(a) shows the distribution of the difference between the seven and one particle ground state energies, $E_{7}(\mathbf{k})-E_{1}(\mathbf{k})$, for $H^{0}_{135}+H^{1}_{HK}$. 
As expected, there is no difference between the energies to within numerical error.
(b) shows the distribution of the difference between the six and two particle ground state energies, $E_{6}(\mathbf{k})-E_{2}(\mathbf{k})$, for $H^{0}_{135}+H^{1}_{HK}$. 
The maximum particle hole symmetry breaking is of order $\lambda^2/U_{1}\sim 0.2$.
(c) shows the distribution of the difference between the five and three particle ground state energies, $E_{5}(\mathbf{k})-E_{3}(\mathbf{k})$, for $H^{0}_{135}+H^{1}_{HK}$. 
(d) shows the distribution of the difference between the seven and one particle ground state energies, $E_{7}(\mathbf{k})-E_{1}(\mathbf{k})$, for $H^{0}_{135}+H^{1}_{HK}+H^{2}_{HK}$. 
As expected, to within numerical error, there is no particle-hole symmetry breaking.
(e) shows the distribution of the difference between the six and two particle ground state energies, $E_{6}(\mathbf{k})-E_{2}(\mathbf{k})$, for $H^{0}_{135}+H^{1}_{HK}+H^{2}_{HK}$. 
Here, $\lambda^2_{1}/(U_{1}+2U_{1})\sim 0.01$.
(f) shows the distribution of the difference between the five and three particle ground state energies, $E_{5}(\mathbf{k})-E_{3}(\mathbf{k})$, for $H^{0}_{130}+H^{1}_{HK}+H^{2}_{HK}$. 
Here, $\lambda^2_{1}/(U_{1}+2U_{2})\sim 0.01$.
(g) shows the distribution of the difference between the seven and one particle ground state energies, $E_{7}(\mathbf{k})-E_{1}(\mathbf{k})$, for $H^{0}_{130}+H^{1}_{HK}+H^{2}_{HK}+H^{3}_{HK}$. 
As expected, to within numerical error, there is no particle-hole symmetry breaking.
(h) shows the distribution of the difference between the six and two particle ground state energies, $E_{6}(\mathbf{k})-E_{2}(\mathbf{k})$, for $H^{0}_{130}+H^{1}_{HK}+H^{3}_{HK}+H^{2}_{HK}$. 
Here, $\lambda^2_{1}/(U_{1}+2U_{2}+U_{3})\sim 0.01$.
(i) shows the distribution of the difference between the five and three particle ground state energies, $E_{5}(\mathbf{k})-E_{3}(\mathbf{k})$, for $H^{0}_{135}+H^{1}_{HK}+H^{2}_{HK}+H^{3}_{HK}$. 
}

\label{fig:ParticleHoleBrillouinZone130}
\end{figure*}

\subsection{Space Group $P4_2/mbc1^\prime$ (\# 135)}\label{subsec:ph135}
The single-particle Hamiltonian $\mathcal{H}_{135}^{0}(\mathbf{k})$ at each $\mathbf{k}$ for our models in space group $P4_2/mbc1^\prime$ (\# 135) is given in \Cref{eq:135 Hamiltonian}. 
Although the spectrum of $\mathcal{H}_{135}^{0}(\mathbf{k})$ is symmetric about zero, there is in general no particle-hole symmetry in the sense of an operator on Fock space. 
We note that particle-hole symmetry can in general be represented on Fock space as
\begin{equation}\label{eq:phgeneral}
Pc_{\mathbf{k}i}P^{-1} = \sum_j A^{ij}c^\dag_{\mathbf{k}j},
\end{equation}
where $i,j$ index the spin and orbital degrees of freedom, and $A^{ij}$ is a matrix. 
Due to the canonical anticommutation relations, for $P$ to be a symmetry of a single-particle Hamiltonian it must anticommute with all symmetric hopping terms, and commute with all antisymmetric hopping terms. 
To see this, we note that for single particle Hamiltonians of the form
\begin{equation}
\mathcal{H}^0(\mathbf{k}) = \sum_{ij}c^\dag_{\mathbf{k}i}h_{\mathbf{k}ij}c_{\mathbf{k}j}\label{eq:generalspham}
\end{equation}
we have
\begin{align}
P\mathcal{H}^0(\mathbf{k})P^{-1} &= \sum_{iji'j'} A^{*ii'}c_{\mathbf{k}i'}h_{\mathbf{k}ij}A^{jj'}c^\dag_{\mathbf{k}} \\
&= -\sum_{iji'j'}c^\dag_{\mathbf{k}j'}A^{\dag i'i} h_{\mathbf{k}ij}A^{jj'}c_{\mathbf{k}i'} + \sum_i h_{\mathbf{k}ii}.\label{eq:spph}
\end{align}
Provided the trace of $h_{\mathbf{k}ij}$ is zero---which is the case for all Hamiltonians we consider in this work---particle-hole symmetry then requires
\begin{equation}
\sum_{i'j'} A^{\dag i'i} h_{\mathbf{k}ij}A^{jj'} = -h_{\mathbf{k}ji},
\end{equation}
obtained by equating \Cref{eq:generalspham,eq:spph}.

No such matrix $A$ exists for the Hamiltonian $\mathcal{H}_{135}^{0}(\mathbf{k})$ in \Cref{eq:135 Hamiltonian} if all hopping amplitudes are nonzero. 
However, if we choose $A=\mu_y\tau_y\sigma_y$, then $P$ commutes with every term in \Cref{eq:135 Hamiltonian} except the hopping with amplitude $t_2^\prime$. 
Thus, the single-particle Hamiltonian is particle-hole symmetric for $t_2^\prime=0$.

Let us turn now to the seven nontrivial orbital HK terms of the form
\begin{equation}
\mathcal{H}_{HK}^{N}(\mathbf{k}) = \sum_{a=2}^8U_a N_a n_{\mathbf{k}i}N^a_{ij}n_{\mathbf{k}j}
\end{equation}
where $U_a$ are positive interaction strength parameters and the matrices $N^a$ are given in Table~\ref{table:HKNterms}. 
Under the action of $P$ defined in \Cref{eq:phgeneral} with $A=\mu_y\tau_y\sigma_y$ we have
\begin{equation}
P\mathcal{H}_{HK}^{N}(\mathbf{k})P^{-1} = H_{HK}^N - \sum_{a=1}^{8}U_a \sum_{i} n_{\mathbf{k}i}.
\end{equation}

Finally, we note that the symmetry-allowed terms of the form
\begin{equation}
\mathcal{H}^{F}_{HK}(\mathbf{k}) = \sum_{b=1}^{3}U'_b \left(\sum_{ij}c^\dag_{\mathbf{k}i}F^b_{ij}c_{\mathbf{k}j}\right)^2
\end{equation}
defined in Equations~\eqref{eq:135f1}--\eqref{eq:135f3} are manifestly invariant under $P$ defined in \Cref{eq:phgeneral} with $A=\mu_y\tau_y\sigma_y$.

Putting this all together, we see then that if the chemical potential
\begin{equation}\label{eq:135musym}
\mu_0 = \frac{1}{2}\sum_{a=1}^{8}U_a \sum_{i}n_{\mathbf{k}i},
\end{equation}
then the Hamiltonian $\mathcal{H}_{135}^{0}(\mathbf{k})+\mathcal{H}_{HK}^{N}(\mathbf{k})+\mathcal{H}^F_{HK}(\mathbf{k}) -\mu_0 \sum_{i} n_{\mathbf{k}i}$ will be particle-hole symmetric if $t_2^\prime=0$.

In particular, this means that when $t_2^\prime=0$ and with chemical potential given in \Cref{eq:135musym}, we expect based on our arguments in Sec.~\ref{sec:orbital-hk} that the ground state at every $\mathbf{k}$ consists of states with four particles. 
Furthermore, we note that since $t_2^\prime \ll \sum_{a=2}^{8} U_a$ for the models we consider, we expect that particle-hole symmetry breaking is weak at every $\mathbf{k}$. 
To verify this, we show in Fig.~\ref{fig:ParticleHole}(a--c) the lowest energy $E_n(\mathbf{k})$ in the $n$-particle subspace at each $\mathbf{k}$ for the models in space group $P4_2/mbc1^\prime$ (\# 135) considered in the main text. 
We see that in all cases, the four-particle energy $E_4(\mathbf{k})$ is the lowest energy, confirming that for the parameter values considered in the text and for all $\mathbf{k}$, the lowest energy state half-fills (in this case four electrons per unit cell) every $\mathbf{k}$ point. 
Furthermore, in Fig.~\ref{fig:ParticleHoleBrillouinZone135} we show the distribution of energy differences $E_{4-n}(\mathbf{k})-E_n(\mathbf{k})$ sampled over the whole Brillouin zone, allowing us to verify that particle-hole symmetry is only weakly broken for the parameter values analyzed in the text.

\subsection{Space Group $P4/ncc1^\prime$ (\# 130)}

We can perform a similar analysis for our HK models in space group $P4/ncc1^\prime$ (\# 130). 
The single-particle Hamiltonian $\mathcal{H}_{130}^{0}$ was defined in \Cref{eq:130noninteractingH}. 
As in Sec.~\ref{subsec:ph135}, this single-particle Hamiltonian is not in general particle-hole symmetric. 
However, when $\lambda_1=0$, $\mathcal{H}_{130}^{0}$ is particle-hole symmetric at every $\mathbf{k}$, with $P$ defined in \Cref{eq:phgeneral} and $A=\mu_y\tau_z\sigma_y$. 
Furthermore, just as in the previous section, all HK interactions are invariant under this particle-hole transformation as well provided the chemical potential $\mu$ is given by \Cref{eq:135musym}. 

Thus, when $\lambda_1=0$ and with chemical potential given in \Cref{eq:135musym}, we expect based on our arguments in Sec.~\ref{sec:orbital-hk} that the ground state at every $\mathbf{k}$ consists of states with four particles. 
Furthermore, we note that since $\lambda_1 \ll \sum_{a=2}^{8} U_a$ for the models we consider, we expect that particle-hole symmetry breaking is weak at every $\mathbf{k}$. 
To verify this, we show in Fig.~\ref{fig:ParticleHole}(d--f) the lowest energy $E_n(\mathbf{k})$ in the $n$-particle subspace at each $\mathbf{k}$ for the models in space group $P4/ncc1^\prime$ (\# 130) considered in the main text. 
We see that in all cases, the four-particle energy $E_4(\mathbf{k})$ is the lowest energy, confirming that for the parameter values considered in the text and for all $\mathbf{k}$, the lowest energy state half-fills (in this case four electrons per unit cell) every $\mathbf{k}$ point. 
Furthermore, in Fig.~\ref{fig:ParticleHoleBrillouinZone130} we show the distribution of energy differences $E_{4-n}(\mathbf{k})-E_n(\mathbf{k})$ sampled over the whole Brillouin zone, allowing us to verify that particle-hole symmetry is only weakly broken for the parameter values analyzed in the text.

\subsection{Space group $P2_1/c1^\prime$ (\# 14)}
Lastly, we consider the HK models in space group $P2_1/c1^\prime$ (\# 14) analyzed in Sec.~\ref{sec:SG14}. 
Here we are more fortunate: the single-particle Hamiltonian \Cref{eq:sg14ham} is particle-hole symmetric for all values of the hopping parameters. 
The particle-hole symmetry operation is given by \Cref{eq:phgeneral} with $A=\tau_y\sigma_x$. 
Since the HK interaction terms are all particle-hole symmetric as well (provided the chemical potential is appropriately chosen), no special care is needed to deduce that the ground state for the HK models in Sec.~\ref{sec:SG14} at half filling consists of two particles at every $\mathbf{k}$.
Nevertheless, we show in \Cref{fig:ParticleHole} that, for the parameter values considered here, for all $\mathbf{k}$, the lowest energy state half-fills (in this case four electrons per unit cell) every $\mathbf{k}$ point.

\section{Ground states at $\Gamma$ and $A$ for \sg135 and \sgb}
\label{sec:groundstates}
We provide here for reference the exact numerical ground states of the system at the $\Gamma$ and $A$ points for our interacting models of space group \sg135 and space group \sgb.
\subsection{Space group \sg135}
For the space group \sg135 Hamiltonian with all HK interactions considered in \Cref{subsec:splitting}:
\begin{equation}
    H=H^{0}_{135}+H_{HK}^{1}+H_{HK}^{2}+H_{HK}^{3},
\end{equation}
the ground state at the $A$ point is given by
\begin{align}
    \ket{GS}_{A}&=\alpha_{1}(\ket{AA\uparrow;AA\downarrow;BA\uparrow;BA\downarrow}+\ket{AB\uparrow;AB\downarrow;BB\uparrow;BB\downarrow}) \\
    &+\alpha_{2}(\ket{AA\uparrow;AA\downarrow;BB\uparrow;BB\downarrow}+\ket{AB\uparrow;AB\downarrow;BA\uparrow;BA\downarrow}) \nonumber \\
    &+\alpha_{3}(\ket{AA\uparrow;AB\downarrow;BA\uparrow;BB\downarrow}+\ket{AA\downarrow;AB\uparrow;BA\downarrow;BB\uparrow}) \nonumber \\
    &+\alpha_{4}(\ket{AA\uparrow;AB\downarrow;BA\downarrow;BB\uparrow}+\ket{AA\downarrow;AB\uparrow;BA\uparrow;BB\downarrow}) \nonumber
\end{align}
with
\begin{align}
    &\alpha_{1}=0.07774353109915988, \nonumber \\
    &\alpha_{2}=0.12444952353294593, \nonumber\\
    &\alpha_{3}=0.3117239903386059, \nonumber \\
    &\alpha_{4}=0.6174920350191032, \nonumber.
\end{align}
The ground state at the $\Gamma$ point is given by:
\begin{align}
    &\alpha_{1}(\ket{AA\uparrow;AA\downarrow;AB\uparrow;AB\downarrow}+\ket{BA\uparrow;BA\downarrow;BB\uparrow;BB\downarrow}) \\
&\alpha_{2}\bigg[\ket{AA\uparrow;AA\downarrow;AB\uparrow;BB\downarrow}+\ket{AA\uparrow;BA\downarrow;BB\uparrow;BB\downarrow} \nonumber \\
&+\ket{AB\uparrow;BA\uparrow;BA\downarrow;BB\downarrow}+\ket{AA\uparrow;AB\uparrow;AB\downarrow;BA\downarrow} \nonumber \\
&-\big[\ket{AA\uparrow;AA\downarrow;AB\downarrow;BB\uparrow}+\ket{AB\downarrow;BA\uparrow;BA\downarrow;BB\uparrow} \nonumber \\
&+\ket{AA\downarrow;BA\uparrow;BB\uparrow;BB\downarrow}+\ket{AA\downarrow;AB\uparrow;AB\downarrow;BA\uparrow})\big]\bigg] \nonumber \\
&\alpha_{3}(\ket{AB\uparrow;AB\downarrow;BB\uparrow;BB\downarrow}+\ket{AA\uparrow;AA\downarrow;BA\uparrow;BA\downarrow}) \nonumber \\
&\alpha_{4}\bigg[\big[(\ket{AA\uparrow;AB\downarrow;BA\uparrow;BA\downarrow}+\ket{AA\uparrow;AA\downarrow;BA\uparrow;BB\downarrow} \nonumber \\
&+\ket{AA\uparrow;AB\downarrow;BB\uparrow;BB\downarrow}+\ket{AB\uparrow;AB\downarrow;BA\uparrow;BB\downarrow})\big] \nonumber \\
&-\big[(\ket{AA\downarrow;AB\uparrow;BB\uparrow;BB\downarrow}+\ket{AB\uparrow;AB\downarrow;BA\downarrow;BB\uparrow} \nonumber \\&+\ket{AA\uparrow;AA\downarrow;BA\downarrow;BB\uparrow}+\ket{AA\downarrow;AB\uparrow;BA\uparrow;BA\downarrow})\big]\bigg] \nonumber \\
&\alpha_{5}(\ket{AB\uparrow;AB\downarrow;BA\uparrow;BA\downarrow}+\ket{AA\uparrow;AA\downarrow;BB\uparrow;BB\downarrow}) \nonumber \\
&\alpha_{6}(\ket{AA\uparrow;AB\uparrow;BA\downarrow;BB\downarrow}+\ket{AA\downarrow;AB\downarrow;BA\uparrow;BB\uparrow}) \nonumber \\
&\alpha_{7}(\ket{AA\downarrow;AB\uparrow;BA\downarrow;BB\uparrow}+\ket{AA\uparrow;AB\downarrow;BA\uparrow;BB\downarrow}) \nonumber \\
&\alpha_{8}(\ket{AA\downarrow;AB\uparrow;BA\uparrow;BB\downarrow}+\ket{AA\uparrow;AB\downarrow;BA\downarrow;BB\uparrow}) \nonumber 
\end{align}
with

\begin{align}
&\alpha_{1}=-0.015244493568331597,\nonumber \\
&\alpha_{2}=0.08320294092507857,\nonumber \\
&\alpha_{3}=-0.07514072175059767,\nonumber \\
&\alpha_{4}=0.002815471666648206,\nonumber \\
&\alpha_{5}=-0.14266271433303265,\nonumber \\
&\alpha_{6}=0.057070353250615435,\nonumber \\
&\alpha_{7}=-0.2813020658633854,\nonumber \\
&\alpha_{8}=-0.603040898093542,\nonumber
\end{align}

\subsection{Space group \sgb}
\label{sec:sg130}
For the interacting Hamiltonian invariant under space group \sgb given in \Cref{eq:130fullinteractingH},
\begin{equation}
    H=H^{130}_{0}+H^{1}_{HK}+H^{2}_{HK}+H^{3}_{HK},
\end{equation}
the ground state at the $A$ point is the same as in space group \sg135 and is given by
\begin{align}
\ket{GS}_{A}=&\alpha_{1}(\ket{AA\uparrow;AA\downarrow;BA\uparrow;BA\downarrow}+\ket{AB\uparrow;AB\downarrow;BB\uparrow;BB\downarrow}) \\
&\alpha_{2}(\ket{AA\uparrow;AA\downarrow;BB\uparrow;BB\downarrow}+\ket{AB\uparrow;AB\downarrow;BA\uparrow;BA\downarrow}) \nonumber \\
&\alpha_{3}(\ket{AA\uparrow;AB\downarrow;BA\uparrow;BB\downarrow}+\ket{AA\downarrow;AB\uparrow;BA\downarrow;BB\uparrow}) \nonumber \\
&\alpha_{4}(\ket{AA\uparrow;AB\downarrow;BA\downarrow;BB\uparrow}+\ket{AA\downarrow;AB\uparrow;BA\uparrow;BB\downarrow}) \nonumber,
\end{align}
with the amplitudes
\begin{align}
&\alpha_{1}=-0.07774353109915998\nonumber \\
&\alpha_{2}=-0.1244495235329457\nonumber \\
&\alpha_{3}=-0.31172399033860637\nonumber \\
&\alpha_{4}=-0.6174920350191034\nonumber.
\end{align}

At the $\Gamma$ point, the ground state is given by:
\begin{align}
\ket{GS}_{\Gamma}&=\alpha_{1}(\ket{AA\uparrow;AA\downarrow;AB\uparrow;AB\downarrow}+\ket{BA\uparrow;BA\downarrow;BB\uparrow;BB\downarrow}) \\
&\alpha_{2}(\ket{AA\uparrow;BA\uparrow;BA\downarrow;BB\downarrow}+\ket{AA\uparrow;AA\downarrow;AB\uparrow;BA\downarrow}+\nonumber \\& \ket{AA\uparrow;AA\downarrow;AB\downarrow;BA\uparrow}+\ket{AA\downarrow;BA\uparrow;BA\downarrow;BB\uparrow}) \nonumber \\
&\alpha_{3}(\ket{AA\uparrow;AA\downarrow;AB\uparrow;BB\downarrow}+\ket{AB\uparrow;BA\uparrow;BA\downarrow;BB\downarrow}) \nonumber \\
&\alpha_{4}(\ket{AB\downarrow;BA\uparrow;BA\downarrow;BB\uparrow}+\ket{AA\uparrow;AA\downarrow;AB\downarrow;BB\uparrow}) \nonumber \\
&\alpha_{5}(\ket{AA\uparrow;AA\downarrow;BA\uparrow;BA\downarrow}+\ket{AB\uparrow;AB\downarrow;BB\uparrow;BB\downarrow}) \nonumber \\
&\alpha_{6}(\ket{AB\uparrow;AB\downarrow;BA\uparrow;BB\downarrow}+\ket{AA\uparrow;AA\downarrow;BA\uparrow;BB\downarrow}) \nonumber \\
&\alpha_{7}(\ket{AB\uparrow;AB\downarrow;BA\downarrow;BB\uparrow}+\ket{AA\uparrow;AA\downarrow;BA\downarrow;BB\uparrow}) \nonumber \\
&\alpha_{8}(\ket{AA\uparrow;AA\downarrow;BB\uparrow;BB\downarrow}+\ket{AB\uparrow;AB\downarrow;BA\uparrow;BA\downarrow}) \nonumber \\
&\alpha_{9}(\ket{AA\uparrow;AB\uparrow;AB\downarrow;BA\downarrow}+\ket{AA\uparrow;BA\downarrow;BB\uparrow;BB\downarrow}) \nonumber \\
&\alpha_{10}(\ket{AB\uparrow;BA\downarrow;BB\uparrow;BB\downarrow}+\ket{AA\uparrow;AB\uparrow;AB\downarrow;BB\downarrow}\nonumber \\&+\ket{AA\downarrow;AB\uparrow;AB\downarrow;BB\uparrow}+\ket{AB\downarrow;BA\uparrow;BB\uparrow;BB\downarrow}) \nonumber \\
&\alpha_{11}(\ket{AA\downarrow;AB\downarrow;BA\uparrow;BB\uparrow}+\ket{AA\uparrow;AB\uparrow;BA\downarrow;BB\downarrow}) \nonumber \\
&\alpha_{12}(\ket{AA\uparrow;AB\downarrow;BA\uparrow;BA\downarrow}+\ket{AA\uparrow;AB\downarrow;BB\uparrow;BB\downarrow}) \nonumber \\
&\alpha_{13}(\ket{AA\uparrow;AB\downarrow;BA\uparrow;BB\downarrow}+\ket{AA\downarrow;AB\uparrow;BA\downarrow;BB\uparrow}) \nonumber \\
&\alpha_{14}(\ket{AA\uparrow;AB\downarrow;BA\downarrow;BB\uparrow}) \nonumber \\
&\alpha_{15}(\ket{AA\downarrow;AB\uparrow;AB\downarrow;BA\uparrow}+\ket{AA\downarrow;BA\uparrow;BB\uparrow;BB\downarrow}) \nonumber \\
&\alpha_{16}(\ket{AA\downarrow;AB\uparrow;BA\uparrow;BA\downarrow}+\ket{AA\downarrow;AB\uparrow;BB\uparrow;BB\downarrow}) \nonumber \\
&\alpha_{17}(\ket{AA\downarrow;AB\uparrow;BA\uparrow;BB\downarrow}), \nonumber
\end{align}
with the amplitudes
\begin{align}
&\alpha_{1}=0.014249658673534686\nonumber \\
&\alpha_{2}=-0.0012444934053119539 i\nonumber \\
&\alpha_{3}=-0.0831672982330461+0.008862561108818907i\nonumber \\
&\alpha_{4}=0.0831672982330454+0.008862561108818754i\nonumber \\
&\alpha_{5}=0.07544667047624203\nonumber \\
&\alpha_{6}=0.0028493813995047884+0.000858765428111011i\nonumber \\
&\alpha_{7}=-0.002849381399504752+0.0008587654281110387i\nonumber \\
&\alpha_{8}=0.14149493379607864\nonumber \\
&\alpha_{9}=-0.08316729823304608-0.0088625611088189i\nonumber \\
&\alpha_{10}=0.0012444934053119391i\nonumber \\
&\alpha_{11}=-0.05320252170928138\nonumber \\
&\alpha_{12}=0.0028493813995046618-0.0008587654281109971i\nonumber \\
&\alpha_{13}=0.2816876681657167\nonumber \\
&\alpha_{14}=0.6031003841472453-0.012345362536891847i\nonumber \\
&\alpha_{15}=0.08316729823304603-0.008862561108818893i\nonumber \\
&\alpha_{16}=-0.002849381399504741-0.0008587654281110387i\nonumber \\
&\alpha_{17}=0.6031003841472453+0.012345362536891696i\nonumber
\end{align}

\twocolumngrid

\bibliography{refs}

\end{document}